\documentclass[aps,preprint,superscriptaddress]{revtex4} % preprint option provides increased line spacing
\usepackage[caption=false]{subfig}
\usepackage[final]{graphicx}
\usepackage{float}
\usepackage{setspace}  % Include this for explicit line spacing control
\usepackage{bm}
\usepackage{adjustbox}

\usepackage{array} % put this in the preamble

\usepackage{eqnarray,amsmath}
\usepackage{tabularx}
\usepackage{txfonts}
\usepackage{cleveref}
\usepackage{slashed}
\usepackage{graphicx,pstricks}
\usepackage{amsmath,amssymb}
\usepackage{amsfonts}
\newcommand{\hypgeo}[2]{%
  {\vphantom{F}}_{#1}\kern-\scriptspace F_{#2}%
}

\usepackage[english]{babel}
\usepackage{afterpage}
\usepackage{graphicx}% Include figure files
\usepackage{dcolumn}% Align table columns on decimal point
\usepackage{bm}% bold math
\usepackage{color}
\usepackage{slashed}%slashed D, k, A, etc.
\usepackage{latexsym}
\newcommand{\ii}{\textrm {i}}
\newcommand{\ee}{\textrm {e}}
\newcommand{\ga}[1]{\gamma^{#1}}

\newcommand{\unit}{\mathbb{I}}

\newcommand{\de}[1]{\partial_{#1}}

\newcommand{\br}[1]{\left( #1 \right)}

\newcommand{\brt}{\br{t}}

\newcommand{\bet}[1]{\left \lvert #1 \right \rvert}
\newcommand{\va}{E_0}
\newcommand{\ttau}{\br{\frac{t}{\tau}}}

\newcommand{\ttanh}[1]{\textrm{tanh}\br{#1}}
\newcommand{\tsech}[1]{\textrm{sech}\br{#1}}
\newcommand{\Tanh}{\ttanh{\frac{t}{\tau}}}

\newcommand{\Hyper}[1]{ {}_2 \mathcal F_1 \br{#1}}

\newcommand{\be}[1]{\begin{equation} #1 \end{equation}}

\newcommand{\ma}[1]{\begin{pmatrix} #1 \end{pmatrix}}

\newcommand{\dd}{d}  %intgration measure       
\newcommand{\op}[1]{\hat{#1}} %operator
\newcommand{\bra}[1]{\langle{#1}\vert} %bra vector
\newcommand{\ket}[1]{\vert{#1}\rangle} %ket vector
\usepackage{natbib}
\begin{document}
\title{Pair Production  in time-dependent Electric field at Finite times}
\author{Deepak Sah\footnote{Corresponding author.\\E-mail address: deepakk@rrcat.gov.in(Deepak).}}
\author{Manoranjan P. Singh}
\affiliation{Theory and Simulations Lab,Theoretical and Computational Physics Section,  Raja Ramanna Centre for Advanced Technology,  Indore-452013, India}
\affiliation{Homi Bhabha National Institute, Training School Complex, Anushakti Nagar, Mumbai 400094, India}

%\date{\today}
%%%%%%
\begin{abstract}
We investigate the finite-time behavior of pair production from the vacuum by a time-dependent Sauter-pulsed electric field. By examining the temporal behavior of the single-particle distribution function, we observe oscillatory patterns in the longitudinal momentum spectrum at finite times. These oscillations arise due to quantum interference effects resulting from the various dynamical processes/channels leading to the creation of the (quasi-)particle with a given momentum. Furthermore, we derive a simplified approximate analytical expression for the distribution function at finite times, which allows us to explain the origin and behavior of these oscillations. The role of the vacuum polarization function and its counterterm is also discussed in this regard.
At finite times, the transverse momentum spectrum peaks at a nonzero value of transverse momentum, indicating the role of multiphoton transitions in quasiparticle creation.
We explicitly relate the quasiparticle distribution function to the measurable probability of real pair production when the field is switched off at a finite time, establishing the physical relevance of the observed finite-time structures. This work bridges the quasiparticle description with operationally measurable quantities, offering insights relevant for upcoming high-intensity laser experiments aimed at observing nonperturbative pair production phenomena.
%The transverse momentum spectrum peaks at the nonzero value of the transverse momentum at finite times, which indicates the role of multiphoton transitions in the creation of quasiparticles.  
%We further establish the physical significance of the finite-time distribution by demonstrating  its equivalence to the probability of real pair production when the field is switched off, \(f(\bm{p}, T) = \mathcal{W}(\bm{p}, T)\). This result bridges the quasiparticle description with operationally measurable quantities.
%To understand the physical significance of the quasiparticle distribution function at finite time can be related to the  real pair of production at finite time. we showed that the  probabiltiy of pair production real particle is equal to the quasiparticle distribution function at finite time $T$ where electric field switched off.

\end{abstract}
%\pacs{}
%\keywords{Particle Production, Dynamical Schwinger effect, Strong-Field QED}
\maketitle
%\tableofcontents

%%%%%%%%%%%%%%%%%%%%%%%%%%%%%%%%%%%%%%%%%%%%%%%%%%%%%%%%%%%%%%%%%%%%%%%%%%%%%%%%%%%%%
\section{Introduction}
The concept of pair production in an electromagnetic field has its roots in the mid-1920s after the invention of quantum mechanics, with the formulation of the relativistic wave equation for electrons by Paul Dirac in 1928 \cite{Dirac:1928hu}. The Dirac Sea model was proposed to explain the enigma of negative energy solutions. F. Sauter's work in 1931 demonstrated that strong electric fields can lead to pair creation through tunneling with exponential suppression \cite{Sauter:1931zz}. This paved the way for quantum field theory, recognizing the vacuum as a polarizable medium influenced by constant fluctuations.
In 1936, W. Heisenberg and H. Euler further explored the peculiarities of the Dirac equation, revealing non-linear modifications in Maxwell's equations due to the interaction of electromagnetic fields with the electron vacuum loop ~\cite{Heisenberg:1936nmg}. J. Schwinger's groundbreaking work in 1951 precisely calculated the imaginary part of the one-loop effective Lagrangian in the presence of a static electric field \cite{Schwinger:1951nm}. 
 As a result of his seminal work, the phenomenon of vacuum pair creation by electric fields has since become widely known as the Schwinger effect, and it is also famously referred to as the Sauter-Schwinger effect in recognition of F. Sauter's prior work on solving the Dirac equation in the presence of an electric field \cite{Sauter:1932gsa}. 
 \par
 Schwinger's pioneering calculation opened up new avenues of research in quantum field theory and has profoundly impacted our understanding of particle physics in the presence of strong fields.
This extraordinary property of quantum vacuum producing spontaneous particle-antiparticle pairs has far-reaching implications for understanding the generation of particle-antiparticle pairs in the presence of a strong electric field \cite{Greiner1985}; particle creation in the expanding universe\cite{Parker:1968mv}; black hole evaporation as a result of Hawking radiation \cite{Hawking:1975vcx,Parikh:1999mf}; and Unruh radiation, in which particle production is seen by an accelerating observer\cite{Unruh:1976db,Kim:2016dmm}.
 The study of generating electron-positron pairs through a spatially constant electrical background field was extended to the electric field with various time dependences. In the 1970s, researchers explored the occurrence of pair production from the vacuum in the presence of an oscillating time-dependent electric field \cite{Popov:1971iga, Marinov:1977gq}. Their investigations revealed different qualitative behaviors for this process under various interaction regimes. The interaction regimes can be distinguished by the value of the dimensionless Keldysh parameter, \( \gamma = \frac{m \omega}{|e| E_0} \), where \( E_0 \) is the field amplitude, \( \omega \) is the field frequency, \( e \) is the electron charge, and \( m \) is the electron mass~\cite{Keldysh:1965ojf}. When $\gamma >> 1 $, the process probability exhibits perturbative power-law scaling with field intensity. Instead, for $ \gamma << 1 $, it exhibits a manifestly non-perturbative exponential dependence on $\frac{1}{E_0}$, similar to the case of a constant electric field \cite{Nikishov:1969tt}. 
 \par
 Particle production in a spatially homogeneous single pulse of an electric field has been explored  \cite{Narozhnyi:1970uv, Nikishov:2001ps}, and methods for tackling particle creation in an arbitrary time-dependent electric field have been developed \cite{Narozhnyi:1970uv, Brezin:1970xf, Bagrov:1975xt}. Because a single pulse of an electric field is an idealized form of an electric field created by two colliding laser beams, particle generation in an alternating electric field has been explored for a more realistic scenario \cite{Mostepanenko:1974im,osti_4400849}.
However, it was found that the mass of the created particle, via the Schwinger mechanism, exponentially suppresses the pair generation rate, necessitating a very strong electric field to observe this phenomenon. Therefore, this event has not yet been the subject of any experimental observations. This makes it unclear how well the theoretical prediction captures the physics of pair production.
\par
Nevertheless, the study of pair production in strong electric fields has attracted sustained interest from theoreticians in recent years due to the extraordinary progress in the development of the ultra-intense lasers technique and the strong-field QED experimental studies that are planned at upcoming high-intensity laser facilities, such as the European X-Ray Free-Electron Laser \cite{TESLA:2000tvm}, the Extreme-Light Infrastructure \cite{Heinzl:2009bmy,Dunne:2008kc}, the Exawatt Center for Extreme Light Studies \cite{korzhimanovetal.2023}. The electric field strength is fast getting closer to the critical value $E_{\text{c}}$.
\par
Additionally, it is suggested that the Schwinger mechanism can be indirectly tested in the condensed matter system of a single monolayer of graphene, where the electrons are roughly characterized by the massless pseudo-relativistic Dirac equation  \cite{Allor:2007ei,Klimchitskaya:2013fpa,Fillion-Gourdeau:2015dga}.
\par
Particle production can be viewed as evolving a quantum system from an initial equilibrium configuration to a new final equilibrium configuration via an intermediate non-equilibrium evolution caused by a strong field background.
In intermediate states, when matter fields interact with a time-dependent external field, the classical Hamiltonian loses its time-translation invariance, leading to the evolution of the vacuum state into a state where mixing between positive and negative energy states occurs. Special care is needed in identifying ``positive" and ``negative" energy states in the presence of an external electric field. Thus, energy is no longer a good quantum number, and there is no well-defined quantum number to characterize a ``particle".
However, the standard way of describing particle creation by an external field involves the asymptotic analysis of particle states in the remote past (or in-states, before the external field is switched on) and the distant future (or out-states, long after the interaction with the external field has ended). This method is well understood using the quantum field operator expression in terms of creation and annihilation operators, which are connected to single-particle states in both the past and the future. Then, by relating the two sets of operators in the past and future, one can derive an equation for the S-matrix of the process. This equation, in turn, determines the number of particles produced in the process. 
\par
Various methodologies have been developed to investigate pair production in the presence of strong fields. These include the proper-time method \cite{Schwinger:1951nm,DeWitt:1975ys}, the canonical method \cite{Mottola:1984ar,Tomaras:2001vs,Tomaras:2000ag}, Green’s function techniques \cite{Gavrilov:1996pz}, semiclassical tunneling \cite{Parikh:1999mf}, the Schrödinger-functional approach \cite{Hallin:1994ad}, functional techniques \cite{Fried:2001ur,Avan:2002dn,Ilderton:2014mla}, the Wigner formalism\cite{Zhuang:1995pd,Hebenstreit:2011cr,Hebenstreit:2010cc},mean-field treatments \cite{Smolyansky:1997ji}, and worldline instanton methods \cite{Affleck:1981bma,Kim:2000un}.  
Studies on pair production in intense laser fields \cite{Alkofer:2001ik, Roberts:2002py} have employed analytical and numerical approaches to predict pair-production rates. However, these investigations primarily focus on pair formation at asymptotic times, which alone is insufficient for fully capturing the dynamical aspects of pair production and its time-dependent effects.
Several authors have derived expressions for the number of pairs created after a time $T$, where $T$ exceeds the duration of the electric field pulse or corresponds to a moment after the field has vanished \cite{Gavrilov:1996pz, DiPiazza:2004lsj, 2012CoTPh..57..422M,Ilderton:2021zej,Adorno:2015ibo}. These expressions provide a reliable approximation of the actual pair count.
\par
Understanding pair production dynamics from the initial to final states requires a framework that accounts for transient and non-equilibrium effects. In this context, the distribution function of quasiparticles, governed by the quantum kinetic equation (QKE), offers a suitable description of the time-dependent evolution of pair production \cite{Blaschke:2014fca,Smolyansky:1997ji}. Defining the particle distribution function in the quasiparticle basis is both physically meaningful and intuitive. This approach is grounded in the straightforward physics of how a time-dependent field distorts the time-invariance symmetry of the Hamiltonian, leading to mixing between momentum states of positive and negative energies.
This method is particularly relevant for exploring the complex dynamical behavior of pair production in spatially uniform, time-dependent electric fields. Specifically, it has been demonstrated that the temporal evolution of the quasiparticle distribution function in momentum space provides insights into the asymptotic states of the quantum field and offers a comprehensive description of the process during the interaction between the matter field and the strong field background \cite{Blaschke2016}. It was shown that the distribution function evolves through three distinct temporal stages: the quasi-electron-positron plasma (QEPP) stage, the transient stage, and finally the residual electron-positron plasma (REPP) stage \cite{Banerjee:2018azr,Banerjee:2018fbw}. In examining the pair production process, both the QEPP stage, where quasi-particles are formed, and the REPP stage, where real particles emerge, are crucial \cite{Blaschke:2013ip,Shen:2018lbq}.
While the Schwinger mechanism is primarily known for producing elementary particles in a vacuum under external fields, it can also generate quasiparticle excitations in exotic materials such as graphene \cite{Allor:2007ei,Zubkov:2012ht,Gavrilov:2012jk,Klimchitskaya:2013fpa,Fillion-Gourdeau:2015dga}, which are experimentally more accessible, and in superconductors \cite{Solinas:2020woq}. Moreover, while ultracold atoms in optical lattices have been investigated as simulators for the Schwinger mechanism \cite{Kasper:2015cca,Halimeh:2023lid}.
\par

\newpage
In the present work, we consider the production of electron-positron pairs from the vacuum in a time-varying, spatially uniform pulsed electric field given by $ E \brt =  E_0 \textrm{sech}^2 \ttau$, with amplitude  $E_0$ and pulse duration $\tau$. 
Such background field has received extensive attention in the literature \cite{Narozhnyi:1970uv,Dunne:1998ni, Balantekin:1990aa,Gavrilov:1996pz, Nikishov:2003ig, Levai:2009mn, Klimchitskaya:2013fpa}, with a focus on the asymptotic behavior of the probability of pair production.
To this end, we study the evolution of the single-particle distribution function, $f(\bm{p},t)$, which is rigorously derived by canonical quantization of the Dirac field and subsequent Bogoliubov transformation to a quasi-particle representation \cite{Smolyansky:2016gmp}.
\par
The distribution function exhibits three distinct dynamical stages\cite{Banerjee:2018azr,Otto:2016fdo,Nousch:2016yxy,Smolyansky:2016gmp}. 
By analyzing the temporal evolution of $f(\bm{p},t) $, we find that the emergence and characteristics of these dynamical stages, QEPP, transient, and REPP, are qualitatively and quantitatively influenced by both longitudinal and transverse components of momentum. 
Next, we analyze the time evolution of the longitudinal momentum spectrum (LMS) of created particles. In the tunneling regime ($ \gamma < 1 $), we observe oscillatory behavior in the LMS at time $t > 2\tau$, and this oscillation pattern continuously changes from $t > 2\tau$  to $t < 6\tau$. This oscillatory behavior at finite times clearly illustrates the quantum interference effects arising from various dynamical processes and channels contributing to the creation of (quasi-)particles with a given momentum. It can also be understood in terms of the vacuum polarization function $ u(\bm{p},t)$  and its counterpart, the depolarization function $v(\bm{p},t) $, which encapsulate the underlying particle production dynamics.
We emphasize that the oscillations seen in the LMS  are not artifacts but rather possess significant physical relevance. In the multi-photon regime, we observe that LMS at finite times near $ t=3\tau$ exhibits a multi-modal structure. 
\par
Furthermore, we develop an analytical framework valid for finite times \( t > \tau \). Specifically, we derive an analytical expression for the single-particle distribution function as a power series in the small parameter \( (1 - y) \), where \( y = \frac{1}{2} (1 + \tanh{(\frac{t}{\tau})}) \). The intriguing dynamical features of the momentum distribution function at finite times arise from a function appearing in this expansion. The finite-time behavior of the momentum spectra can be understood in terms of functions that reveal three distinct behaviors, which are discussed in detail in the results section. 
\par
Additionally, we explored how the transverse momentum affects LMS and discovered that the oscillatory behavior depends on the value of transverse momentum at that time. Subsequently, we analyzed the temporal progression of the transverse momentum spectrum(TMS) of generated pairs, providing valuable insights into the pair production process. At finite times, the transverse momentum spectra reach their peak at a nonzero value, reflecting the influence of multi-photon transitions in quasiparticle creation.

%Following the approach of Ref.~\cite{Ilderton:2021zej}, we also consider the probability of pair creation at a finite time $T$ within the Dirac model, where the electric field is switched off at $t = T$. In this context, the quasiparticle distribution function $f(\bm{p},T)$ can be directly related to the probability of producing real, observable electron-positron pairs. This establishes a clear physical interpretation: the quasiparticle distribution at a finite time corresponds to the actual number of particles that would be measured if the field were turned off at that instant. Hence, the study of $f(\bm{p},t)$ not only captures the real-time dynamics of particle creation, including oscillatory structures from quantum interference, but also connects the formal quasiparticle description to an operationally measurable particle count.  

%Furthermore, 

Crucially, following the approach of Ref. ~\cite{Ilderton:2021zej}, we relate the quasiparticle distribution function \( f(\bm{p}, t) \) to an experimentally meaningful quantity: the probability \( \mathcal{W}(\bm{p}, T) \) of producing real electron-positron pairs when the electric field is switched off at a finite time \( t = T \). We demonstrate explicitly that \( f(\bm{p}, T) = \mathcal{W}(\bm{p}, T) \), thereby bridging the quasiparticle description with an operationally measurable particle count. This equivalence ensures that the oscillatory structures and momentum spectra analyzed in this work correspond to observable features in pulsed-field experiments.

%%%%

%%%Crucially, following the approach of Ref.~\cite{Ilderton:2021zej}, we relate the quasiparticle distribution function f(p,t) to an experimentally meaningful quantity: the probability W(p,T) of producing real electron–positron pairs when the electric field is switched off at a finite time t=T. We demonstrate explicitly that f(p,T)=W(p,T), thereby bridging the quasiparticle description with an operationally measurable particle number. This equivalence ensures that the oscillatory structures and momentum spectra analyzed in this work correspond to observable features in pulsed-field experiments.

\par
The preliminary results of this work were presented in Refs. \cite{Sah:2023udt, Sah:2024qcs}. After the report in Ref. \cite{Sah:2023udt} appeared, one of the authors of Ref. \cite{Diez:2022ywi} (C.Kohlfürst, private communication) informed us about their findings, which exhibit some overlap with the results presented here. In this manuscript, we compare and discuss our findings with those of Ref. \cite{Diez:2022ywi} whenever applicable.
%The preliminary results of this work were presented in Refs.~\cite{Sah:2023udt,Sah:2024qcs}. Following the report in Ref.~\cite{Sah:2023udt}, one of the authors of Ref.~\cite{Diez:2022ywi} (C. Kohlfürst, private communication) informed us of overlapping findings. In this manuscript, we discuss and compare our results with those of Ref.~\cite{Diez:2022ywi} where relevant.

\par
%This article is organized as follows: In Sec.II detailed the theoretical formulation is given. This largely follows the derivation from \cite{Grib1988,Grib1972}. In Sec. III, we present expressions for the particle momentum distribution function using the exact analytical solution for the mode function in the case of a Sauter-pulsed electric field. Results are discussed in Sec. IV. The article is concluded in Sec. V.

This article is organized as follows: In Sec. II detailed the theoretical formulation is presented. This largely follows the derivation from ~\cite{Grib1988,Grib1972}. In Sec. III, we present expressions for the particle momentum distribution function using the exact analytical solution for the mode function in the case of a Sauter-pulsed electric field. Results are discussed in Sec. IV. In Sec. V, we provide a physical interpretation linking the quasiparticle distribution to measurable pair production. The article is concluded in Sec. VI.
\newline 
Throughout the paper, we use natural units and set $ \hslash = c = m = 1 $, and express all variables in terms of the electron mass units.
%%%%%%%%
\section{Theory}
In this section, we review the canonical quantization method within the context of a time-dependent electric field, as discussed in \cite{Grib1972}. This approach offers valuable insights into pair production phenomena, including detailed information on particle number and momentum distribution at each moment. To begin, we establish solutions for the single-particle scenario governed by the Dirac equation in the presence of such a field. We start by writing the Dirac equation for a particle in an electromagnetic field, which takes the following form:
\begin{equation}
\bigl(\ii  \gamma^\mu \partial_\mu -e \gamma^\mu A_\mu -m \bigr) \Psi(\bm x ,t)=0,
\label{t1}
\end{equation} 
 where $A^\mu$ is the four-vector potential,  $\Psi(\bm x ,t)$ is a four-component spinor. For Dirac gamma matrices $\gamma^\mu$ we chose the Weyl basis \cite{Ryder:1985wq}
\begin{equation}
\ga{0}  = \ma{\unit & 0 \\ 0 & -\unit}, \quad 
\ga{i} = \ma{0 & -\sigma^i \\ \sigma^i & 0}, \quad (i = 1,2,3),
\end{equation}
where $\unit$ is the identity matrix and $\sigma^i$ are the Pauli matrices. The Dirac gamma matrices $\gamma^\mu$ satisfy the anti-commutation relations:
 \begin{equation}
     \{\ga{\mu}, \ga{\nu}  \} = 2 g^{\mu \nu},
 \end{equation}
  with the metric tensor, 
\begin{align}
     g^{\mu \nu} = \text{diag}(1, -1,-1,-1).
\end{align}
Four coupled differential equations result from the Dirac equation for the spinor, and it is typically challenging to find precise analytical solutions in the presence of external fields.
Feynman and Gell-Mann were able to circumvent this difficulty by taking into account a two-component form of the Dirac equation  \cite{Feynman:1958ty}.
Accordingly, we turn this equation into a second-order differential equation by assuming the existence of a bispinor $\chi(\bm{x},t)$ such that
\begin{equation}
\Psi(\bm x ,t)=\bigl(\ii  \gamma^\nu \partial_\nu -e \gamma^\nu A_\nu + m \bigr) \chi(\bm x ,t).
\label{t2}
\end{equation}
and inserting Eq.~\eqref{t2} into Eq.~\eqref{t1}, it follows that $\chi(\bm x ,t)$ satisfies the 
second-order Dirac equation
%quadratic Dirac equation
\begin{equation}
    \left[ (i \partial_\mu - e A_\mu)^2 - \frac{e}{2} \sigma^{\mu \nu} \mathcal{F}_{\mu \nu} - m^2 \right] \chi(\bm{x} ,t) = 0,
    \label{5}
\end{equation}
where $\mathcal{F}_{\mu \nu} = \partial_\mu  A_\nu - \partial_\nu  A_\mu  $ is the field strength tensor, $\sigma^{\mu \nu}  =  \frac{i}{2} [\gamma^\mu ,\gamma^\nu]$ is the commutator of the gamma matrices, and
$\chi(\bm x ,t)$ is a four component spinor. 
\par
In general, $\mathcal{F}_{\mu \nu}$  is space and time dependent. However, to simplify the discussion,we consider the case where the electromagnetic field tensor is $ \mathcal{F}^{\mu 0} =(0,{\bm E}(t))\equiv(0,0,0, E(t))$ with $E(t)$ linearly polarized  time-dependent quasi-classical spatially uniform  electric field along the $z$-axis and the corresponding  four-vector potential $A^\mu(\bm x ,t)=(0,{\bm A}(t))\equiv(0,0,0, A(t))$, with an arbitrary $A(t)$ such that 
 $E(t)=-\frac{\dd A(t)}{\dd t}.$
\newline
Then, now Eq.\eqref{5} can be simplified to
\begin{equation}
\left[\partial_{\mu}\partial^{\mu} + e^2 A^2(t) + 2 i A(t)  \partial_3 - i e \partial_t A(t) \gamma^0 \gamma^3 + m^2\right] \chi(\bm x ,t) = 0.
\label{t3}
\end{equation}
Spatial homogeneity implies that the solutions of the form
\begin{equation}
\chi(\bm x ,t)=\ee^{\ii \bm{p} \cdot {\bm x}}\chi_{\bm{p}}( t),
\label{t4}
\end{equation}
where $\chi_{\bm{p}}( t)$ is independent of the position ${\bm x}$ and we label it by  momentum of a particle ${\bm p}$. 
\newline
Subsequently, Eq.~\eqref{t3} reduces to
\begin{equation}
\left(\partial_t^2 + \ii e E(t)\gamma^0\gamma^3 + \omega^2(\bm{p}, t)\right)\chi_{\bm{p}}(t) = 0,
\label{t5}
\end{equation}
where the time-dependent frequency is
\begin{equation}
    \omega(\bm{p}, t) = \sqrt{\epsilon_\perp^2(p_\perp) + P^2(p_\parallel, t)}.
\end{equation}
Here, the transverse energy is given by \(\epsilon_{\perp}(p_\perp) = \sqrt{m^2 + p_\perp^2}\), and the longitudinal kinetic momentum is \(P(p_\parallel, t) = p_\parallel - eA(t)\). The longitudinal momentum component \(p_\parallel \equiv p_3\) is directed along the external electric field, while the transverse momentum magnitude \(p_\perp = \sqrt{p_1^2 + p_2^2}\) corresponds to the vector \(\bm{p}_\perp \), which lies in the plane perpendicular to the field.
\par
We now expand the function $\chi_{\bm{p}}(t)$ in the basis of eigenvectors of $\gamma^0 \gamma^3.$ The matrix representation of $\gamma^0 \gamma^3$ is
\be{
\ga{0} \ga{3} = \ma{\unit & 0 \\ 0 & -\unit} \ma{0 & \sigma^3 \\ -\sigma^3 & 0} = \ma{0 & \sigma^3 \\ \sigma^3 & 0},
}
making it easy to recognize the eigenvector. They are given by
\begin{equation}
R_1=\begin{pmatrix}
1 \\ 0 \\ 0 \\ 0
\end{pmatrix},\quad
R_2=\begin{pmatrix}
0 \\ 0 \\ 0 \\ 1
\end{pmatrix},\quad
R_3=\begin{pmatrix}
0 \\ 1 \\ 0 \\ 0
\end{pmatrix},\quad
R_4=\begin{pmatrix}
0 \\ 0 \\ 1 \\ 0
\end{pmatrix}.
\label{t9}
\end{equation}
There are two doubly degenerate eigenvectors for $\gamma^0\gamma^3$: $R_1$ and $R_2$ with eigenvalue $``1"$, and $R_3$ and $R_4$ with eigenvalue $``-1"$. However, selecting one from each pair suffices \cite{Grib1988}. Now, we will seek the solutions of Eq.~\eqref{t5} in the form
\begin{equation}
\chi_{\bm{p}}(t)\equiv\chi_{\bm{p}  r}(t)= \psi_{{\bm p}}(t) R_r,
\label{t8}
\end{equation}
where $\gamma^0\gamma^3 R_r=R_r.$ 
\newline
Solving a differential equation for a scalar function $\psi_{\bm p}(t)$ simplifies the problem,
\begin{equation}
\Bigl(\partial_t^2+\ii e E(t)+\omega^2({\bm p},t)\Bigr) \psi_{\bm{p}}( t)=0.
\label{a9}
\end{equation}
Let's examine the resulting solutions. According to Eq.~\eqref{a9}, in a vanishing electric field as $t \rightarrow -\infty $, $ \omega(\bm{p},t)$ becomes independent of time, denoted $\omega(\bm{p}) = \sqrt{m^2 + \bm{p}^2}$ . The scalar function $\psi_{\bm{p}}(t)$ then satisfies the asymptotic equation,
\begin{equation}
\Bigl(\partial_t^2+\omega^2 (\bm{p})\Bigr) \psi_{\bm{p}}(t)=0.
\label{a10}
\end{equation}
There are two linearly independent solutions to this time-dependent harmonic  oscillator equation, which correspond to energy
$\pm \omega({\bm p})$. In what follows, we will label these solutions with superscripts $\lambda=+$ and $\lambda=-$, respectively.
The solutions to Eq.~\eqref{a10}, are clearly given by plane waves. Therefore,
\begin{equation}
\psi_{\bm{p}}^{(\lambda)}(t)\underset{t\rightarrow -\infty}{\sim}\ee^{-\ii \lambda  \omega({\bm p}) t}.
\label{a11}
\end{equation}
These solutions will be interpreted as describing an electron ($\lambda=+$), and its antiparticle, i.e., a positron ($\lambda=-$).
\newline
Finally, the corresponding solutions of Eq.~\eqref{t3} have the form,
\begin{equation}
\chi_{{\bm p} r}^{(\lambda)}(\bm x ,t)=\ee^{\ii {\bm p}\cdot {\bm x}}\psi_{\bm{p}}^{(\lambda)}(t) R_r,
\label{a12}
\end{equation}
Those of the Dirac equation, however, are derived 
\begin{align}
\Psi_{{\bm p}r}^{(\lambda)}(\bm x ,t)=\Bigl[\ii\gamma^0 \partial_t+\gamma^i p_i-e\gamma^3 A(t) +m\Bigr]\psi_{\bm{p}}^{(\lambda)}(t)R_r \ee^{\ii {\bm p}\cdot {\bm x}},
\label{a13}
\end{align}
where the spinor solutions, $\Psi_{{\bm p}r}^{(\lambda)}(\bm x ,t)$ are normalized according to the product:
 \begin{equation}
      \int \dd^3{\bm x} [\Psi_{{\bm p}r}^{(\lambda)}(\bm x ,t)]^\dagger \Psi_{{\bm p'}r'}^{(\lambda')}(\bm x ,t) = (2\pi)^3 \delta({\bm p}-{\bm p}') \delta_{rr'} \delta_{\lambda \lambda'}.
 \end{equation}
Hence, the newly constructed eigenstates of the Dirac equation representing an electron or positron in a time-dependent electric field provide a complete and orthonormal relation
%%%\sum_{\lambda=\pm}\sum_{r=\pm}
\begin{equation}
\sum_{\lambda}\sum_{r}\int\frac{\dd^3{\bm p}}{(2\pi)^3}\,\Psi_{{\bm p}r}^{(\lambda)}(\bm x ,t)[\Psi_{{\bm p}r}^{(\lambda)}( \bm x',t)]^\dagger=\delta({\bm x}-{\bm x}').
\label{a18}
\end{equation}
\par
The Dirac fermion field operator $\hat{\Psi}(\bm x,t)$ in the framework of second quantization is  written in the form 
\begin{equation}
\hat{\Psi}(\bm x ,t)=\sum_{r}\int\frac{\dd^3 {\bm p}}{(2\pi)^3}\Bigl(\Psi_{{\bm p}r}^{(+)}(\bm x ,t)\op{b}_{{\bm p}r}+\Psi_{-{\bm p}r}^{(-)}(\bm x ,t)\op{d}_{{\bm p}r}^\dagger\Bigr),
\label{a19}
\end{equation}
where $\Psi^{(\lambda)}_{{\bm p } r}(\bm{x},t)$  are the single-particle solutions of the Dirac equation, whereas $\op{b}_{{\bm p}r}$ and $\op{d}_{{\bm p}r}^\dagger$ are the electron annihilation and positron creation operator respectively. The operators satisfy the usual fermionic anti-commutation relations,
%These operators define the initial vacuum state  (or in-state) at $t \rightarrow  -\infty$ through the condition that $\op{b}_{{\bm p}r} | 0_{in}> = 0 $ and $\op{d_{{-\bm p}r} | 0_{in}> = 0 $ %%%%
\begin{equation}
\{ \hat{b}_{{\bm p}r},\op{b}_{{\bm p}'r'}^\dagger \}=\{\hat{d}_{{\bm p}r},\op{d}_{{\bm p}'r'}^\dagger\}=\delta({\bm p}-{\bm p}')\delta_{r r'}.
\label{a20}
\end{equation}
Then the  $\hat{\Psi}(\bm{x},t)$ field operator also satisfies the anti-commutation relation,
\begin{equation}
     \{ \hat{\Psi}_n( \bm x ,t) , \hat{\Psi}_m^\dagger( \bm x',t) \}  = (2 \pi)^3 \delta({\bm x}-{\bm x}')\delta_{m n}.
\end{equation}
Now, the Hamiltonian can be calculated from the energy-momentum tensor, which yields
\begin{equation}
\op{H}(t)= \ii \int \hat{\Psi}^{\dagger}(\bm{x},t)\, \dot{\hat{\Psi}}(\bm{x},t)\, d^3 \bm{x},
\label{E-T}
\end{equation}
where \(\dot{\hat{\Psi}}(\bm{x},t) = \partial_t \hat{\Psi}(\bm{x},t)\). 
\newline
The diagonal and off-diagonal parts of the Hamiltonian are given as
\begin{widetext}
\begin{align}
\op{H}_{diag}(t)= \ii \sum_{r}\int\frac{\dd^3{\bm p}}{(2\pi)^3}\Bigl[\varepsilon^{(++)}_{\bm p}(t)\op{b}_{{\bm p}r}^\dagger\op{b}_{{\bm p}r}+
\varepsilon^{(--)}_{\bm p}(t)\op{d}_{-{\bm p}r}\op{d}_{-{\bm p}r}^\dagger\Bigr],
\label{a23a}
\end{align}
\begin{align}
\op{H}_{offdiag}(t)= \ii \sum_{r}\int\frac{\dd^3{\bm p}}{(2\pi)^3}\Bigl[
\varepsilon^{(+-)}_{\bm p}(t)\op{b}_{{\bm p}r}^\dagger\op{d}_{-{\bm p}r}^\dagger
+\varepsilon^{(-+)}_{\bm p}(t)\op{d}_{-{\bm p}r}\op{b}_{{\bm p}r} \Bigr],
\label{a23b}
\end{align}
\end{widetext}
where the factors $\varepsilon_{\bm p}^{(\lambda \lambda')}(t)$ are expressed as
\begin{widetext}
\begin{equation}
\varepsilon_{\bm p}^{(\lambda \lambda')}(t)=\left\{ 
\begin{array}{ll}
- (p_\parallel -e A(t)) - 2 \epsilon^2_\perp(p_\perp)  \textrm{Im} \Bigl([\psi_{\bm p}^{(\lambda)}(t)]^* \dot{\psi}_{\bm p}^{(\lambda)}(t) \Bigr)  &\quad{\mbox{if}}\,\lambda=\lambda',\\
\displaystyle 
\ii \epsilon^2_\perp(p_\perp)  \Bigl( [\psi_{\bm p}^{(\lambda)}(t)]^* \dot{\psi}_{\bm p}^{(\lambda')}(t) - [\dot{\psi}_{\bm p}^{(\lambda)}(t) ]^*  \psi_{\bm p}^{(\lambda')}(t) \Bigr)
&\quad{\mbox{if}}\,\,\lambda\neq\lambda'.
\end{array}
\right.
\label{a24}
\end{equation}
\end{widetext}
As the above Hamiltonian has non-vanishing off-diagonal terms, the positive and negative energy modes mix, and thus, a clear interpretation in terms of particles and antiparticles is difficult.
\par
To calculate the spectrum, the Hamiltonian is diagonalized through a basis transformation to the quasi-particle representation using the new time-dependent operators $\op{B}_{{\bm p}r}(t)$ and $\op{D}_{{\bm p}r}(t)$. The relation between the  $\op{b}_{{\bm p}r},\op{d}_{{\bm p}r} $and  $\op{B}_{{\bm p}r}(t) ,\op{D}_{{\bm p}r}(t)$ operators is given by the Bogoliubov transformation \cite{Grib1972,Schmidt:1998vi}
\begin{align}
\op{B}_{{\bm p}r}(t)&=\alpha_{\bm p}(t)\op{b}_{{\bm p}r}+\beta_{\bm p}(t)\op{d}^\dagger_{-{\bm p}r},\label{a26a}\\
\op{D}_{{\bm p}r}(t)&=\alpha_{-{\bm p}}(t)\op{d}_{{\bm p}r}-\beta_{-{\bm p}}(t)\op{b}_{-{\bm p}r}^\dagger,
\label{a26b}
\end{align}
with the condition
\begin{equation}
|\alpha_{\bm p}(t)|^2+|\beta_{\bm p}(t)|^2=1.
\label{a27}
\end{equation}
Here, the operators $\op{B}_{{\bm p}r}(t)$ and $\op{D}_{{\bm p}r}(t)$ describe the creation and annihilation of quasiparticles at time $t$ with respect to the instantaneous vacuum state \cite{Schmidt:1998vi}. The substitution of Eqs.~\eqref{a26a} and \eqref{a26b} into Eq.~\eqref{a19} results in a new representation of the field operator
\begin{equation}
\op{\Psi}(\bm x ,t)=\sum_{r}\int\frac{\dd^3{\bm p}}{(2\pi)^3}\Bigl[\Phi_{{\bm p}r}^{(+)}(\bm x ,t)\op{B}_{{\bm p}r}(t)+\Phi_{-{\bm p}r}^{(-)}(\bm x ,t)
\op{D}_{{\bm p}r}^\dagger(t)\Bigr],
\label{a29}
\end{equation}
with the spinors $\Phi_{{\bm p}r}^{(\lambda)}(\bm x ,t)$ such that
\begin{align}
\Phi_{{\bm p}r}^{(+)}(\bm x ,t)&=\alpha_{\bm p}^*(t)\Psi_{{\bm p}r}^{(+)}(\bm x ,t)+\beta_{\bm p}^*(t)\Psi_{{\bm p}r}^{(-)}(\bm x ,t), \label{a30}\\
\Phi_{{\bm p}r}^{(-)}(\bm x ,t)&=\alpha_{\bm p}(t)\Psi_{{\bm p}r}^{(-)}(\bm x ,t)-\beta_{\bm p}(t)\Psi_{{\bm p}r}^{(+)}(\bm x ,t).\label{a31}
\end{align}
It follows from this that $\Phi_{{\bm p}r}^{(\lambda)}(\bm x, t)$  have a spin structure similar to that of $\Psi_{{\bm p}r}^{(\lambda)}(\bm x, t)$ in Eq.~\eqref{a13},
\begin{align}
\Phi_{{\bm p}r}^{(\lambda)}(\bm x ,t)&=\Bigl[\ii\gamma^0 \partial_t- {\bm p}\cdot{\bm \gamma}+ eA(t)\gamma^3 + m\Bigr]\ee^{\ii {\bm p}\cdot {\bm x}}\phi_{\bm p}^{(\lambda)}(t)R_r.
\label{a32}
\end{align}
The function $\phi_{\bm{p}}^{(\lambda)}(t)$ is the mode function in the quasi-particle representation, chosen according to the ansatz.
\begin{equation}
\phi_{\bm p}^{(\lambda)}(t)=\frac{e^{ -\ii \lambda \Theta_{\bm p}(t_0,t)}}{\sqrt{2\omega(\bm {p},t)\bigl[\omega(\bm {p},t)-\lambda P(p_\parallel,t)\bigr]}},
\label{a35a}
\end{equation}
with the dynamical phase 
\begin{align}
     \Theta_{\bm p}(t_0,t) &= \int^{t}_{t_0} \omega(\bm{p},t') dt'.
     \label{a35b1}
\end{align}
The lower limit $t_0$ of the integral in Eq.~\eqref{a35b1} is not fully determined as it only fixes an arbitrary phase. 
Note that the functions $\phi_{\bm p}^{(\pm)}(t)$  are chosen such that they coincide with the mode functions $\psi_{\bm p}^{(\pm)}(t)$ in the case of a vanishing electric field \cite{Smolyansky:1997ji,Kluger:1998bm}.
\newline
Now, combining Eqs.~\eqref{a13},~\eqref{a30} to ~\eqref{a32} , we obtain that
\begin{align}
\psi_{\bm p}^{(+)}(t)=\alpha_{\bm p}(t)&\phi_{\bm p}^{(+)}(t)-\beta_{\bm p}^*(t)\phi_{\bm p}^{(-)}(t), \label{a33}\\
\psi_{\bm p}^{(-)}(t)=\beta_{\bm p}(t)&\phi_{\bm p}^{(+)}(t) +\alpha_{\bm p}^*(t)\phi_{\bm p}^{(-)}(t),\label{a34}
\end{align}
 and the coefficients $\alpha_{\bm p}(t)$ and $\beta_{\bm p}(t)$ are given by
\begin{align}
\alpha_{\bm p}(t) = i \phi_{\bm p}^{(-)}(t)  \epsilon_\perp(p_\perp)     (\partial_t - i \omega(\bm {p},t) ) \psi_{\bm p}^{(+)}(t),
\label{a35b}
\end{align}
\begin{align}
\beta_{\bm p}(t) = -i \phi_{\bm p}^{(+)}(t) \epsilon_\perp(p_\perp)    (\partial_t + i \omega(\bm {p},t) ) \psi_{\bm p}^{(+)}(t).
\label{a36}
\end{align}
From the above equations, if we know $\psi_{\bm p}(t)$ from a solution of differential Eq.~\eqref{a9} for a specific electric field, we can find out the Bogoliubov transformation coefficients.
%%%%%%%%%%%%%%%%%%%%%%%%%%%%
\par
We now define the  occupation number of electrons in the given eigenmode $\bm{p} r$ of the fermionic field using the time-dependent creation and annihilation operators for the initial vacuum state as 
\begin{align}
f_r(\bm{p},t)&=  \bra{0_{in}}\op{B}_{{\bm p}r}^\dagger(t)\op{B}_{{\bm p}r}(t)\ket{0_{in}},
\label{a28.1}
\end{align}
Similarly, the occupation number of the positron is given by
\begin{align}
\bar{f}_r(-\bm{p},t)&= \bra{0_{in}}\op{D}_{-{\bm p}r}^\dagger(t)\op{D}_{-{\bm p}r}(t)\ket{0_{in}}.
\label{a28.2}
\end{align}
%%%%
Because of the  charge conjugation invariance,
 \begin{align}
f_r(\bm{p},t)&=\bar{f}_r(-\bm{p},t).
\label{a29.1}
\end{align}
%%%%
In the quasi-particle representation,  $f_r(\bm{p},t)$ and $\bar{f}_r(-\bm{p},t)$ will act as single-particle distribution functions \cite{Schmidt:1998vi}.
%%%%%
Since the Hamiltonian Eq.\eqref{E-T} does not have any spin-dependent terms, the spin index$(r)$ is dropped, and the distribution function 
\begin{align}
f(\bm{p},t) &= 2 |\beta_{\bm p}(t)|^2.
\label{a43}
\end{align}
%%%%
where the factor $``2"$ corresponds to the spin degree of freedom.
One approach to find $f(\bm{p},t)$ involves solving Eq.~\eqref{a9} for $\psi_{\bm{p}}(t)$ and subsequently using it to determine the Bogoliubov coefficients, which are essential for obtaining the single-particle distribution function.
%%%%%%%%%%%%%%%%%%%%%%%%%%%%%%%%%%%%%%%%%%%%%%%%%%%

%%%%%%%%%%%%%%%%%%%%%%%%%%%
\section{Pair Production in Sauter-pulse electric field}
\label{Sauterpulse}
\begin{figure}[t]
\begin{center}
{\includegraphics[width = 2.8543219808012in]{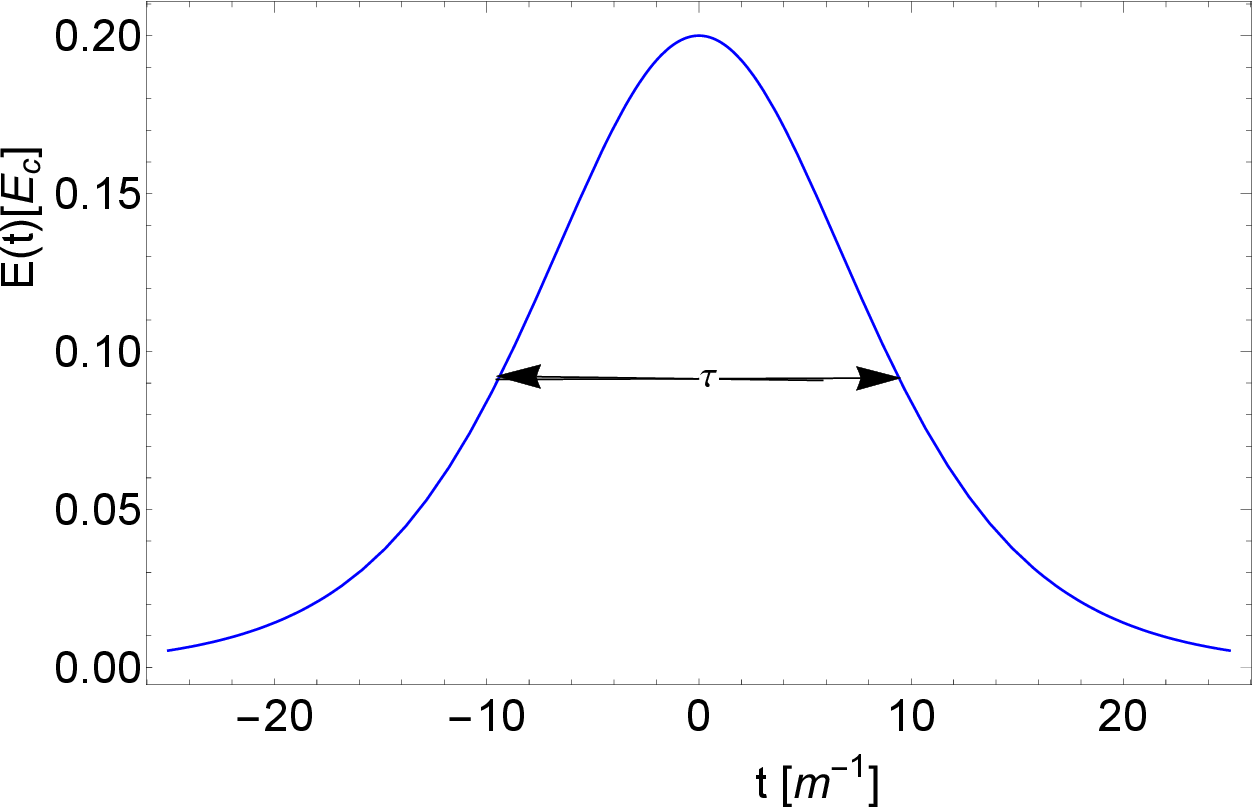}
\includegraphics[width = 2.8543219808012in]{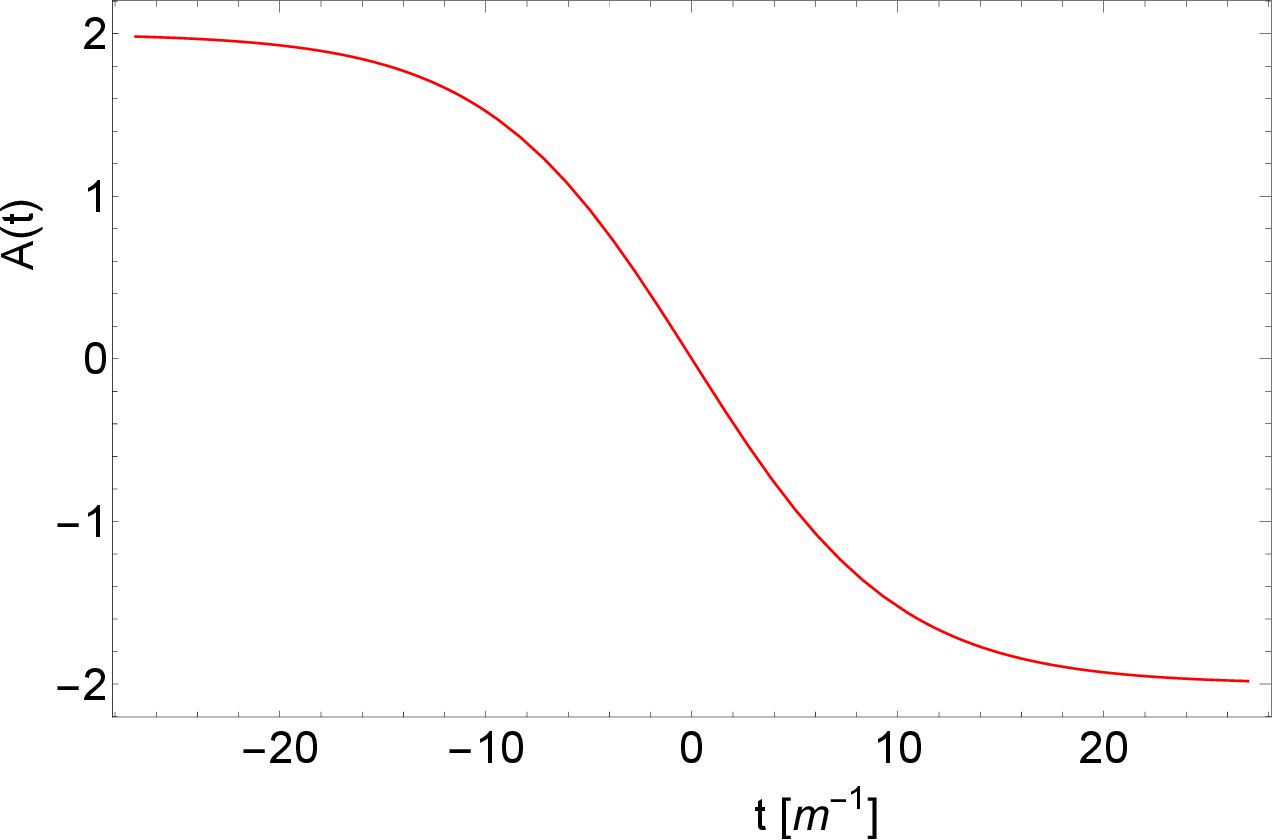}
}
\caption{Temporal profile of the electric field  (left) and the associated vector potential with the choice $A(t=0) =0$ (right).
The field parameters are $E_0 = 0.2 E_c $ and $ \tau = 10 [m^{-1}]$ and all the units are taken in the electron mass unit.}
   	\label{fig:ea}
\end{center}
\end{figure}

A spatially uniform external field background is a common approximation of the electromagnetic field near the focal region of two counter-propagating laser pulses along the z-axis, generating a standing wave \cite{Nakajima:2004tn,Blaschke:2008wf}. In general, the pair-production process takes place close to the electric field maximum (comparable to the critical field limit), where the magnetic field vanishes. Despite the fact that laser fields typically include many optical cycles, in this case, we examine a relatively simple model of the external field made up of the Sauter profile, which can be thought of as an extremely short laser pulse.
%%%%%
\begin{equation}
    E \brt =  E_0 \textrm{sech}^2 \ttau ,
\label{ef}
\end{equation}
where $\tau $ is the  duration of pulse, and $E_0$ is field strength. 
This electric field exponentially goes to zero for $ |t | >> \tau.$ In the limit of $ \tau \rightarrow \infty$ the electric field becomes homogeneous in time. We can choose a gauge in which $A_0 = 0$ and the vector  potential associated with the electric field is $ (0,0, A(t)= -
\int dt E(t)).$
After the integration, we find the Sauter-type gauge potential,

\begin{equation}
    A \brt = -E_0   \tau  \Tanh,
\label{vp}
\end{equation}

%%%%
with $ A(t=0) =0.$ The left panel of figure~\ref{fig:ea} shows the temporal profile of the electric field given by Eq.\eqref{ef}. Its peak height is attained at $t = 0$, its half height is reached at $t = \pm 0.81 \tau $, and at $ t = \pm \tau  $, the pulse amplitude has already dropped to $ 41 \%$ of its peak height, followed by a further dramatic reduction well below $10 \% $ at  $t = \pm 2 \tau  $.
%%%%%%%%%%%%%%%%%%%%%%%%%%%%%%
\par
Originally, Sauter explored the Dirac equation within an inhomogeneous scalar potential $V(z) = V_0 \tsech{\frac{z}{d}}^2$ \cite{Sauter:1932gsa}. Since the problem is essentially reduced to solving a one-dimensional differential equation, the potential can also be addressed with a similar functional dependence on time, as given by Eq.~\eqref{ef}. In 1970, the vacuum instability in a time-dependent Sauter-like electric field was initially examined by N. B. Narozhny et al. \cite{Narozhnyi:1970uv}. Subsequently, many researchers revisited this topic, finding it practical to test various approaches, including approximate methods, in the specific context being considered. Examples include Refs.\cite{Hebenstreit:2010vz,Kohlfurst:2012rb,Dabrowski:2016tsx,Gelis:2015kya,Bialynicki-Birula:2011lzb}, and the references therein. However, for the sake of completeness, we reproduce the essential steps and find out the expression for the single-particle distribution function, $f(\bm{p},t)$, utilizing the solution of the differential equation.
%%%% % % % % % % % % % % %   % % % % % % % % % % % % % % % % % % % %
%%%%%
\par
Now, in the presence of an external Sauter-pulse field, the equation of motion for the mode function  Eq.~\eqref{a9} reads
\begin{equation}
\Bigl(\partial_t^2 +\ii \va \textrm{sech}^2 \ttau +\omega^2(\bm{p},t)\Bigr) \psi_{\bm p}(t)=0,
\label{a9.1}
\end{equation}
where we have skipped the irrelevant index $\lambda,$ of the mode function $\psi_{\bm p}(t).$
This equation is solved analytically by converting it into a hypergeometric differential equation. By changing the time variable to
\begin{align}
    y &= \frac{1}{2} \left(1 + \Tanh\right), \quad \text{with} \quad y \in [0,1],
    \label{newt}
\end{align}
so that
\begin{align}
    P(p_\parallel, y) &= p_\parallel + e E_0 \tau (2y - 1), \\
    \omega^2(p_\parallel, y) &= \epsilon^2(p_\perp) + P^2(p_\parallel,y).
\end{align}
Additionally, we compute
\begin{align}
     \de{t} = (\de{t} y) \de{y} = \frac{2}{\tau} y (1-y) \de{y}, \\
      \de{t}^2 =\frac{4}{\tau^2 } y (1-y) \de{y} y (1-y).
\end{align}
Equation \eqref{a9.1} can be rewritten as follows
%Now, Eq.\eqref{a9.1}  can be rewritten as 
\begin{align}
    \left[\frac{4}{\tau^2} y \br{1-y} \de{y} y \br{1-y} \de{y} + \omega^2(\bm{p},y) + 4 \ii \va y \br{1-y} \right]\psi_{\bm p}(y)  = 0.
\label{37}
\end{align}
Further, by using  the following ansatz 
\begin{align}
    \psi_{\bm{p}}(y)  = y^{k} \br{1-y}^{l} \eta_{\bm p}(y)  
\label{eq2_7:9}
\end{align}
where
\[
k = \frac{-\ii \tau \omega_0}{2}, \quad l = \frac{\ii \tau \omega_1}{2},
\]
and we define
\begin{align}
    P_0 = P(p_\parallel, y = 0) = p_\parallel - e E_0 \tau, \quad P_1 = P(p_\parallel, y = 1) = p_\parallel + e E_0 \tau,
\end{align}

\begin{align}
    \omega_0 = \omega(\bm{p}, y = 0) = \sqrt{\epsilon_{\perp}^2 (p_\perp) + P_0^2}, \quad \omega_1 = \omega(\bm{p}, y = 1) = \sqrt{\epsilon_{\perp}^2 (p_\perp) + P_1^2}.
\end{align}

Inserting this ansatz into Eq.~\eqref{37}, we have  $\eta_{\bm p}(y)$  satisfying  the following hypergeometric differential equation \cite{abramowitz}
%with, 
%$k = \frac{- \ii \tau \omega_0}{2}, $
%$l = \frac{\ii \tau \omega_1}{2},$ $\omega_0= \sqrt{\epsilon_{\perp}^2 +P_0^2},$  $\omega_1 =\sqrt{\epsilon_{\perp}^2 +P_1^2},$ 
%$ P_0 = \br{p_\parallel -e E_0 \tau},$ $ P_1= \br{p_\parallel + e E_0 \tau},$ 
%\newline
%in  Eq.~\eqref{37}, we have  $\eta_{\bm p}(y)$  satisfying  the following hypergeometric differential equation \cite{abramowitz}
 \be{
\br{y\br{1-y} \de{y}^2 + \br{c -\br{a+b+1}y} \de{y} -ab} \eta_{\bm p}(y)   = 0,
\label{eq2_7:10}
}
Here,
\begin{align}
    a &= -\ii \va \tau^2 - \frac{\ii \tau \omega_0}{2} + \frac{\ii \tau \omega_1}{2} = i \zeta_1, \notag \\
b &= 1+\ii \va \tau^2 - \frac{\ii \tau \omega_0}{2} + \frac{\ii \tau \omega_1}{2} = 1 + i \zeta_2 ,\label{eq2_7:3} \\
c &= 1-\ii \tau \omega_0 = 1 + i \zeta_3. \notag
\end{align}
The two linearly independent solutions of Eq.\eqref{eq2_7:10} are 
 %$ \eta_{\bm p}^{(\pm)}(y) $ which are analytic in the neighborhood of the singular point $ y=0$ 
\be{
\eta_{\bm p}^{(+)}(y)   =  \Hyper{a,b,c;y},
}
\be{
\eta_{\bm p}^{(-)}(y)   = y^{1-c} (1-y)^{c-a-b} \Hyper{1-a,1-b,2-c;y},
}
with $\Hyper{a,b,c;y}$ denoting the Gauss-hypergeometric function\cite{abramowitz}.
\newline
To get the mode functions $\psi_{\bm p}^{(\pm)}(y)  $ we have to resubstitute 
$\eta_{\bm p}^{(\pm)}(y)$ in the ansatz Eq.\eqref{eq2_7:9}
\begin{align}
		 \psi^{(+)}_{\bm{p}}(y)   = N^{(+)}(\bm{p})  y^{k} \br{1-y}^{l}\Hyper{a,b,c;y},\label{sol1} 
	\end{align}
    \begin{align}
		 \psi^{(-)}_{\bm{p}}(y)   = N^{(-)}(\bm{p})   y^{-k} \br{1-y}^{-l} \Hyper{1-a,1-b,2-c;y},
   \label{sol2}
	\end{align}
%%%
with  $ N^{(\pm)}(\bm{p})$  being normalization factors.
\newline
We chose  the normalization factors such that the mode functions  $\psi^{(+)}_{\bm{p}}(y)$ and the functions $\phi_p^{(\pm)}(y)$ coincide for $y \rightarrow 0:$
\begin{align}
     \psi^{(\pm)}_{\bm p} (y) =  \phi^{(\pm)}_{\bm p} (y) \quad \text{for} \quad y \rightarrow 0.
\end{align}
The  asymptotic  behavior of the mode function for $y\rightarrow 0$ is given by:
 \begin{align}
      \psi^{(+)}_{\bm{p}}(y)   \stackrel{y \to 0}{=}   N^{(+)}(\bm{p}) e^{k \ln{(y)}} ,
        \label{normal1}
 \end{align}
\begin{align}
      \psi^{(-)}_{\bm{p}}(y)   \stackrel{y \to 0}{=}   N^{(-)}(\bm{p}) e^{-k \ln{(y)}} .
        \label{normal2}
 \end{align}
 The asymptotic behavior of the adiabatic mode functions:
 \begin{equation}
\phi_{\bm p}^{(\pm)}(y)=\frac{e^{ \mp \ii  \Theta_{\bm p}(y_0,y)}}{\sqrt{2\omega(\bm {p},y)\bigl(\omega(\bm {p},y)\mp P(p_\parallel,y)\bigr)}},
\label{phi_y}
\end{equation}
 with
\begin{align}
    \Theta_{\bm p}(y_0,y)= \frac{\tau}{2} \int^y_{y_0} dy' \frac{\omega ( \bm{p},y')}{y' \br{1-y'}},
\end{align}
is determined in order to fix the value of normalization constants $N^{(\pm)}(\bm{p}).$

One needs to integrate the phase carefully, since the \( 1/y \) factor leads to a divergence at \( y = 0 \). 
We follow the limiting integration procedure described in Ref.~\cite{Hebenstreit:2010vz}. 
The dynamical phase \( \Theta_{\bm{p}}(y_0,y) \) diverges as \( y \rightarrow 0 \). 
To isolate this divergence, the phase can be split into two parts:
\begin{align}
    \Theta_{\bm{p}}(y_0,y) \stackrel{y \to 0}{=} \frac{\tau}{2} \omega_0 \ln{y} + \Tilde{\Theta}_{\bm{p}}(y_0,y=0).
\end{align}
The first term captures the divergent behavior, while \( \Tilde{\Theta}_{\bm{p}}(y_0,y) \) represents the regular, finite part. 
This regular term depends on a reference point \( y_0 \not= \{0,1\} \), which is introduced to regulate the lower limit of the integral in the phase expression.
\newline
So that the asymptotic behavior of the adiabatic mode functions $\phi_{\bm p}^{(\pm)}(y)$ reads 
\begin{align}
     \phi_{\bm p}^{(\pm)}(y)=\frac{e^{ \mp \ii \Tilde{\Theta}_{\bm p} (y_0,0)}}{\sqrt{2\omega_0\bigl(\omega_0 \mp P_0\bigr)}} e^{ \mp \ii \frac{\tau}{2} \omega_0 \ln{(y)} }.
     \label{phi_asy}
\end{align}
The normalization constants $N^{(\pm)}(\bm{p}) $ are accordingly given by:
\begin{align}
      N^{(\pm)}(\bm{p}) = \frac{e^{ \mp \ii \Tilde{\Theta}_{\bm p} (y_0,0)}  }{\sqrt{2 \omega_0\bigl(\omega_0 \mp P_0\bigr)}}.
      \end{align}

%%%%%%%%%%%%%%%%%%%%%%%%%%%%%%%%%%%%%%%%%%%%%%%%%%%%%%%%%%%%%%%%%%%%%%%%%%%%%%%%%%%%%%%%%%%%%%%%%%%%%%%%%%%%%%%%%%%%%One-particle distribution function as %function%%%%%%%%%%%%%%%%%%%%%%%%%%%%%%%%%%%%%%%%%%%%%%%%%%%%%%%%%%%%%%%%%%%%%%%%%%%%%%%%%%%%%%
\subsection{Particle distribution function}
\label{PDF}
To obtain the single-particle distribution function using the Bogoliubov coefficient $\beta(\bm{p},t)$ in terms of the new time variable $y$. This transformation yields
\begin{align}
    | \beta_{\bm{p}}(y) |^2 =\frac{\epsilon^2_\perp(p_\perp)}{ 2 \omega(\bm{p},y) \bigl(\omega(\bm{p},y)- P(p_\parallel,y ) \bigr)} 
    \left|\Biggl(  \frac{2}{\tau} y (1-y) \partial_y  + \ii \omega(\bm{p},y) \Biggr)\psi^{(+)}_{\bm{p}}(y) \right|^2.
\end{align}
Using Eq.~\eqref{sol1}, the analytical expression for the single-particle distribution function in terms of the transformed time variable $y$ can be written as 
\begin{align}
      f(\bm{p},y)  &= |N^{(+)}(\bm{p})|^2  \Biggl(  1 + \frac{ P(p_\parallel,y )}{\omega(\bm p ,y)}  \Biggr) 
        \bigg| \frac{2}{\tau} y (1-y)  \frac{a b}{c} f_1 + \ii \biggl(\omega(\bm p ,y)- (1-y) \omega_0 -y \omega_1 \biggr) f_2 \bigg|^2,
\label{59.1}
\end{align}
where   $|N^{(+)}(\bm{p})|^2 = \frac{1}{{2 \omega_0\bigl(\omega_0 - P_0\bigr)}} ,f_1 = \Hyper{1+a,1+b,1+c;y}$, and $f_2 =\Hyper{a,b,c;y}.$
\newline
This expression for the time-dependent single-particle distribution function provides insight into pair production at various dynamical stages. In this context, we focus on analyzing the evolution of the distribution function from the initial QEPP stage to the final REPP stage, passing through the transient stage~\cite{Banerjee:2018azr}. Within this framework, the distribution function encodes the spectral information of the produced particles, and its behavior is discussed in Sec.~\ref{Result}. Additionally, in Sec.~\ref{Approximate analytical}, we derive a simple approximate expression for the distribution function, valid for \(t \gg \tau\).
\par
It is important to note that \(y\) is a transformed time variable used to obtain the analytic solution of the differential equation. This, in turn, leads to an analytical relation for the distribution function, which depends on time \(t\). However, we employ the \(y\)-variable as time to derive an approximate relation for the distribution function. Plots of the momentum distribution function, or momentum spectrum, will be shown for different time values (\(t\)-values), which will allow us to compare our results with those from the literature wherever possible.
%%%%%%%%%%%%%%%%%%%%%%%%%%%%%%%%%%%%%%%%%%%%%%%%%%%%%%%%%%%%%%%%%%%%%%%%%%%%%%%%%%%%%%%%%%correlation function
%%%%%%%%%%%%%%%%%%%%%%%%%%%%%%%%%%%%%%%%%%%%%%%%%%%%%%%%%%%%%%%%%%%%%%%%%%%%%%%%%%%%%%%%%%%%%%%%%%%%%%%%%%%%%%%%%%%%%%%%%%%%%%%
\subsection{Correlation function}
\label{CF}
The quantum vacuum in the presence of an external electric field may be described by a complex correlation function. It is defined as,
\begin{align}
\mathcal{C}(\bm{p},t)&=\bra{0_{in}} \op{D}_{-{\bm p}r}^\dagger(t) \op{B}_{{\bm p}r}^\dagger(t)\ket{0_{in}}= 2  \alpha_{\bm p}^*(t) \beta_{\bm p}(t),
\label{a45}
\end{align}
\begin{align}
\mathcal{C}^*(\bm{p},t)&=\bra{0_{in}} \op{D}_{-{\bm p}r}(t) \op{B}_{{\bm p}r}(t)\ket{0_{in}}= 2   \beta_{\bm p}^*(t) \alpha_{\bm p}(t).
\label{a45.1}
\end{align}
 As can easily be seen, this function $\mathcal{C}(\bm{p},t)$, consisting of creation operators for a particle and an anti-particle with the opposite momentum, describes the process of production of $e^-e^+$ pair.
%%%%%%%%%%%%%%%%%%%%%%%%%%%%%%%%%%%%%%%%%%%
\par
In several research articles \cite{Schmidt:1998vi, Blaschke2016,Fedotov:2010ue}, the particle-antiparticle correlation function is redefined by incorporating the slowly varying component of the time-dependent creation and annihilation operators in adiabatic number basis:
\begin{align}
   \op{\mathcal{B}}_{{\bm p}r}(t) = \op{B}_{{\bm p}r}(t) \ee^{-\ii\Theta_{\bm{p}} (t_0,t)}, \\
     \op{\mathcal{D}}_{-{\bm p}r}(t) = \op{D}_{-{\bm p}r}(t) \ee^{-\ii \Theta_{\bm{p}} (t_0,t)}.
    \end{align}
So that, 
\begin{align}
\mathcal{C} (\bm{p},t)&=\bra{0_{in}} \op{\mathcal{D}}_{-{\bm p}r}^\dagger(t) \op{\mathcal{B}}_{{\bm p}r}^\dagger(t)\ket{0_{in}},\nonumber\\
&=  \ee^{ 2 \ii \Theta_{\bm{p}} (t_0,t)} \bra{0_{in}} \op{D}_{-{\bm p}r}^\dagger(t) \op{B}_{{\bm p}r}^\dagger(t)\ket{0_{in}},  \nonumber\\
 &= 2  \alpha_{\bm p}^*(t) \beta_{\bm{p}}(t) \ee^{ 2 \ii \Theta_{\bm{p}} (t_0,t)}.
\label{cf2}
\end{align}
Using the above relation, the pair correlation function can be calculated using the Bogoliubov coefficients (Eq.\eqref{a35b},\eqref{a36}) and the mode function $\psi_{\bm p}^{(+)}(y)$( Eq.~\eqref{sol1}:
\begin{align}
    \mathcal{C}( \bm{p}, t ) &= |N^{(+)}(\bm{p})|^2  
    \frac{\epsilon^2_\perp (p_\perp)}{\omega(\bm p,y)}  \Biggl[ \frac{4}{\tau^2}  y^2 (1-y)^2 \left|\frac{a b}{c} f_1 \right|^2  - \Bigl(\omega^2(\bm{p},y) - (y \omega_1 + (1-y) \omega_0)^2 \Bigr) |f_2|^2   
     \nonumber \\
      &
    - \left(2 y (1-y) (\omega_1 - \omega_0 -2 E_0 \tau) \right)\left( (y \omega_1  + (1-y) \omega_0 ) \Re{(\frac{b}{c} f_1 f^*_2)} +\ii \omega(\bm p,y) \Im{(\frac{b}{c} f_1 f^*_2)} \right)\Biggr].
    \end{align}

Vacuum polarization effects play a fundamental role in the process of pair production, influencing the dynamics of particle creation in strong fields. These effects are captured by the correlation function \(\mathcal{C}(\bm{p}, t)\), which can be decomposed into its real and imaginary parts to describe different aspects of the process ~\cite{Banerjee:2018fbw,Smolyansky:2019dqd}. Specifically, the real part of \(\mathcal{C}(\bm{p}, t)\) is denoted as \(u(\bm{p}, t) = \Re\left(\mathcal{C}(\bm{p}, t)\right)\), while the imaginary part is given by \(v(\bm{p}, t) = \Im\left(\mathcal{C}(\bm{p}, t)\right)\). 
These components are crucial for understanding the underlying mechanisms of pair production and its dependence on time and momentum.
\newline
Therefore,
%%%%%%
 \begin{align}
   u( \bm{p}, t ) &=
   |N^{(+)}(\bm{p})|^2  
    \frac{\epsilon^2_\perp (p_\perp)}{\omega(\bm p,y)}  \Biggl( \frac{4}{\tau^2}  y^2 (1-y)^2 \left|\frac{a b}{c}f_1 \right|^2 - (\omega^2(\bm{p},y) - (y \omega_1 + (1-y) \omega_0)^2) |f_2|^2   
     \nonumber \\
      &
    - 2 y (1-y) (\omega_1 - \omega_0 -2 E_0 \tau) (y \omega_1  + (1-y) \omega_0 ) \Re{(\frac{b}{c} f_1 f^*_2)} \Biggr),
    \end{align}
   \begin{align}
   v( \bm{p}, t ) &=   2 \ii y (1-y) \epsilon^2_\perp (p_\perp) |N^{(+)}(\bm{p})|^2 
    (2 E_0 \tau + \omega_0 - \omega_1)  \Im{(\frac{b}{c} f_1 f^*_2)}.
    \end{align}
The function $u(\bm{p}, t)$ depicts vacuum polarization effects and pair production phenomena. The function $v(\bm{p}, t)$ serves as a counterterm to pair production, effectively representing pair annihilation in the vacuum excitation process. A comprehensive discussion of these insights is presented in detail in Sec.~\ref{VP}.
%%%%%%%%%%%%%%%%%%%%%%%%%%%%%%%%%%%%%%%%%%%%%%%%%%%%%%%%%%%%%%%%%%%%%%%%%%%%%%%%%%%%%%%%%%%%%%%%%%%%%%%%%%%%%%%%%%%%%%%%%%%%%%%%%%%%%%%%%%%%%%%%%%%%%%%%%%%%%%%%%%%%%%%%%%%%%%%%%%%%%%%%%%%%%%%%%%%%%%%%%%%%%%%%%%%%%%%%%%%%%%%%%%%%%%%
\subsection{Approximate analytical expression for the distribution function at finite time}
\label{Approximate analytical}
%To investigate the behavior of the distribution functions in the limit $t >> \tau,$To analyze this limit, we  approximate  the Gamma  and Gauss-hypergeomtric  functions governing $f(\bm{p},y)$ we employ approximations based on the suitable expressions of the Gamma and the Gauss-hypergeometric functions~\cite{abramowitz}. These approximations enable us to deduce a simplified analytical expression for particle distribution functions.
%%%%
To investigate the behavior of the distribution functions in the asymptotic limit $t \gg \tau$, we analyze the governing expressions for $f(\bm{p}, y)$, which involve the Gamma and Gauss-hypergeometric functions. In this limit, we employ suitable asymptotic approximations for these special functions, as given in standard references~\cite{abramowitz}. These approximations allow us to derive a simplified analytical expression for the particle distribution functions, capturing their leading behavior at late times.
\par
First, we start with approximating the Gauss-hypergeometric function as $y \to 1$. It is crucial to ensure smooth convergence towards the limit of $\Hyper{a,b,c;y\rightarrow 1}$. This task is complicated by the intricate nature of the parameters $a$, $b$, and $c$ in this specific context, which makes it essential to exercise caution when dealing with this limit. Therefore, it is beneficial to transform the argument by substituting $y$ with $(1-y)$. This transition can be achieved using the following mathematical identity
\begin{multline}
\Hyper{a,b,c;z} = \frac{\Gamma \br{c} \Gamma \br{c-a-b}}{\Gamma \br{c-a} \Gamma \br{c-b}} \Hyper{a,b,a+b-c+1;1-z} 
\\  + \br{1-z}^{c-a-b} \frac{\Gamma \br{c} \Gamma \br{a+b-c}}{\Gamma \br{a} \Gamma \br{b}} \Hyper{c-a, c-b, c-a-b+1; 1-z}.
\label{1.66}
\end{multline}
where $\bet{\textrm{arg}(1- z )} < \pi$ for the identity to hold \cite{abramowitz}.
\newline
In general, the Gauss-hypergeometric function can be expressed as a power series: 
\begin{align}
    \Hyper{a,b,c;z}  &=  1 +  \frac{ a b }{c } z +  \frac{ a(a+1)  b(b+1) }{c (c+1) } \frac{z^2}{2!} +  \frac{ a(a+1)(a+2)   b(b+1)(b+2) }{c (c+1)(c+2) } \frac{z^3}{3!}+...
    \label{1.69}
 \end{align}
 The series continues with additional terms involving higher powers of $z.$ Each term in the series involves the parameters $a, b,$ and $c$ as well as the variable $z$ raised to a specific power \cite{abramowitz}.
 \par
 Using the above relations Eq.~\eqref{1.66} and ~\eqref{1.69}, we approximate the Gauss-hypergeometric functions $f_1$ and $f_2$ present in the Eq.\eqref{59.1} as follows:
 \begin{align}
 f_1  &=   \Biggl( \frac{ c \Gamma ( c) \Gamma (c-a-b-1) }{\Gamma (c-a) \Gamma (c-b)} \Biggr) \Biggl( 1 +  \frac{ (a+1) (b+1) }{(2+a+b-c) } (1-y) 
  \nonumber \\
 & +  \frac{ (a+1)(a+2)  (b+1)(b+2) }{(2+a+b-c) (2+a+b-c+1) } \frac{(1-y)^2}{2!} + ...  \Biggr)
  \nonumber \\
 &+   (1-y)^{(c-a-b-1)}  
  (a+b-c)   \Biggl(\frac{ c \Gamma ( c) \Gamma (a+b-c) }{ a \Gamma (a)  b \Gamma (b)}  \Biggr) \Biggl( 1 +  \frac{ (c-a) (c-b) }{(c-a-b) } (1-y)  
  \nonumber \\
  &+  \frac{ (c-a)(c-a+1)  (c-b)(c-b+1) }{(c-a-b) (c-a-b+1) } \frac{(1-y)^2}{2!} + ...   \Biggr).
  \label{hypfun1}
\end{align}
Similarly, 
\begin{align}
   f_2  &= \frac{ \Gamma (c) \Gamma (c-a-b) }{\Gamma (c-a) \Gamma (c-b)} \Biggl( 1 +  \frac{ a b}{1+a+b-c}(1-y) +...\Biggr)
   \nonumber \\ & +  (1-y)^{(c-a-b)}  
  \frac{ \Gamma (c) \Gamma (a+b-c) }{\Gamma (a) \Gamma (b)} \Biggl(1 + \frac{(c-a)(c-b)}{(1+c-a-b)} (1-y) + ... \Biggr).
  \label{hypfun2}
  \end{align}
Note that here $y$ is a transformed time variable that evolves according to Eq.~\eqref{newt}.
\par
As time progresses, especially in the case with significant time intervals such as $t > \tau$, it becomes evident that \( (1 - y) \) approaches zero. Consequently, the dominant contribution to the particle distribution function as $y \rightarrow 1 $  originates from the zeroth-order term, which is independent of $(1-y).$ As discussed in many literature for deriving the asymptotic expression for the particle distribution function, the Gauss-hypergeometric functions are truncated up to zeroth order only\cite{Hebenstreit:2010vz,Gavrilov:1996pz}.
\par
Building on the earlier discussion (see Sec.~\ref{PDF}), our focus is primarily on understanding the dynamics of the particle distribution function over finite time rather than just its asymptotic trends. We want to examine the distribution function, $f(\bm{p}, y)$, in the vicinity of $y \rightarrow 1.$ In that case, it is necessary to incorporate other higher-order terms in the expression of the asymptotic particle distribution function. It means expanding the time-dependent distribution function defined in Eq.~\eqref{59.1} as a power series in the  variable $(1-y)$ as:
\begin{align}
      f(\bm{p},y) \approx \mathrm{C}_0 (\bm{p},y) + (1-y) \mathrm{C}_1(\bm{p},y) + (1-y)^2 \mathrm{C}_2(\bm{p},y) +...+ (1-y)^n \mathrm{C}_n(\bm{p},y).
      \label{SERIESF}
\end{align}
%%%%%%%%%%
To compute an approximate expression for the distribution function that depends on finite time, we can truncate the power series of the Gauss-hypergeometric functions $ f_1$ and $f_2$ given by  Eqs.~\eqref{hypfun1} and \eqref{hypfun2}  respectively, up to a specific order. The order of truncation will depend on the desired accuracy and the characteristics of the finite-time behavior under consideration.
%\newline
%%%%%%%%%%%%%%%%%%%%%%first term%%%%
\par
Let's start by approximating the different terms present in the particle distribution Eq.~\eqref{hypfun1} and \eqref{hypfun2}, considering only up to the second-order terms and neglecting higher-order terms.

Let us begin by approximating the various terms appearing in the particle distribution functions
Eqs.~\eqref{hypfun1} and \eqref{hypfun2}, retaining contributions up to second order and neglecting
higher-order terms. Accordingly, we obtain
%\newline
%%Therefore,
\begin{align}
    \frac{2y}{\tau}(1-y)  \frac{ab}{c} f_1 &\approx \frac{2}{\tau}y(a+b-c)  \Gamma_2 (1-y)^{(c-a-b)} +(1-y) \frac{2 y }{\tau}
\Biggl(a b \Gamma_1 -  (c-a)(c-b)  \Gamma_2 (1-y)^{(c-a-b)}
\Biggr) \nonumber \\
  & + (1-y)^2 \frac{2y}{\tau}\Biggl(  \Gamma_1 \frac{a(1+a) b(1+b)}{(2+a+b-c)}  +  \Gamma_2 (1-y)^{(c-a-b)} \frac{(c-a)(c-b)(c-a+1)(c-b+1)}{(a+b-c-1)}\Biggr),
\end{align}
where we have defined
\begin{align}
     \Gamma_1  &=\frac{ \Gamma ( c) \Gamma (c-a-b-1) }{\Gamma (c-a) \Gamma (c-b)}, \\
   \Gamma_2 &=\frac{ \Gamma ( c) \Gamma (a+b-c) }{\Gamma (a) \Gamma (b)}.
\end{align}

Here, $\Gamma(\cdot)$ denotes the Gamma function, while $\Gamma_1$ and $\Gamma_2$ are shorthand
coefficients constructed from ratios of Gamma functions.

%%Here, $\Gamma ()$ denotes  the Gamma function, whereas $\Gamma_1$ and $\Gamma_2$ are shorthand coefficients constructed from Gamma functions.
  Similarly,
 \begin{align}
      \Bigl(\omega(\bm{p},y) - (1-y) \omega_0- y \omega_1 \Bigr) f_2 \approx (1-y) \Biggl((\omega_1 - \omega_0) - \frac{2 E_0 \tau }{\omega_1} P_1 \Biggr) \Biggl(\Gamma_1 (c-a-b-1)  + e^{- \ii \tau\omega_1 \ln{(1-y)}} \Gamma_2\Biggr)
      \nonumber \\
      + (1-y)^2 \Biggl( \Gamma_1 (c-a-b-1)  \Biggl(  \frac{ 2 E_0^2 \tau^2  \epsilon_\perp^2(p_\perp) }{\omega_1^3}  +     \frac{ a b ( \omega_1(\omega_1 - \omega_0) - 2 P_1 E_0 \tau )}{ \omega_1(1+a+b-c)} \Biggl)
      \nonumber \\
      + \Gamma_2  e^{-\ii \tau \omega_1 \ln{(1-y)}} \Biggl( \frac{ 2 E_0^2 \tau^2 \epsilon_\perp^2(p_\perp) }{\omega_1^3} + \Bigl((\omega_1 - \omega_0) - \frac{2 E_0 \tau }{\omega_1} P_1 \Bigr)  \frac{(c-a)(c-b)}{(1+c-a-b)} \Biggr)\Biggr).
 \end{align}

 Additionally, we can approximate 
    \begin{align}
   1 + \frac{P(p_\parallel,y)}{\omega(\bm{p},y)}   &\approx \Omega_0 + (1-y) \Omega_1  + (1-y)^2 \Omega_2,   
\end{align}
with
 \begin{align}
     \Omega_0 &= 1 + \frac{P_1}{\omega_1} ,\\
     \Omega_1 &=  \frac{-2 E_0 \tau}{ \omega_1^3} \epsilon^2_\perp (p_\perp),\\
     \Omega_2 &= \frac{-6 E_0^2 \tau^2}{\omega_1^5} P_1 \epsilon^2_\perp(p_\perp).
\end{align}

We also approximate the cross term $\Gamma_1 \Gamma^{*}_2$, which appears in the expression for the distribution function. The detailed calculation of this term using the approximation for the Gamma function is provided in Appendix~\ref{Approximate function appendix}. Here, we simply write final expression for the $\Gamma_1 \Gamma^{*}_2$ :
\begin{align}
    \Gamma_1 \Gamma^{*}_2 &\approx \left| \Gamma_1 \Gamma^{*}_2 \right| e^{\ii \varrho},
\end{align}

where the modulus and phase are given by

\begin{align}
    \left| \Gamma_1 \Gamma^{*}_2 \right| &= \frac{\omega_0}{2 \omega_1} \cdot \frac{\exp\left[\frac{\pi \tau}{2} (\omega_0 - \omega_1 + 2 E_0 \tau)\right]}{ \sinh(\pi \tau \omega_0)\sqrt{1 + \omega_1^2 \tau^2}} 
    \sqrt{\frac{(\omega_0 + 2 E_0 \tau)^2 - \omega_1^2}{\omega_1^2 - (2 E_0 \tau - \omega_0)^2}},
\end{align}
and
\begin{align}
    \varrho &= \frac{\tau}{2} (\omega_0 + \omega_1 - 2 E_0 \tau) \ln\left[\tau (\omega_0 + \omega_1 - 2 E_0 \tau)\right] 
    + (-\omega_0 + \omega_1 - 2 E_0 \tau) \ln\left[\tau (\omega_0 - \omega_1 + 2 E_0 \tau)\right] \nonumber \\
    &\quad + (-\omega_0 + \omega_1 + 2 E_0 \tau) \ln\left[\tau (-\omega_0 + \omega_1 + 2 E_0 \tau)\right] 
    + (\omega_0 + \omega_1 + 2 E_0 \tau) \ln\left[\tau (\omega_0 + \omega_1 + 2 E_0 \tau)\right] \nonumber \\
    &\quad + \pi - \tan^{-1}(\tau \omega_1) - 2 \tau \omega_1 \ln(2 \omega_1 \tau).
\end{align}
%%%%%%%%%%%%%%%%
Therefore, the approximate expression for the particle distribution function :
\begin{align}
f(\bm{p},y) \approx &\, |N^{(+)}(\bm{p})|^2    \left( \Omega_0 + (1-y) \Omega_1 + (1-y)^2 \Omega_2 \right) \Bigg| 
\frac{2y}{\tau} (a+b-c) \Gamma_2 e^{- \ii \tau \omega_1 \ln(1-y)} \nonumber \\
&+ (1-y) \Bigg[ 
  \Gamma_1 \left( \frac{2y}{\tau} a b + 
  \ii \frac{c-a-b-1}{\omega_1} (\omega_1^2 - \omega_1 \omega_0 - 2 E_0 \tau P_1) \right) \nonumber \\
&\qquad\quad + \Gamma_2 e^{-\ii \tau \omega_1 \ln(1-y)} 
\left( \frac{2y}{\tau} (c-a)(b-c) + \frac{\ii}{\omega_1} (\omega_1^2 - \omega_1 \omega_0 - 2 E_0 \tau P_1) \right) 
\Bigg] \nonumber \\
&+ (1-y)^2 \Bigg[ 
\Gamma_1 \Bigg( 
  \frac{2y}{\tau} ab \frac{(1+a)(1+b)}{2+a+b-c} \nonumber \\
&\qquad\qquad\quad + \frac{\ii}{\omega_1} 
\left( \frac{2 E_0^2 \tau^2 \epsilon_\perp^2(p_\perp)}{\omega_1^2}(c-a-b-1) - ab(\omega_1^2 - \omega_1 \omega_0 - 2 E_0 \tau P_1) \right) 
\Bigg) \nonumber \\
&\qquad + \Gamma_2 e^{-\ii \tau \omega_1 \ln(1-y)} \Bigg(
\frac{2 E_0^2 \tau^2 \epsilon_\perp^2(p_\perp)}{\omega_1^3} 
+ \frac{(c-a)(c-b)}{\omega_1 (a+b-c-1)} \nonumber \\
&\qquad\qquad\qquad \times \left( \frac{2y}{\tau} \omega_1 (c-a+1)(c-b+1) - (\omega_1^2 - \omega_1 \omega_0 - 2 E_0 \tau P_1) \right) 
\Bigg) \Bigg]
\Bigg|^2.
\label{apppdf1}
\end{align}

%%%%%%%%%%%%%%%%%%%%%%%%%%%%%%%%%%%%%%%%%%%%%%%%%%%%%%%%%%%%%%%%%%%%%%%%%%%%%%%%%%%%%%
%%%%%%%%%%%%%%%16nov
To explore the behavior of the distribution function at finite times $( t > \tau )$, we aim to express $ f(\bm{p},y) $ in a series involving $(1-y)$, as discussed previously (refer to Eq. \eqref{SERIESF}). We can then consider truncating higher-order terms to simplify the analysis while still capturing essential features.
In this context, we focus exclusively on terms up to order $ (1-y)^2$, disregarding higher-order terms in Eq. \eqref{apppdf1}. This approach is sufficient to explain the interesting results discussed later.
\par
After performing extensive calculations, we derive a simplified expression for the particle distribution function in terms of the small parameter $(1 - y)$, up to second order. A detailed evaluation of $f(\bm{p}, y)$ for better clarity is provided in Appendix \ref{Approximate function appendix}. This corresponding expression is given by:
%%%%%%%%%%%%%%%%%%%%%%%%%%%%%%%%%%%%%%%%%%%%%%%%%%%%%%%%%%%%%A detailed evaluation of \( f(\bm{p}, y) \) for better clarity is provided in Appendix~\ref{Approximate function appendix}. The corresponding expression is given by:
%%%%%%%%%%%%%%%%%
\begin{align}
        f(\bm{p},y) &\approx |N^{(+)}(\bm{p})|^2 \Bigl( \mathrm{C}_0(\bm{p},y) +  (1-y) \mathrm{C}_1(\bm{p},y)+ (1-y)^2 \mathrm{C}_2(\bm{p},y) \Bigr),
   \label{appdisfun}
\end{align}
%%%
where
\begin{align}
     \mathrm{C}_0(\bm{p},y) &=  4  \Omega_0 \omega_1^2 y^2 |\Gamma_2|^2,
\label{pdfC0}
\end{align}
%%%%%%%%%%%%%%%%%%%%%%%%%%%%%%%%%%%%%%%%%%%%%%%%%%%%%%%%%%%%%%%%%%%%%%%%%%%%%%%%%%%%%%%%%%%%%%%%%%%%%%%%%%%%%%%%%%%%%%%%%%%%%%%%%%%%%%%%%%%%%%%%%%%%%%%
\begin{figure}[t]
\begin{center}
\includegraphics[width = 2.03905in]{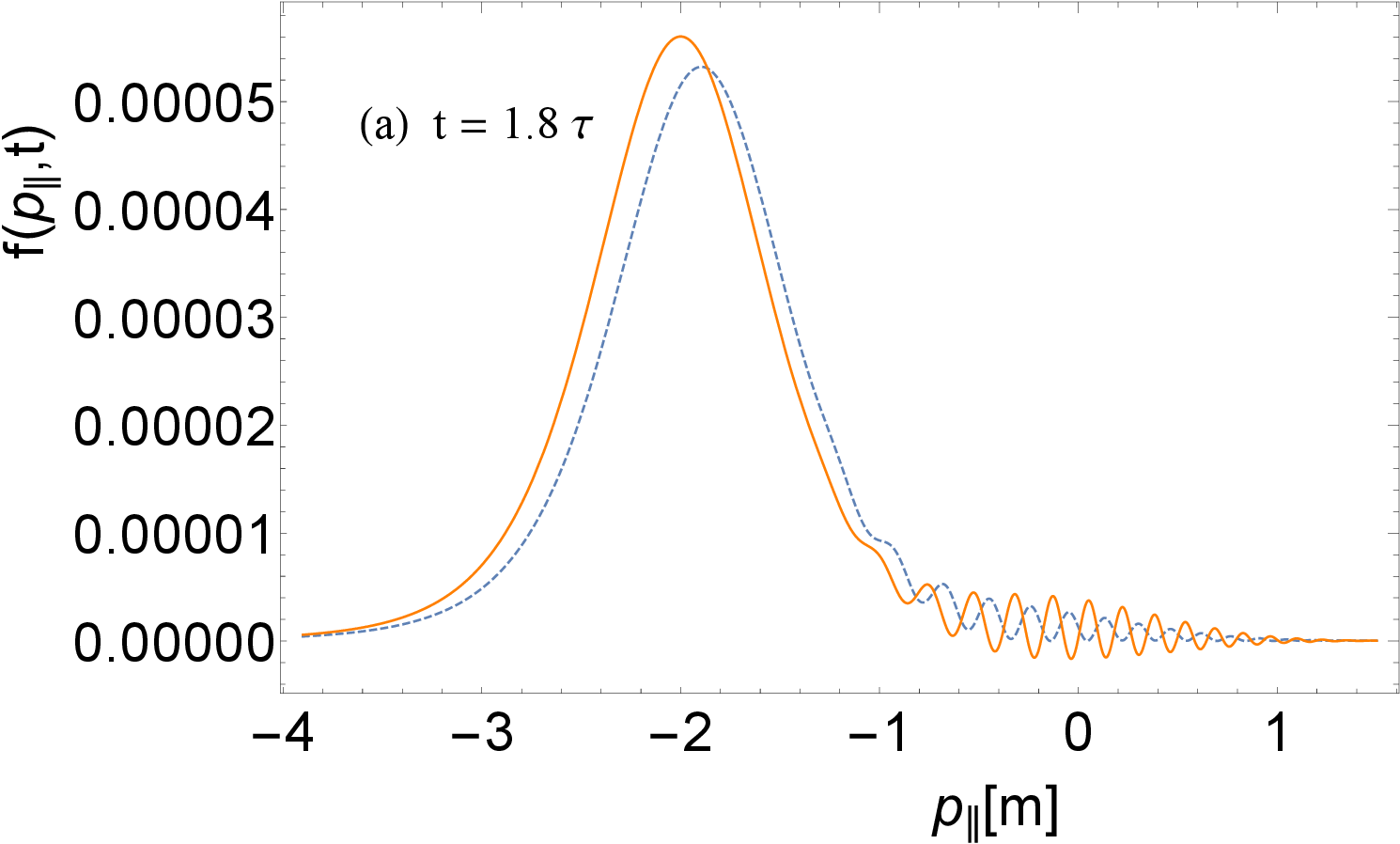}
\includegraphics[width = 2.03905in]{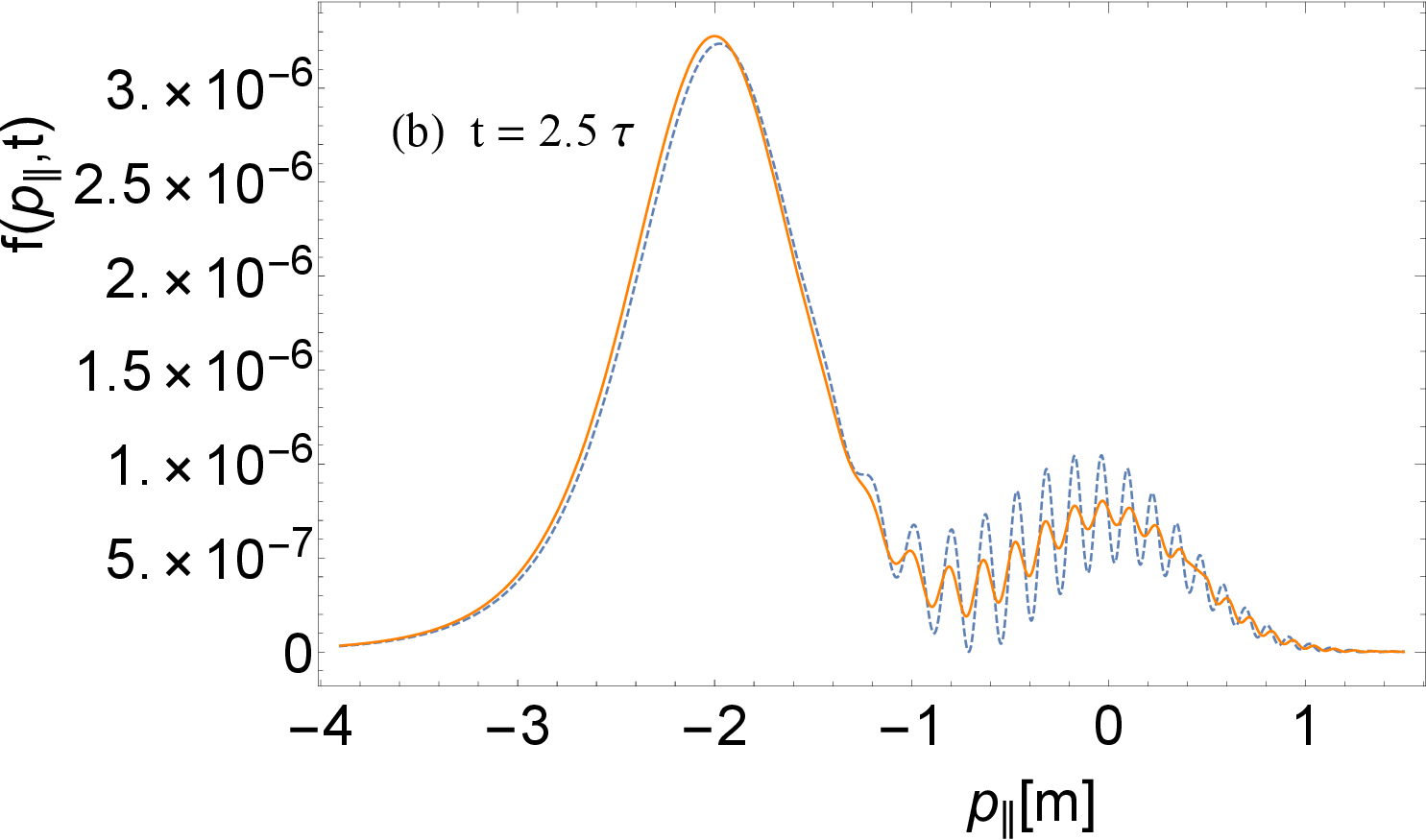}
\includegraphics[width = 2.03905in]{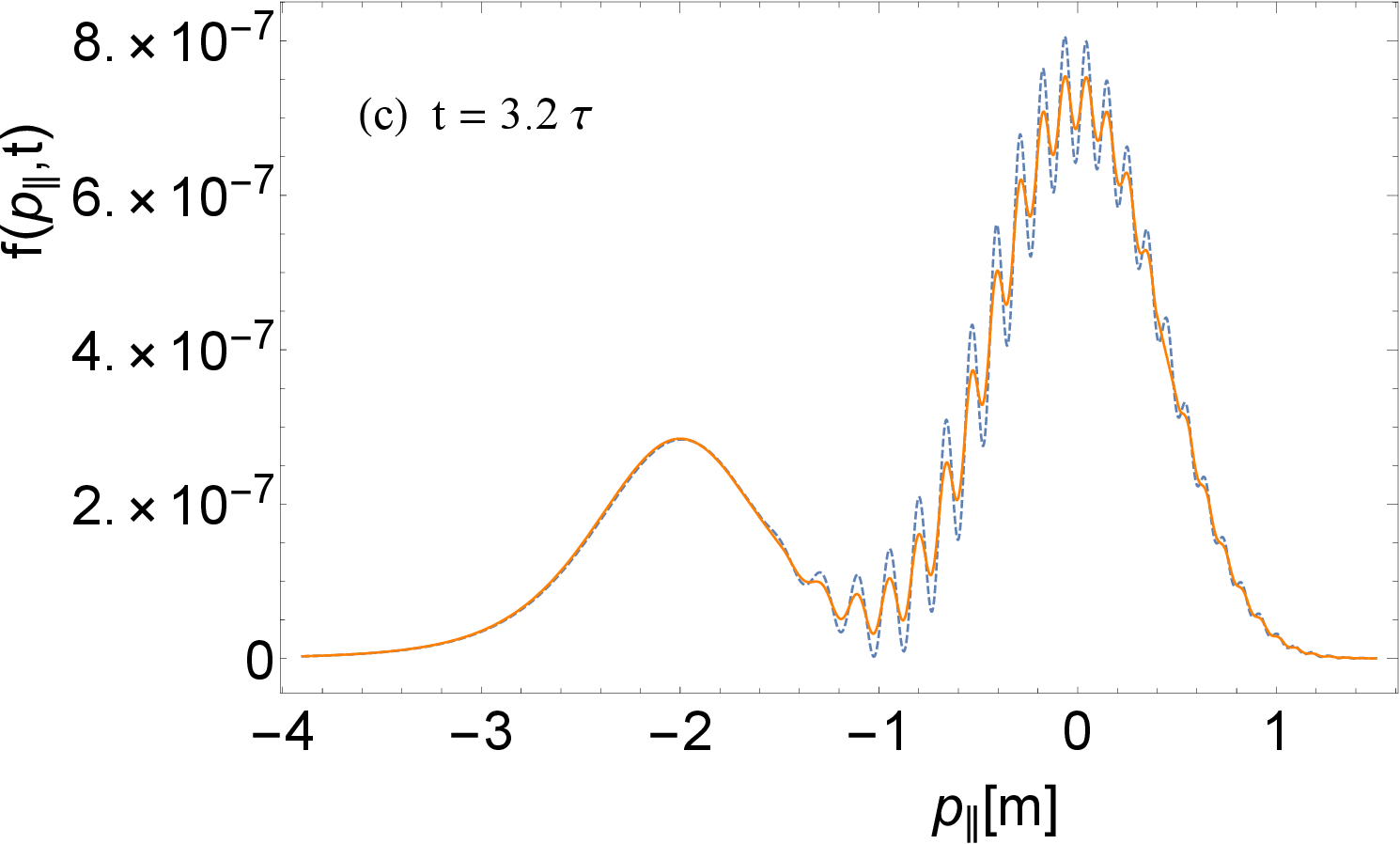}
\caption{The longitudinal momentum distribution function, $f(p_\parallel,t)$, as a function of $p_\parallel$ for different times with the field parameter $E_0=0.2 E_{c}$ and $ \tau =10 [m^{-1}].$ The value of transverse momentum is taken to be zero. Here, the dashed blue line represents the exact result, while the solid orange line is from the approximate calculation. All units are in electron mass units.}
\label{app_exact}
\end{center}
\end{figure}
\begin{align}
    \mathrm{C}_1(\bm{p},y) &=  4 y  \Omega_0  |\Gamma_1 \Gamma^{*}_2| E_0 \tau  (P_1 -\omega_1 )   \cos{(\Upsilon)} +4 y  |\Gamma_2|^2   (2 \Omega_0 ( \omega_1^2 -\omega_0 \omega_1 +(\omega_1 - P_1) E_0 \tau ) +  \Omega_1 \omega_1^2 y),
    \label{pdfC1}
\end{align}
\begin{align}
    \mathrm{C}_2 (\bm{p},y)   &= \frac{4 E_0^2 \tau^2}{\omega_1^2} \Omega_0 (P_1 - \omega_1)^2 |\Gamma_1|^2 + |\Gamma_2|^2 \Biggl[8 \Omega_1 y (2 E_0 \tau(\omega_1 -P_1) - \omega_0 \omega_1   + \omega_1^2) 
  \nonumber   \\ 
     &+ 4 \Omega_2 y^2 \omega_1^2  + \frac{\Omega_0 }{4 \omega_1^2 ( 1 + \tau^2 \omega_1^2)}\sigma_0 \Biggr]  + \frac{|\Gamma_1 \Gamma^{*}_2| ( \cos{(\Upsilon)} \sigma_1 + \sin{(\Upsilon)} \sigma_2  )}{2 \omega_1^2 (4 + \omega_1^2 \tau^2)} .
     \label{pdfC2}
\end{align}

where we define,
\begin{align}
\Upsilon &= \mathrm{\varrho} + \tau \omega_1 \ln{(1-y)}, \\
\sigma_0 &=\bigg(16 (\omega_1^2 + E_0 (\omega_1 - P_1) \tau)^2 (1 + \omega_1^2 \tau^2) -   \nonumber\\
&4 \Big(-8 (\omega_0 - \omega_1) \omega_1^2 (\omega_1 + P_1) + 
16 E_0 \omega_1^2 (\omega_1 + P_1) \tau +   \nonumber\\
&\big(8 E_0^2 (-\omega_1^2 + P_1^2) - (\omega_0 - \omega_1) \omega_1^3 (\omega_0^2 + 
\omega_1 (7 \omega_1 + 8 P_1))\big) \tau^2 +   \nonumber\\
&2 E_0 \omega_1^2 (\omega_0 + 3 \omega_1) (2 \omega_1^2 - \omega_0 P_1 + 3 \omega_1 P_1) \tau^3 +   \nonumber\\
&4 E_0^2 \omega_1^2 \big((\omega_0 - 3 \omega_1) \omega_1 + 2 \omega_1 P_1 + 2 P_1^2\big) \tau^4 + 
8 E_0^3 \omega_1^2 P_1 \tau^5\Big) y +   \nonumber\\
&\omega_1^2 \Big(32 \omega_1 (\omega_0 + \omega_1) + 
64 E_0 \omega_1 \tau + (\omega_0 + \omega_1)^2 (\omega_0^2 - 
14 \omega_0 \omega_1 + 25 \omega_1^2) \tau^2 +   \nonumber\\
&16 E_0 \omega_1^2 (2 \omega_0 + 3 \omega_1) \tau^3 - 
\big(\omega_1^2 (\omega_0 + \omega_1)^4 + 
8 E_0^2 (\omega_0^2 - 6 \omega_0 \omega_1 - 3 \omega_1^2)\big) \tau^4 +   \nonumber\\
&64 E_0^3 \omega_1 \tau^5 + 
8 E_0^2 \big(2 E_0^2 + \omega_1^2 (\omega_0 + \omega_1)^2\big) \tau^6 - 
16 E_0^4 \omega_1^2 \tau^8\Big) y^2\bigg),
\end{align}

\begin{align}
 \sigma_1 &= 16 E_0^3 \Omega_0 \omega_1^2 \tau^5 y (P_1 (4 + \omega_1^2 \tau^2) - 
     4 \omega_1 y)  -16 E_0^4 \Omega_0 \omega_1^4 \tau^8 y^2 +  \Omega_0 \omega_1^2 (-\omega_0 + 
     \omega_1) y (2 (-8 (\omega_1 + P_1)  \nonumber   \\ 
     &+ (\omega_0 - 
           \omega_1)^2 \omega_1 \tau^2) (4 + \omega_1^2 \tau^2) +  \omega_1 (32 - 
        4 (\omega_0 - \omega_1) (4 \omega_0 + \omega_1) \tau^2 + (\omega_0 - 
           \omega_1)^3 \omega_1 \tau^4) y) \nonumber   \\ 
     & + 8 E_0^2 \Omega_0 \tau^2 (-2 (\omega_1 - P_1)^2 (4 + \omega_1^2 \tau^2) - (4 + \omega_1^2 \tau^2) (-2 P_1^2 +  \omega_1^2 (2 + \omega_1 (-\omega_0 + \omega_1) \tau^2)) y  \nonumber   \\ 
     & +  \omega_1^3 \tau^2 (-8 \omega_0 + 
        6 \omega_1 + (\omega_0 - \omega_1)^2 \omega_1 \tau^2) y^2) + 
  4 E_0 \omega_1^2 \tau (-4 \Omega_0 (\omega_1 - P_1) (4 + 
        \omega_1^2 \tau^2) 
        \nonumber   \\ 
     &  - (4 + 
        \omega_1^2 \tau^2) (4 \Omega_1 (\omega_1 - P_1) - 
        8 \Omega_0 (\omega_1 + P_1) + 
        \Omega_0 (\omega_0 - \omega_1)^2 P_1 \tau^2) y
    \nonumber   \\ 
     &  - 4 \Omega_0 \omega_1 (4 + (-\omega_0^2 + \omega_1^2) \tau^2) y^2),   \\ 
    \sigma_2 &= 2 (4 + \omega_1^2 \tau^2) (\omega_1^3 (2 + \omega_1) - E_0 \omega_1^2 (\omega_1 + 3 P_1) \tau 
     \nonumber \\
     & + 4 E_0^2 (\omega_1^2 + \omega_1 (-2 + P_1) - P_1^2) \tau^2 + 
     4 E_0^3 (-\omega_1 + P_1) \tau^3)  \nonumber \\
     & + \omega_0^4 \omega_1^2 \tau^2 y - 
  \omega_1^2 (\omega_1 - 2 E_0 \tau) (12 \omega_1 + 8 E_0 \tau + 
     3 \omega_1^3 \tau^2 + 2 E_0 \omega_1^2 \tau^3 - 4 E_0^2 \omega_1 \tau^4 + 
     8 E_0^3 \tau^5) y  \nonumber \\
     &+ 4 \omega_0 \omega_1^2 (4 + \omega_1^2 \tau^2) (-2 + \omega_1 y) + 
  2 \omega_0^2 ((\omega_1 (2 + \omega_1) + E_0 (\omega_1 - P_1) \tau) (4 + 
        \omega_1^2 \tau^2) \nonumber \\
        & -y \omega_1^2 (10 + 3 \omega_1^2 \tau^2 - 2 E_0 \omega_1 \tau^3 + 
        4 E_0^2 \tau^4) ) .
        \label{sigm2}
\end{align}
Also note that $\mathrm{C}_0(\bm{p},y)$, $\mathrm{C}_1(\bm{p},y)$, and $\mathrm{C}_2(\bm{p},y)$ are independent of the function of $(1-y)$.
%%%%%%%
%
\par
To verify the accuracy of the approximate expression of the distribution function Eq.~\eqref{appdisfun}, we will directly compare it with the exact relation \eqref{59.1}. For this purpose, we plot the approximate relation \eqref{appdisfun} against the exact one for $t > \tau$, as shown in Fig.\ref{app_exact}. This comparison is crucial for analytically investigating the momentum distribution function at finite time (see Sec. \ref{ApprLMSresult}).
The parameters describing the electric field are $E_0 = 0.2 E_c$ and $\tau = 10 [m^{-1}]$. The transverse momentum is assumed to be zero.
%%%%%%%%
In Fig. \ref{app_exact}(a), we observe a small discrepancy between the exact and approximate results, which appears to depend on $t$. Consequently, the approximate result provides a highly accurate prediction for $t \ge 2 \tau$. The mean relative error of the approximate result remains below $0.2\%$ at $t = 2.5 \tau$ (see Fig. \ref{app_exact}(b)). As time progresses, both approaches yield nearly identical results, as illustrated in Fig. \ref{app_exact}(c) for $t = 3.2 \tau$.  
Based on the Eq.~\eqref{appdisfun}, we investigate the approximate momentum distribution function, considering both longitudinal and transverse momentum components. These components are expressed in terms of elementary functions, as detailed in \cref{sec:Longitudinal momentum,sec:TMS}. 
%%%%
\section{Results and Discussion}
\label{Result}
%%%%%%%%%%%%%%%%%%%%
\subsection{Temporal evolution of particle distribution }
\label{temporal}
\begin{figure}[t]
\begin{center}
\includegraphics[width = 5.05in]{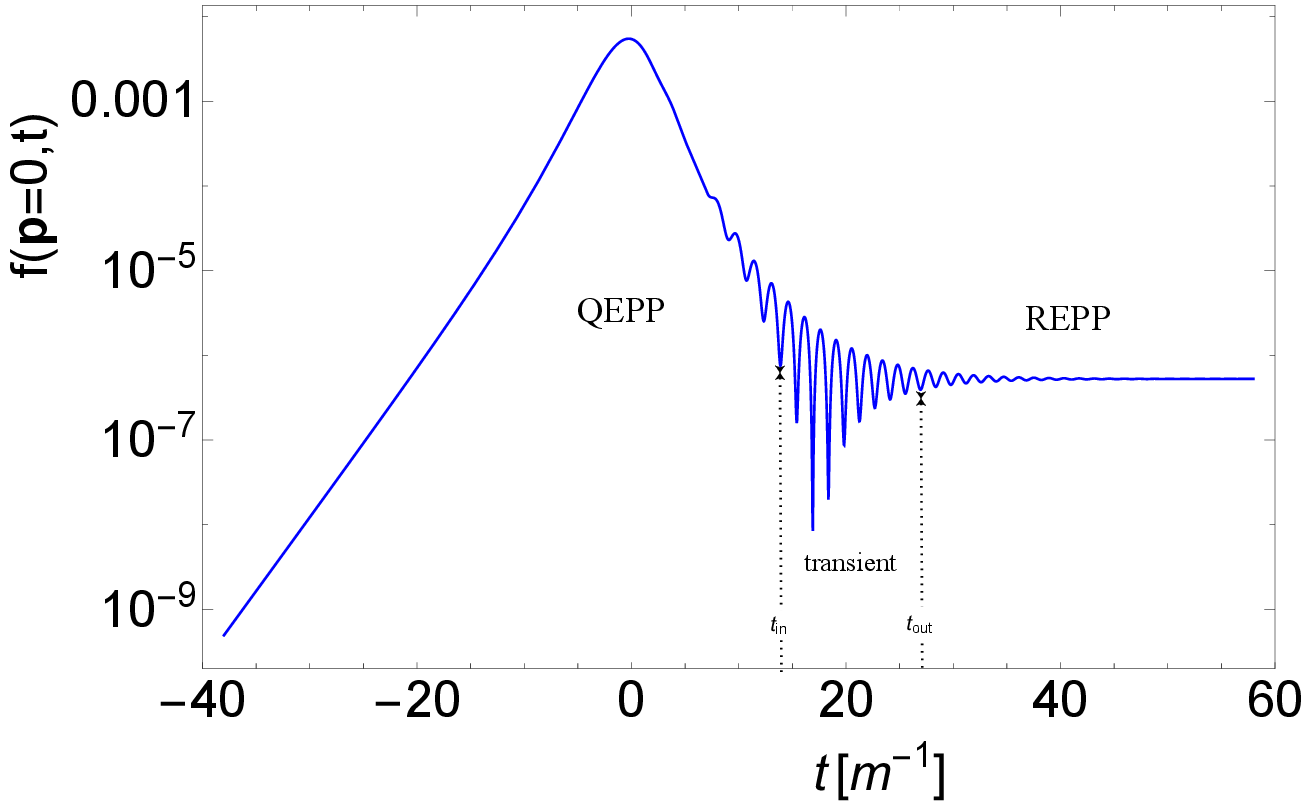}
\caption{Evolution of quasi-particle distribution function for Sauter-pulsed field for longitudinal momentum $p_\parallel = 0$ and transverse momentum $p_\perp =0.$
The field parameters are $E_0=0.2 E_{c}$ and $ \tau=10 [m^{-1}]$, and all the measurements are taken in electron mass units.}
   	\label{longi_trans}
\end{center}
\end{figure}
%%%%%%
Quantum vacuum becomes unstable under the action of the external electric field. As a consequence, virtual particle-antiparticle pairs are created in an off-shell configuration. These virtual charged particles are accelerated by the electric field to enough energy to become real particles in an on-shell mass configuration. During the action of the external force, pair annihilation processes occur simultaneously with the pair creation, giving rise to a dynamical quasiparticle plasma \cite{Blaschke:2014fca}. This results in different states of in and out-vacuums. 
The most complete description of vacuum pair creation from in-state to out-state is given by the single-particle distribution function. 
\par
The time evolution of the single-particle distribution function shows that electron-positron plasma(EPP) excited from vacuum passes through three different temporal stages: the QEPP stage, the transient stage, and the final REPP in the out-state as indicated by the author in \cite{Blaschke:2011is,Banerjee:2018azr}. This is illustrated in Figure~\ref {longi_trans}. The initial stage QEPP and the final REPP stage are separated by the fast oscillation of EPP by the transient stage. The transient stage is considered to begin at time $t_{in}$, where the oscillation of $f(\bm{p},t)$ first reaches the REPP level. The time $t_{out}$, marking the end of the transient stage, is defined as when the average level of the oscillating $f(\bm{p},t)$ hits the REPP level. After this point, the REPP stage begins.
 At the REPP stage, quasi-particles become independent, and real particle-antiparticles are observed with a lesser value off than that at the electric field maximum at $t = 0.$ Each of the three stages contributes to various physical effects, including vacuum polarization effects \cite{Smolyansky:2010as}, the emission of annihilation photons from the focal point of colliding laser beams \cite{Blaschke:2005hs,Blaschke:2010vs,Blaschke:2011af,Blaschke:2011is}, birefringence effects \cite{DiPiazza:2006pr}, and other secondary phenomena. To accurately estimate the contributions of these stages to observable effects, a detailed analysis of each period of the EPP's evolution is quite helpful.
\par
\begin{figure}[t]
\begin{center}
{
\includegraphics[width = 3.15in]{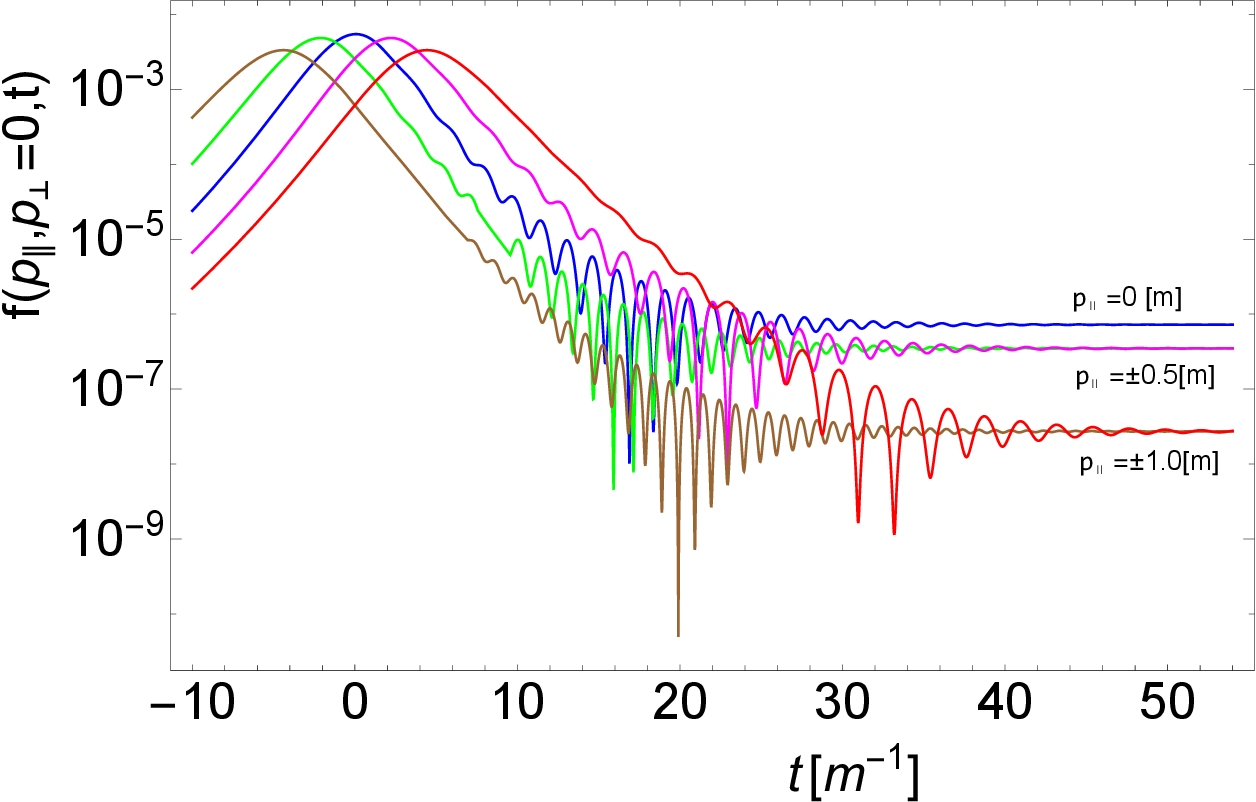}
\includegraphics[width = 3.15in]{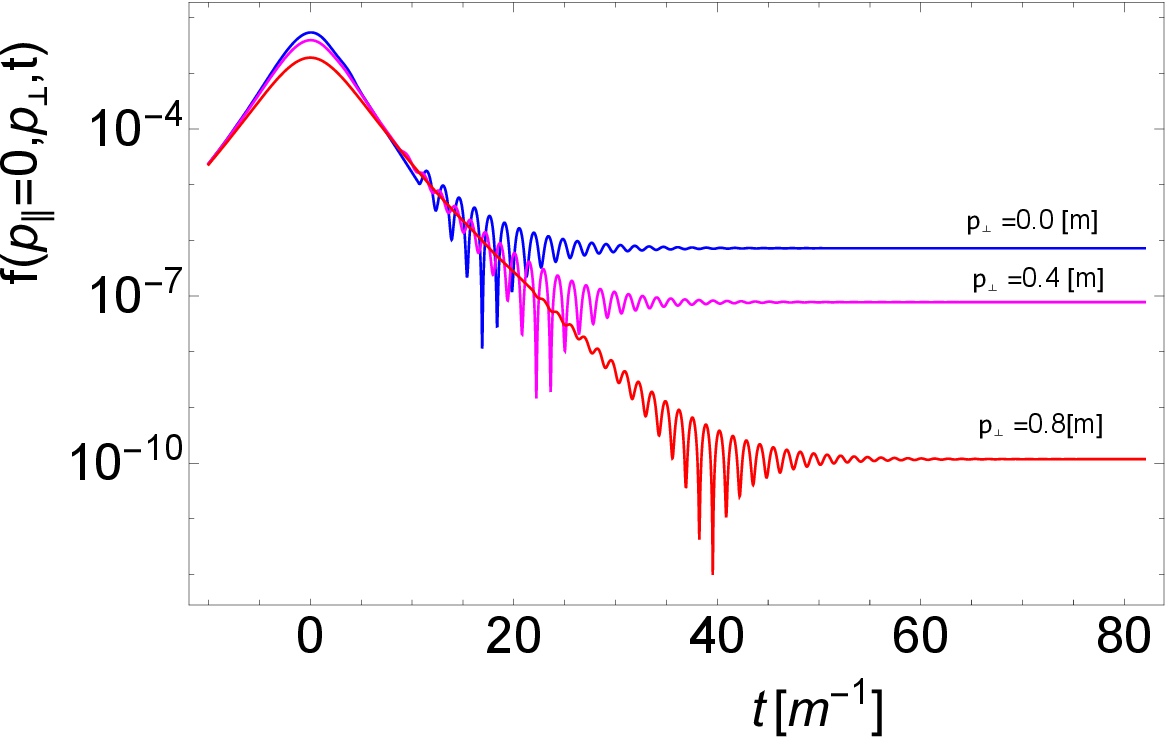}
}
\caption{The relationship between momentum and the transient domain's time of occurrence. \textbf{Left panel:} for longitudinal momentum. \textbf{Right panel:} for transverse momentum.  The field parameters are $E_0=0.2 E_c$ and $ \tau =10 [m^{-1}]$ . All units are in electron mass units.}
   	\label{pos_neg}
\end{center}
\end{figure}

Since the particle distribution function depends on momentum $p_\parallel = p_3$ parallel to the direction of the electric field and the modulus of the transverse momentum $p_\perp = \sqrt{ p_1^2 + p_2^2}.$ Due to the rotation symmetry of the problem about the field axis, one may parameterize the momentum vector as $ \bm{p} = (p_1,0,p_3)$ with the transverse component, $p_\perp = p_1$, and longitudinal component $p_\parallel = p_3.$  Furthermore, the quasienergy $\omega(\bm{p},t)$, the transverse energy $\epsilon_\perp(p_\perp)$
and the longitudinal quasi-momentum $P(p_\parallel,t)$ are defined as
\begin{align}
     \omega(\bm{p},t) &= \sqrt{\epsilon^2_\perp(p_\perp) + P^2(p_\parallel,t) },\\
     \epsilon_\perp(p_\perp) &= \sqrt{m^2 + p^2_\perp},\\
     P(p_\parallel,t) &= (p_\parallel -e A(t)).
\end{align}
%%%%
Figure~\ref{pos_neg} illustrates how the temporal evolution of the distribution function depends on the momentum value. As depicted in the left panel of Fig.~\ref{pos_neg}, the particle distribution function demonstrates a consistent increase over time within the region of QEPP. Notably, it showcases a higher value for negative longitudinal momentum than positive longitudinal momentum. On closer inspection, it becomes evident that the distribution function reaches its peak precisely when the longitudinal quasi-momentum $P(p_\parallel,t) = 0 $. This particular peak occurrence is influenced by the specific choice of $p_{\parallel}$ value. After $t=0$, where the electric field reaches its maximum value distribution function decreases. After decreasing to a certain value, it shows rapid oscillations and thus transitions from QEPP to the transient region. One observes a gradual narrowing and a disappearance of the fluctuations in the transient regime for higher $p_\parallel$ values. The transient region appears later for a higher $p_\parallel$ value. We confirm this by quantifying the period of the transient stage, $t_{in}$ and $t_{out}$ as shown in Table 1. The transient stage starts nearly after $ t \approx \tau$ (see Table 1). Afterward, it enters the REPP stage, during which the distribution function $f_{\text{out}}$ becomes constant. $f_{\text{out}}$ reaches its maximum when $p_{\parallel} = 0$ and remains the same for positive and negative values of $p_{\parallel}$.
%%%%%%%%%%%%%%%%
%%%%\newline
%%%%%

\begin{table}[h!]
\centering
\begin{adjustbox}{width=\textwidth}
\begin{tabular}{|c|c|c|c|c|}
  \hline
  $p_\parallel~[m]$ & $t_{\text{in}}~[m^{-1}]$ & $f(p_\parallel , t_{\text{in}})~(\times 10^{-7})$ & $t_{\text{out}}~[m^{-1}]$ & $f(p_\parallel , t_{\text{out}})~(\times 10^{-7})$ \\ 
  \hline
  0.00 & 12.57 & 7.203 & 24.038 & 3.309 \\
  \hline
  0.25 & 13.37 & 5.576 & 24.803 & 3.112 \\
  \hline
  0.50 & 13.78 & 2.242 & 24.320 & 1.599 \\ 
  \hline
  0.75 & 14.53 & 0.776 & 25.799 & 0.7033 \\
  \hline
  1.00 & 16.76 & 0.174 & 26.866 & 0.1082 \\
  \hline
\end{tabular}
\end{adjustbox}
\caption{Transient region times labeled by longitudinal momentum $p_\parallel$.The transverse momentum is taken to be zero, and all units are in electron mass. The field parameters are \(E_{0}=0.2E_{c}\) and \(\tau=10\) [\(m^{-1}\)].}
\label{tab:transient-times}
\end{table}

The time interval during which a strong oscillation emerges in the evolution of the distribution function is referred to as $t_{in}$, and $t_{out}$ marks the point where this vigorous oscillation diminishes, leading to a residual distribution function. 
%The transverse momentum is taken to be zero, and all units are in electron mass. The field parameters are $E_0 = 0.2 E_c$ and $\tau = 10~[m^{-1}]$.
%\end{center}
%%%%\label{table} Transient region time labeled by longitudinal momentum $p_\parallel$. The time interval during which a strong oscillation emerges in the evolution of the distribution function is referred to as $t_{in}$, and $t_{out}$ marks the point where this vigorous oscillation diminishes, leading to a residual distribution function. The transverse momentum is taken to be zero, and all units are in electron mass. The field parameters are $E_0 = 0.2 E_c$ and $\tau = 10~[m^{-1}]$.
%%%%%%%%%%%%%\newline
%\newline
\par
Next, we examine the influence of transverse momentum on temporal stages, as depicted in the right panel of Fig. \ref{pos_neg}. We observe intriguing behavior where the distribution function slowly approaches the REPP stage with a residual value $ f(p_\parallel=0, p_\perp, t > t_{out}) $, which is lowest compared to small transverse momentum values $ p_\perp $. 
The behavior of these stages is primarily determined by the double quasi-energy $2\omega(\bm{p},t)$. Increasing the transverse momentum value, and thus the transverse energy $\epsilon_\perp(p_\perp)$, increases the dynamical energy gap $2 \omega(\bm{p}, t)$. Consequently, achieving the on-shell mass configuration takes a longer time, see the right panel of Fig. \ref{pos_neg}.
%%%%%%%%%%%%%%%%%%%%%%%%%%%%%%%%%%%%%%%%%%%%%%%%%%%%%%%%%%%%%%%%%%%%%%%%%%%%%%%%%%%%%%%%%%%%%%%%%%%%%%%%%%%%%%%%%%%%%%%%%%%%%%%%%%%%%%%%%%%%%%%%%%%%%%%%%%%%%%%%%%%%%%%%%%%%%%%%%%%%%%%%%%%%%%%%%%%%%%%%%%%%%%%%%%%%%%%%%%%%%%%%%%%%%%%%%%%%%%%%%%%%%%%%%%%%%%%%%%%%%%%%%%%%%%%%%%%%%%%%%%%%%%%%%%%%%%%%%%%%%%%%%%%%%%%%%%%%%%%%%%%%%%%%%%%%%%%%%%%%%%%%%%
\subsection{Longitudinal Momentum Spectrum}
\label{sec:Longitudinal momentum}

In this subsection, we discuss the longitudinal momentum spectrum using the distribution function given in Eq.~\eqref{59.1}, with \( p_\perp = 0 \). The longitudinal momentum distribution function of the produced pairs at any time is denoted as \( f(p_\parallel, t) \).

%In this subsection, we discuss the momentum distribution function as defined in Eq.~\eqref{59.1}. Specifically, we focus on the longitudinal momentum spectra by setting $p_\perp = 0$ and denote the longitudinal momentum distribution function of the produced pairs at any time as $f(p_\parallel, t)$.\newline
\begin{figure}[t]
\begin{center}
{
\includegraphics[width = 2.015in]{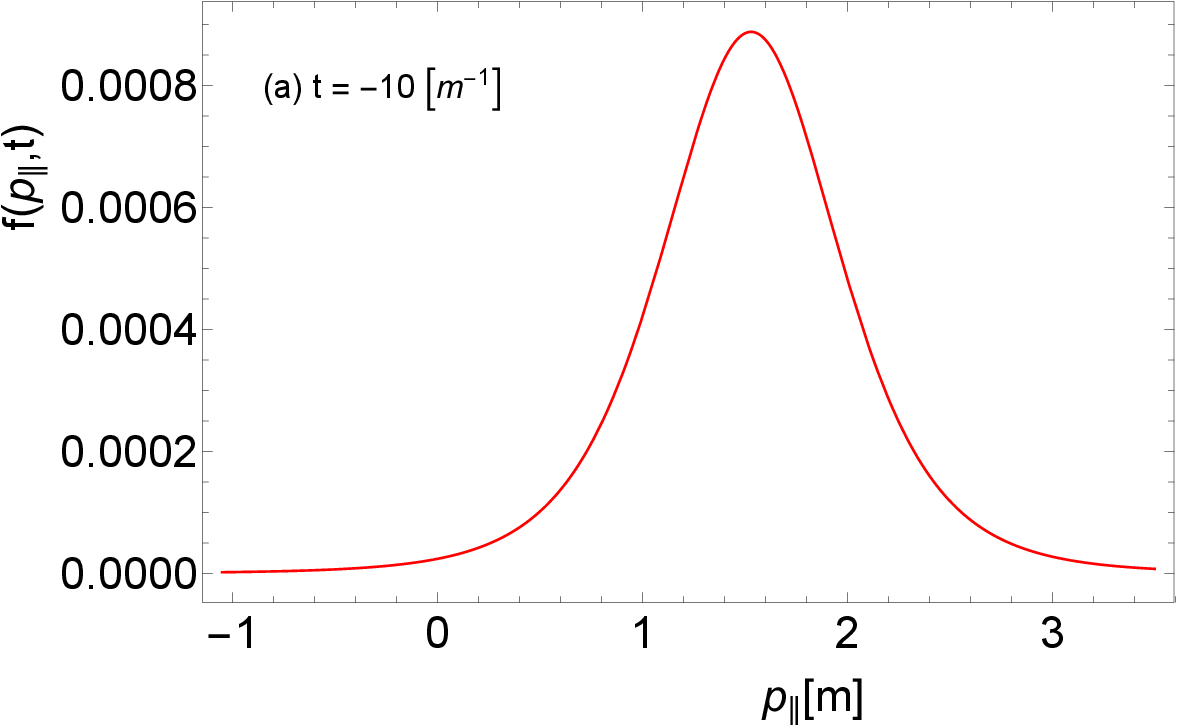}
\includegraphics[width = 2.015in]{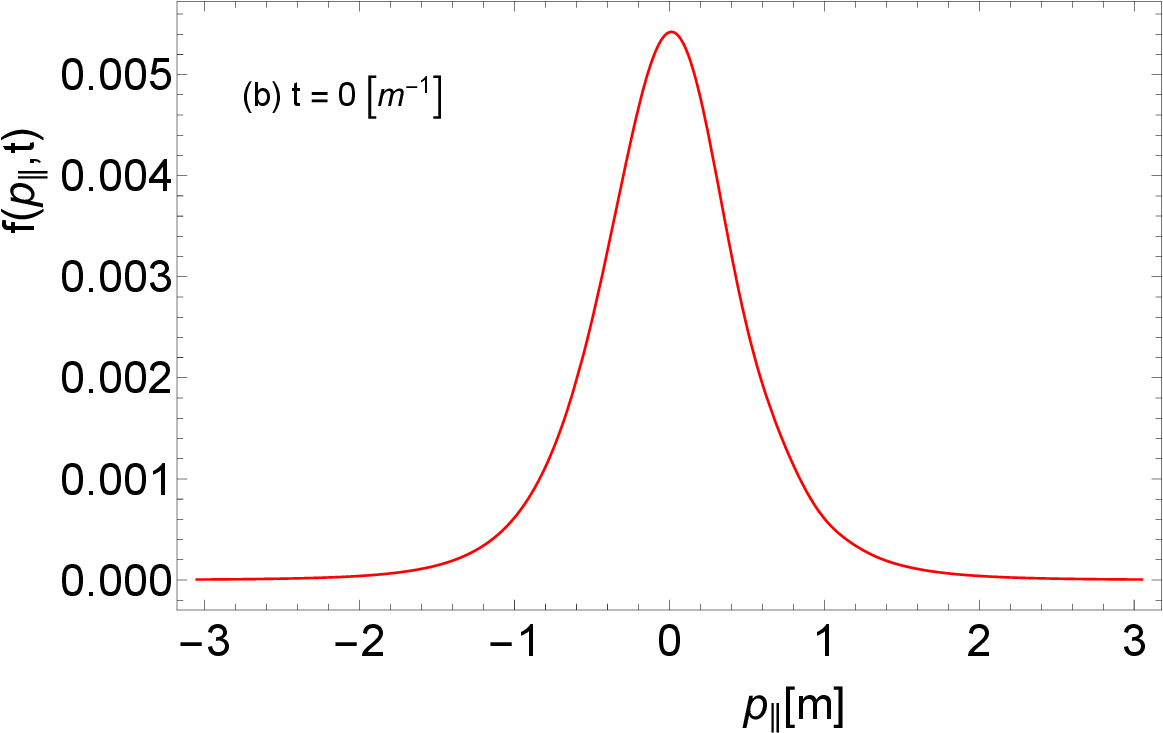}
\includegraphics[width = 2.015in]{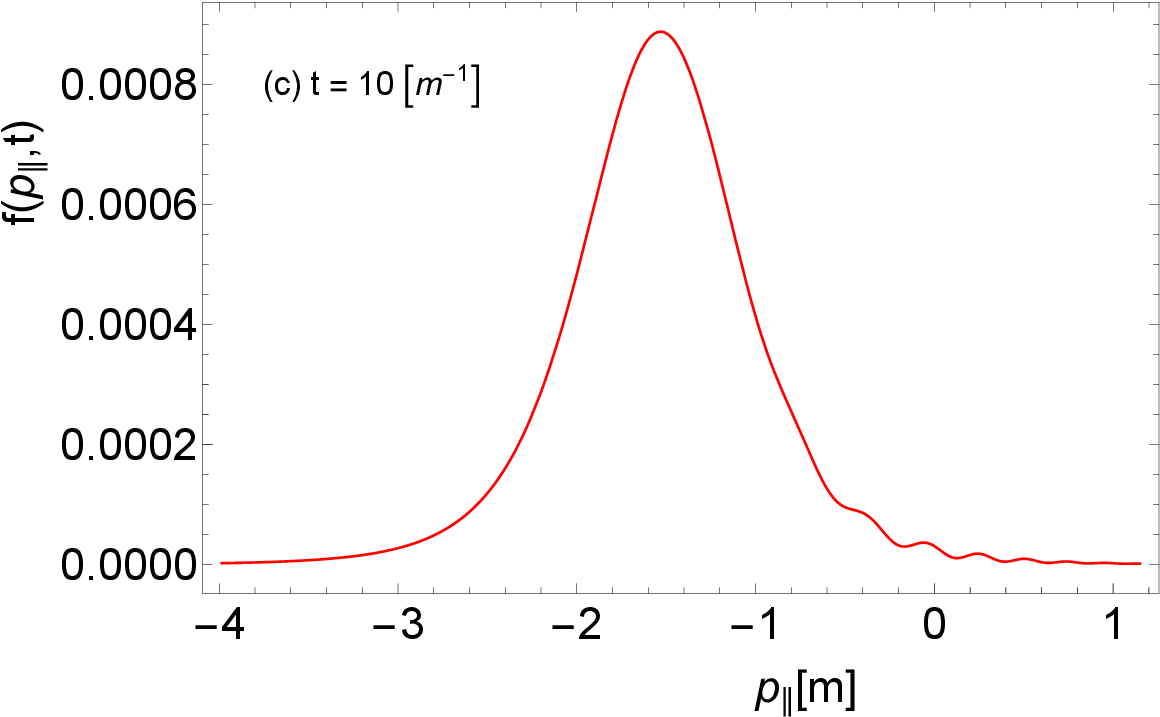}
\includegraphics[width = 2.015in]{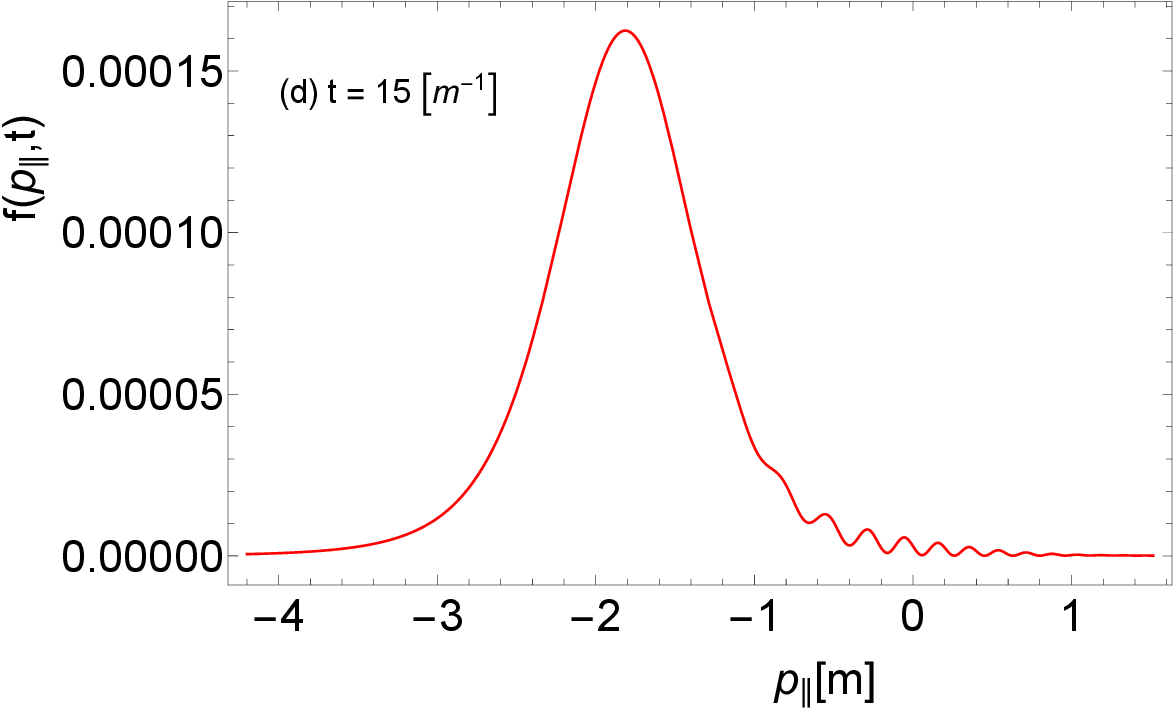}
\includegraphics[width = 2.015in]{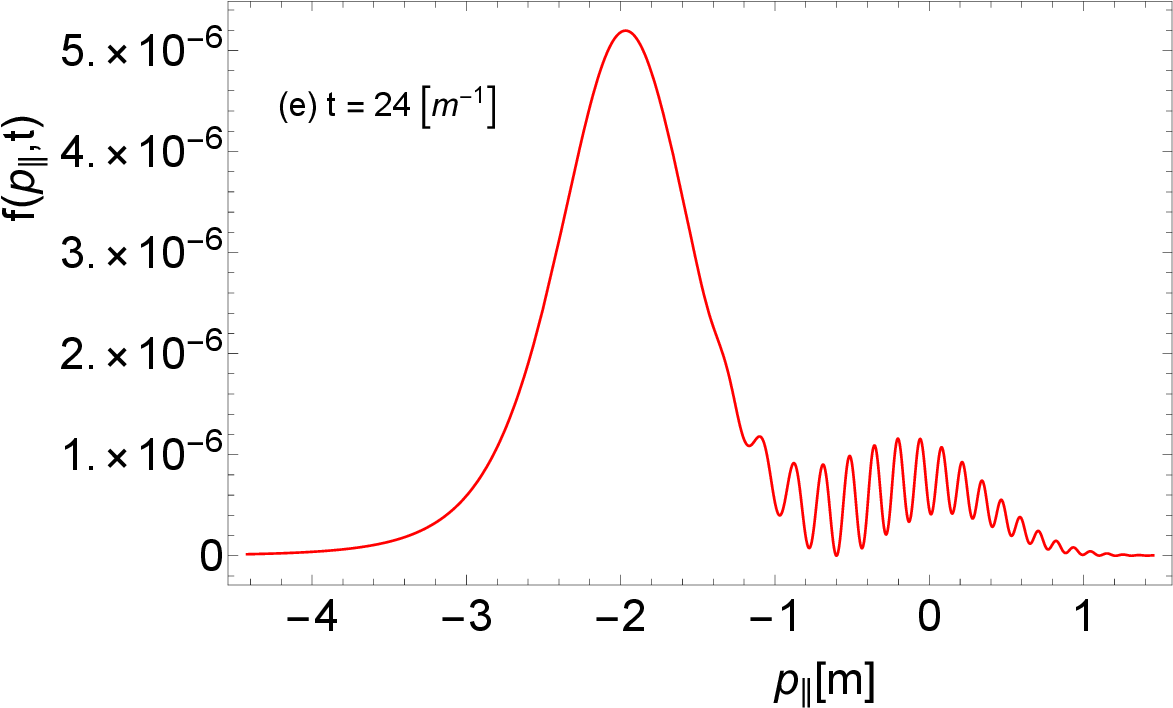}
\includegraphics[width = 2.015in]{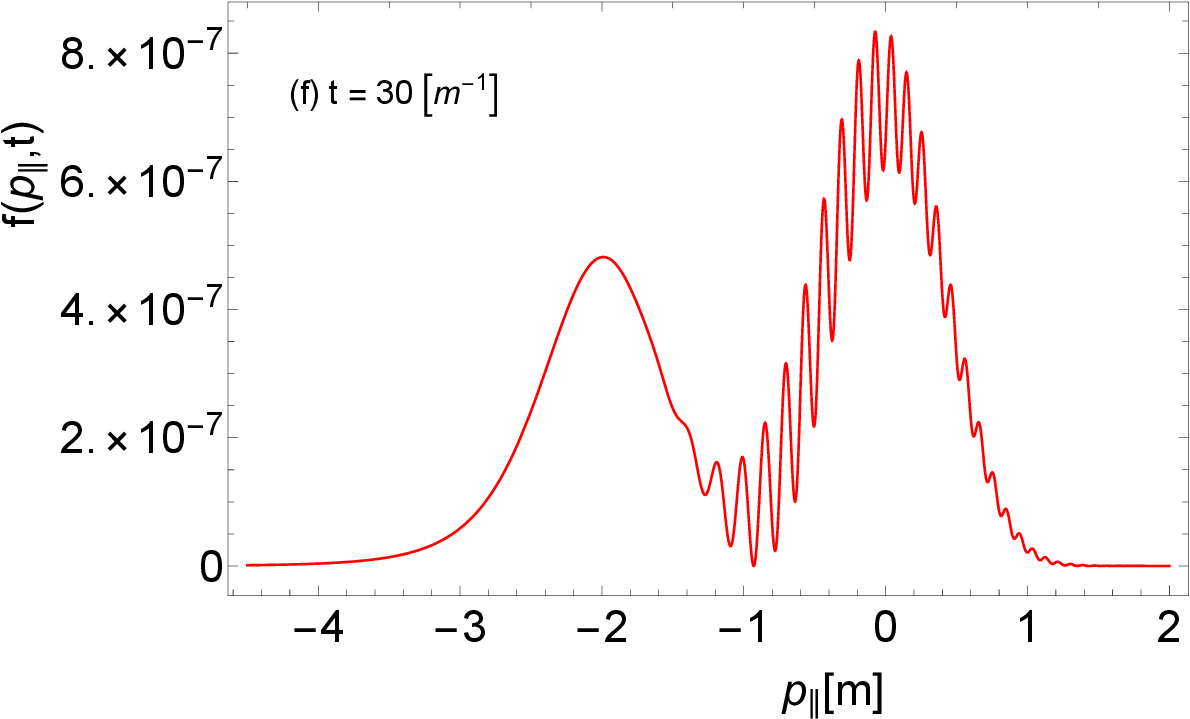}
\includegraphics[width = 2.015in]{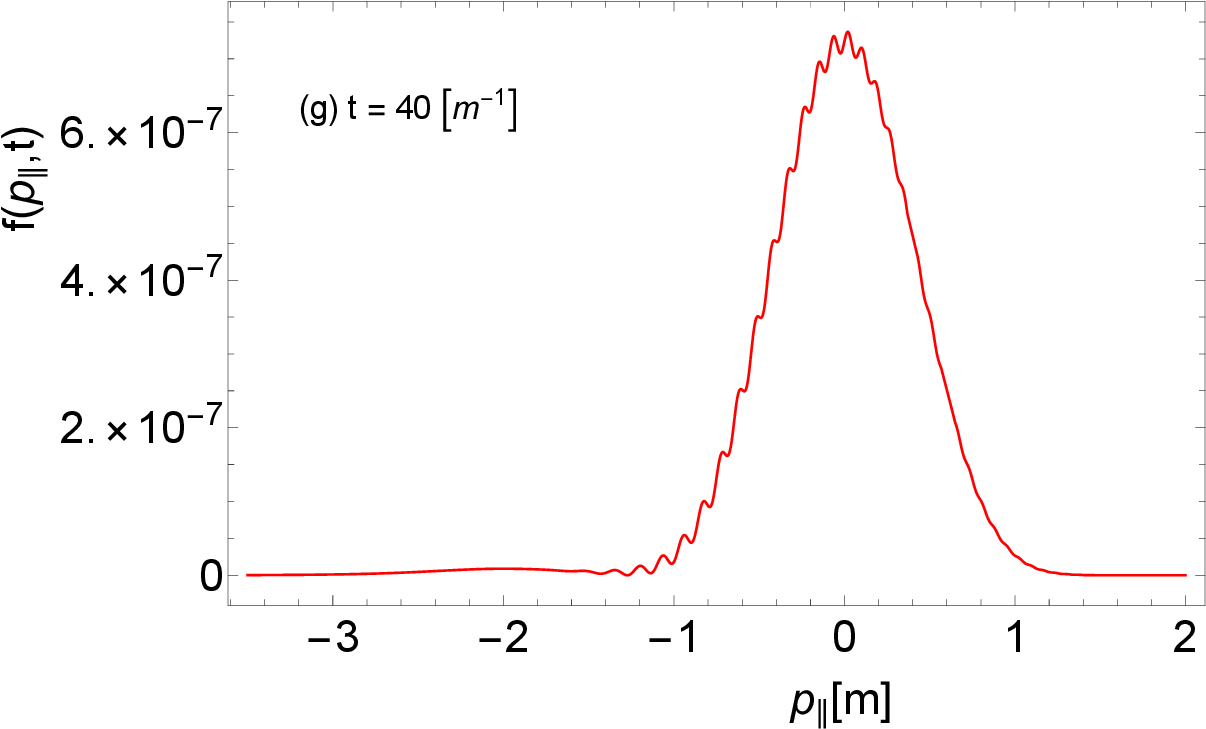}
\includegraphics[width = 2.015in]{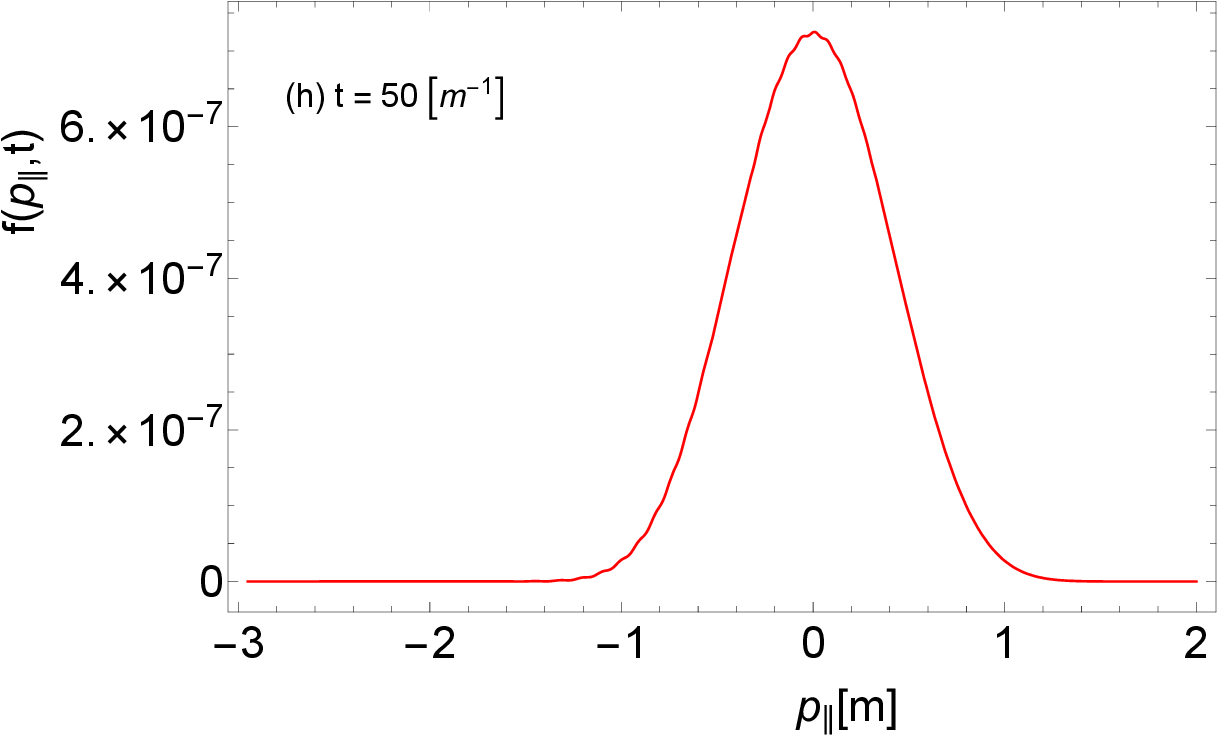}
\includegraphics[width = 2.015in]{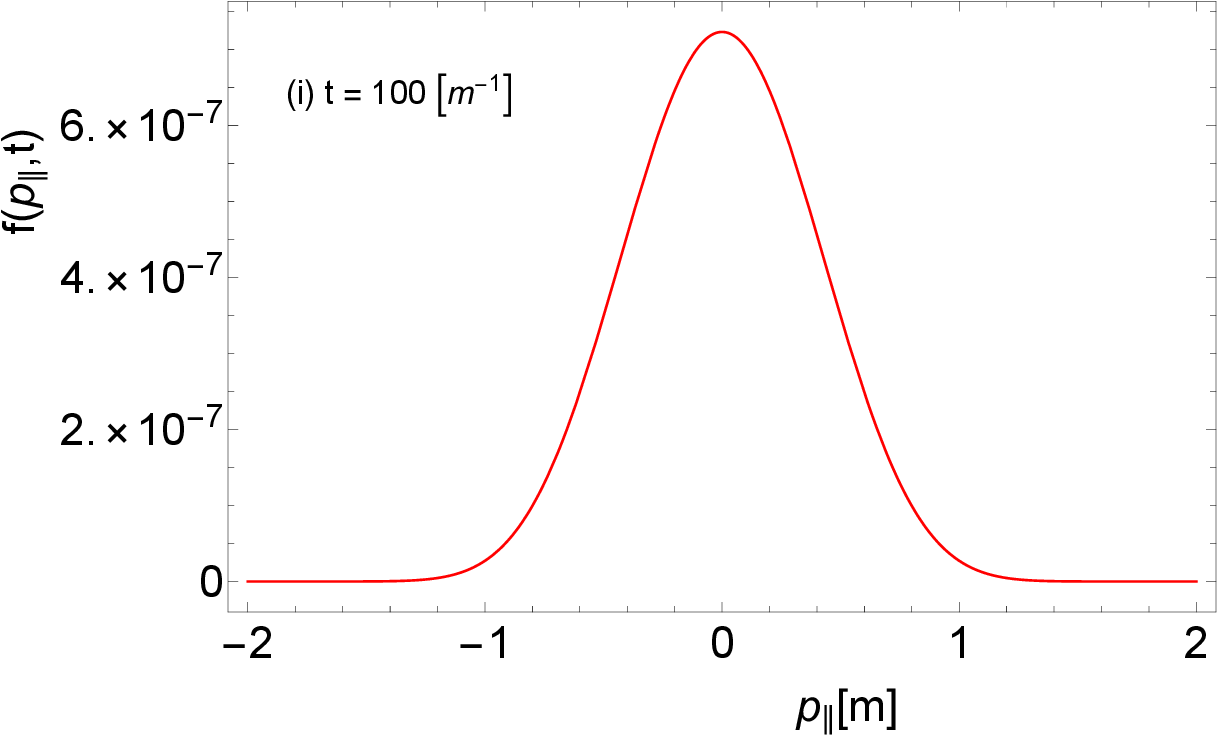}
}
\caption{LMS of created particles at different times. The value of transverse momentum is considered to be zero, and all units are in electron mass units. The field parameters are  $E_0=0.2 E_c$ and $ \tau =10 [m^{-1}].$}
   	\label{fig:6.1}
\end{center}
\end{figure}
%%%%%%%
From Figure \ref{fig:6.1}, we see how particle creation proceeds. At early times, $t = -10 [m^{-1}]$  in the QEPP region, when the electric field is increasing, the created particle shows a smooth unimodal-like structure of momentum spectrum with a peak at $p_\parallel \sim 2 [m].$ 
The electric field propels the newly created particles towards the negative $z$-direction. The movement of the distribution function's peak from $ p_\parallel =2$ $( ~ e E_0 \tau) $ to $-2$ $( ~ - e E_0 \tau) $ can be understood through the concept of longitudinal quasi-momentum, $P(p_\parallel,t)$ as depicted in the time evolution shown in Figs. \ref{fig:6.1}(a) to (c). This momentum shift implies that the longitudinal momentum distribution of the generated particles is expected to span a range of $\Delta p_\parallel = 2$, with the spread determined by the electric field parameters.
At $ t=0$ (electric field is maximum), the momentum distribution function, $f(p_\parallel=0,t =0)   \approx 5 \times 10^{-3}$, follows the tendency $ E^2(t)/8$ \cite{Schmidt:1998zh}. 
For $t > 0$, a reduction in the field strength with a decrease in an intriguing longitudinal momentum distribution function. It rapidly drops to $99.2\%$ of its maximum value, driven by its reliance on the strength of the electric field at $t = 10 [m^{-1}]$, as illustrated in Fig.~\ref{fig:6.1}(c). Within a narrow range of $-1 < p_\parallel < 1$, the typically smooth unimodal profile of the longitudinal momentum spectra is disrupted, as depicted in Fig.~\ref{fig:6.1}(d).
\par
Near \( t \approx 2 \tau \), when the electric field's magnitude decreases to approximately $93\%$ of its maximum, a secondary peak appears in the LMS. This peak emerges around $p_\parallel = 0$ and is accompanied by oscillations within a confined range of $p_\parallel$, as shown in Fig. \ref{fig:6.1}(e). Over time, as the electric field weakens, this secondary peak grows near $p_\parallel = 0$, while simultaneously, the dominant peak begins to diminish.  
At  $t = 3\tau $, the small peak becomes dominant and exhibits oscillatory behavior near the beginning of the REPP stage, as shown in Fig.~\ref{fig:6.1}(f). In this stage, oscillations occur within a narrow range of longitudinal momentum as the electric field decreases to approximately one-hundredth of its maximum magnitude. Intriguingly, these oscillations exhibit asymmetry, with their amplitude more pronounced for negative longitudinal momentum compared to positive longitudinal momentum. Within Fig.~\ref{fig:6.1}(f), the Gaussian bump centered around $p_\parallel \approx -2 [m]$ arises from particles generated during the early stages of the process. Conversely, the dominant peak, characterized by the onset of oscillations, comprises particles formed at later instances, which have encountered relatively less acceleration since their inception.
\par
The oscillations in the LMS originate from quantum interference effects, which arise due to the interplay of dynamical processes (or channels) leading to the creation of particles with a given momentum. These time-evolving processes involve inter-band dynamics within the particle momentum representation. To understand this, one can consider multiple channels within momentum space that facilitate particle tunneling \cite{Sah:2024qcs}. As time progresses, distinct scenarios emerge:  
(i) At time \( t \), a particle can tunnel directly, acquiring a momentum \( p_\parallel' \).  
(ii) In the early stages, a particle with lower momentum (\( p_\parallel < p_\parallel' \)) can tunnel first and subsequently accelerate to reach \( p_\parallel' \).  
(iii) Conversely, a particle initially possessing higher momentum (\( p_\parallel > p_\parallel' \)) can tunnel at time \( t_1 \) and then decelerate to reach \( p_\parallel' \).  When the probability amplitudes of these processes are summed, they exhibit quantum interference effects at some time \( t \). 
%However, at asymptotic times, phase coherence is lost due to averaging over particle paths, resulting in a smooth unimodal structure in the LMS, devoid of quantum interference effects.%%%as elucidated in reference \cite{Keldysh2016}.
\par
Around $ t \approx 4 \tau $, a minor peak at $ p_\parallel = -2[m] $ is nearly diminished. Only the dominant peak at $ p_\parallel = 0 $ persists, with a faint onset of oscillatory behavior superimposed on a Gaussian-like structure, as depicted in Figs. \ref{fig:6.1}(g) to (h). The oscillations gradually fade away by $ t = 50 [m^{-1}] $.
\par
In the asymptotic region $(t = 10 \tau),$ the LMS lacks oscillatory structure for a Sauter pulse, instead forming a single-peaked unimodal structure were reported in  \cite{Banerjee:2018azr, Dumlu:2011rr}. This holds true at asymptotic times (\( t \to \infty \)), consistent with Fig.~\ref{fig:6.1}(i).
%%%%%%%%%%%
\par
To the best of our knowledge, the evolution of LMS has only been discussed in ref.~\cite{Banerjee:2018azr}. However, in Ref. \cite{Banerjee:2018azr}, the pulse duration of the Sauter pulse was significantly longer than what is considered here. Due to this larger pulse duration, the quantum interference effects on the momentum spectrum at finite times were not evident in their study. These interference effects were explicitly analyzed only in a recent paper by Diez et al.\cite{Diez:2022ywi}.
In Ref.~\cite{Diez:2022ywi}, the authors identified the formation time scale based on quantum interference signatures in the spectra for the field parameters \( E_0 = 0.4 E_c \) and \( \tau = 40 [m^{-1}] \). However, we present results on the dynamical aspects of pair production, providing a broader perspective on the time evolution of the process. Additionally, we discuss the phenomenon in the multiphoton region, which is further explored later in \cref{sec:LMS multi-photon}.

%%%%%%%%%%%%%%%%%%%%%%%%%%
%%%%%%%%%%%%Furthermore, we can mark some time scale related to the quantum signature that appears in the form of oscillation in LMS of created particles in the REPP stage. Based on the occurrence of a secondary peak, we quantify a three-time scale $(i)$ $t_{cp}$, presence of the secondary peak or time after that central peak build-up, $(ii)$ $t_{sep},$ Dominated peak represented by central peak or time corresponding after that two peaks are clearly separated, and $(ii)$ $t_{dis},$ Oscillation in central peak disappears or time after that primary peak (or left side peak) vanished.
%%%
\par
Certainly, we can identify specific time scales associated with the quantum signature that manifest as oscillations within the LMS of created pairs. Drawing from the emergence of a secondary peak, we define three distinct time scales: $(i)$ $t_{cp}$ (central peak formation): This time scale is characterized by the emergence of the secondary peak and the beginning of its development, $(ii)$ $t_{sep}$ (peak separation): At this time, the central peak becomes dominant or the time after the two peaks become distinctly separated,  $(iii)$ $t_{dis}$ (disappearance of oscillation): This indicates the time when the oscillations within the central peak fade away or the time after the primary peak (or the left-side peak) ceases to exist. By identifying and quantifying these time scales, we can better characterize the intricate quantum dynamics reflected in the LMS during the REPP stage.
%%%%%%%%%%%%%%%%%%%%%%%%
\begin{table}[h!]
\centering
\begin{tabular}{ |c|c|c|c| } 
  \hline
  $E_0~[E_{c}]$ & $t_{cp}~[m^{-1}]$ & $t_{sep}~[m^{-1}]$ & $t_{dis.}~[m^{-1}]$ \\ 
  \hline
   0.1 & 47 & 57 & 70  \\
  \hline
   0.2 & 22 & 30 & 50  \\
  \hline
   0.3 & 15 & 20 & 35  \\
  \hline
   0.4 & 9  & 15 & 28  \\
  \hline
   0.5 & 6  & 12 & 23  \\
  \hline
\end{tabular}
\caption{\label{table2} Effect of electric field strength $E_0$ on the time scales related to the formation of a central peak. 
$t_{cp}$ denotes the appearance of a central peak, $t_{sep}$ the time when two peaks become distinctly separated, and $t_{dis.}$ the disappearance of the oscillation in LMS.}
\end{table}

%%%
The time scales governing the process are influenced by the electric field strength $E_0$, as shown in Table II. A clear pattern emerges—these time scales consistently follow a specific trend as $E_0$ increases. Notably, for a fixed pulse duration \( \tau \), key events occur earlier in time with increasing $E_0$. 
%In general, these time scales can be quantified in terms of the electric field parameters, specifically $E_0$ and $\tau.$ Although a detailed analysis of time scales is beyond the scope of this paper, they can be expressed as:  
%%%%%%%%%\begin{equation} t_{cp} =  0.1698 \, [m^{3.1}] \tau^{1.102} {E_0}^{-1.478}, \\ t_{sep} = 0.456 \, [m^{2.2}]\tau^{1.065} {E_0}^{-1.059}, \\ t_{dis} = 1.831 \, [m^{1.6}]\tau^{0.871} {E_0}^{-0.830}.\end{equation}  These relations highlight the nonlinear dependence of each timescale on both the pulse duration and the electric field strength.  Our analysis is specifically applicable to Sauter pulsed electric fields within the range \( 5 < \tau < 30 \) and \( 0.1 < E_0 < 0.5 \).  The time scales found in ref.~\cite{Diez:2022ywi} differ from the scaling we have identified in our analysis. 
%%%%%%%%%%%%%%%%%%%%%%%%%%%%%%%%%%%%%%%%%%%%%%%%%%%%%%
\par
\begin{figure}[t]
\begin{center}
{
\includegraphics[width = 2.015in]{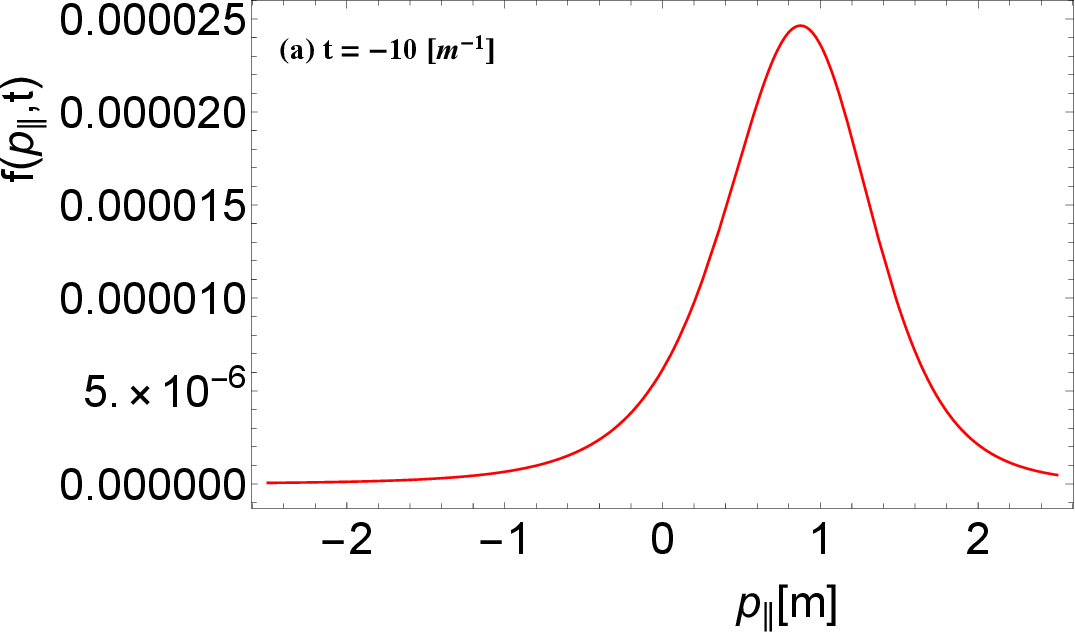}
\includegraphics[width = 2.015in]{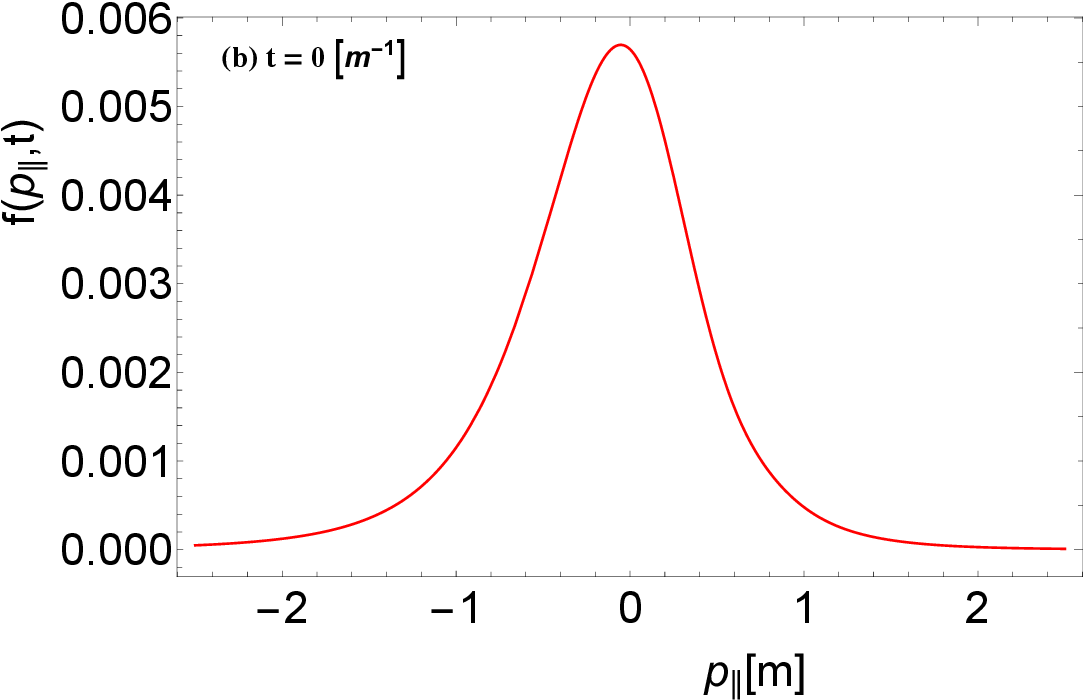}
\includegraphics[width = 2.015in]{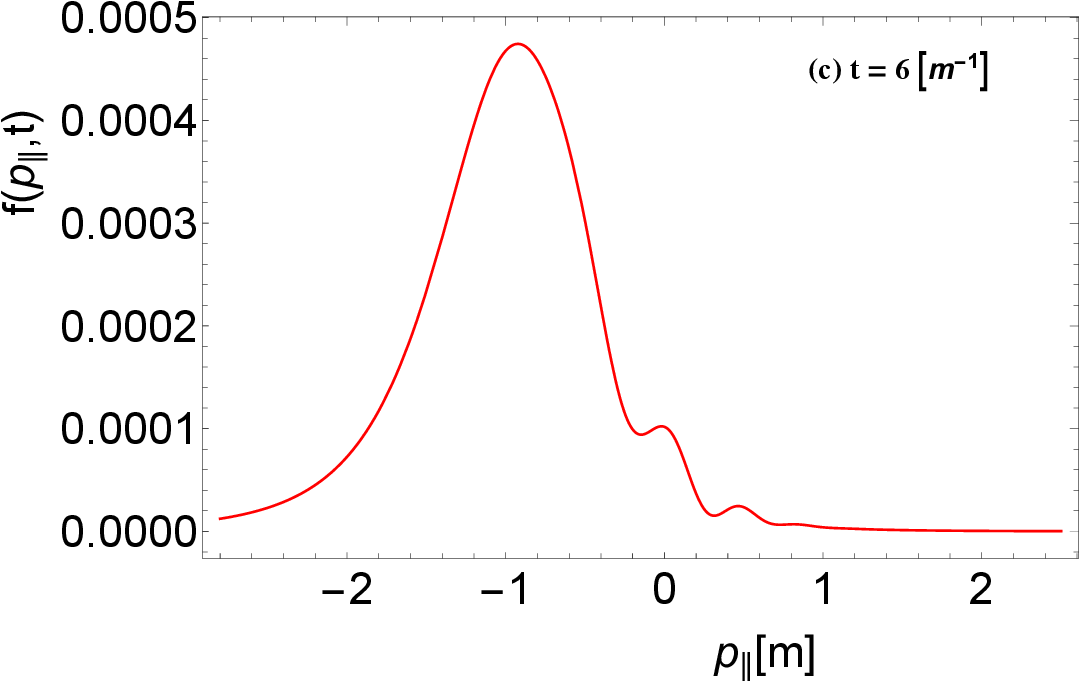}
\includegraphics[width = 2.015in]{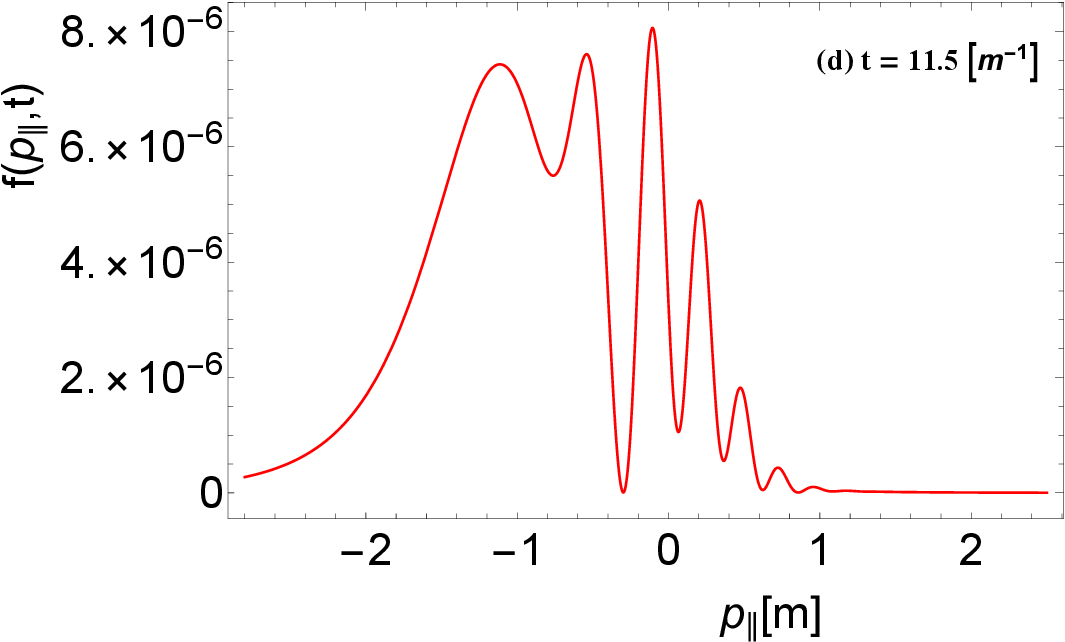}
\includegraphics[width = 2.015in]{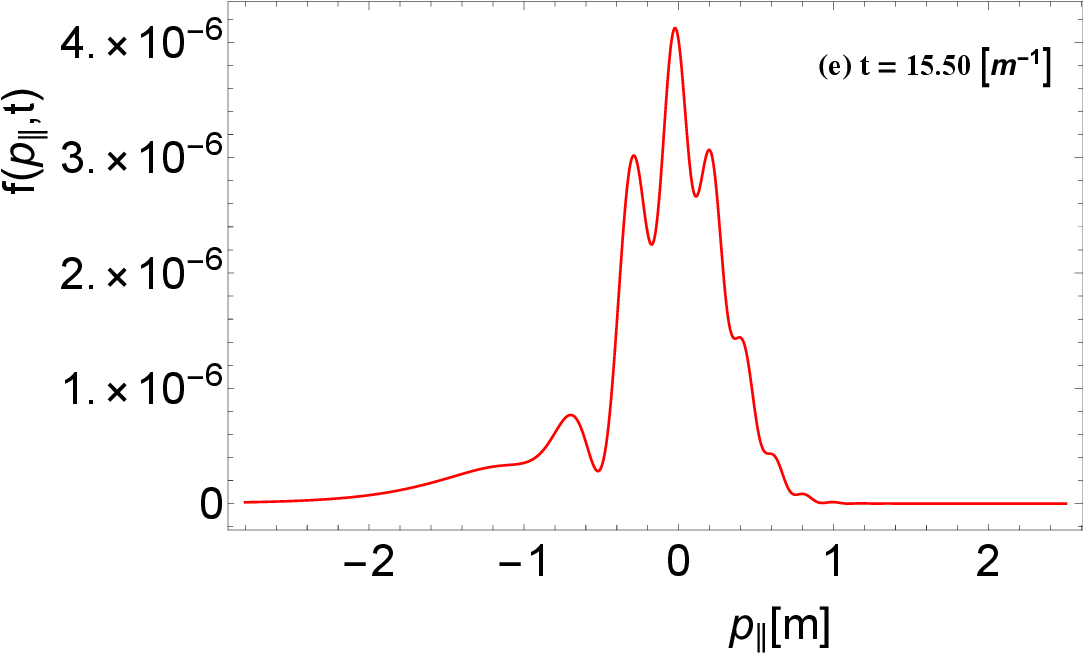}
\includegraphics[width = 2.015in]{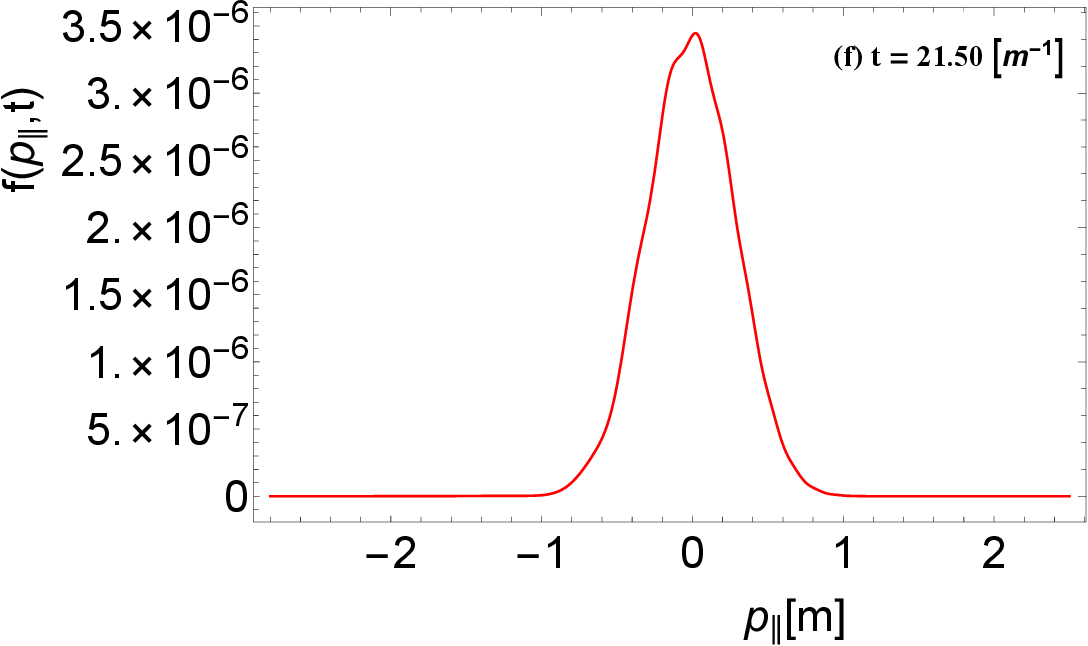}
}
\caption{LMS of created particles at different times. The value of transverse momentum is considered to be zero, and all units are in the electron mass unit.The field parameters are  $E_0=0.2 E_c$ and $ \tau = 5 [m^{-1}].$}
   	\label{fig:6.2}
\end{center}
\end{figure}
\begin{figure}[t]
\begin{center}
{
\includegraphics[width = 2.015in]{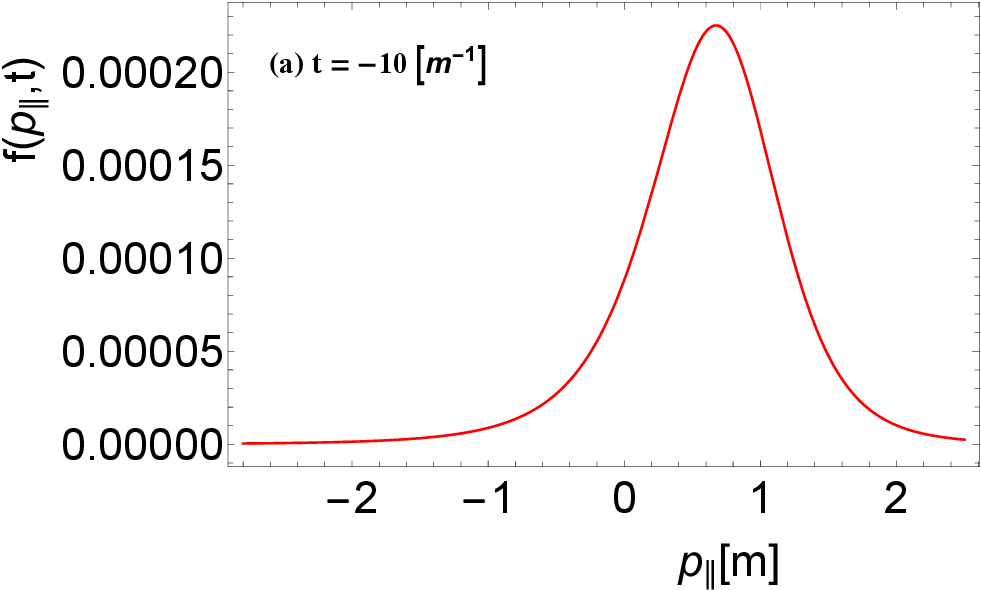}
\includegraphics[width = 2.015in]{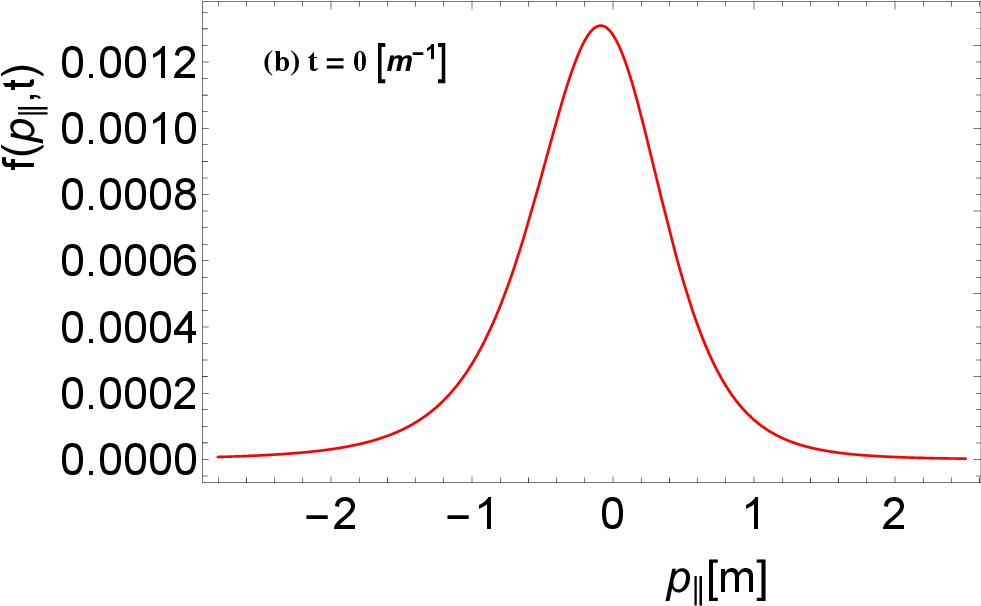}
\includegraphics[width = 2.015in]{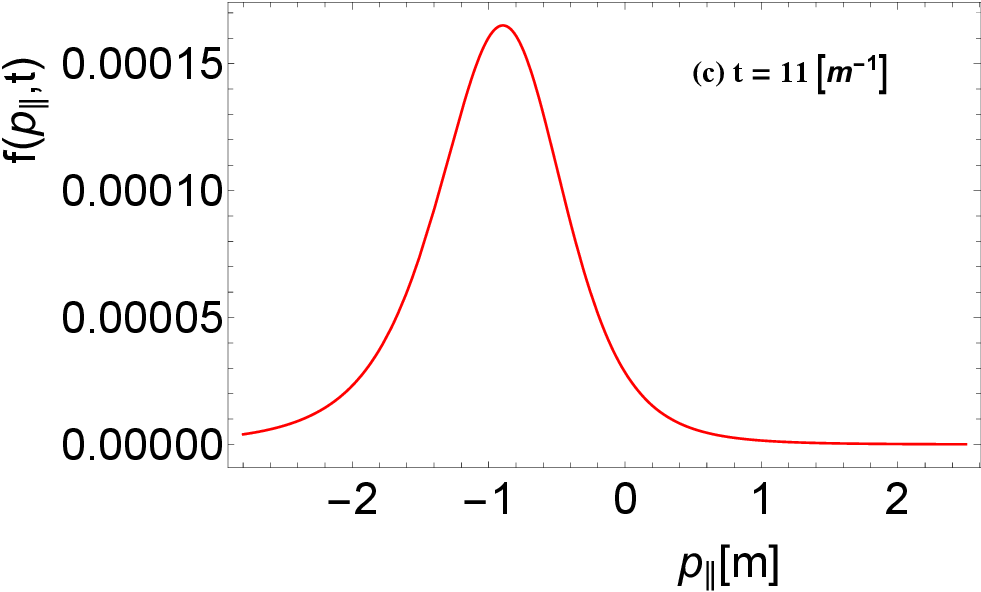}
\includegraphics[width = 2.015in]{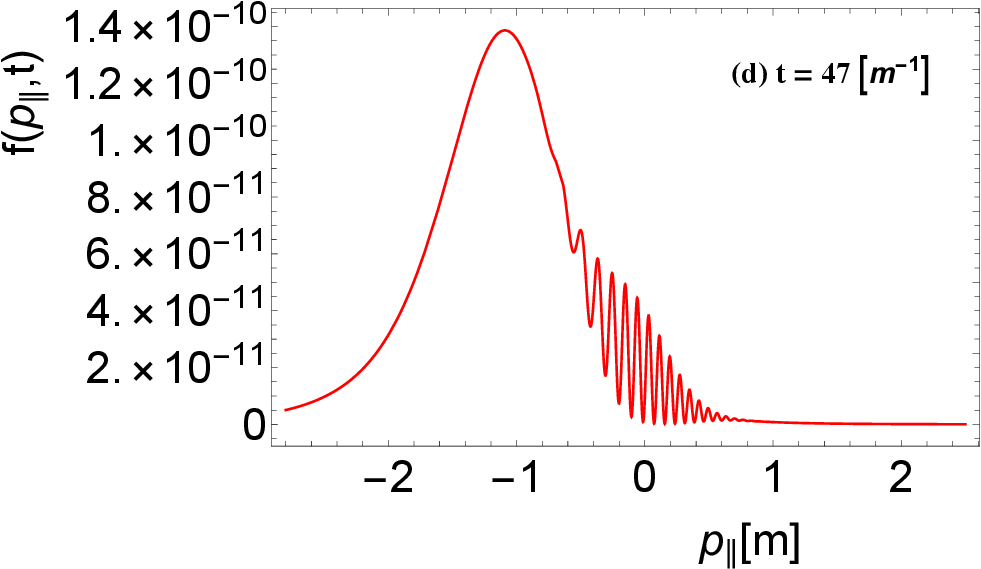}
\includegraphics[width = 2.015in]{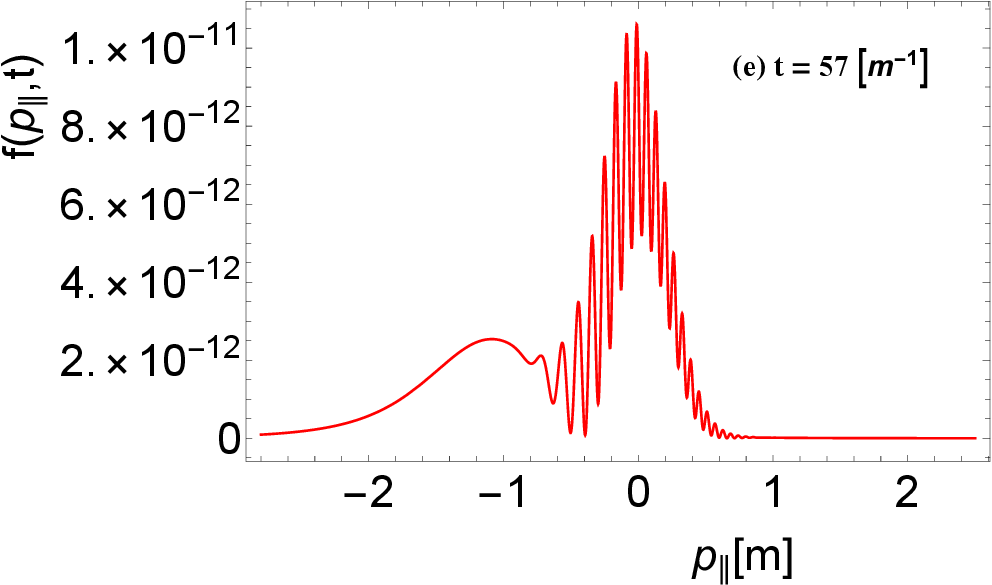}
\includegraphics[width = 2.015in]{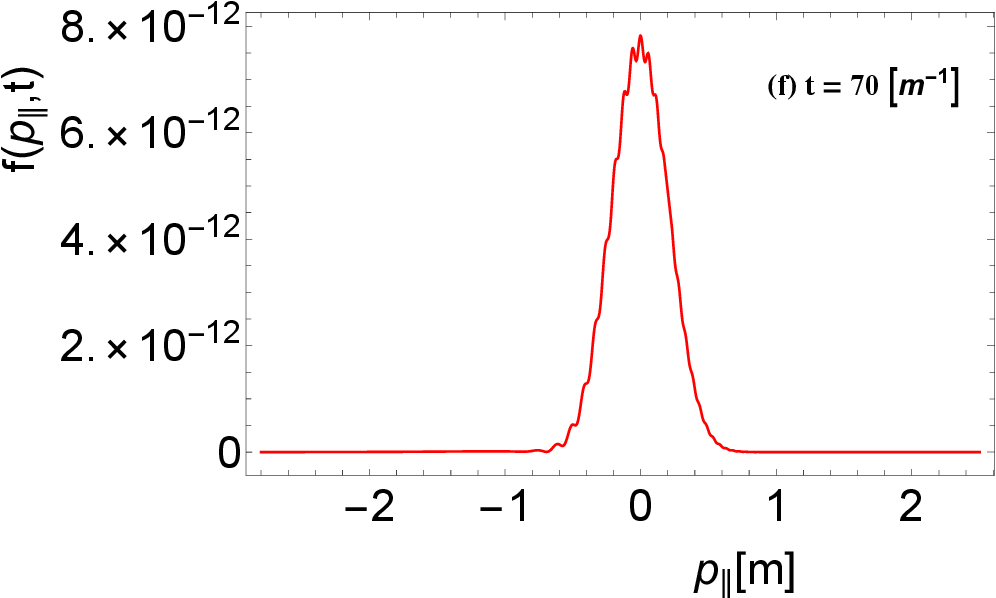}
}
\caption{LMS of created particles at different times. The value of transverse momentum is considered to be zero, and all units are in the electron mass unit.The field parameters are  $E_0=0.1 E_c$ and $ \tau = 10 [m^{-1}].$}
   	\label{fig:6.3}
\end{center}
\end{figure}

We have not yet discussed the Keldysh parameter, \( \gamma \), which classifies the pair production process as either a multi-photon or tunneling mechanism. Additionally, we are investigating whether the oscillatory behavior explicitly depends on \( \gamma \). For this study, we select two different combinations of \( E_0 \) and \( \tau \) to vary \( \gamma \) and examine its behavior over a finite time interval. Specifically, we set \( \gamma = 1 \), corresponding to two distinct parameter configurations \( (E_0, \tau) \) associated with the Sauter pulse electric field. It is known that \( \gamma = 1 \) represents an intermediate regime involving both tunneling and multi-photon processes, which are inherently non-perturbative \cite{Avetissian:2002ucr}. This approach allows us to demonstrate how the same \( \gamma \) value can lead to qualitatively different outcomes in the time evolution of the pair production process.  
Figure \ref{fig:6.2} shows LMS for the Sauter pulse electric field with $ E_0 = 0.2 E_c $ and $ \tau = 5 \, [m^{-1}] $, while Fig.\ref{fig:6.3} shows LMS for  field parameter  $ E_0 = 0.1 E_c $ and $ \tau = 10 \, [m^{-1}] $.
From Figs.~\ref{fig:6.2} (a) to (c) we can see that due to the electric field smooth unimodal peaked profile accelerated towards negative $z-$ direction and after the pulse duration $t = 6 [m^{-1}] $ we observed smooth profile becomes deformed near the region  $-0.5<p_\parallel< 0.5.$ At $t = 11.5 [m^{-1}]$, the interference effect is observed, due to which the smooth unimodal peak structure becomes a multi-modal structure as shown in Figs.\ref{fig:6.2}(d) and \ref{fig:6.2}(e). That interference effect nearly disappears at $ t  \approx 4 \tau.$ 
\par
Comparing Figs. \ref{fig:6.2}(d), (e) and \ref{fig:6.3}(d), (e), clear structural differences emerge. These differences give rise to two qualitatively distinct behaviors at finite times under the condition \(\gamma = 1\). Notably, the primary factor influencing this behavior at finite times is the pulse duration $\tau$.
%\par Depending on the classification of short and long pulses, we can outline two distinct behaviors: (i) Long Pulse Behavior ($\tau > 8$): In this regime, a bi-modal peaked structure is seen in  LMS. One of these profiles forms at the origin, potentially with the onset of oscillations.(ii) Short Pulse Behavior ($\tau < 8$): In contrast, shorter pulses lead to a unique outcome. Initially, a smooth unimodal structure experiences a splitting near the REPP region. This leads to the formation of a multi-modal Gaussian structure, which eventually converges into a single-peaked Gaussian profile as time progresses towards the asymptotic state.
 \newline
%%%%%%%%%%%%%%%%%%%%%%%%%%%%%%%%%%%%%%%%%%%%%%%%%%%%%%%%%%%%%%%%%%%%%%%%%%%%%%%%%%%%%%%%%%%%%%%%%%%%%%%%%%%%%%%%%%%%%%%%%%%%%%%%%%%%%%%%%%%%%%%%%%%%%%%%%%%%%%%%%%%%%%%%%%%%%%%%%%%%%%%%%%%%%%%%%%%%%%%%%%%%%%%%%%%%%%%%%%%%%%%%%%%%%%%%%%%%%%%%%%%%%%%%%%%%%%%%%%%%%%%%%%%%%%%%%%%%%%%%%%%%%%%%%%%%%%%%%%%%%%%%%%%%%%%%%%%%%%%%%%%%%%%%%%%%%%%%%%%%%%%%%%%%%%%%%%%%%%%%%%%%%%%%%%%%%%%%%%%%%%%%%%%%%%%%%%%%%%%%%%%%%%%%%%%%%%%%%%%%%%%%%%%%%%%%%%%%%%%%%%%%%%%%%%%%%%%%%%%%%%%%%%%%%%%%%%%%%%%%%%%%%%%%%%%%%%%%%%%%%%%%%%%%%%%%%%%%%%%%%%%%%%%%%%%%%%%%%%%%%%%%%%%%%%%%%%%%%%%%%%%%%%%%%%%%%%%%%%%%%%%%%%%%%%%%%%%%%%%%%%%%%%%%%%%%%%%%%%%%%%%%%%%%%%%%%%%%%%%%%%%%%%%%%%%%%%%%%%%%%%%%%%%%
\subsubsection{Approximate expression for longitudinal momentum distribution function }
\label{ApprLMSresult}
In this subsection, we aim to analyze the oscillatory behavior of the LMS in the late-time limit, specifically for $ t > 2 \tau $. We utilize the previously derived approximate analytical expression for the distribution function Eq.\eqref{appdisfun}. By setting $ p_\perp = 0 $, we derive an approximate expression that depends solely on the longitudinal momentum and the transformed time variable $y$. The longitudinal momentum distribution function $f(p_\parallel,y)$ is then expressed as a series expansion in powers of $(1-y)$ up to the second order and neglecting higher-order terms:
%%%
\begin{align}
   f(p_\parallel,y) &=|N^{(+)} (p_\parallel)|^2 \Bigl( \mathrm{C}_0 (p_\parallel,y) +  (1-y) \mathrm{C}_1 (p_\parallel,y)   + (1-y)^2 \mathrm{C}_2(p_\parallel,y) \Bigr).
\label{appPDFlong}
\end{align} 
\par
Following Eqs.~\eqref{pdfC0}, \eqref{pdfC1}, and \eqref{pdfC2}, it is evident that for $ t \gg 2 \tau $, the coefficients of $(1-y)$ and $(1-y)^2$ can be further approximated by retaining only the dominant contributions and disregarding the others, as follows:
\begin{align}
     \mathrm{C}_0 (p_\parallel,y) & \approx   4 \omega_1 y^2  ( \omega_1+ P_1 ) |\Gamma_1|^2,
      \label{apppdfc0}
\end{align}
\begin{align}
  \mathrm{C}_1 (p_\parallel,y)&\approx -4 y|\Gamma_1 \Gamma^{*}_2|\frac{ E_0 \tau }{\omega_1} \cos{(\Upsilon)},
  \label{apppdfc11}
  \end{align}
 %%%%%
 \begin{align}
     \mathrm{C}_2 (p_\parallel,y) &\approx  \frac{4 E_0^2 \tau^2 (\omega_1 -P_1)}{\omega_1^2}|\Gamma_1|^2 + 4 |\Gamma_1 \Gamma^{*}_2|  \cos{(\Upsilon)} \Biggl[  \omega_0 ( 1  +  3 P_1 (\omega_1 + P_1)) - \omega_1 ( 3 P_1(1+\omega_1) +1) 
      \nonumber \\ &
     +  3 E_0 \tau ((\omega_1 + P_1)^2 -1 )
     +  2 E_0 \tau^2\frac{  (\tau \omega_1^4 - 2 E_0 P_1 (4 + 3 \tau^2 \omega_1^2))}{\omega_1^3 ( 4 + \tau^2 \omega_1^2)}  \Biggr].
    \label{apppdfc22}
\end{align}
%%%%%%%%%%%%%%%
\begin{figure}[t]
\begin{center}
{
\includegraphics[width = 1.52015in]{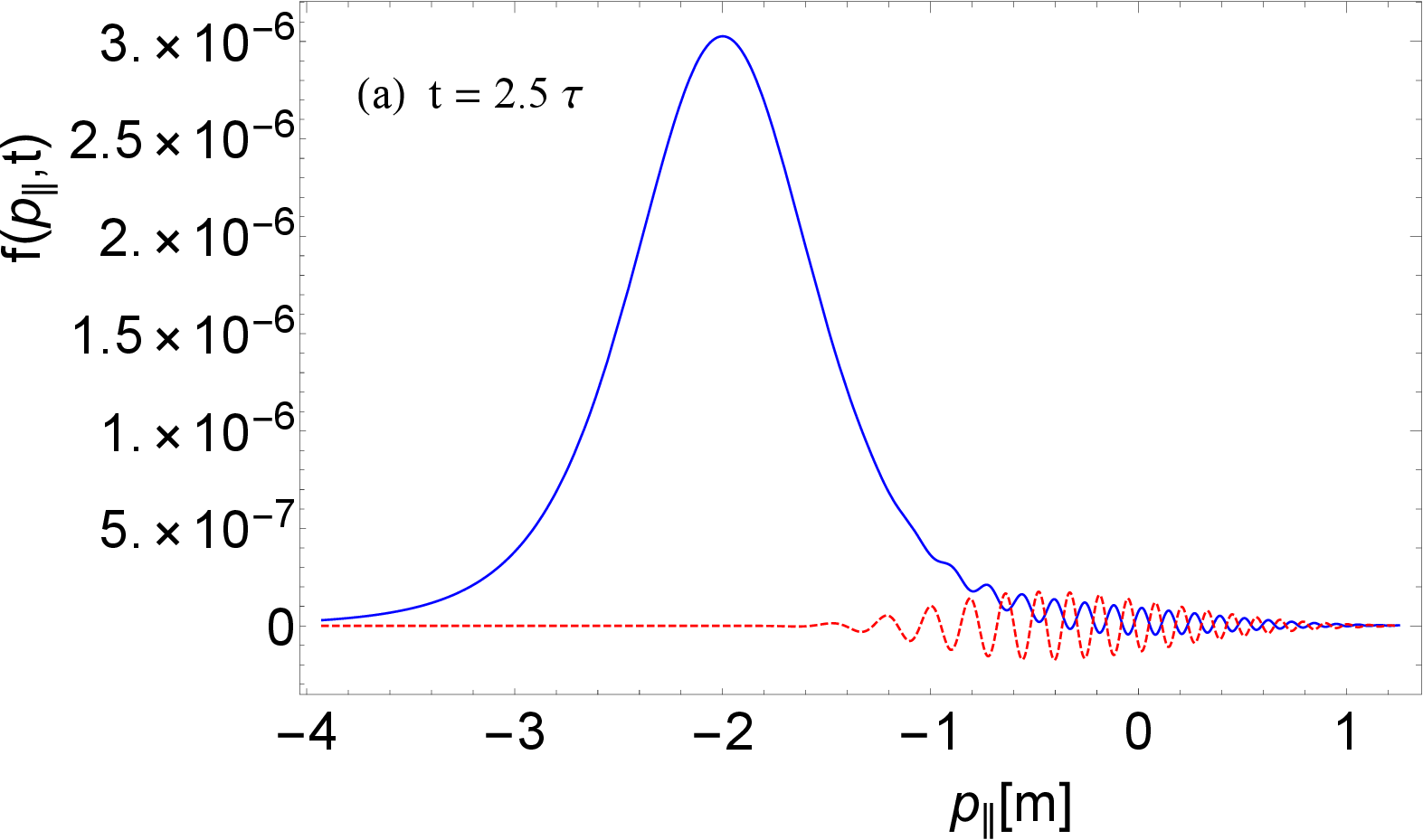}
\includegraphics[width = 1.52015in]{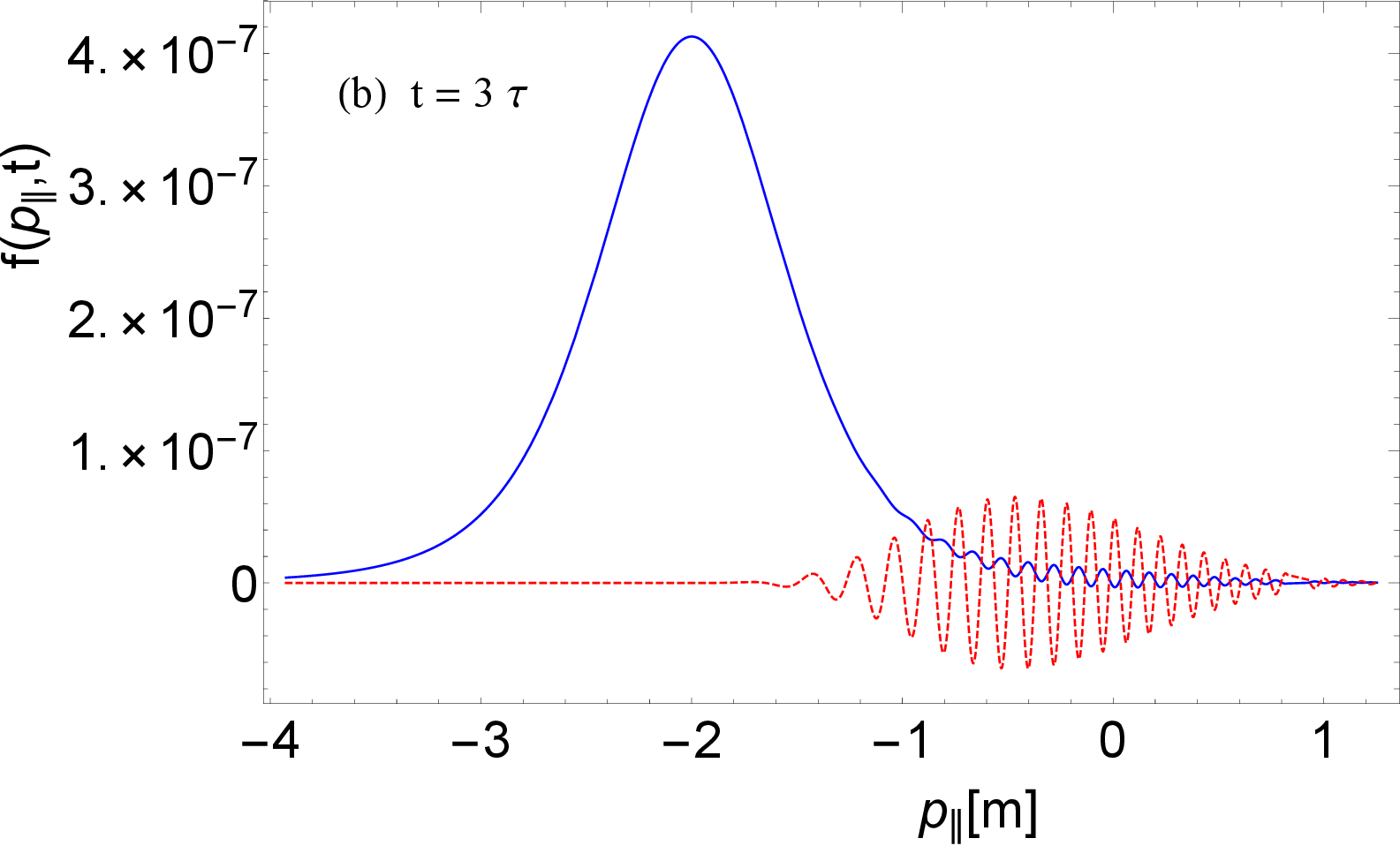}
\includegraphics[width = 1.52015in]{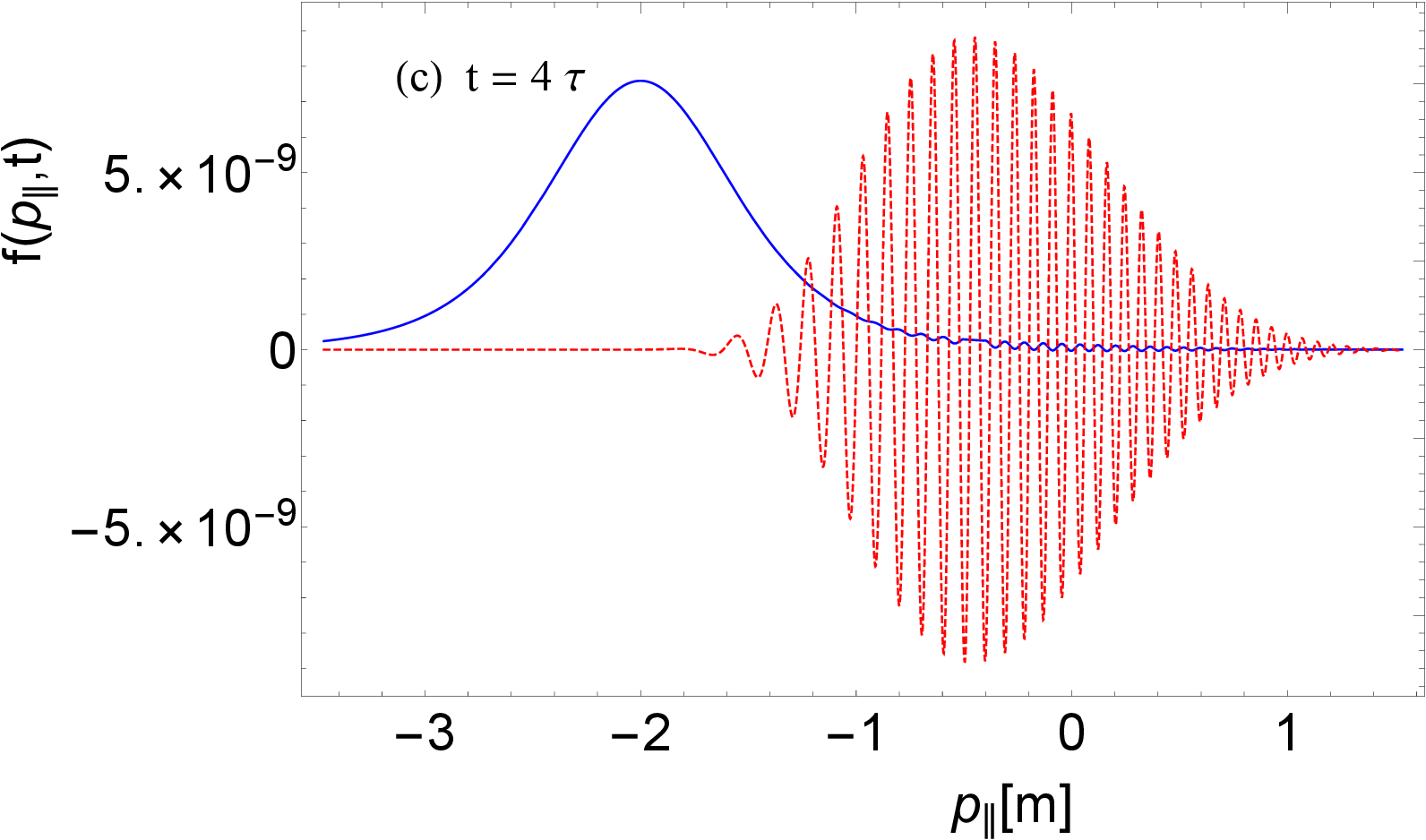}
\includegraphics[width = 1.52015in]{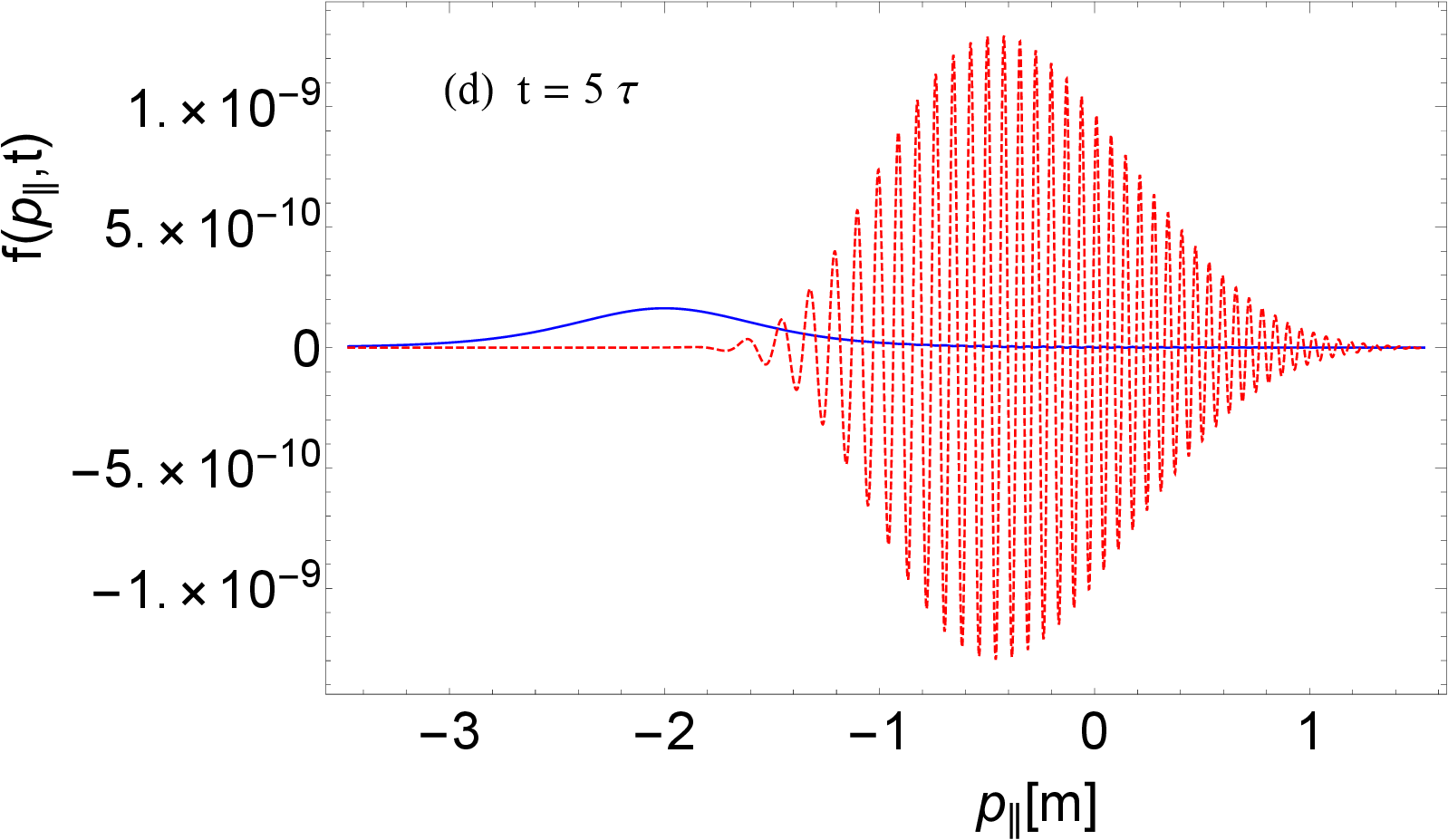}
}
\caption{ Time evolution of the first and second terms present in the longitudinal momentum distribution function $f(p_\parallel, t)$ (see Eq.~\eqref{appPDFlong}) as a function of $p_\parallel$. Red curve: $(1-y) \mathrm{C}_1$ and Blue curve: $(1-y)^2 \mathrm{C}_2$. The field parameters are $E_0 = 0.2 E_c$ and $\tau = 10 \, [m^{-1}]$.
}
\label{FIG:8}
\end{center}
\end{figure}
%%%%%%%%%%%%%%%%

\begin{figure}[t]
\begin{center}
\includegraphics[width = 3.3015in]{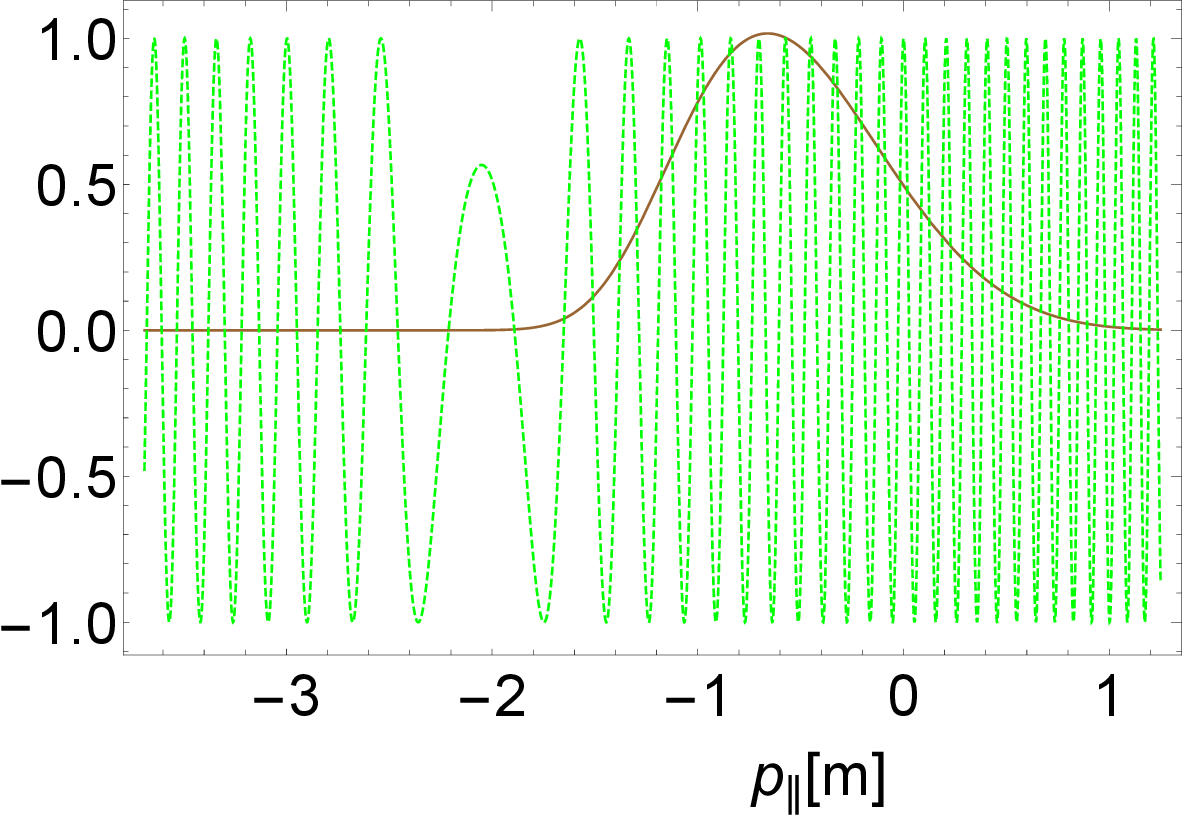}
\caption{The individual components of $\mathrm{C}_1(p_\parallel, y)$ as defined in \eqref{apppdfc11}, as a function of longitudinal momentum, depicted for $y \simeq 1$, $E_0 = 0.2 E_c$, and $\tau = 10 [m^{-1}]$. The green dashed curve represents $\cos{(\Upsilon)}$, and the brown curve represents the amplitude function ($\mathrm{A}_{C1}$).
}
\label{first_cos}
\end{center}
\end{figure}

In Figure \ref{FIG:8}, we illustrate the first-order term \((1-y)\mathrm{C}_1(p_\parallel,y)\) and the second-order term \((1-y)^2\mathrm{C}_2(p_\parallel,y)\). These terms are plotted as functions of longitudinal momentum at different times, allowing for a comparison of their behavior over time. 
Fig. \ref{FIG:8}(a), we observe that during earlier times, the term $(1-y)^2\mathrm{C}_2(p_\parallel,y)$ is more pronounced compared to $(1-y)\mathrm{C}_1(p_\parallel,y)$. The profile of $(1-y)^2\mathrm{C}_2(p_\parallel,y)$ shows a prominent peak around $p_\parallel \simeq -2 [m]$, largely influenced by $|\Gamma_1|^2$ (refer to Eq. \eqref{apppdfc22}), which peaks near $p_\parallel \approx -E_0 \tau$ depending on parameters $E_0$ and $\tau$. The oscillations observed within $-1 < p_\parallel < 1$ stem from the  $\cos{(\Upsilon)}$ in the expression for $(1-y)^2\mathrm{C}_2(p_\parallel,y)$, contributing to this oscillatory pattern.
%Additionally, it depends on both $p_\parallel$ and $t$, contributing to the modulation of these oscillations within the specified range of $p_\parallel$ as seen in Fig.\ref{FIG:8}.
Over time, the influence of the first-order term $(1-y)\mathrm{C}_1(p_\parallel, y)$ gradually increases compared to earlier times. This term exhibits oscillations within a Gaussian envelope, with the oscillations decaying away from $ p_\parallel \approx 0 $ (see Figs.~\ref{FIG:8}(b) and (c)).
To understand the oscillations and their dependence on  $ p_\parallel$ value and the time variable $y$, the argument $ \Upsilon $ of the cosine function is approximated as follows:
%%%%%%%%%%%%
\begin{align}
      \cos{(\Upsilon)} &= \cos{\left[K_0  + (p_\parallel+ E_0 \tau)(K_1 + K_2(p_\parallel+ E_0 \tau))\right]},
      \label{apxphase}
\end{align}
where we define 
\begin{align}
    K_0 &= \pi - \tan^{-1}(\tau) - 2\tau \ln{(2)} + \tau \ln(1 - y) 
     \nonumber \\
     &+ \frac{1}{2} \tau \left(   
     4 \ln{(2)} ( 1 +  2 E_0 \tau)  + 2 \ln{(E_0 \tau)} ( 1 +  E_0 \tau)  +\sqrt{1 + 4 E_0^2 \tau^2}  
      \right),
\end{align}

\begin{align}
K_1 &= \frac{E_0 \tau^2}{\sqrt{1 + 4 E_0^2 \tau^2}} \ln\left( \frac{\sqrt{1 + 4 E_0^2 \tau^2} - 1 }{\sqrt{1 + 4 E_0^2 \tau^2} + 1} \right),
\end{align}

\begin{align}
   K_2 &= \frac{\tau}{4} \left[ 
   2 - \frac{2}{1 + \tau^2} - \frac{2}{1 + 4 E_0^2 \tau^2}  + 
   2 \ln(E_0 \tau) + 2 \ln(1 - y) \right. \nonumber \\
   & \quad \left. + \frac{1}{(1 + 4 E_0^2 \tau^2)^{3/2}} \ln\left( \frac{\sqrt{1 + 4 E_0^2 \tau^2} - 1}{\sqrt{1 + 4 E_0^2 \tau^2} + 1} \right)   \right],
\end{align}
up to quadratic term and neglecting other higher-order.
\newline
According to Eq. \eqref{apxphase}, the time-dependent oscillation frequency depends on $ p_\parallel $,and the field parameters $ (E_0, \tau) $. Near $ p_\parallel = -E_0 \tau $, the cosine oscillation flattens out, and its behavior near this point changes over time. However, regular oscillations occur at other points (see Fig. \ref{first_cos}).
\par
%In addition to the cosine function within $\mathrm{C}_1(p_\parallel, y)$, the coefficient function $\mathrm{A}_{C1}$ also significantly influences the observed oscillatory behavior  in $\mathrm{C}_1(p_\parallel, y)$
In addition to the cosine function within \( \mathrm{C}_1(p_\parallel, y) \), the coefficient function \( \mathrm{A}_{C1} \) also plays a significant role in influencing the observed oscillatory behavior in \( \mathrm{C}_1(p_\parallel, y) \).
\begin{align}
 \mathrm{A}_{C1} &= \frac{4 |\Gamma_1 \Gamma^{*}_2| E_0 \tau }{\sqrt{1 + (p_\parallel + E_0 \tau)^2} }  .
\label{f_cos}
  \end{align}
%%%%%%%%
The combined effect of $\cos(\Upsilon)$ and $\mathrm{A}_{C1}$ is observed in the spectra, we compare their behaviors by plotting them for the late-time limit, as shown in Fig.\ref{first_cos}. The function $\mathrm{A}_{C1}$ can be approximated as a Gaussian-like profile:
\begin{align}
 \mathrm{A}_{C1}&\approx    \frac{ 2 E_0 \tau e^{\frac{\pi \tau }{2} ( 2 E_0 \tau - \sqrt{1+ (p_\parallel- E_0 \tau)^2 } - \sqrt{1+ (p_\parallel + E_0 \tau)^2})}}{(1 + (p_\parallel + E_0 \tau)^2) \sqrt{ 1  + \tau^2 (1 + (p_\parallel + E_0 \tau)^2)}} \Biggl(1 + (p_\parallel - E_0 \tau)^2
\nonumber \\ &
+  \sqrt{ 1 + (p_\parallel -E_0 \tau)^2} (-p_\parallel + E_0 \tau) \Biggr),
\label{f_cos2}
  \end{align}
  \begin{align}
    \frac{ \partial{\mathrm{A}_{C1}}}{\partial {p_\parallel}} \Bigg|_{E_0=0.2,\tau=10}  =0 .
\end{align}
This leads us to find that the maxima occur at  $ p_\parallel \approx -0.6$. The function exhibits a nearly Gaussian-like structure.
%%%%%%%%%%%%%%%%%%%%%%%%%%%%%%%
%%%%%%%first derivative to find out the maximum and minimum 
Figure \ref{first_cos} illustrates that the function $\mathrm{C}_1(p_\parallel,y)$ combines a smooth Gaussian-like envelope, $\mathrm{A}_{C1}$, with an oscillatory function, $\cos(\Upsilon)$. This function displays a pattern where the amplitude decreases smoothly as one moves away from the peak of the Gaussian envelope. The Gaussian envelope ensures a gradual decline in the amplitude of function $\mathrm{C}_1(p_\parallel,y)$, while the oscillatory component introduces periodic variations. 
%The frequency and amplitude of these oscillations depend on field parameters $(E_0, \tau)$, and $p_\parallel$.
\par
At $t = 4 \tau$, the magnitude of the first-order term $(1-y)\mathrm{C}_1(p_\parallel,y)$ dominates over the second-order term $(1-y)^2\mathrm{C}_2(p_\parallel,y)$ as seen in Fig.\ref{FIG:8}(c), and the second-order term diminishes at $t = 5 \tau$ (see Fig.\ref{FIG:8}(d)). It's worth noting that both orders have a $(1-y)$ factor, which contributes to the distribution function. However, this contribution gradually decreases as time progresses, becoming less important in the late-time limit $y \simeq 1$. At $t \gg 2 \tau$, the onset of oscillations in the Gaussian-like structure can be explained by the zeroth and first-order terms of the power series expansion of $f(p_\parallel,y)$. 
\par
\begin{figure}[t]
\begin{center}
{
\includegraphics[width = 2.03015in]{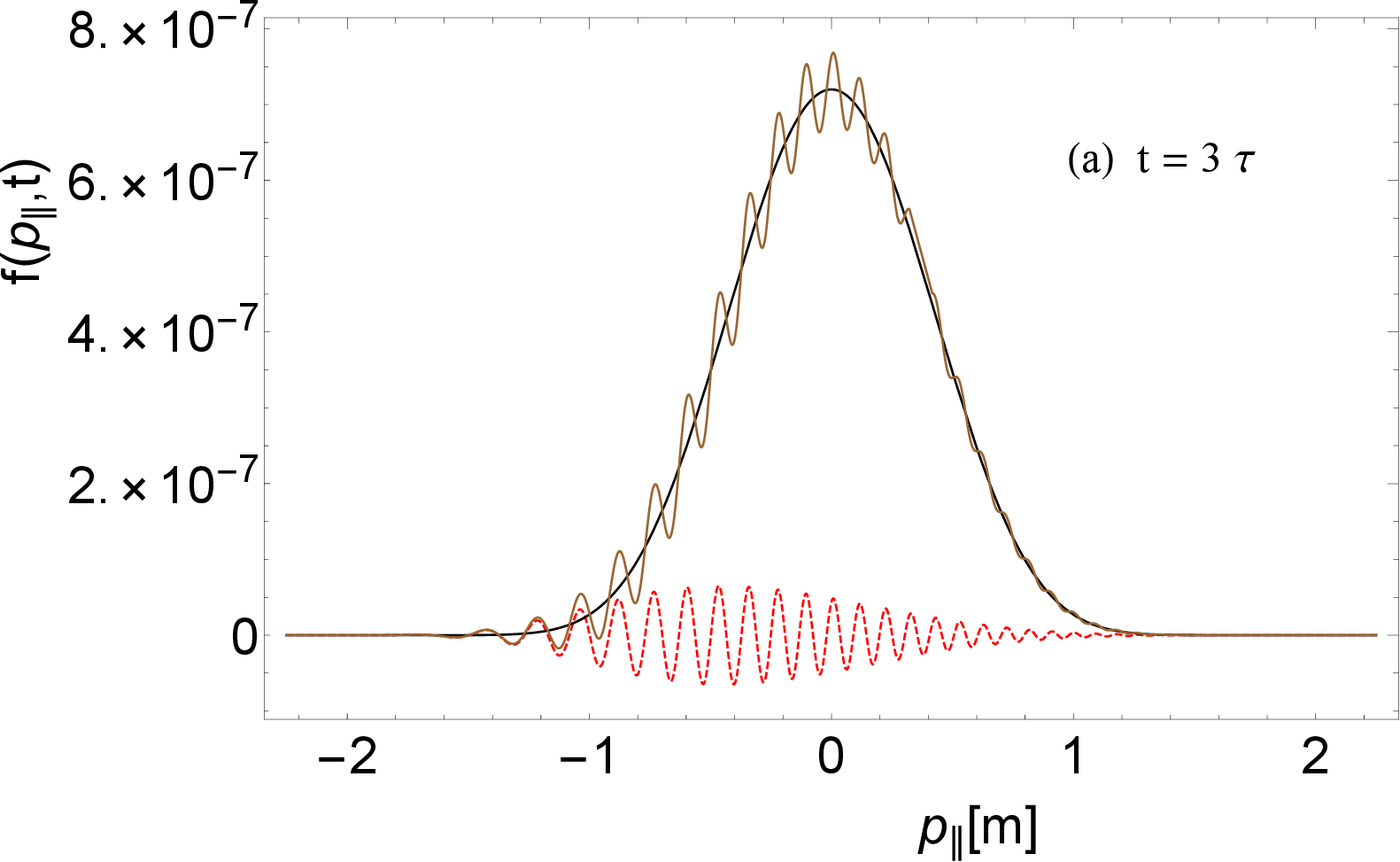}
\includegraphics[width = 2.03015in]{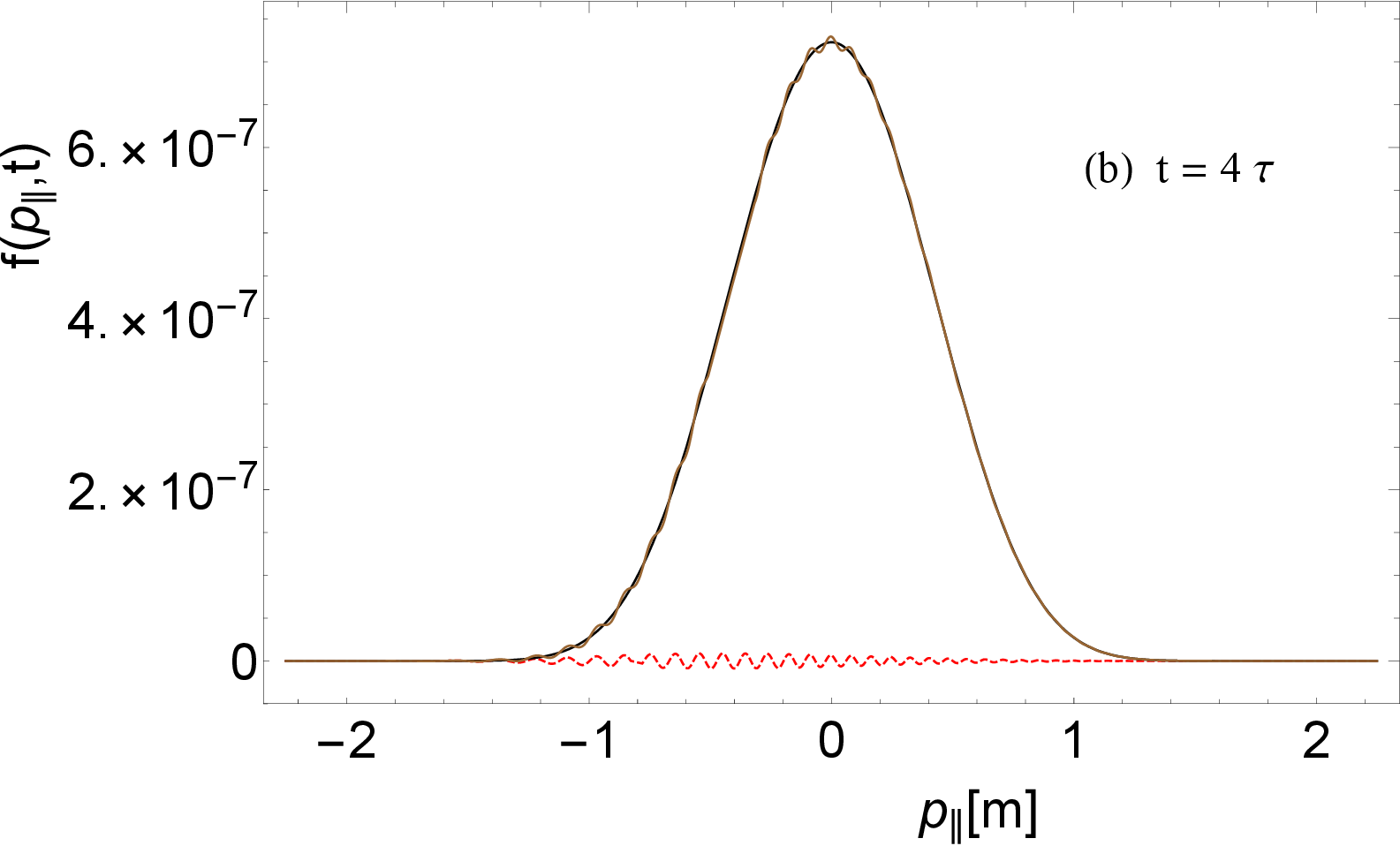}
\includegraphics[width = 2.03015in]{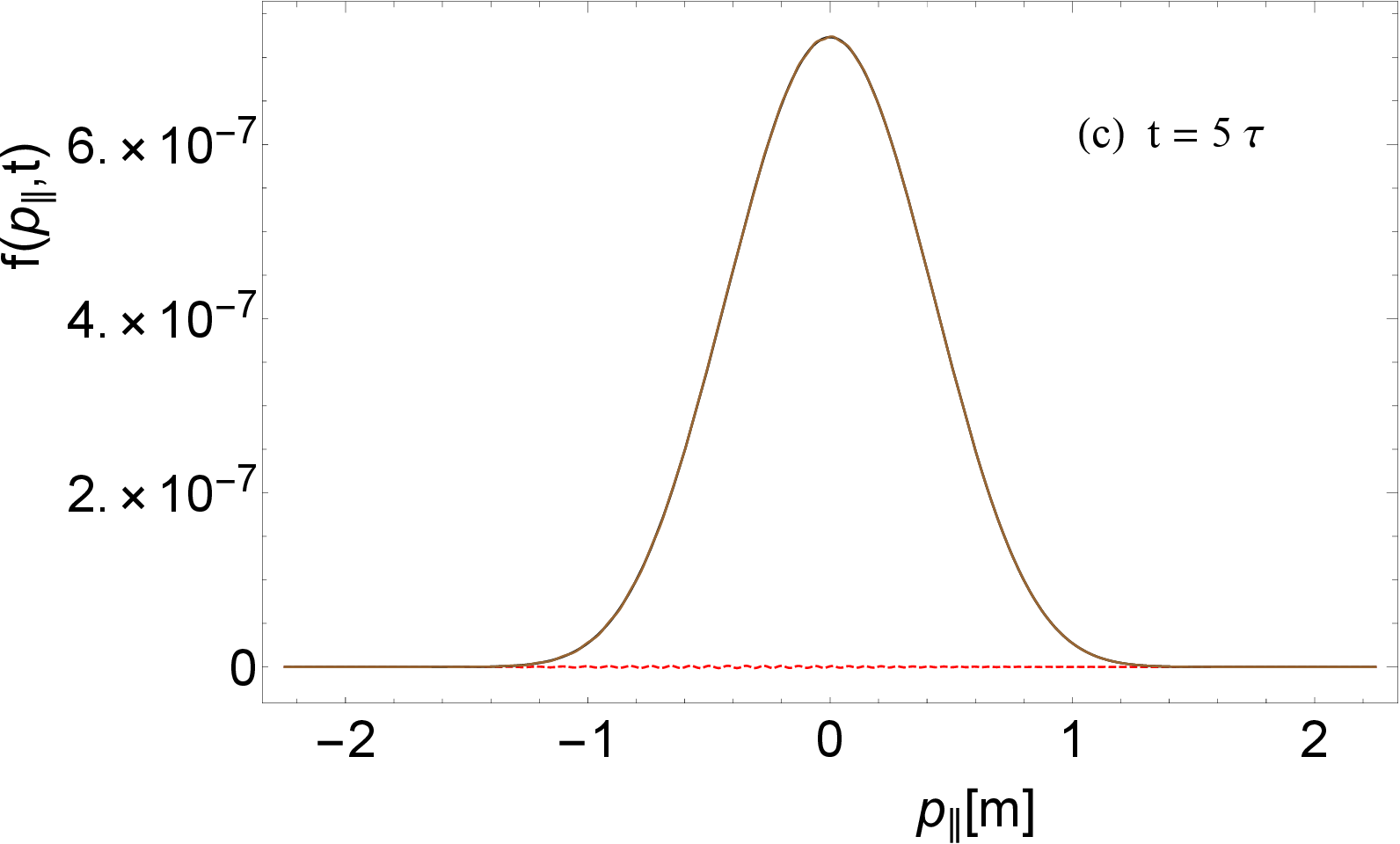}
}
\caption{The zeroth and first-order terms defined in  Eq.\eqref{appPDFlong} as a function of the longitudinal momentum at different times. Black curve: $\mathrm{C}_0(p_\parallel,y)$ and  Red curve : $(1-y) \mathrm{C}_1(p_\parallel,y),$ Brown curve: $\left(\mathrm{C}_0(p_\parallel,y) +(1-y) \mathrm{C}_1(p_\parallel,y) \right).$
The field parameters are  $E_0=0.2 E_c$ and $ \tau = 10 [m^{-1}].$}
\label{zero_first}
\end{center}
\end{figure}

In figure \ref{zero_first}, the comparison of $\mathrm{C}_1(p_\parallel,y)$, $\mathrm{C}_0(p_\parallel,y)$, and $\mathrm{C}_0(p_\parallel,y) + (1-y)\mathrm{C}_1(p_\parallel,y)$ is shown to identify which order is responsible for the oscillatory features of LMS at $t \geq 3 \tau$. At $t = 3 \tau$, $\mathrm{C}_0(p_\parallel,y)$ exhibits a smooth single peak located at $p_\parallel = 0$, while $(1-y)\mathrm{C}_1(p_\parallel,y)$ shows oscillatory behavior within the small range $-1 < p_\parallel < 1$. The combined behavior resembles a smooth Gaussian structure with onset oscillations contributing to the distribution function, as confirmed by Fig. \ref{zero_first}(a). As time progresses, the oscillation amplitude decreases due to the presence of the $(1-y)$ factor in the first-order term, reducing its impact (see Fig.~\ref{zero_first}(b)). Eventually, $\mathrm{C}_0(p_\parallel,y)$ dominates and explains the distribution function's behavior at asymptotic times as shown in Fig.\ref{zero_first}(c).
%%%%%%%%%%%%%%%%%%%%%%%%%%%%%%%%%%%%%%%%%%%%%%%%%%%%%%%%%%%%%%
\par
On careful examination of the individual components of the longitudinal momentum distribution function in Eq.\eqref{appPDFlong}, we observe that during electron-positron formation at $t > 2\tau$, the second-order term $\bigl((1-y)^2 \mathrm{C}_2(p_\parallel,y) \bigr)$ is responsible for the primary peak located at $p_\parallel \approx -2 $ (see Fig.\ref{fig:6.1}). As time progresses, this primary peak diminishes, and a secondary peak at $p_\parallel = 0$ starts to build up due to the first-order term ($(1-y) \mathrm{C}_1(p_\parallel,y)$) and zeroth-order term $\mathrm{C}_0(p_\parallel,y)$, which are responsible for the onset oscillation on that peak.
As time progresses towards infinity, $\mathrm{C}_1(p_\parallel,y)$ leads to suppression. Consequently, we observe only a secondary peak due to the dominance of the $\mathrm{C}_0(p_\parallel,y)$ zeroth-order term( see Fig.\ref{fig:6.1}). Since the component $(1-y) \mathrm{C}_1(p_\parallel,y)$ represents an oscillatory finite function whose magnitude depends on time, the magnitude of this function plays a crucial role in determining the dynamics of $f(p_\parallel,t)$ in $p_\parallel$-space.
%at finite times.
\par

In the limit \( y \rightarrow 1 \) ( or $t \rightarrow \infty$), we find an asymptotic expression for the distribution function.
At this stage, the second and first-order terms become negligible, leaving only the zeroth-order term $\mathrm{C}_0(p_\parallel,  y\rightarrow 1)$ surviving:
\begin{align}
   f(p_\parallel,y \rightarrow 1 ) &= |N^{(+)}(p_\parallel)|^2 \mathrm{C}_0(p_\parallel, y \rightarrow 1) ,\nonumber \\
   &= 4 \omega_1 |N^{(+)}(p_\parallel)|^2 (\omega_1 + P_1) |\Gamma_1|^2,
\end{align}
We can write the expression for the distribution function at asymptotic times as follows:
\begin{align}
   f(p_\parallel,y \rightarrow 1 ) &= \frac{ 2 \sinh\left(\frac{\pi \tau ( 2 E_0 \tau + \omega_0 - \omega_1)}{2}\right) \sinh\left(\frac{\pi \tau ( 2 E_0 \tau - \omega_0 + \omega_1 )}{2}\right)}{\sinh(\pi \tau \omega_0) \sinh(\pi \tau \omega_1)}.
\end{align}
%%%%%%%%%%%%%%%%%%%%%%%%%%%%
The function $f(p_\parallel,y \rightarrow 1 )$ aligns with earlier findings on the asymptotic particle distribution function \cite{Taya:2014taa, Bialynicki-Birula:2011lzb}.
%%%%%%%%%%%%%%%%%%%%
By approximating the $\omega_1 $ and $\omega_0$ as series expansions around  $p_\parallel = 0 $  up to quadratic order and neglecting higher order and further simplifying, we  get
\begin{align} f(p_\parallel) &\approx 2 \exp{ \left\{2\pi \tau \left( E_0 \tau - \sqrt{1 + E_0^2 \tau^2} \right) - \pi \frac{p_\parallel^2 \tau}{\left(1 + E_0^2 \tau^2\right)^{3/2}}\right\}}. \label{Z0gauss_fixed} \end{align}
%%%%%
%\par
\newline

Next, we analyze the qualitative picture of LMS at finite time for the Keldysh parameter $\gamma = 1$, as discussed in the previous subsection. This analysis uses two different configurations of parameters, $E_0$ and $\tau$, to illustrate their impact on the finite-time behavior of LMS.
We observe that LMS exhibits multi-peaks for $\tau = 5 \, [m^{-1}]$ and a bimodal Gaussian-like profile with onset oscillations for $\tau = 8 \, [m^{-1}]$, as depicted in Fig.~\ref{fig:6.3}. This trend can be understood through the presence of $\cos(\Upsilon)$ in the first and second-order terms of the approximate distribution function relation \eqref{appPDFlong}. We approximate the argument $\Upsilon$ using a series expansion around $p_\parallel = 0$ as follows:
\begin{align}
	 \Upsilon &\approx \pi - \tan^{-1} (\frac{\tau}{\gamma}\sqrt{1+\gamma^2})) + \frac{\tau}{\gamma} \ln{\Bigl(  \frac{\sqrt{1+ \gamma^2} +1}{\sqrt{1+ \gamma^2} -1}  \Bigr)} + \sqrt{1+ \gamma^2} \ln{\Bigl(  \frac{\gamma^2 (1-y)}{1+ \gamma^2}  \Bigr)} \nonumber \\
	 & +  p_\parallel\frac{\tau}{\sqrt{1+\gamma^2} (\tau^2 + \gamma^2 (1+\tau^2))} \left(\ln{\biggl(\frac{(1-y)}{1+\gamma^2} \biggr)} (\gamma^2(1+\tau^2) + \tau^2) - \gamma^2\right)
	 \nonumber \\
	 & + p_\parallel^2  \sqrt{1+ \gamma^2} \gamma \tau \Biggl( \frac{\gamma^2 \tau^2 (2 + \gamma^2 -\gamma^4) - \gamma^6}{(\gamma^2 (1+\tau^2) + \tau^2)^2} -2 + \frac{\gamma^2}{2 (1+\gamma^2)} \ln{\Bigl(  \frac{\gamma^2 (1-y)}{1+ \gamma^2}  \Bigr)} \Biggr).
	 \label{gamma_1}
\end{align}
Eq.~\eqref{gamma_1} indicates that the cosine function's argument depends on the pulse duration $\tau$ and the Keldysh parameter $\gamma$. When $\gamma$ is fixed, the frequency of momentum oscillations is determined by $\tau$ alone. As $\tau$ decreases, these oscillations occur less frequently, leading to visible patterns in the momentum spectrum, as illustrated in Fig. \ref{fig:6.2}. According to Eq. \eqref{gamma_1}, reducing $\tau$ results in fewer oscillations, which manifests as a multi-peak structure in the spectrum.
%%%%%%%%%%%%%%%%%%%%

\par
We next examine how these interference phenomena manifest in the multiphoton regime $(\gamma >> 1)$, where distinct spectral features emerge.
\subsubsection{LMS in the multi-photon regime}
\label{sec:LMS multi-photon}
\begin{figure}[t]
\begin{center}
{
\includegraphics[width = 2.015in]{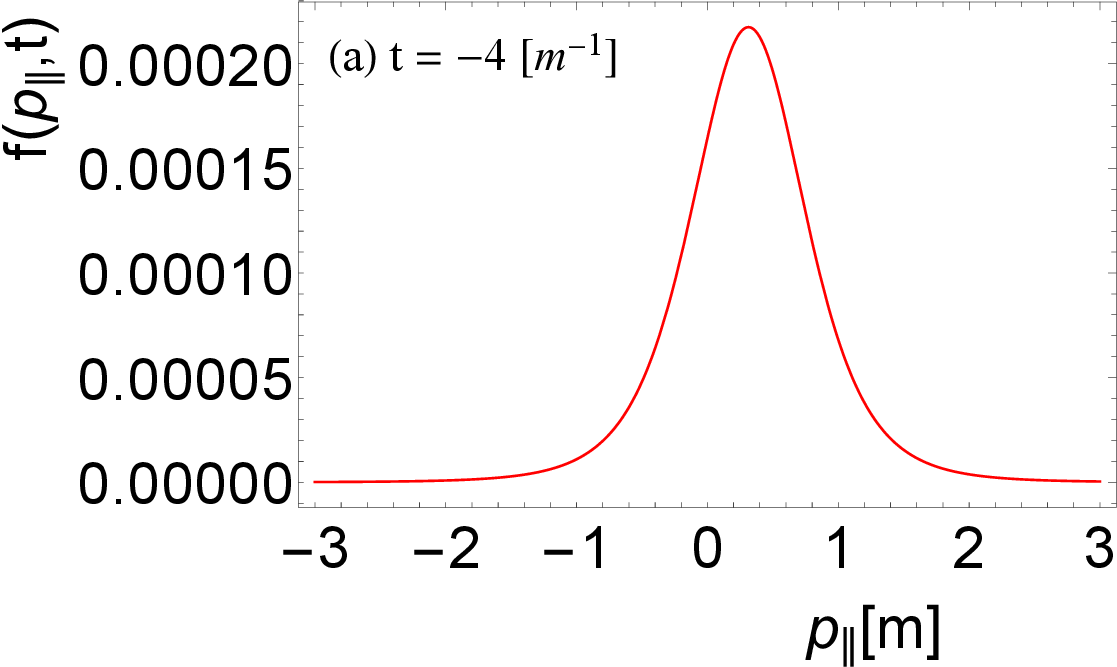}
\includegraphics[width = 2.015in]{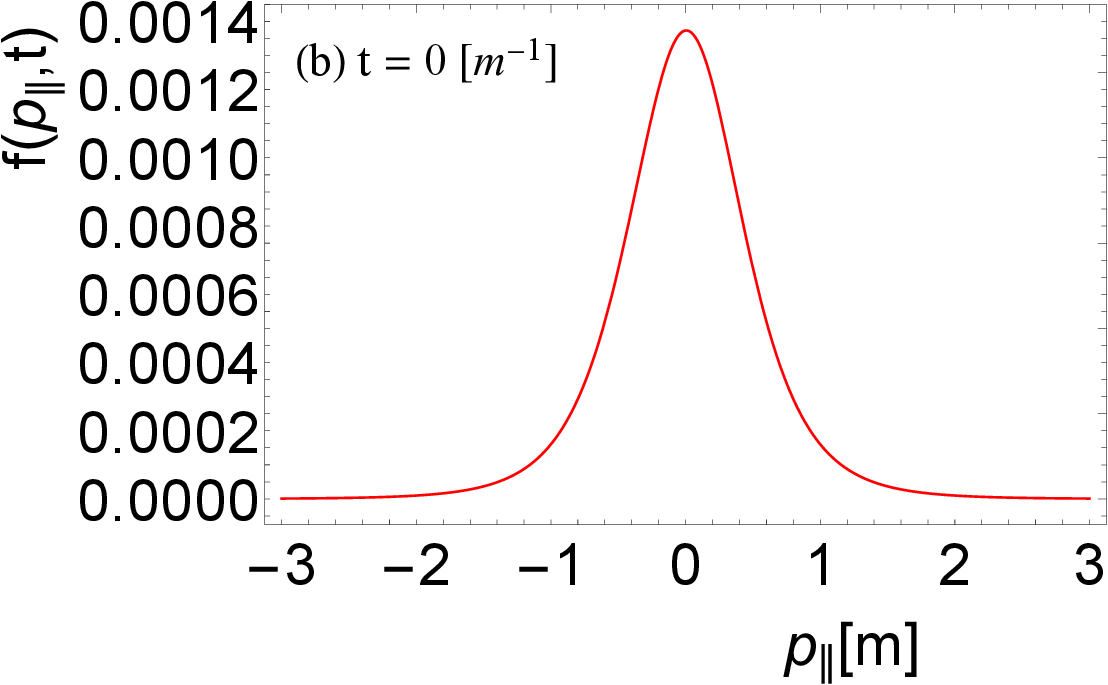}
\includegraphics[width = 2.015in]{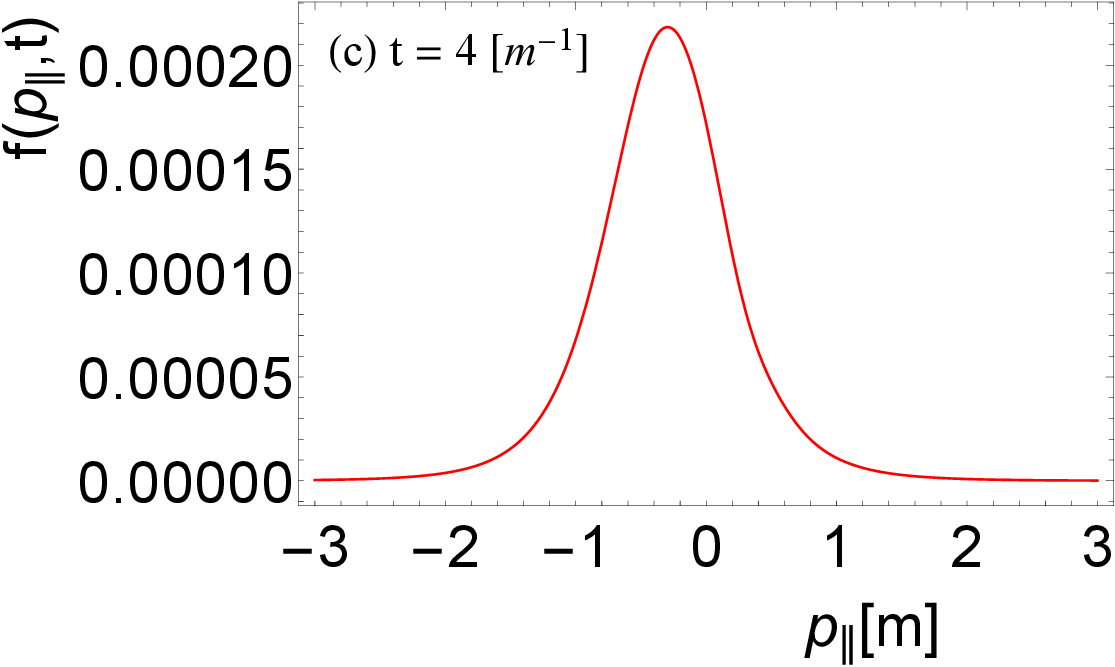}
\includegraphics[width = 2.015in]{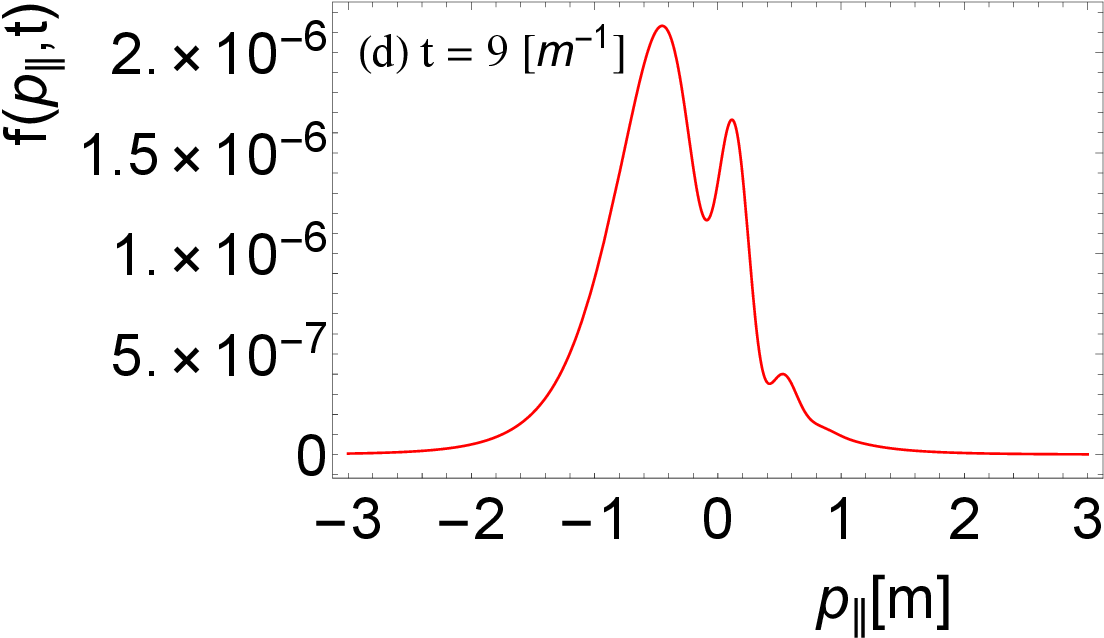}
\includegraphics[width = 2.015in]{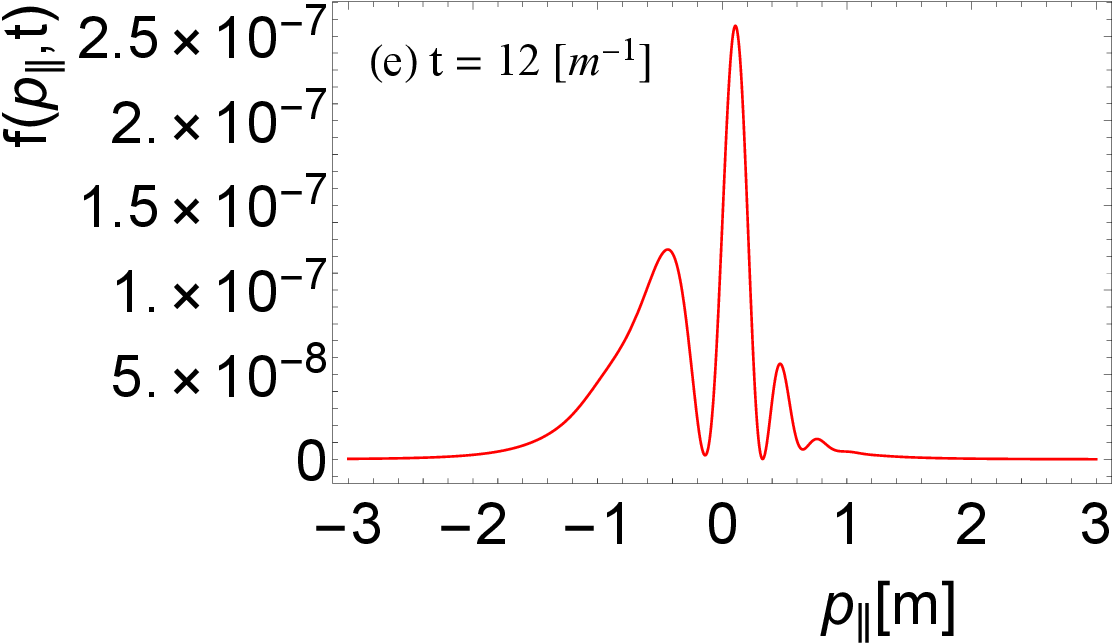}
\includegraphics[width = 2.015in]{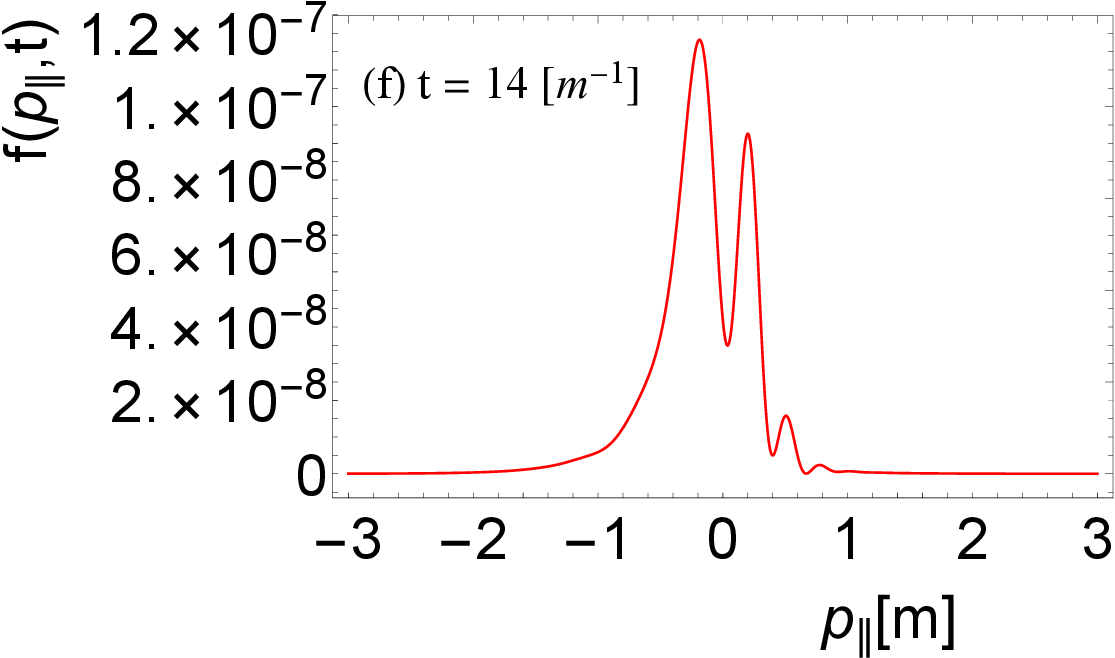}
\includegraphics[width = 2.015in]{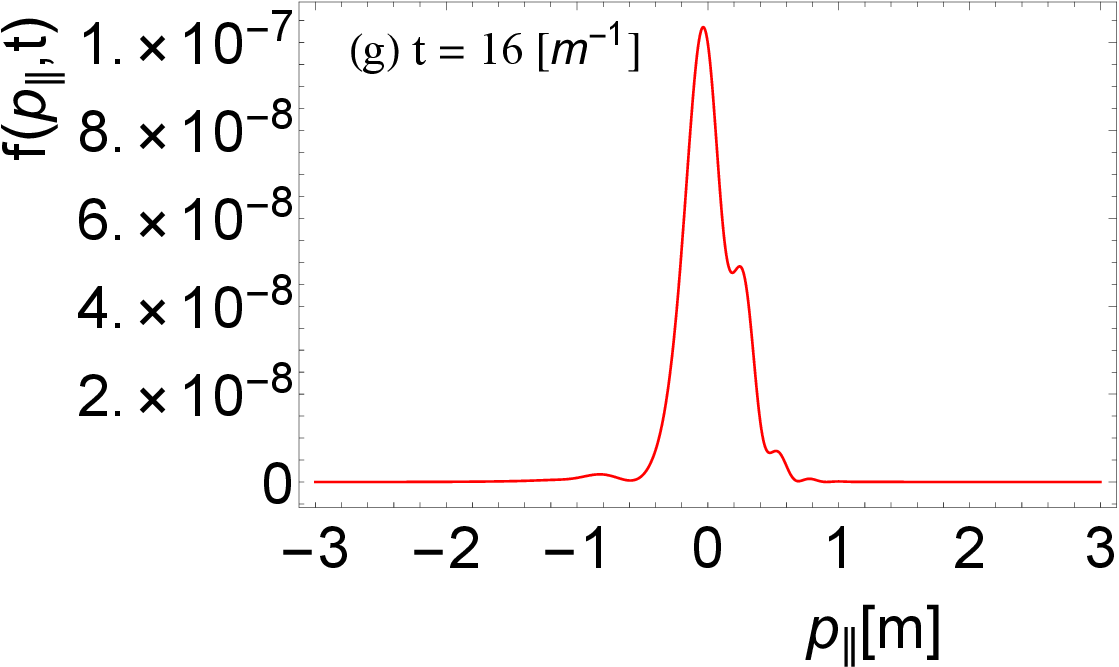}
\includegraphics[width = 2.015in]{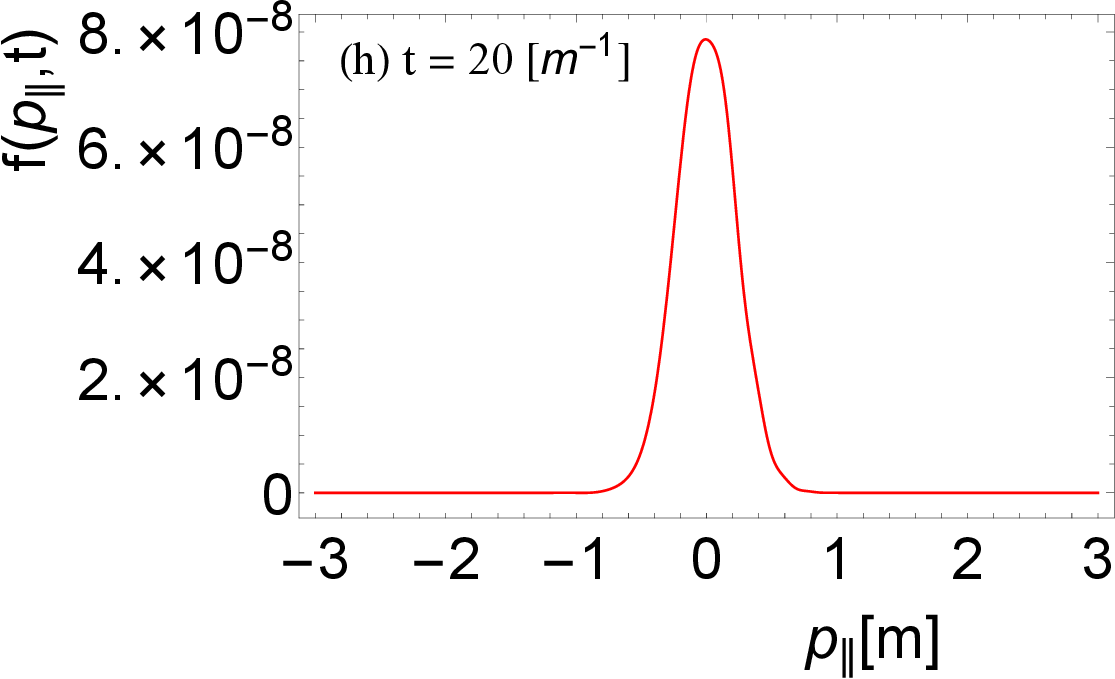}
\includegraphics[width = 2.015in]{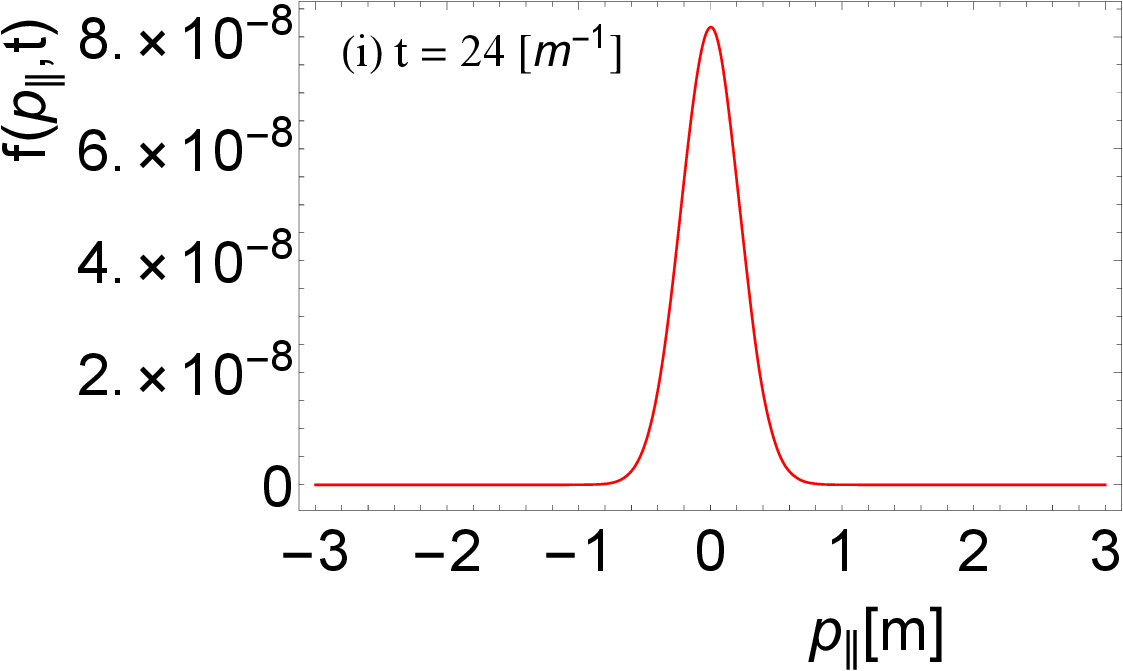}
}
\caption{LMS of created particles at different times. The value of transverse momentum is considered to be zero, and all units are in the electron mass unit. The field parameters are  $E_0=0.1 E_c$ and $ \tau =4 [m^{-1}].$}
   	\label{multi_photon}
\end{center}
\end{figure}

In this subsection, we explored the LMS of the created particle in the multiphoton regime. We choose the parameters of the laser pulse in such a way that the Keldysh parameter, $\gamma >> 1.$
Figure \ref{multi_photon} shows LMS for short pulse-duration $\tau = 4[m^{-1}] $ and $E_0 = 0.1 E_c.$ The Keldysh parameter, in this case, is close to $2.5$, i.e., $\gamma >> 1$ corresponds to $n^{th}$ order perturbation theory, with $n$ being the minimum number of photons to be absorbed in order to overcome the threshold energy for pair creation $n \omega > 2m.$ In the early time of the creation of pairs at $t = -4 [m^{-1}]$ spectrum has a unimodal Gaussian-like profile peak at $p_\parallel \simeq -0.4 [m]$, and as time proceeds, we see shifting of peak $p_\parallel \simeq - e E_0 \tau $ to $p_\parallel \simeq e E_0 \tau $ due to the action of force and its peak value shows the maximum value for $t = 0$, i.e., $f(p_\parallel=0,t=0) = 1.4 \times 10^{-3}$ as seen in figure \ref{multi_photon}
(a-c). However, the previously smooth and unimodal spectrum now exhibits slight modulation at $t = 9[m^{-1}].$  However, at $t= 12 [m^{-1}]$ spectrum has a quad-modal profile, as seen in Fig.\ref{multi_photon}(e). The central peak, which is located at $p_\parallel =0$, is much more prominent than the two unequal peaks at $p_\parallel \simeq \pm 0.4[m]$ and another very small peak at $p_\parallel \simeq 0.7[m].$
Fig.~\ref{multi_photon}(f-g)  shows the merging of those multimodal peaks as a result of the fading of different peaks that occur and the single smooth Gaussian peaks observed at $t > 5\tau $  as shown in figure  \ref{multi_photon} (h-i).
\par
An interesting qualitative contrast becomes evident when comparing the previous scenario with $\gamma = 0.5$ to the current situation. The presented Figs.\ref{multi_photon} illustrates this distinct behavior. Notably, a multi-modal pattern emerges at a specific moment, characterized by the presence of more than two peaks, occurring approximately at $t=3\tau$.
%%%%
%%%%
%%%%%%%%%%%%%%%%%%%%%%%%%%%%%%%%%%%%%%%%%%%%%%
%%%%%%%%%%%
\par
\begin{figure}
    \centering
    \includegraphics[width = 3.015in]{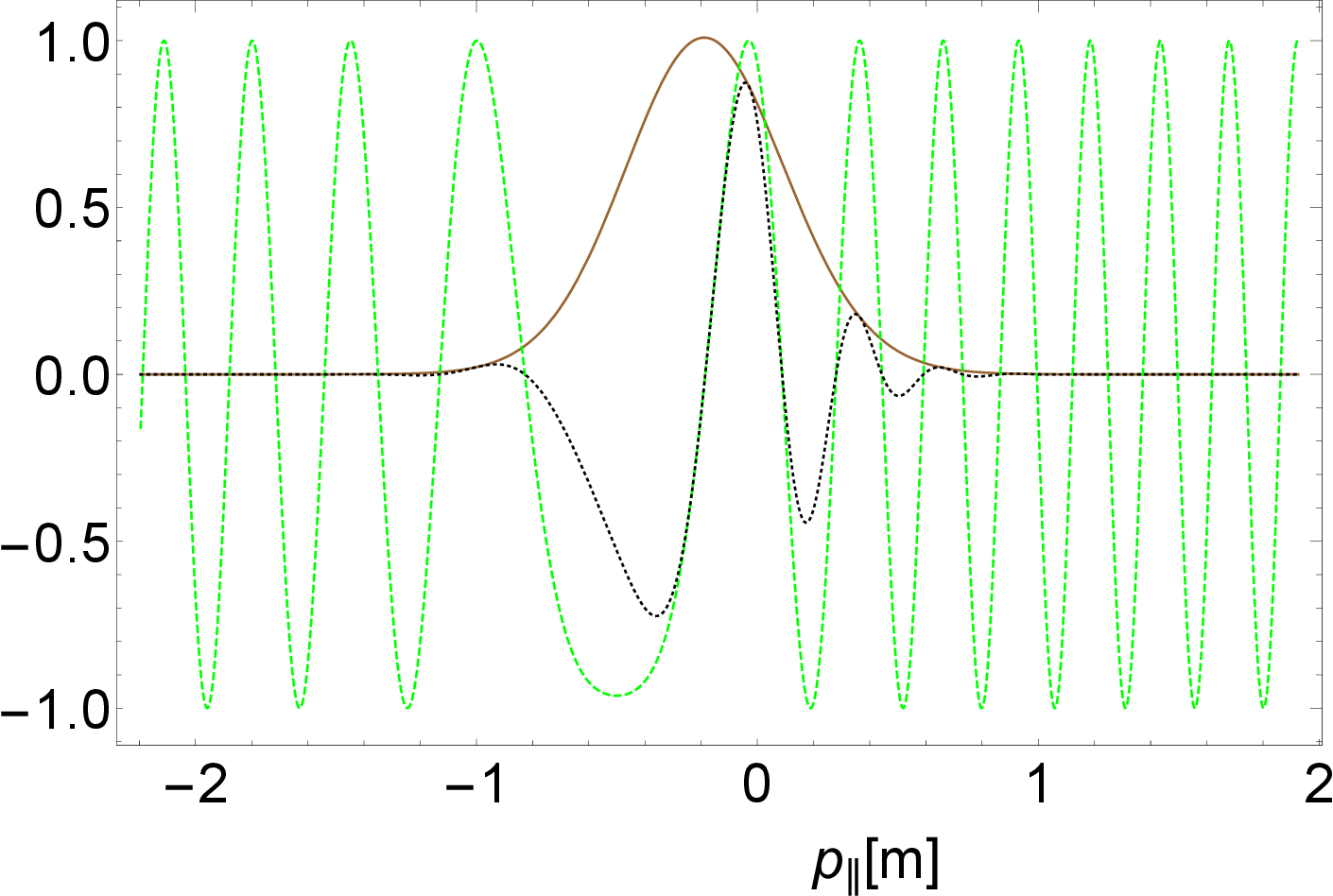}
    \caption{The first order term, $\mathrm{C}_1(p_\parallel,y)$, and its individual components as defined in Eq.\eqref{multi}, as a function of longitudinal momentum, depicted for $y \simeq 1$, $E_0 = 0.1 E_c$, and $\tau = 4 \, [m^{-1}]$. The green dashed curve represents $\cos{(\Upsilon)}$, the brown curve represents the amplitude function ($\mathrm{A}_{C1}$), and the black dotted curve represents $\mathrm{C}_1(p_\parallel,y)$.
}
    \label{mult_first_cos}
\end{figure}

As discussed in the previous section, the characteristic features of the LMS can be mathematically explained using a simple approximate distribution function, as shown in Eq. \eqref{appPDFlong}. A similar analysis can help us understand the behavior of the multiple peaks observed in the LMS at finite time. In the multi-photon case, where $E_0 \tau < 1$, the first-order term becomes 
\begin{align}
      \mathrm{C}_1 (p_\parallel,y) & \approx  \mathrm{A}_{C1}\cos{(\Upsilon)},
      \label{multi}
\end{align}
with

\begin{align}  \mathrm{A}_{C1}&\approx \frac{1}{2} \exp\left\{ \frac{\pi \tau}{2} \left( E_0 \tau (2 - E_0 \tau) - 2 \right) + p_\parallel^2 (1 - E_0^2 \tau^2) \right\}[ 4 + 2 E_0 \tau (2 - E_0 \tau - 2 E_0^2 \tau^2) ]
 \nonumber \\ 
      &+ p_\parallel^2 \left( -2 + E_0 \tau (12 + 57 E_0 \tau) \right) + 2 p_\parallel \left( -2 - 10 E_0 \tau - 6 E_0^2 \tau^2 + 15 E_0^3 \tau^3 \right), \label{AC1} \end{align}
\begin{align}
     \Upsilon &\approx   E_0 \tau^2 \ln{\Biggl(\frac{E_0 \tau + \sqrt{1 + E_0^2 \tau^2}}{-E_0 \tau + \sqrt{1 + E_0^2 \tau^2}} \Biggr)} +  \tau \sqrt{1 + E_0^2 \tau^2}   \ln{ \Biggl(\frac{(1-y)}{1 + E_0^2 \tau^2} \Biggr)}   
     \nonumber \\ &+  p_\parallel  \frac{E_0 \tau^2}{\sqrt{1+E_0^2 \tau^2}} \ln{\Biggl(\frac{E_0^2 \tau^2 (1-y)}{(1+E_0^2 \tau^2)}\Biggr)}
     + \frac{p_\parallel^2 \tau}{2 (1+ E_0^2 \tau^2)^{3/2}} 
     \Biggl( -2 E_0^2 \tau^2 + \ln{\Biggl( \frac{(1-y)}{(1+ E_0^2 \tau^2)}\Biggr)}   \Biggr).
     \label{APPRCOS_MULTI}
\end{align}
%%%
In Fig. \ref{mult_first_cos}, for the given field parameters, a decrease in the oscillation frequency of the $\cos{(\Upsilon)}$ function is observed compared to the previous case with $\gamma = 0.5$. As the oscillation frequency decreases for smaller values of $\tau$, this is reflected in the spectrum. Near $p_\parallel = -E_0 \tau$, the cosine function behaves irregularly, flattening out and oscillating regularly away from this point.
The function $\mathrm{A}_{C1}$ plays a crucial role here, resembling a Gaussian single-peak structure with its peak located at $p_\parallel = -0.1 [m]$, as represented by the brown curve in Fig. \ref{mult_first_cos}. The width of the Gaussian envelope and its peak location affect the overall behavior of $\mathrm{C}_1(p_\parallel,y)$. Since the field parameters $(E_0, \tau)$ influence the occurrence of the peak in the function $\mathrm{A}_{C1}$, they also affect the behavior of the $\cos{(\Upsilon)}$ function.
Consequently, the presence of oscillations within the envelope depends on the field parameters. The appearance of multiple peaks in the spectrum is the combined effect of the oscillation of the cosine function and the function $\mathrm{A}_{C1}$. As seen in Fig.~\ref{mult_first_cos}, the black dotted curve represents $C_1(p_\parallel,y)$, which shows fewer irregular oscillations. In the long-time limit, the first-order term contribution becomes suppressed, and only the zeroth-order term, $C_0(p_\parallel,y)$, becomes dominant and governs $f(p_\parallel,y \rightarrow 1).$ As a result, the multi-modal pattern gradually dissipates, eventually leading to a singular smooth peak above $t > 5 \tau$. 

%%%%
%%%%%%%%%%%%%%%

%%%%%%%%%%%%%%%%%%%%%%
\subsection{Vacuum Polarisation and depolarisation function }
\label{VP}

\begin{figure}[t]
\begin{center}
{\includegraphics[width = 2.973169493219808012in]{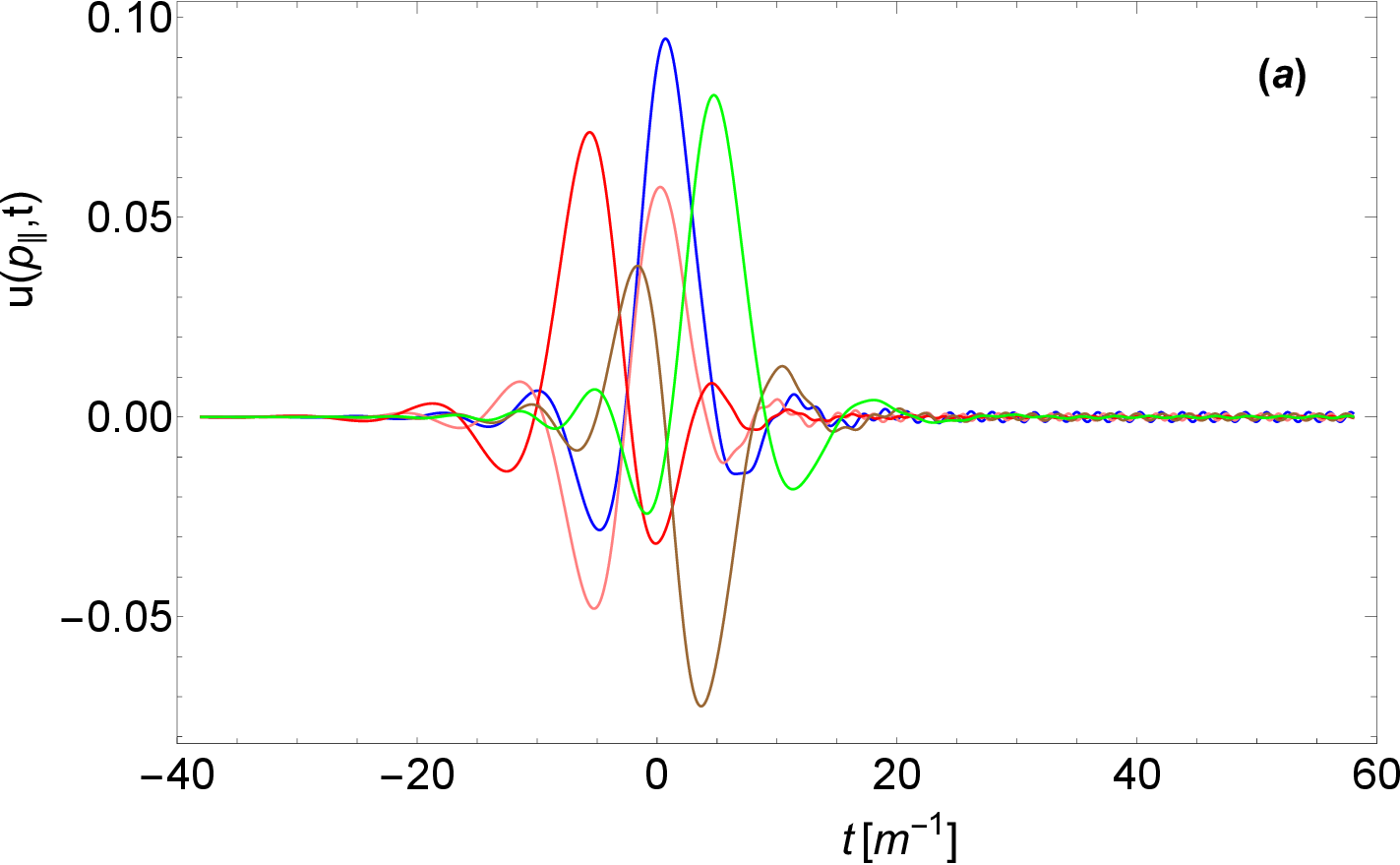}
\includegraphics[width = 2.973169493219808012in]{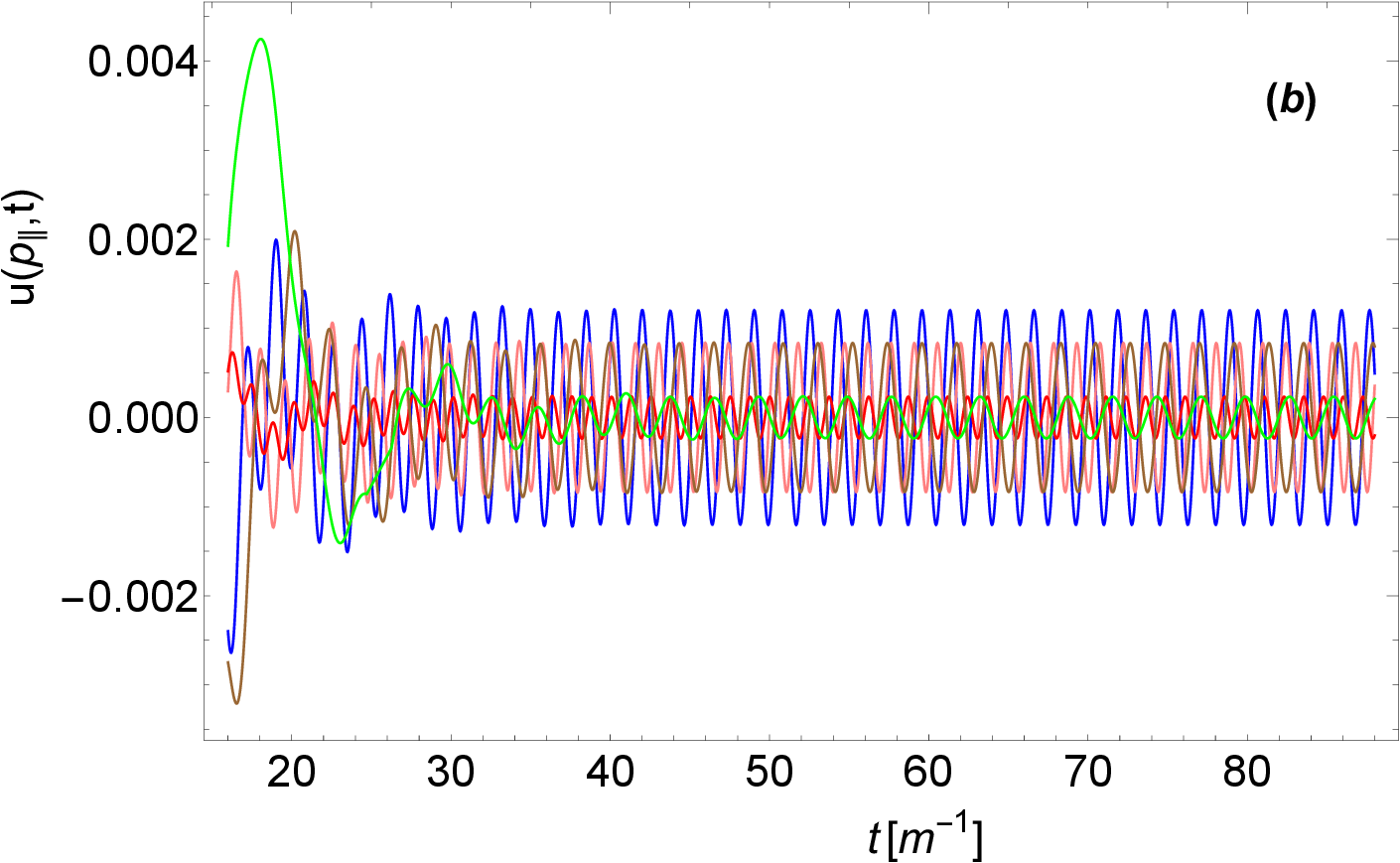}
\includegraphics[width = 2.973169493219808012in]{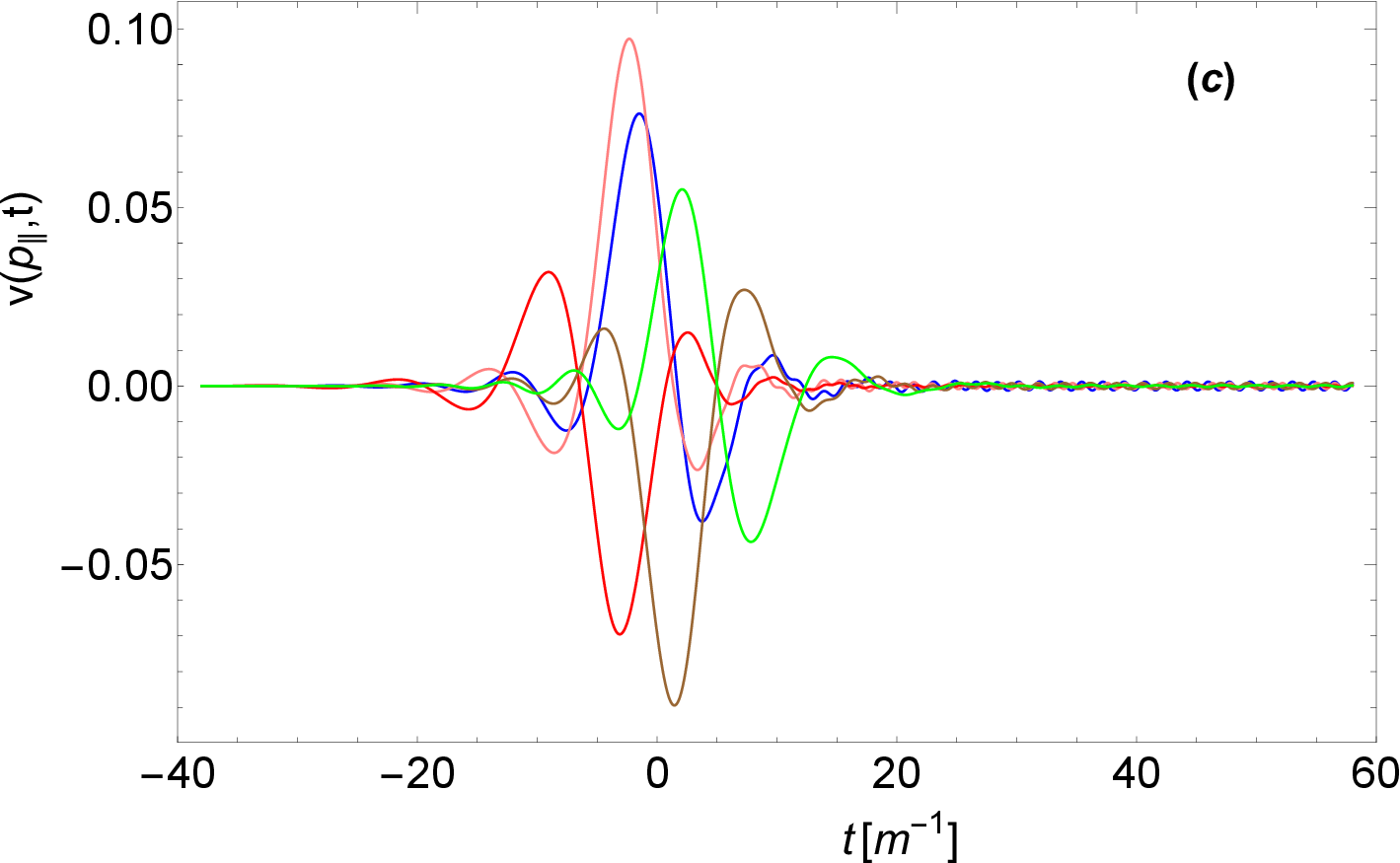}
\includegraphics[width = 2.973169493219808012in]{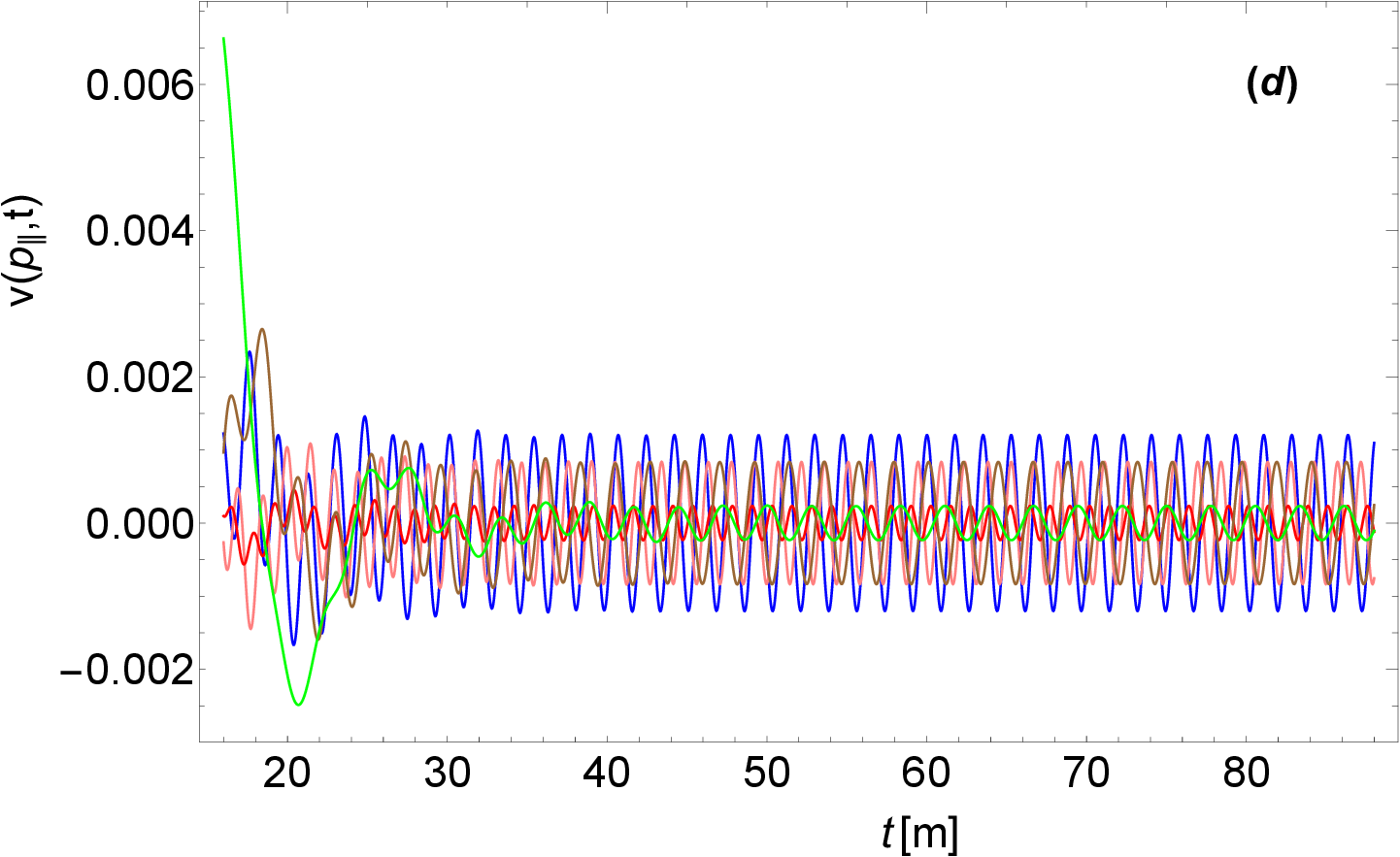}}
    \caption{Evolution of $u(p_\parallel,t)$ and $v(p_\parallel,t)$ for $p_\parallel = 0$ (blue), $0.5$ (pink), $-0.5$ (brown), $1$ (red), and $-1$ (green) with $p_\perp = 0$, $E_0 = 0.2 E_c$, and $\tau = 10 \, [m^{-1}]$. All the measurements are taken in electron mass units.}
\label{fig:uv}
\end{center}
\end{figure}

For a better understanding of the phenomenon of particle creation under a strong electric field, we will also trace the evolution of the vacuum polarization function $u(p_\parallel,t)$ and its counterterm $v(p_\parallel,t)$, which governs the depolarization.
%%%%
Figure \ref{fig:uv} shows the time evolution of $u( p_\parallel,t )$ and $v( p_\parallel,t)$ for different values of $p_\parallel$ at zero transverse momentum. 
Due to pair annihilation being stronger than pair creation in the early times, the depolarisation function $v( p_\parallel,t)$ dominates over $u(p_\parallel,t)$ (See Figs.~\ref{fig:uv}(a) and (c)).
It appears that in the polarization function, there is a sinusoidal function behavior. On the other hand, the depolarization function $v( p_\parallel,t )$ shows an unimodal Gaussian peak structure in its temporal evolution. Both $u(p_\parallel, t)$ and  $v(p_\parallel, t)$ show oscillations with varying amplitudes during the initial transient stage. These oscillations are particularly pronounced when $p_\parallel = 0 [m]$, as evident from Figs.\ref{fig:uv}(a) and (c). As time progresses, irregular oscillations are observed in the transient stage. However, the oscillations become regular and stable as the system enters the REPP stage. Both $u(p_\parallel, t)$ and $v(p_\parallel, t)$ now exhibit regular oscillations centered around the zero value. Moreover, as the momentum value $p_\parallel$ increases, the amplitudes of these oscillations diminish, as shown in Fig.~\ref{fig:uv}(b) and (d). During the REPP stage, one interesting finding is that $u(p_\parallel, t)$ and $v(p_\parallel, t)$ demonstrate balancing behavior characterized by similar oscillatory patterns. This balance results from the formation of real electron-positron pairs.
%%%%%%%%%%%%%%%%%%%%%%%%%%%%%%%%%%%%%%%%%%%%%%%%%%%%%%%%%%%%%%%%%%%%%%%%%%%%%%%%%%%%%%%%%%%%%
%%%%%%%%%%%%%%%%%%%%%%%%%%%%%%%%%%%%%%%%%%%%%%%%%%%%%%%%%%%%%%%%%%%%%%%%
\subsubsection{Momentum Spectra of $u(p_\parallel,t)$ and $v(p_\parallel,t)$ }
\begin{figure}[t]
\begin{center}

{\includegraphics[width = 1.9in]{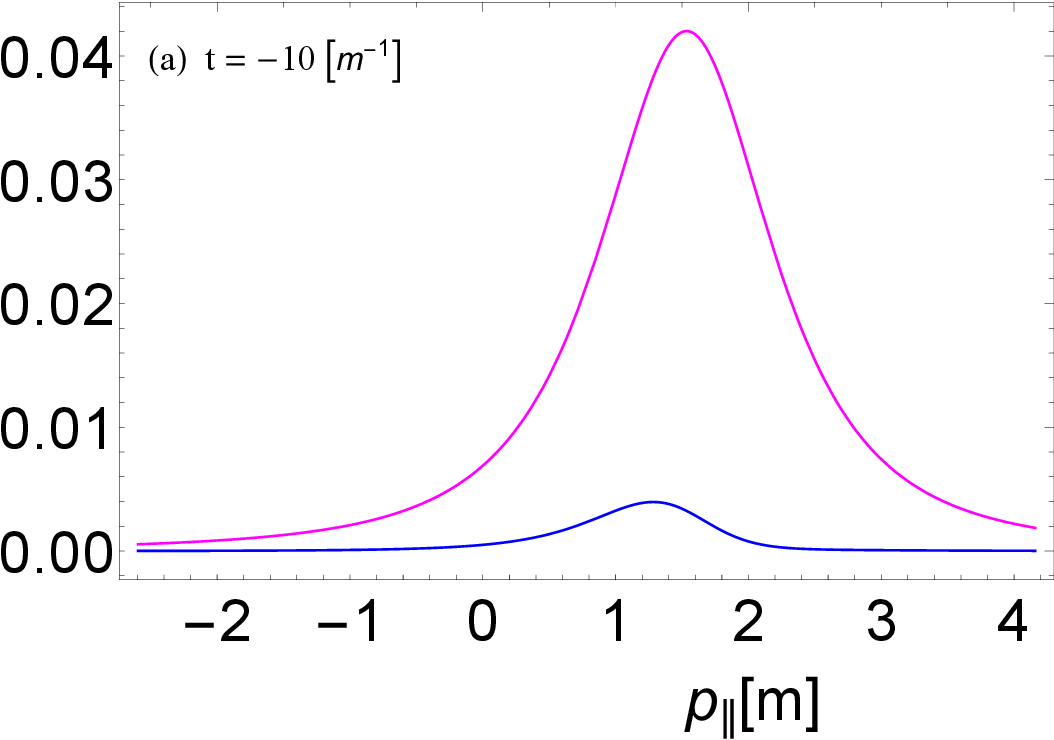}
\includegraphics[width =1.9in]{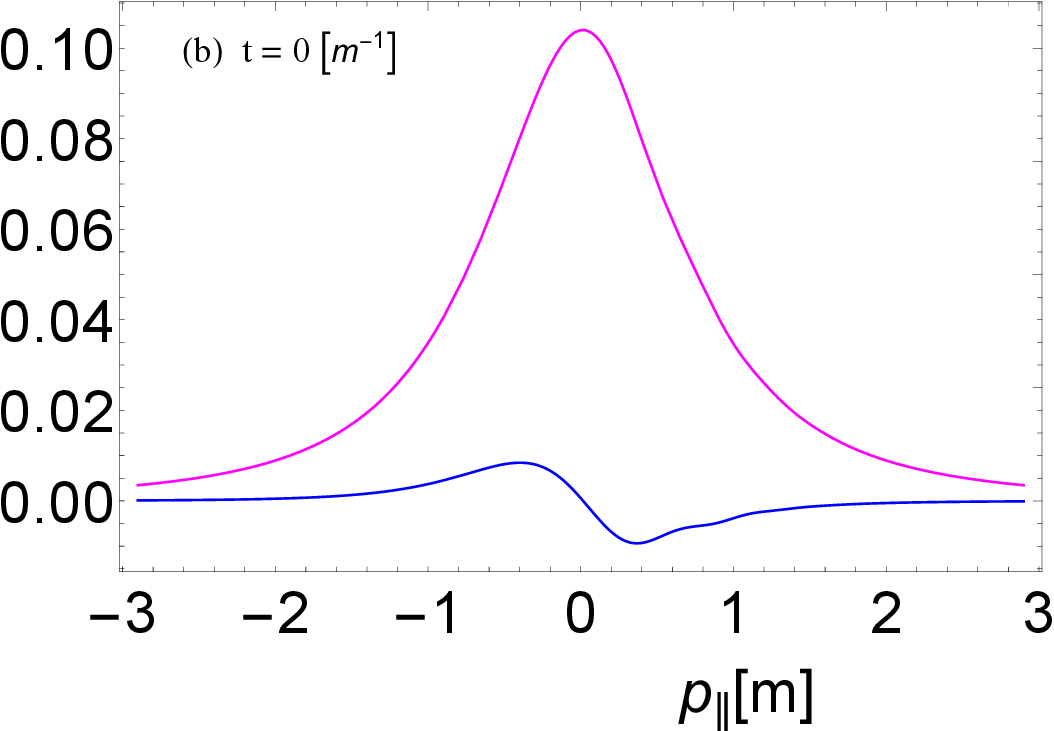}
\includegraphics[width = 1.9in]{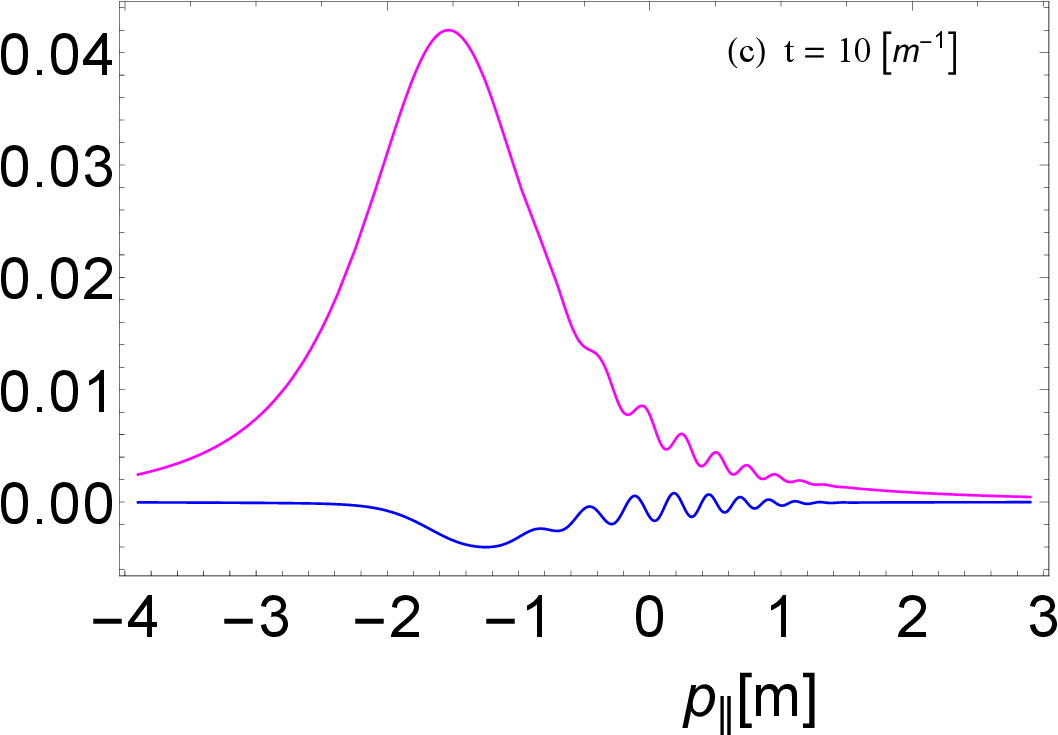}
\includegraphics[width = 1.9in]{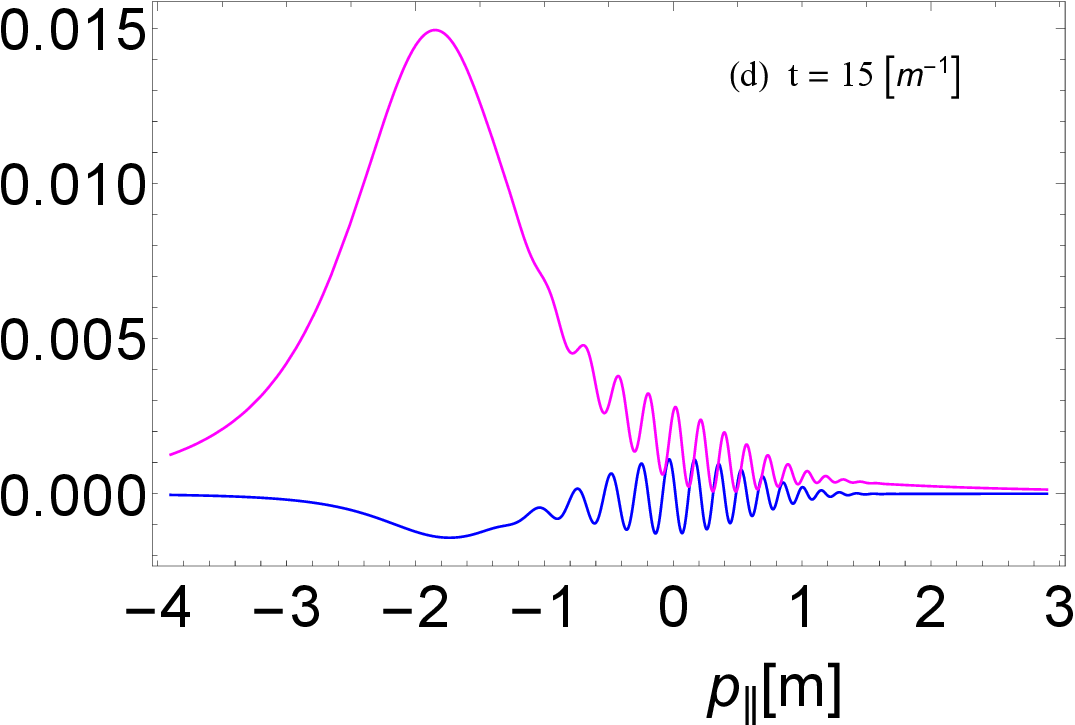}
\includegraphics[width = 1.9in]{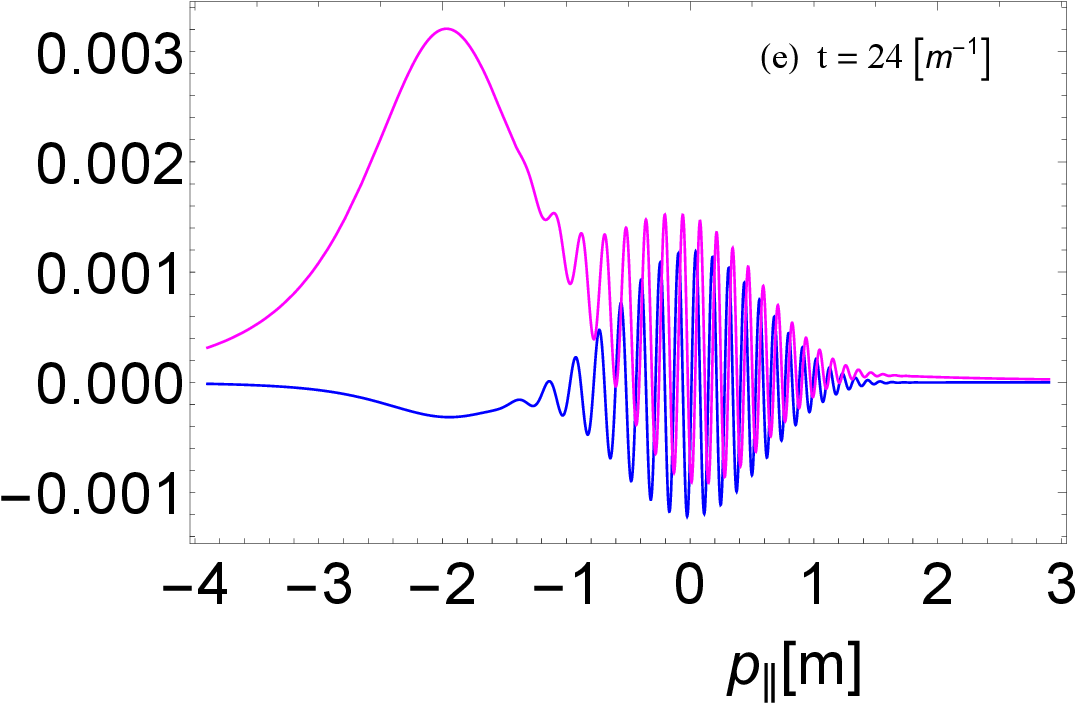}
\includegraphics[width = 1.9in]{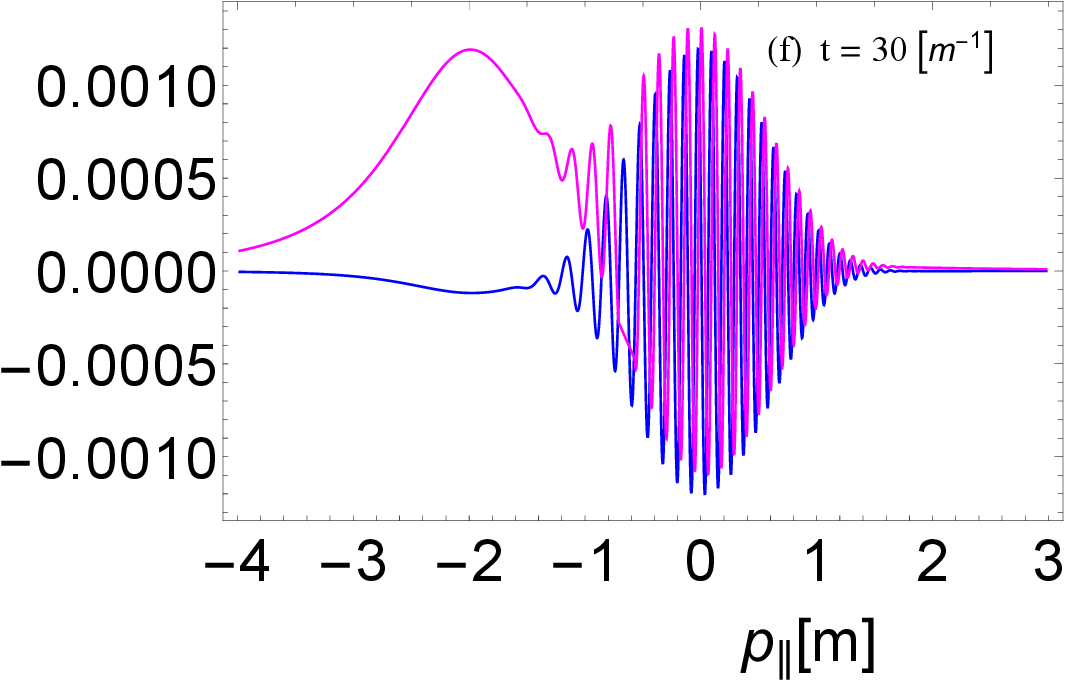}
\includegraphics[width =1.9in]{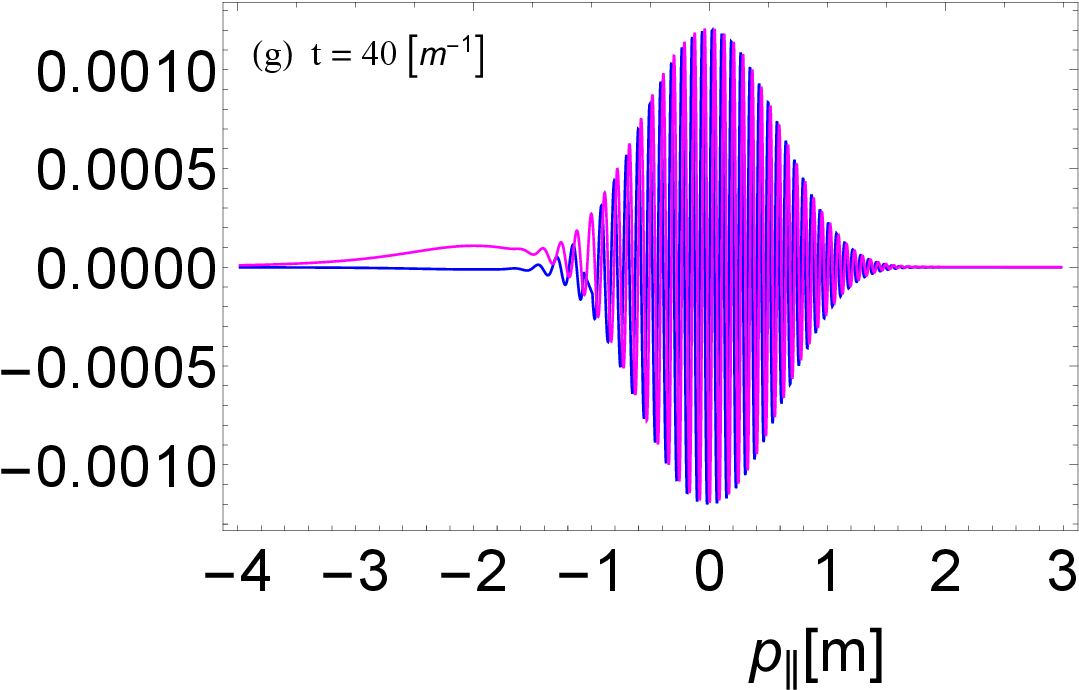}
\includegraphics[width = 1.9in]{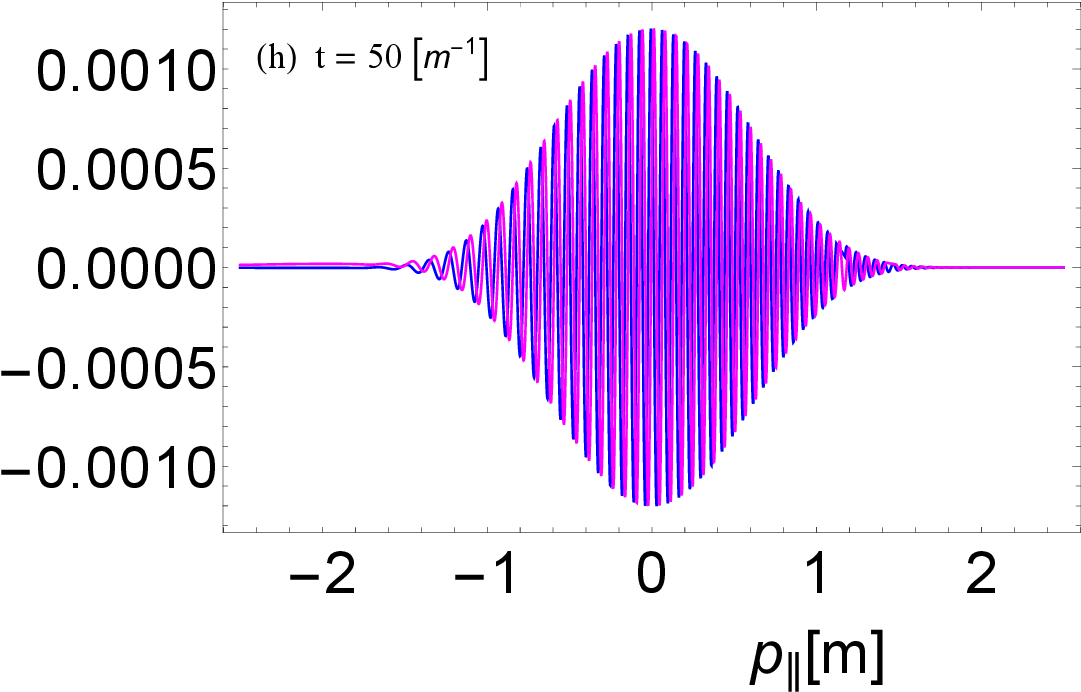}
\includegraphics[width = 1.9in]{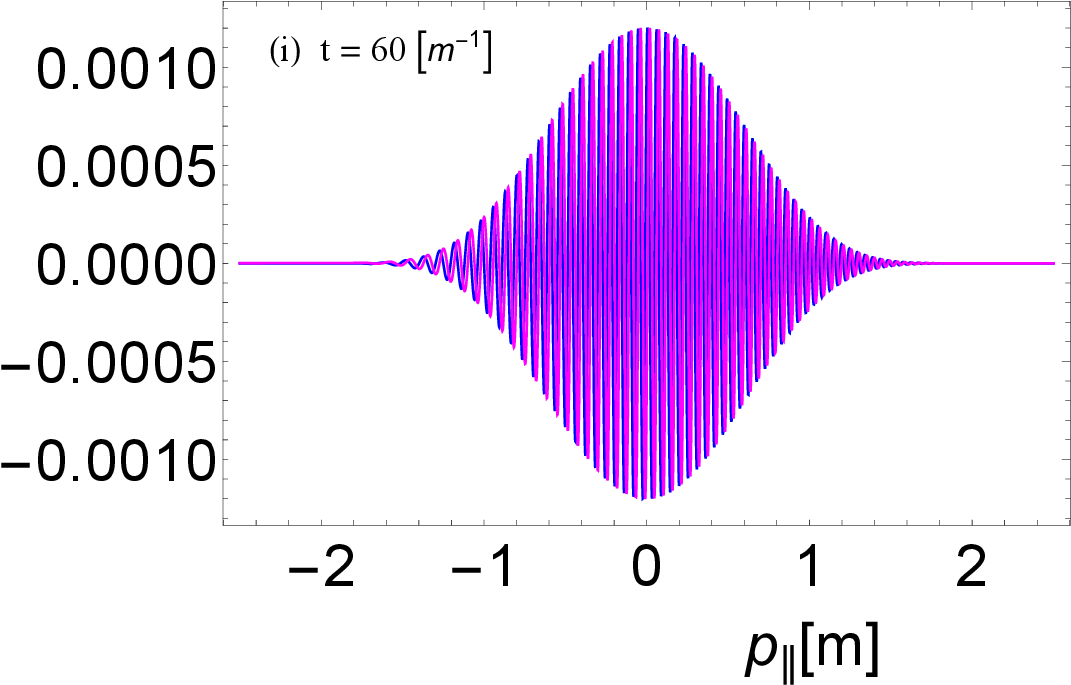}
}
\caption{Momentum spectrum at finite time. Blue curve: polarization function ($u(p_\parallel, t)$) and Magenta curve: de-polarization function ($v(p_\parallel, t)$). The transverse momentum is considered zero, and all units are in electron mass units. The field parameters are $E_0 = 0.2 E_c$ and $\tau = 10 \, [m^{-1}]$.} 
   	\label{fig:LMS_u}
\end{center}
\end{figure}

%%%%%%%%%%%%%%%%%%%%%%%%%%%%%%%%%%%%%%%%%%%%%%%%%%%%%%%%%%%%%%%%
The longitudinal momentum significantly impacts the vacuum polarization function's qualitative traits. To understand its dependence on $p_\parallel$, we comprehensively analyze the momentum spectra of both $u(p_\parallel,t)$ and its associated counter term $v(p_\parallel,t)$.
Figure \ref{fig:LMS_u} shows the momentum spectra of $u(p_\parallel, t)$ and $v(p_\parallel,t).$ In the initial stages of particle formation, specifically at $t < 0$, $u(p_\parallel,t)$ and $v(p_\parallel,t)$ display asymmetric Gaussian peaks with roughly similar profiles. During this time, $v(p_\parallel, t)$ dominates over $u(p_\parallel,t)$. Upon closer examination, it becomes evident that the peaks occur at different positions around $p_\parallel \approx 1$, as illustrated in Fig.~\ref{fig:LMS_u}(a). At $t = 0$, the initial Gaussian-like structure of $u(p_\parallel, t)$ is deformed, becoming bi-modal asymmetric structure. The spectrum $u(p_\parallel, t)$ exhibits a peak at $p_\parallel \approx -0.35 \ [m]$ and a dip at $p_\parallel \approx 0.35 \ [m]$. This indicates a significant asymmetry in the function. On the other hand, the $v(p_\parallel, t)$ spectrum maintains its Gaussian-like unimodal profile, with a peak at $p_\parallel = 0$. During this stage, while the shape of $u(p_\parallel, t)$ undergoes considerable changes and becomes asymmetric, the overall shape of $v(p_\parallel, t)$ remains unchanged. Due to the force factor $ ``e E(t)"$ and corresponding longitudinal quasi-momentum $P(p_\parallel,t) = (p_\parallel - e A(t))$, the spectrum moved to the left side of the origin, and its peak now located at $p_\parallel \approx -2 [m] $ for   $v(p_\parallel,t)$   whereas small dip for  $u(p_\parallel,t)$ with some disruption in the tail ( $ -1< p_\parallel < 1$) as shown in the Figs.\ref{fig:LMS_u}(c-d) near to the transient stage. 
%%%%%%%%%%%%%%%%%%%%%%%%%%%%%%%%%%%%%%%%%%%%%%%%%%%%%%
%%%%%%%%%%%%%%%%%%%%%%%%%%%%%%%%%%%%%%%%%%%%%%%%%%%%%%%%%%
At $t = 24 \ [m^{-1}]$, we observe two distinct structures in the polarization and depolarization functions. The first structure appears at $p_\parallel = -2 \ [m]$ due to earlier particle creation and annihilation events. The second structure, within $-1 < p_\parallel < 1$, exhibits varying amplitude oscillations, with the maximum occurring at zero longitudinal momentum. At this stage, $u(p_\parallel,t)$ and $v(p_\parallel,t)$ compete, as shown in Fig.\ref{fig:LMS_u}(d) and (e).
\par
As we approach the REPP stage, $u(p_\parallel, t)$ and $v(p_\parallel, t)$ exhibit a single oscillating structure resembling a sine or cosine function with varying amplitude. The maximum now appears near $p_\parallel = 0$, dominating compared to the secondary peak (and dip for $u(p_\parallel, t)$) previously present at $p_\parallel \approx -2 \ [m]$. As time progresses further, this left-side structure slowly vanishes, as shown in Figs. \ref{fig:LMS_u}(e) and (f). Eventually, as $A(t)$ reaches a constant value, a balance is achieved between the processes of particle creation and annihilation, as depicted in Fig.\ref{fig:LMS_u}(g).
\par
In the late REPP stage, where the particle distribution function $f(p_\parallel,t)$ is constant, the polarization function $u(p_\parallel,t)$ is balanced by its counterpart $v(p_\parallel, t)$. This results in very regular and rapid oscillations with varying amplitude within the Gaussian envelope, as explicitly shown in Figs.~\ref{fig:LMS_u}(h-i). The overall behavior observed in both $u(p_\parallel, t)$ and $v(p_\parallel, t)$ during this process provides valuable insights into the complex dynamics of particle-antiparticle pair creation and annihilation under the influence of the electric field.
%%%%%%%%%%%%%%%%%%%%
\subsubsection{Examining quantum interference effect in LMS: Role of Vacuum Polarization and Depolarization Functions}
%%%%%%%%%%%%%
During the formation of electron-positron pairs from the quantum vacuum, several processes can take place. Electrons are predominantly created and annihilated simultaneously, resulting in electron acceleration and deceleration throughout the formation of real $e^-e^+$ independent pairs from the virtual $e^-e^+$ pairs.
%%%%%
In these scenarios, the polarization functions $u(p_\parallel, t)$ and its counterpart $v(p_\parallel, t)$ are responsible for the acceleration and deceleration of electrons, respectively. In this context, an electron is typically created in the direction of the external field with positive momentum and is subsequently decelerated. As soon as $p_\parallel < 0$, the electron may be annihilated again. These processes can be understood in terms of $u(p_\parallel, t)$ and $v(p_\parallel, t).$
\par
\begin{figure}[t]
\begin{center}
{
\includegraphics[width = 2.08in]{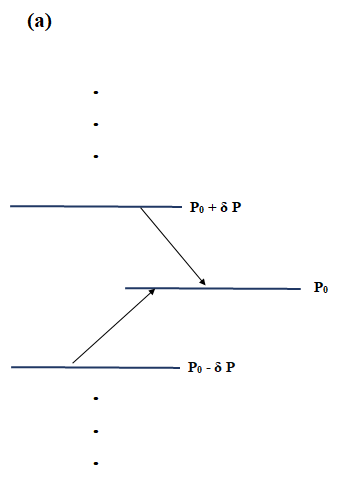}
\includegraphics[width = 2.08in]{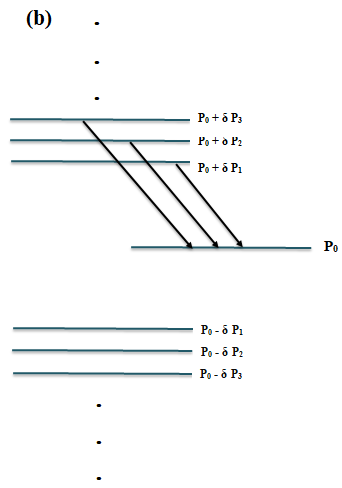}
\includegraphics[width = 2.08in]{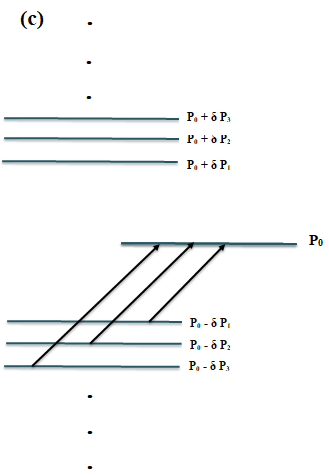}
}
\caption{ A schematic representation of particles occupying different momentum states in various scenarios during particle creation events.
}
\label{MSL}

\end{center}
\end{figure}

 In quasi-momentum space, there are many possibilities that particles can find with specific momentum $p_0$  with time $t_0.$ Suppose particles at a lower momentum level with $(p_0 - \delta p_0)$  in earlier time reach momentum value $p_0$ by acceleration process showed by $u(p_\parallel,t) $ and it is also possible that particles at higher momentum level with $(p_0 + \delta p_0)$ followed by deceleration process ($p_0 - \delta p_0$)  and finally come to momentum value $p_0$  (as shown in figure\ref{MSL}). 
%%%%%%%%%%%%%%%%%%%%%%%%%%%%%%%%%%%%%%%%%%%%%%%%%%%%%%%%%%%%%%%%%%%%%%%%%%%%%%%%%%%%%%%%%%%%%%%%%%%%%%%%%%%%%%%%%%%%%%%%%%%%%%%%%%%%%%%%%%%%%%%%%%%%%%%%%%%%%%%%%%%%%%%%%%%%%%%%%%%%%%%%%%%%%%%%%%%%%%%%%%%%%%%%%%%%%%%%%%%%%%%%%%%%%%%%%%%%%%%%%%%%%%%%%%%%%%%%%%%%%%%%%%%%%%%%%%%%%%%%%%%%%%%%%%%%%%%%%%%%%%%%%%%%%%%%%%%%%%%%%%%%%%%%%%%%%%%%%%%%%%%%%%%%%%%%%%%%%%%%%%%%%%%%%%%%%%%%%%%%%%%%%%%%%%%%%%%%%%%%%%%%%%%%%%%%%%%%%%%%%%%%%%%%%%%%%%%%%%%%%%%%%%%%%%%%%%%%%%%%%%%%%%%%%%%%%%%%%%%%%%%%%%%%%%%%%%%%%%%%%%%%%%%%%%%%%%%%%%%%%%%%%%%%%%%%%%%%%%%%%%%%%%%%%%%%%%%%%%%%%%%%%%%%%%%%%%%%%%%%%%%%%%%%%%%%%%%%%%%%%%%%%%%%%%%%%%%%%%%%%%%%%%%%%%%%%%%%
%\par
At a specific time $t_0$, two possible events can give rise to the quantum interference effect. This expectation is explicitly confirmed in the middle panel of Fig.~\ref{fig:6.1}, where onset-oscillations are observed in a bell-shaped profile in LMS $t \approx 3\tau$. These oscillations in the LMS are observed for a very short duration, during which coherence is maintained in the LMS of $u(p_\parallel, t)$ and $v(p_\parallel, t)$, as evident from Figs.\ref{fig:LMS_u}(e-f). However, these oscillations slowly fade away in the late REPP region around $t = 4\tau$.
The disappearance of oscillations in the particle LMS can be understood in terms of the loss of coherence that was maintained in the LMS of $u(p_\parallel, t)$ and $v(p_\parallel, t)$. After the loss of coherence, the LMS of $u(p_\parallel,t)$ and $v(p_\parallel,t)$ become identical at $t = 6\tau$ (see Fig.\ref{fig:LMS_u} (i)). The envelope of the Gaussian with a cosine or sine behavior results in a smooth unimodal structure particle momentum distribution. 
%%%%%%%%%%%%%%%%%%%
\subsection{Dependence on Transverse momentum}
\begin{figure}[t]
\begin{center}
{\includegraphics[width = 2.54082690187513in]{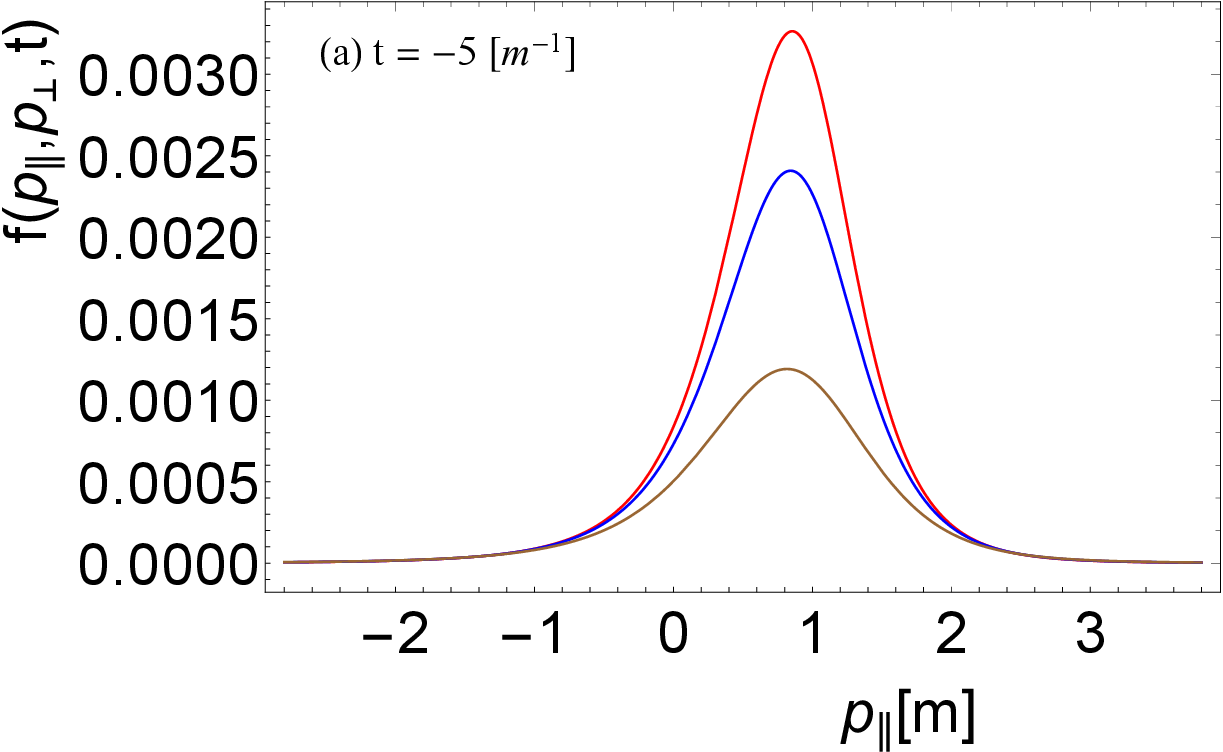}
\includegraphics[width = 2.5408269890187513in]{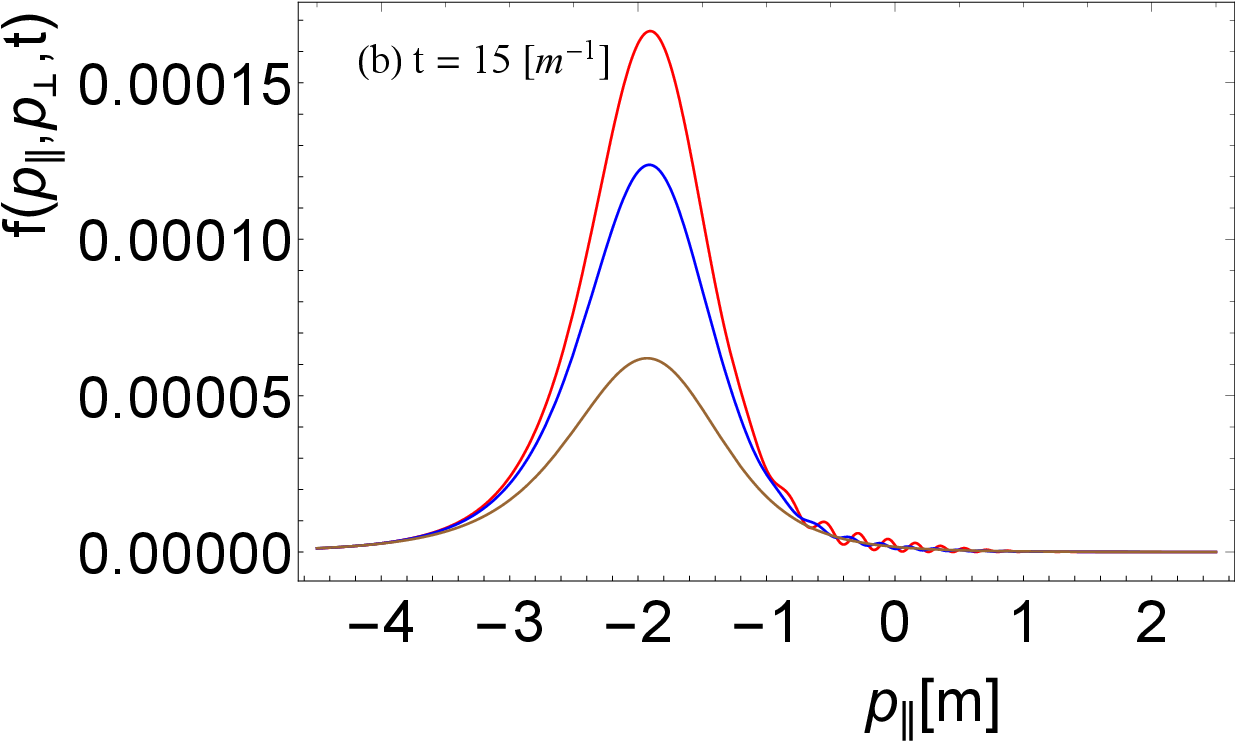}
\includegraphics[width = 2.54082690187513in]{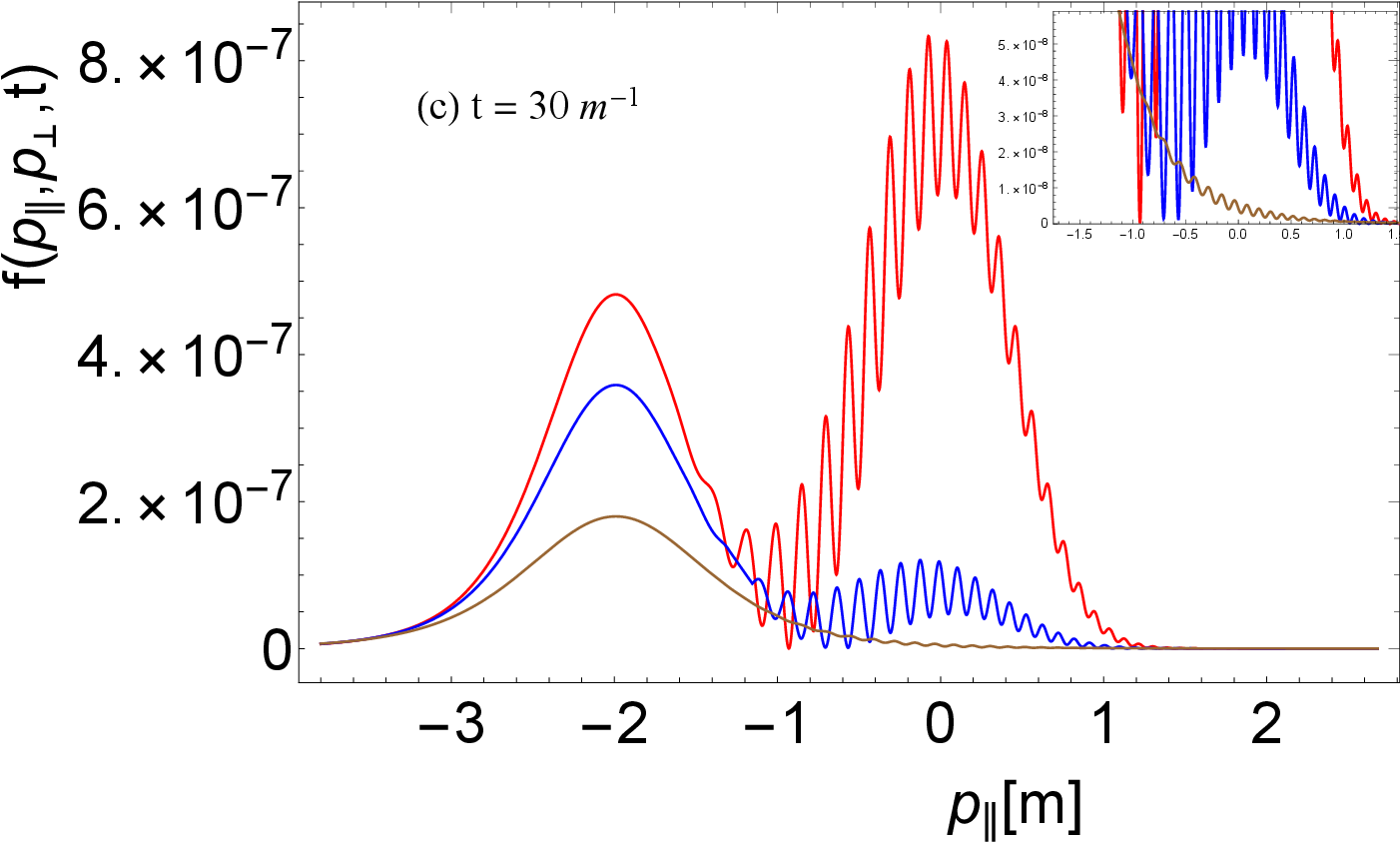}
\includegraphics[width = 2.54082390187513in]{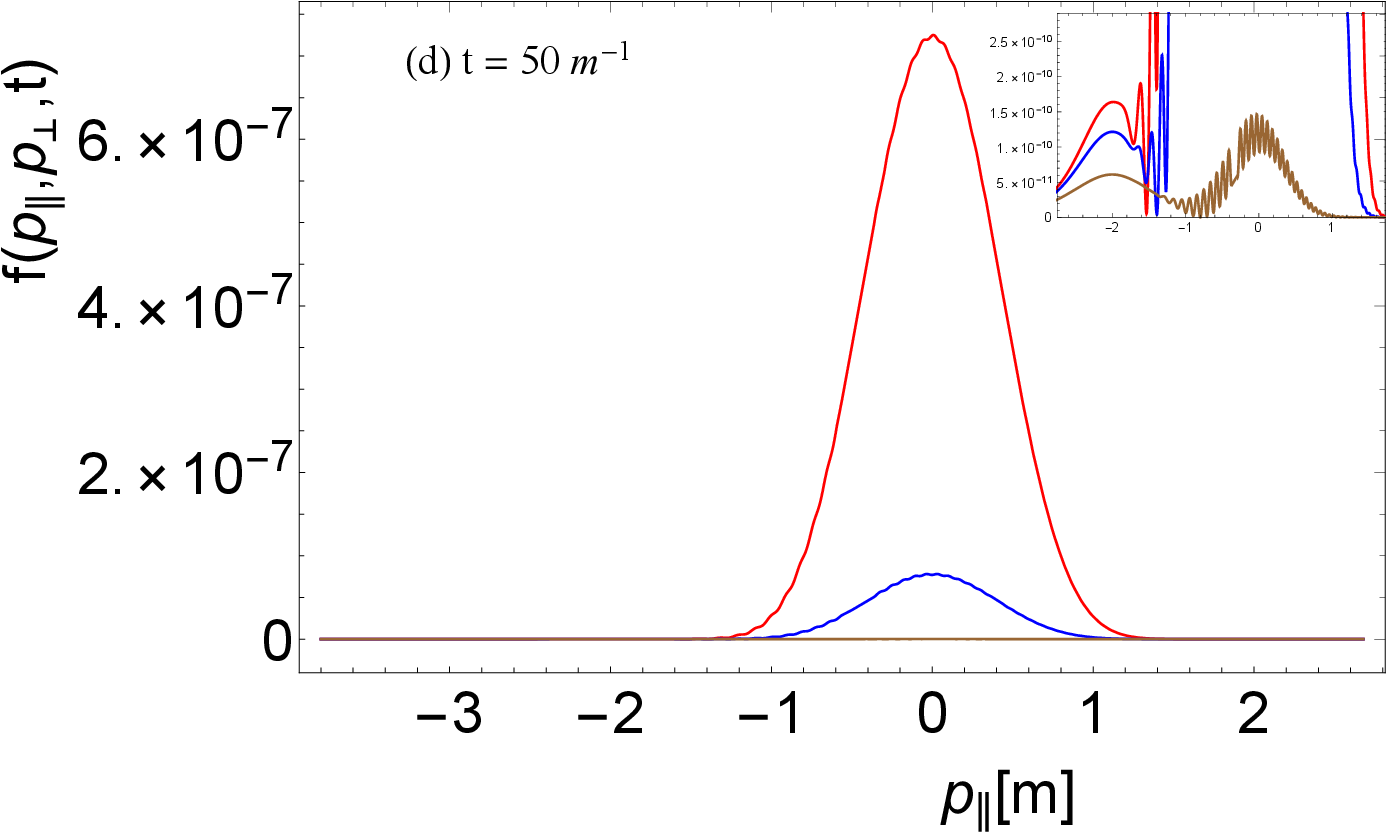}}
\caption{  LMS of created particles for different values of the transverse momentum $p_\perp$ ($p_\perp = 0$ (red), $p_\perp = 0.4$ (blue), and $p_\perp = 0.8$ (brown)). The field parameters are $E_0 = 0.2 E_c$ and $\tau = 10 [m^{-1}]$.
}
\label{fig:6.14}
\end{center}
\end{figure}

%%%%%%%%%%%%%%%%%%%%%%%%%%%%%%%%%%%%%%%%%%%%%%%%%%%%%%%%%%%
In this subsection, we extensively investigate the effect of transverse momentum on 
the LMS. By plotting the time evolution of the LMS for different fixed values of $p_\perp$, the result is displayed in Figure \ref{fig:6.14}. 
%%%%%%%%%%%%%%%%%%%%%%24 NOV 2023 %%%%%%%%%%%%%%%%%%%%%%%%%%%%%%%%%%%%%%%%%
The spectra exhibit a smooth unimodal structure during the initial stages, as depicted in  Fig.~\ref{fig:6.14}(a). Notably, the peak's location shifts, as illustrated in  Fig.\ref{fig:6.14}(b), accordingly with the  $ (p_\parallel - e A(t))= 0$ as explained in section~\ref{sec:Longitudinal momentum}. Within a narrow spectral region for $-1 < p_\parallel < 1,$ some irregular oscillating structure is observed. 
\par
Furthermore, these figures indicate that the peak value of the spectra decreases with the increase in the value of transverse momentum. Fig. \ref{fig:6.14} (c), at $t =30 [m^{-1}]$ spectra shows bimodal distribution. The central peak structure with onset oscillation located at $p_\parallel = 0 $ dominates over another peak at $p_\parallel = -2 [m]$  for a small value of $p_\perp.$
\par
One interesting fact observed was that the central peak magnitude diminished very rapidly compared to other peaks, and that oscillatory structure does not exist for high transverse momentum at this time. The existence of oscillation in the central peak depends on the time taken for particles to reach the on-shell configuration from the initial off-shell state, a process governed by both transverse and longitudinal momenta (see Sec.\ref{temporal}).

%%%%%%%%%%%%%%%%%%%%
\begin{table}[ht]
\centering
\begin{tabular}{ |c|c|c|c| } 
  \hline
  $p_\perp~[m]$ & $t_{cp}~[m^{-1}]$ & $t_{sep}~[m^{-1}]$ & $t_{dis.}~[m^{-1}]$ \\ 
  \hline
   0.00 & 22 & 30 & 50 \\
  \hline
   0.25 & 24 & 32 & 53 \\
  \hline
   0.50 & 31 & 38 & 63 \\
  \hline
   0.75 & 39 & 47 & 71 \\
  \hline
   1.00 & 53 & 60 & 83 \\
  \hline
\end{tabular}
\caption{\label{table3} Effect of transverse momentum $p_\perp$ on the time scales. 
$t_{cp}$ denotes the appearance of a central peak, $t_{sep}$ the time when two peaks become distinctly separated, and $t_{dis.}$ the disappearance of oscillations in LMS.}
\end{table}

%%%%%%
%\caption{ \label{table3} Effect of transverse mode on different time scale. The time scale  $t_{cp}$ appearance of a central peak, $t_{sep} [m^{-1}] $ when two peaks become distinctly separated, and $t_{dis.}$  disappearance of the oscillation in LMS.}
%end{table}

%%%%%%%%%%%%%%%%%%
%%%\newline

Overall, we know that from  ~\cref{temporal} increasing the momentum value particle takes a longer time to reach in REPP stage. So, depending upon the value of $(p_\parallel,p_\perp)$, we can say that different portions of the spectra point to different dynamical stages. The insets of 
Fig.~\ref{fig:6.14} (c) shows only an irregular oscillatory structure in the tail region of the spectra for
$p_\perp = 0.8[m]$, and the central peak structure is not observed. This absence is attributed to the fact that particles are in the QEPP stage for higher transverse values.
At $t = 50 [m^{-1}],$ for $p_\perp = 0$, that oscillation is nearly smooth, but for high value$,p_\perp = 0.8 [m]$ magnitude of peak diminished see Fig.\ref{fig:6.14} (d),  however as shown in insert figure that onset oscillation still observed only magnitude get decreased. From this observation, we can say that the central peak structure does not show the same behavior for a higher value of $p_\perp $. This is exactly what is implied by Fig. \ref{fig:6.14} (d).
As we see that for higher transverse momentum, we still see that central peak with onset oscillation, which means it just reaches the REPP region.
We also observe that for a higher value of $p_\perp$ width of the spectra, $\Delta p_\parallel$ gets decreased, see insert Fig.~\ref{fig:6.14} (d). 
\par
From the occurrence of the oscillation structure, which we previously identified in the time scales discussed in \cref{sec:Longitudinal momentum}, we now observe that the quantum signature also depends on the transverse momentum, as shown in Table 3. Lower values of \(p_\perp\) lead to oscillation patterns becoming visible earlier than those at higher values.

\subsection{Transverse Momentum Spectrum}
\label{sec:TMS}

In the preceding section, we examined how fixing different values of the transverse momentum \( p_\perp \) influences the longitudinal momentum distribution function. We now shift our attention to the transverse momentum distribution itself, focusing on the case where the longitudinal momentum is set to zero, i.e., \( p_\parallel = 0 \). The transverse momentum distribution function of the created pairs at any time \( t \) is denoted by \( f(p_\perp, t) \). 
%%%
\begin{figure}[t]
\begin{center}
{
\includegraphics[width =  1.58802in]{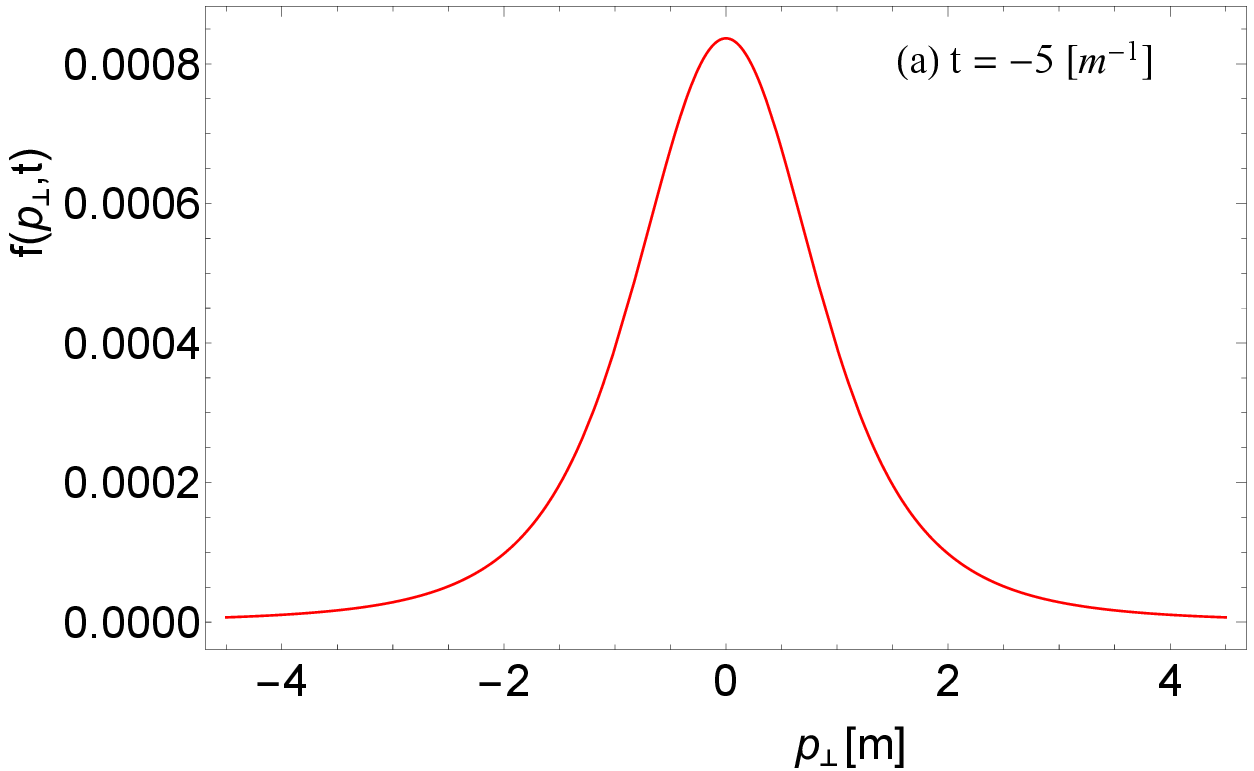}
\includegraphics[width =  1.58802in]{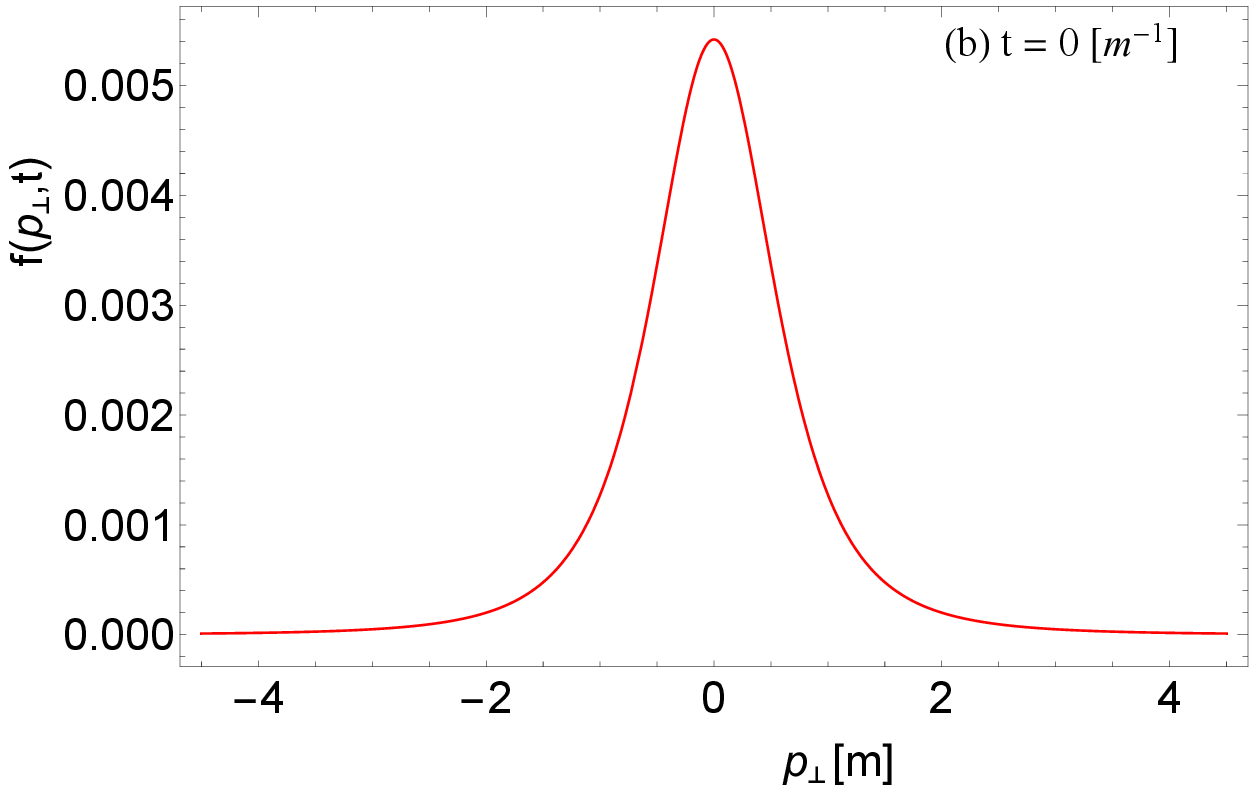}
\includegraphics[width =  1.58802in]{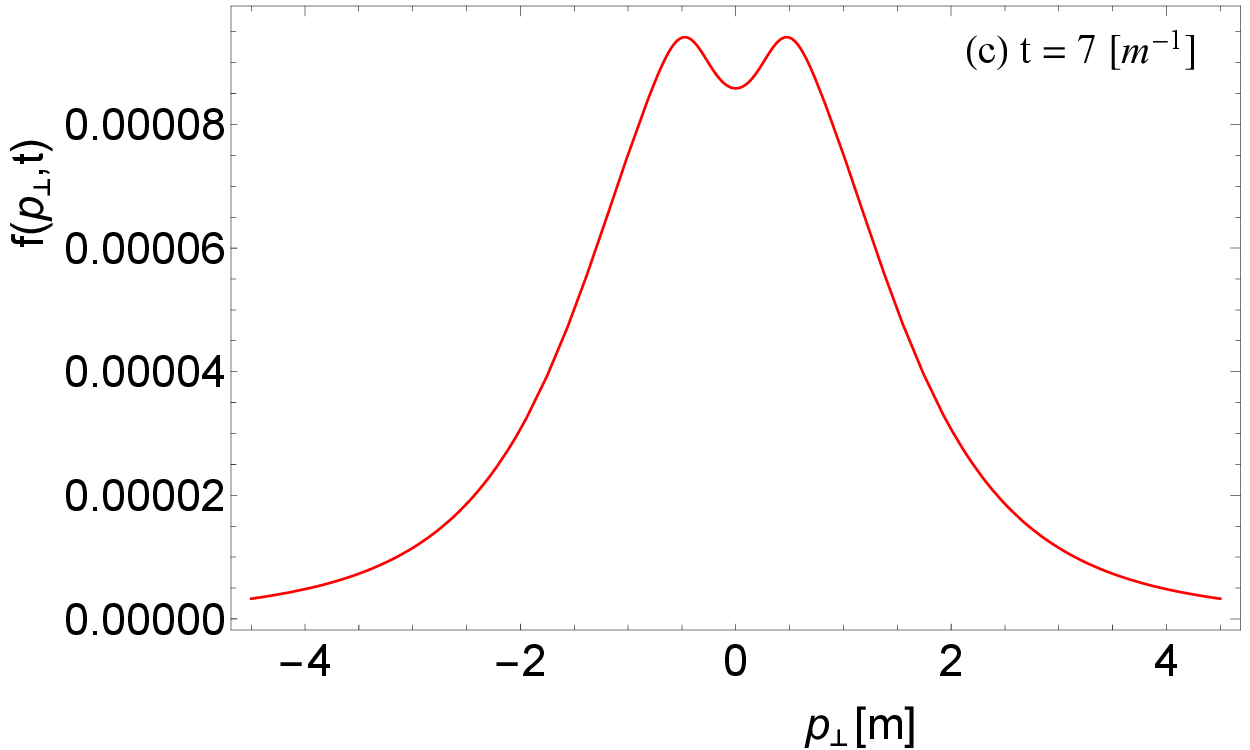}
\includegraphics[width =  1.58802in]{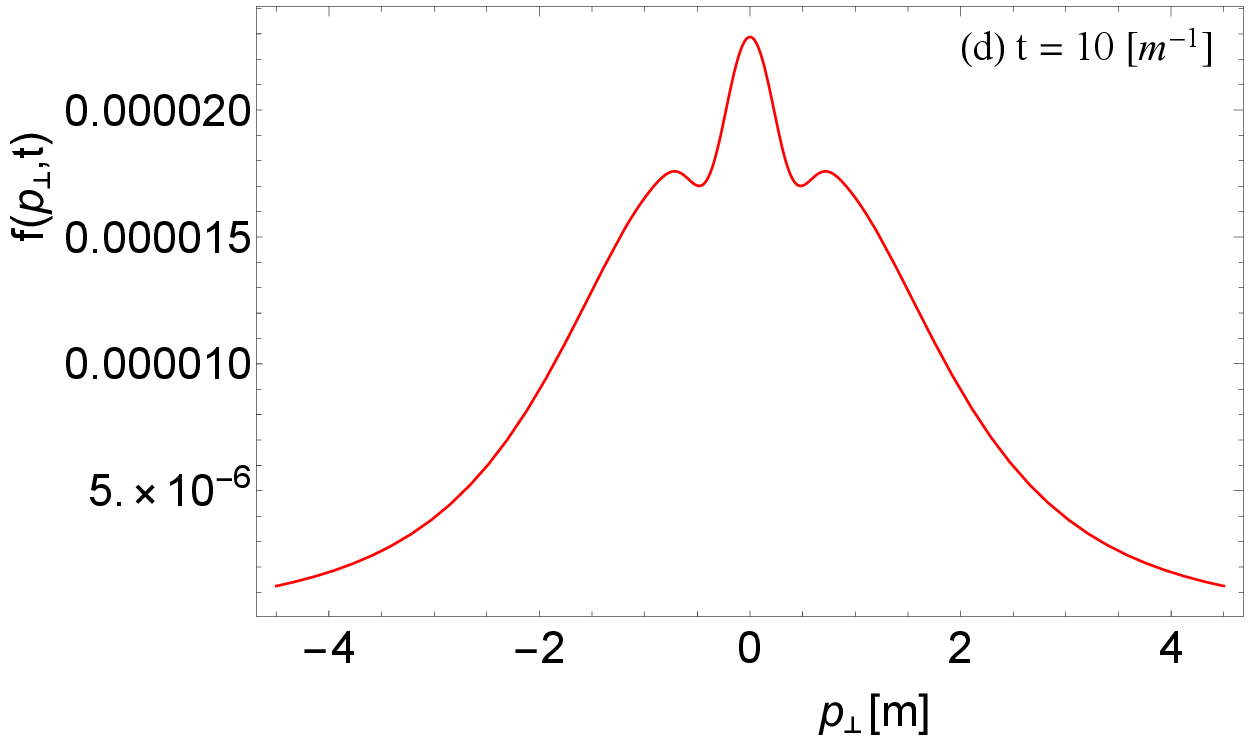}
\includegraphics[width =  1.58802in]{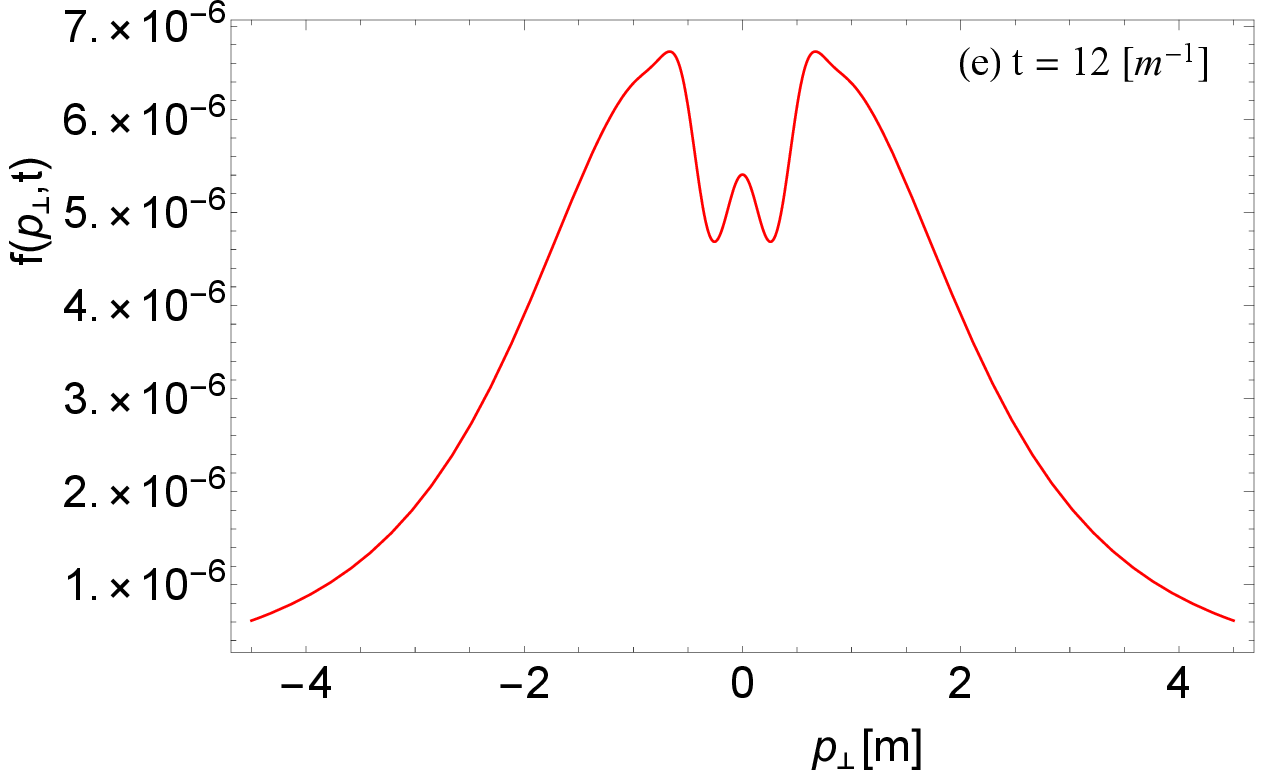}
\includegraphics[width =  1.58802in]{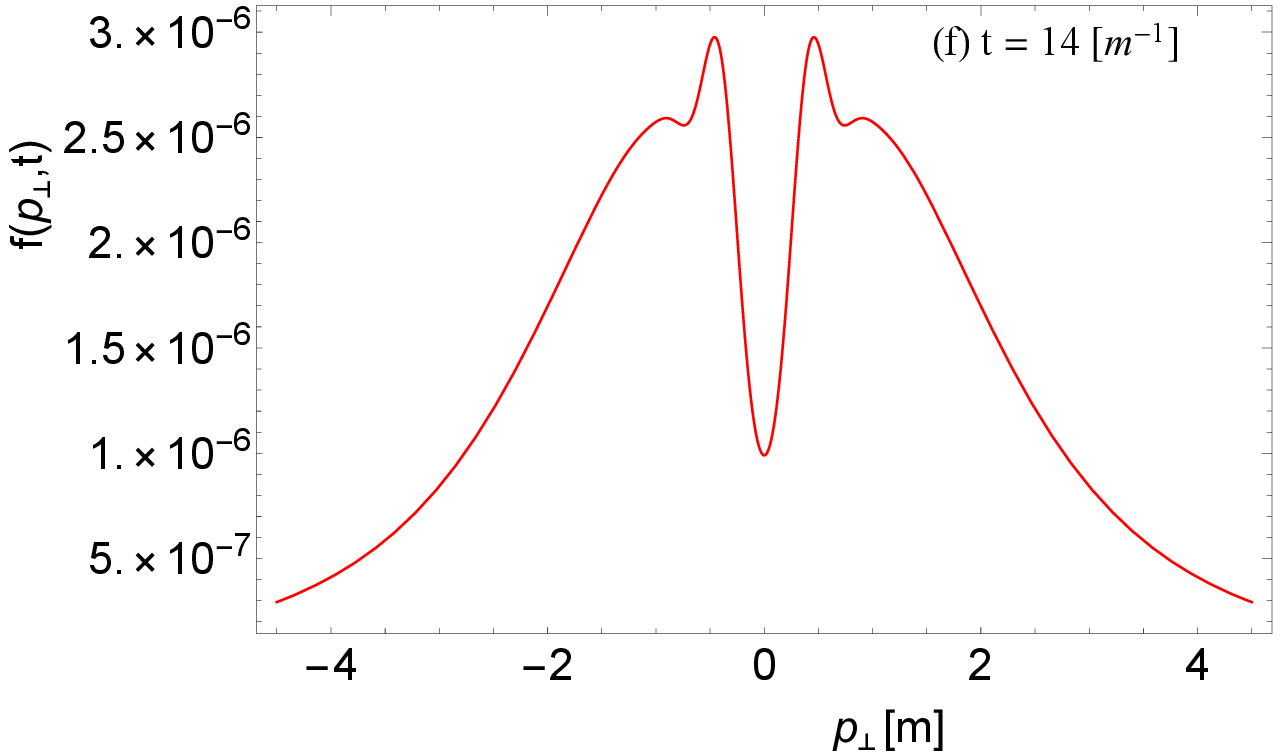}
\includegraphics[width =  1.58802in]{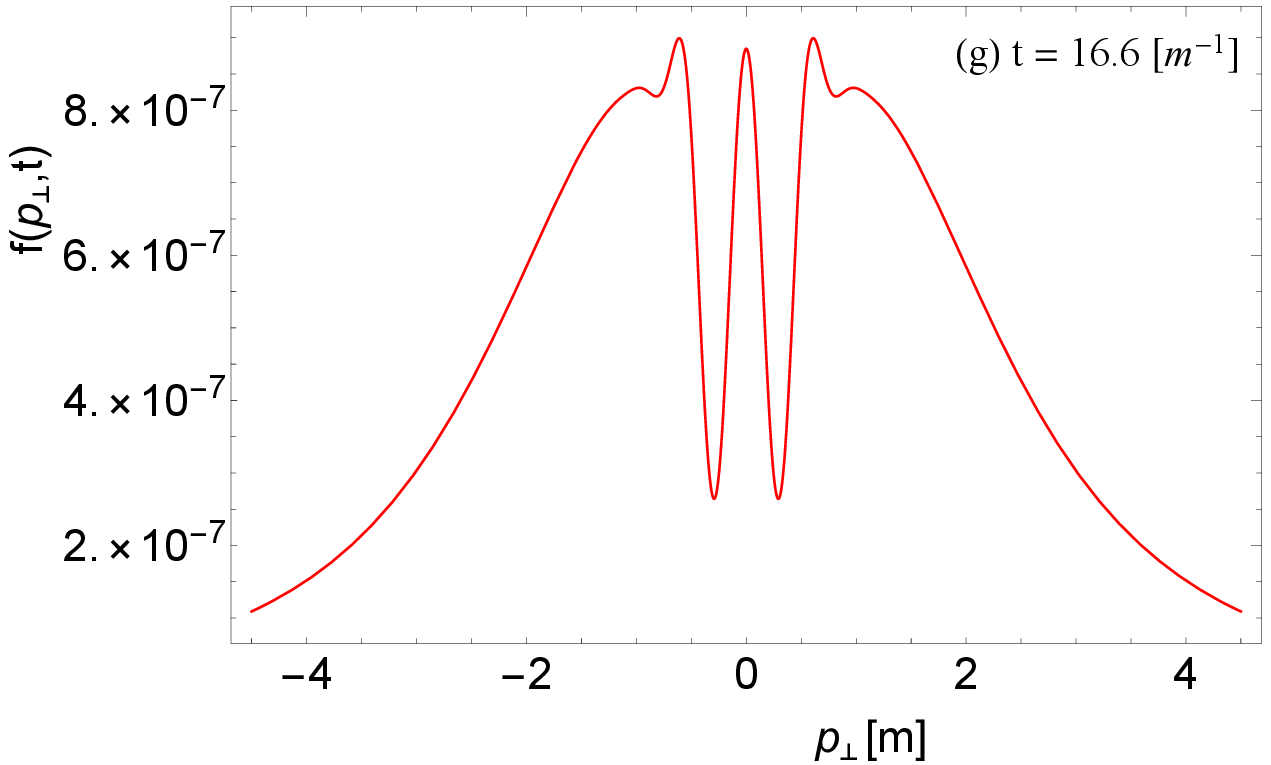}
\includegraphics[width =  1.58802in]{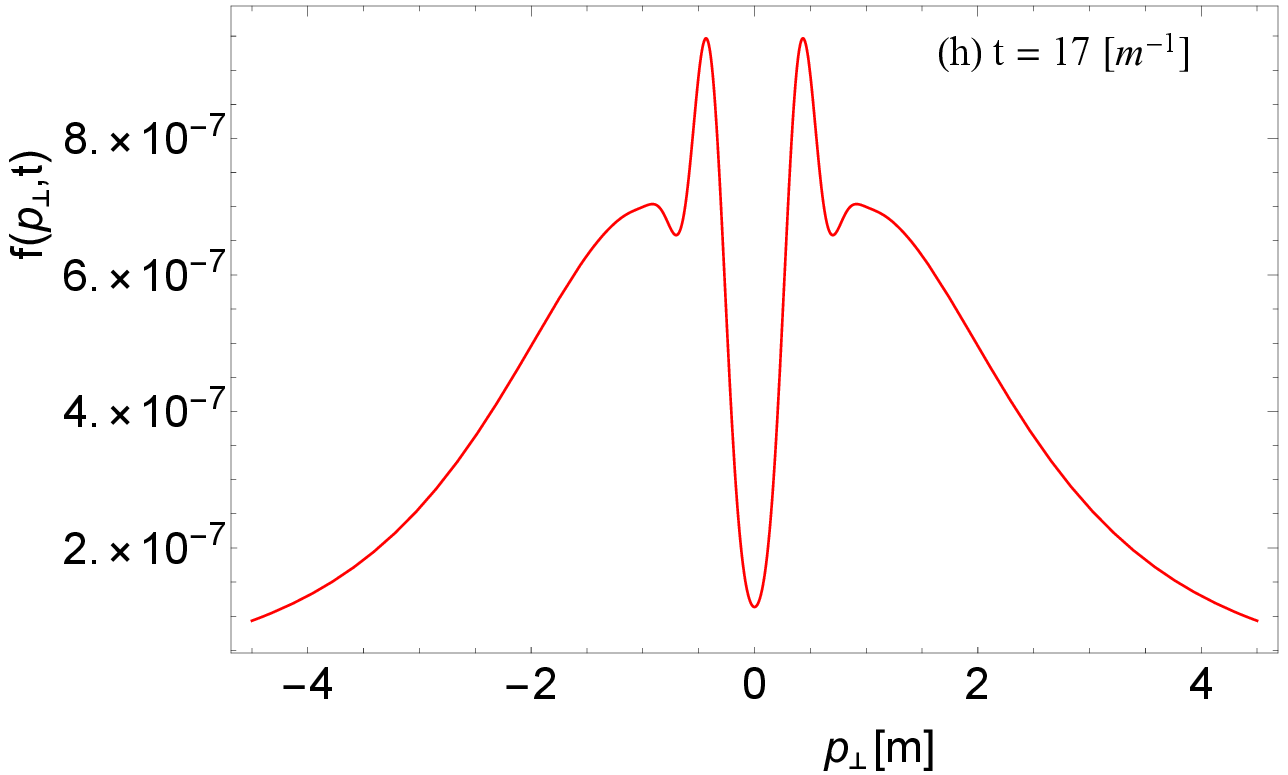}
\includegraphics[width = 1.58802in]{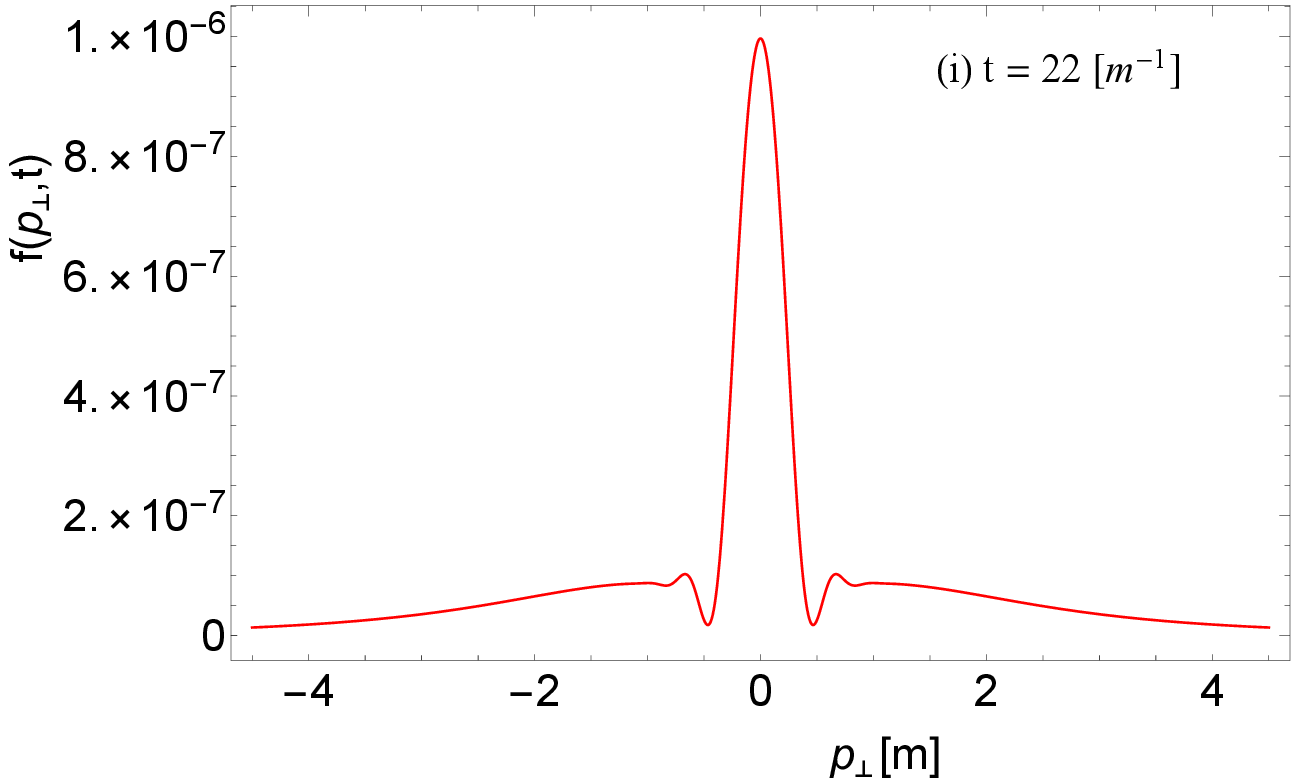}
\includegraphics[width = 1.58802in]{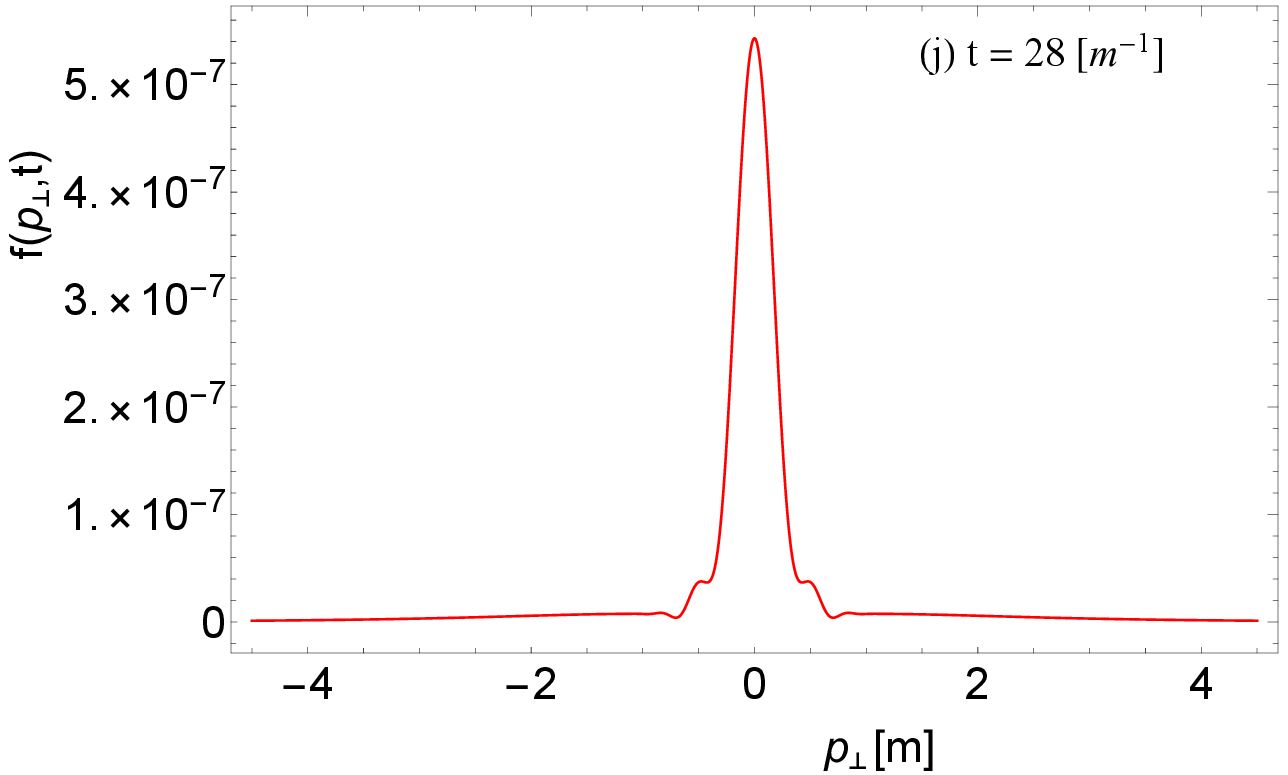}
\includegraphics[width = 1.58802in]{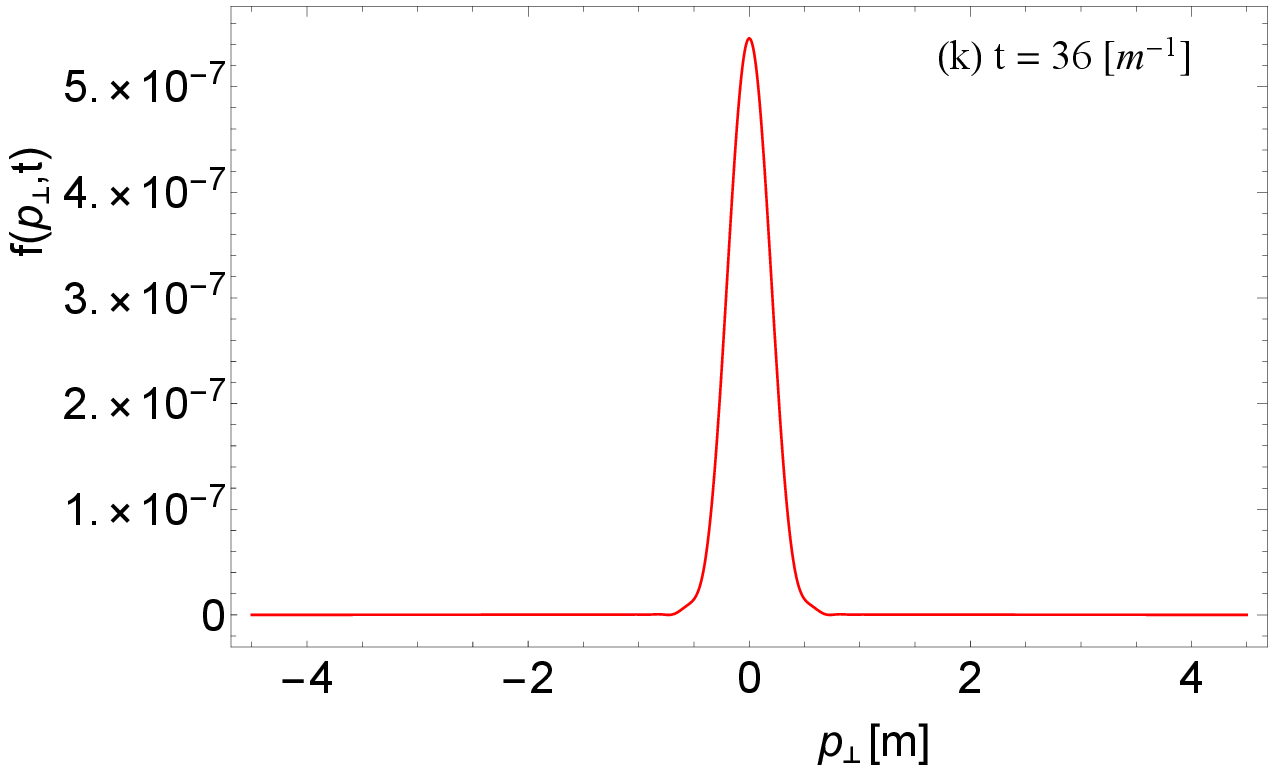}
\includegraphics[width = 1.58802in]{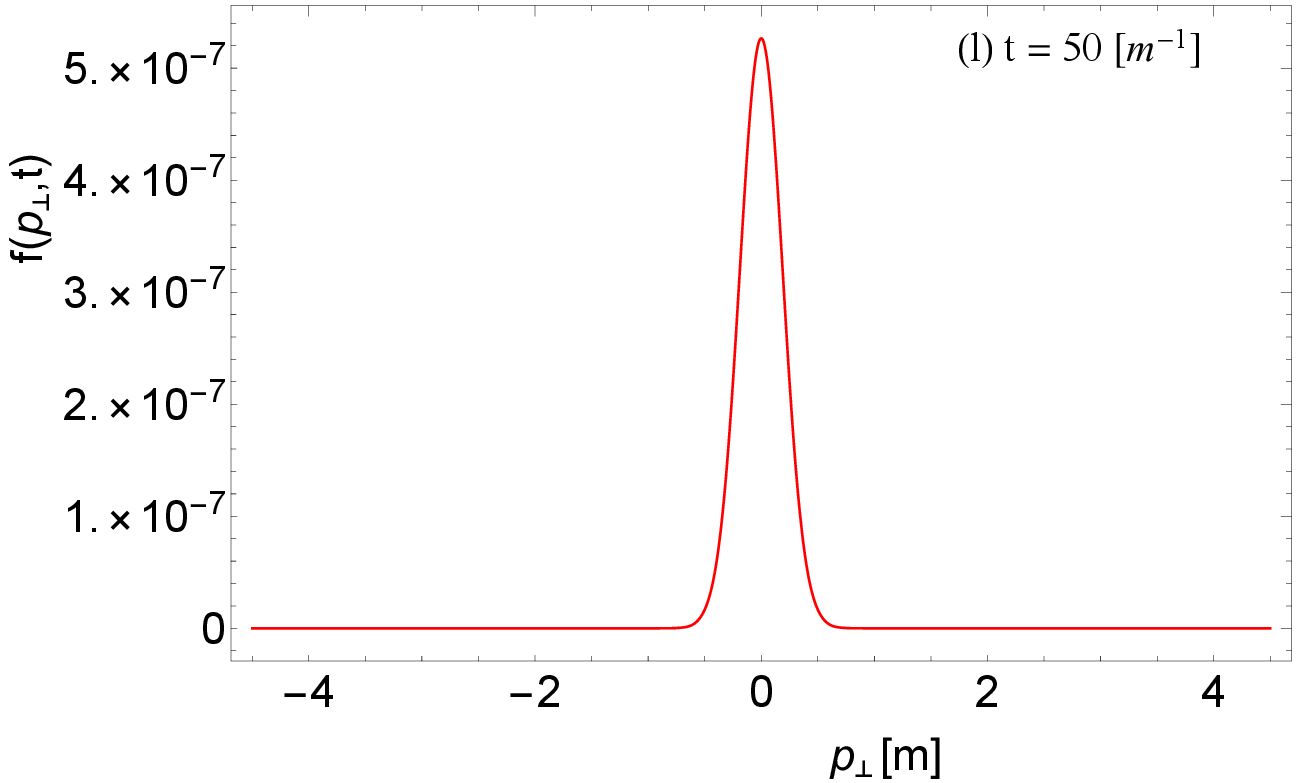}

}
\caption{TMS of created particles in the presence of time-dependent Sauter pulse at different times. 
The longitudinal momentum is set to zero, and all units are in electron mass units.The field parameters are  $E_0=0.2 E_c$ and $ \tau =10 [m^{-1}].$}
   	\label{FG:2}
\end{center}
\end{figure}
\par
Figure \ref{FG:2} shows the time evolution of the TMS for created particles at different times. At the initial time $t = -5 \, [m^{-1}]$, TMS exhibits a smooth, unimodal structure with a peak at $p_\perp = 0$, where the electric field is at $75\%$ of its maximum strength. As the electric field reaches its maximum value at $t = 0$, the spectrum undergoes a dramatic change: both the peak height and width of the spectrum significantly shift, although the peak remains at $p_\perp = 0 $. After the peak reaches its maximum height, it rapidly decreases as the electric field weakens. The single peak structure initially observed at \( p_\perp = 0 \) has disappeared, and new peaks have emerged at \( p_\perp \approx \pm 0.46 \, [m] \) in TMS. As an example, at \( t = 7 \, [m^{-1}] \), TMS exhibits a bimodal profile, as shown in Fig.~\ref{FG:2}(c).
As time progresses, the peaks at \( p_\perp \approx \pm 0.46 \, [m] \) tend to merge, leading to the formation of a peak at \( p_\perp = 0\) with two weakly pronounced peaks, as illustrated in Fig.~\ref{FG:2}(d).
Those weakly pronounced peaks grow over time, while the height of the central peak decreases. As a result, TMS again exhibits a multi-modal structure, and \( f(p_\perp, t) \) reaches its maximum value at non-zero transverse momentum, specifically at \( p_\perp \approx \pm 0.65 \, [m] \), with a small peak still present at \( p_\perp = 0 \), as shown in Fig.~\ref{FG:2}(e).
\par
Overall, TMS displays a substructure (multiple peaks ) profile where the dominant peak is governed by non-zero values of \( p_\perp \). Upon closer inspection, TMS reveals that the peak at \( p_\perp = 0 \) becomes invisible again, and the locations of the dominant side peaks undergo slight modulation at \( t = 14 \, [m^{-1}] \). Specifically, the dominant peak at \( p_\perp = 0.67 [m]\) splits into two weakly pronounced peaks: one at \( p_\perp = 0.45 [m]\) and the other at \( p_\perp = 0.94 [m] \), as depicted in Fig.~\ref{FG:2}(f). As time progresses, the maximum peak at zero transverse momentum reappears. Although this central peak stands out, there are also side peaks (not shown in the figure). These side peaks eventually dominate as the central peak fades, just like in the earlier stages. In this way, the side peaks and the central peak reach the same height, as shown in the figure at \( t = 16.6 [m^{-1}] \), and the spectrum shows a three-peak pattern.
\par
We can say that the spectrum of created particles at a given time exhibits key features, indicating that the spectrum undergoes continuous changes and reveals two distinct structures. One scenario shows a maximum peak at \( p_\perp = 0\), suggesting particles with no transverse momentum have a high probability of creation. The other case displays a maximum peak of the spectrum at a non-zero \( p_\perp \) value. In both cases, there are multiple peaks, showing the complex behavior of the spectrum over time, as seen in figure \ref{FG:2}(a)-(h).
\par
The effect of created pairs tends to have a peak in the momentum distribution function at non-zero transverse momentum, as discussed by Krajewska et al.~\cite{Krajewska:2018lwe}. It is demonstrated that the transverse momentum distribution of real pairs exhibits an off-axis maximum above the one-photon threshold frequency (or high frequency); otherwise, the maximum for the transverse distribution function occurs at \( p_\perp = 0 [m] \) in the presence of an oscillating time-dependent electric field. This study gives the clue that our case of off-axis maxima or multi-modal structure may also be observed due to multiphoton transition processes.
%%%%
To confirm this, we analyze the probability of pair production by calculating the amplitude in the two-level equivalent formalism ~\cite{PhysRevA.64.013414,Krajewska:2018lwe}. We decompose the time-dependent electric field \(E(t)\) via the Fourier transformation into cosine and sine components. This decomposition isolates individual frequency components of the field, allowing us to study their effects on pair production at finite times. 
For a given frequency \(\nu\), the corresponding electric field amplitude $ \tilde{E} (\nu)$ facilitates the evaluation of the particle production spectrum at a specific finite time \(t\) (see \cref{fourier}). 
The spectrum exhibits multiple peaks, forming a multimodal profile. Each peak corresponds to distinct energy or momentum exchanges between the field and the particle pairs, characteristic of multi-photon absorption.
By considering different frequencies \(\nu\) and their respective field amplitudes, we compute the pair production probability for each case. The total probability is obtained by summing the probabilities across different frequencies and then squaring the sum to derive the final result. The resulting spectrum features distinct peaks at various transverse momenta \(p_\perp\). 
%At lower transverse momenta, the peaks are more pronounced, while at higher momenta, the peak heights diminish progressively. This suppression aligns with the multi-photon process, where higher momentum states become less populated as the number of absorbed photons increases. 
The peaks appear at non-zero transverse momenta, with their height and position dependent on the specific parameters of the electric field and the timing of the spectrum evaluation (see appendix \ref{fourier}). The Fourier decomposition approach not only elucidates the origin of multiple peaks in the momentum spectrum but also establishes a connection between these features and the multi-photon process.

After time $t > 2 \tau$, the transverse momentum distribution function has a peak at $p_\perp = 0 \, [m]$, and the behavior of pair production governed by the tunneling process for $\gamma < 1$ appears as shown in Figure \ref{FG:2}(i)-(l).
At $t \approx 22 [m^{-1}]$, the width of the momentum spectra changes with time. The half-width of the $p_\perp-$ distribution is determined by the field strength at that time, which is explicitly confirmed by Fig.~\ref{FG:2}(c)-(I).

%%%%
%the dominant peak is solely governed by zero transverse momentum, and the role of non-zero momentum nearly fades away, though it remains visible as small side peaks as shown in Fig.\ref{}.
%%%%%%%%%%%%%%%%
As the particle reaches the  REPP stage, the spectrum shows a prominent peak at zero transverse momentum, accompanied by weakly pronounced peaks at $p_\perp \approx \pm 0.67 [m]$ and $p_\perp \approx \pm 0.84 [m]$  as seen in the Fig.~\ref{FG:2}(i). 
 As the electric field decays to $98.6\%$ of its maximum value, the width of the $p_\perp$ distribution becomes of the order of $1[m]$, and the substructure in TMS disappears ( see Fig.\ref{FG:2} (j)). At this stage, the TMS shows a smooth Gaussian-like distribution, with the maximum value $f(p_\perp,t)$ occurring at $p_\perp = 0$ as shown in Fig.\ref{FG:2}(k). In the absence of an electric field, the spectrum exhibits a single peak  Gaussian-like structure that remains unchanged, and also the distribution function $f(p_\perp)$ can be well understood by assuming that particle creation is exponentially suppressed with $\exp{(-\frac{m^2 + p_\perp^2}{eE_0})}$ see Fig. \ref{FG:2}(l) momentum spectrum shows a Gaussian distribution. This contrast highlights the dynamic evolution of substructures in the initial stages, contrasting with the smooth Gaussian-like structure observed in the REPP stage.
\newline
Similar kinds of non-trivial features in the transverse momentum distribution are observed, depending on the details of the electric field parameters discussed by Bechler et al.'s work \cite{Bechler:2023kjx}, in which the TMS of generated particles reveals a fascinating characteristic—off-axis maxima above the one-photon threshold frequency (or high frequency). The authors further discuss the significant changes in energy distribution between created particles' longitudinal and transverse motion. This raises questions about the feasibility of pair formation through tunneling theory, given the substantial distribution function values at non-zero transverse momentum but for low frequencies. In our case, we can see that the asymptotic distribution function is described as the Schwinger-like formula, as we see later in \cref{TMSAPPFN}. Also, the transverse momentum distribution function shows a maximum value where the transverse momentum becomes negligible. 
An intriguing aspect is that the tunneling time of particles is altered due to the increase in transverse momentum. This observation is substantiated by the fact that the occurrence of the transient stage is influenced by the momentum values $(p_\parallel, p_\perp)$, as explained in \cref{temporal}.
\par
\begin{figure}[t]
\begin{center}
{
\includegraphics[width =  1.858802in]{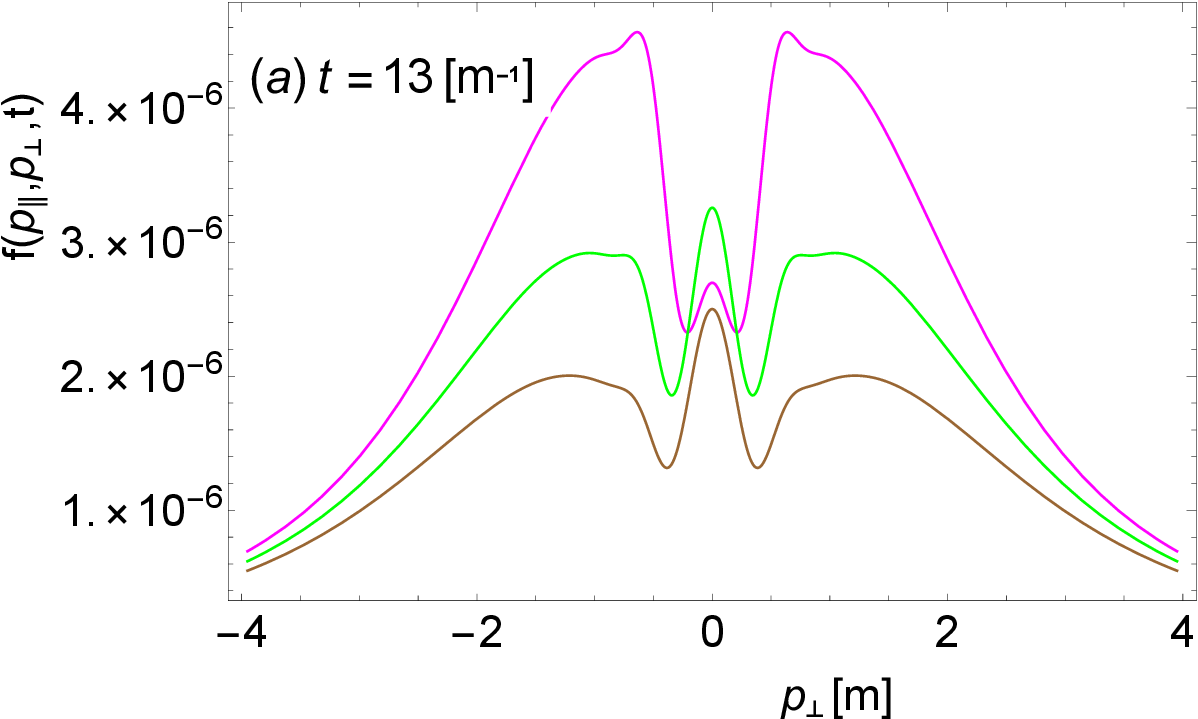}
\includegraphics[width =  1.858802in]{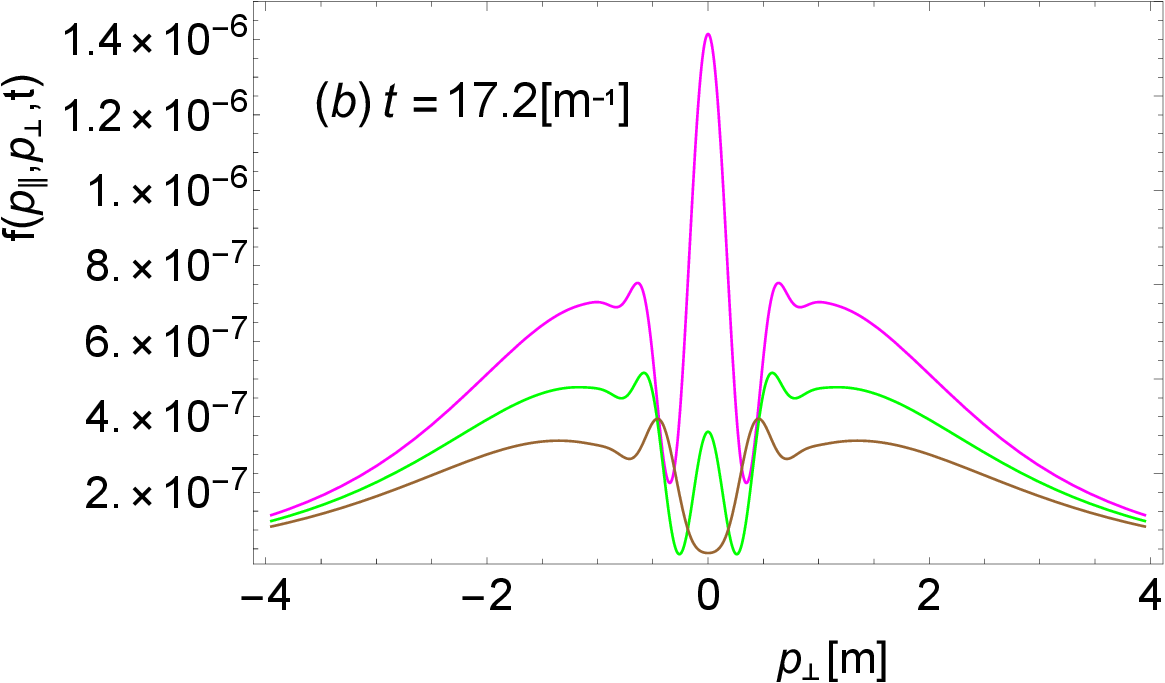}
\includegraphics[width =  1.858802in]{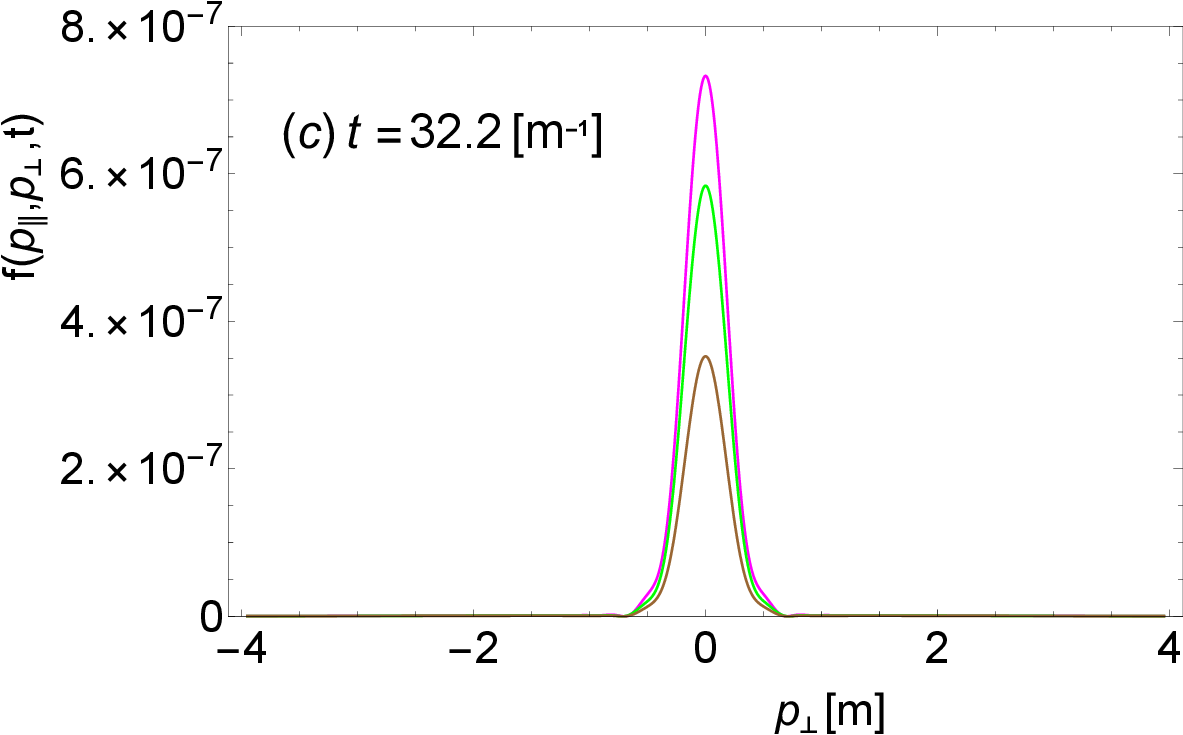}
}
\caption{TMS of created particles for different values of the longitudinal momentum $p_\parallel$ ($p_\parallel = 0.1$ (magenta), $p_\parallel = 0.3$ (green), and $p_\parallel = 0.5$ (brown)). The field parameters are $E_0 = 0.2 E_c$ and $\tau = 10 \, [m^{-1}]$.
}
   	\label{FG:2.a}
\end{center}
\end{figure}

Next, we investigate the impact of longitudinal momentum on TMS at finite time. We observe the interesting feature that different longitudinal values manifest distinct substructures in the spectrum at a given time, as explicitly verified from Figs.~\ref{FG:2.a}(a)-(b). The most common feature observed is that the magnitude of the distribution function decreases for higher longitudinal momentum. Before particles reach the on-shell condition, the spectra of created particles at times $t > 3 \tau$ exhibit a dramatic change in structure, contingent on the value of $p_\parallel.$ For \( p_\parallel = 0.1 [m]\) and \( p_\parallel = 0.3 [m] \), the maximum peak value occurs at vanishing transverse momentum. In contrast, 
%for \( p_\parallel = 0.5 [m] \), the maximum peak is observed at non-vanishing transverse momentum, as shown in Fig.~\ref{FG:2.a}(b).
for \( p_\parallel = 0.5 [m] \), the TMS peak shifts to \( p_\perp \approx 0.6 [m] \) (Fig.~\ref{FG:2.a}(b)), indicating energy sharing between modes.
%At this finite time, we observe the sharing of longitudinal and transverse motion.
At this finite time, we observe a sharing of longitudinal and transverse motion, indicating that the two modes do not evolve independently. A similar interplay between longitudinal and transverse degrees of freedom has been reported in Ref. 
\cite{Majczak:2024hmt,Bechler:2023kjx}.

%within the scattering-matrix description of the dynamical Sauter–Schwinger process.

%We observe a clear sharing between the longitudinal and transverse motion. A similar type of mode sharing between these two degrees of freedom has also been reported in Ref. [Bechler et al., Acta Phys. Pol. A 143, S18 (2023); Majczak et al., Scattering matrix approach to the dynamical Sauter–Schwinger process].
The qualitative change in the shape of the spectrum, due to the influence of longitudinal momentum. Specifically, it requires more time for the quasi-particle to evolve into a real particle with the inclusion of longitudinal momentum when analyzing the TMS of quasi-particles at finite time. In this sense, we can say that the motion of the longitudinal and transverse directions is strongly coupled, and its impact is easily seen in  Figs.\ref{FG:2.a}. 
These concerns can be understood by examining the highly oscillating phase kernel $\cos{(\Theta_{\bm p}(t_0,t))}$ present in QKE ~\cite{Banerjee:2018fbw}. This kernel depends on the quasi-energy $\omega(\bm{p}, t)$, which connects the longitudinal and transverse motions of the created particles.
At the REPP stage, the substructure in the spectrum disappears, and the peak height for higher longitudinal momentum value is consistently lower, shown in Fig.~\ref{FG:2.a}(c). 

%These concerns can be understood through the highly oscillating phase kernel $\cos{\Theta_{\bm p}(t)}$ that present in quantum kinetic equation (QKE) \cite{Banerjee:2018fbw} which depends on the quasi-energy $\omega(p_\parallel, p_\perp, t)$  through which longitudinal and transverse motion of created particles are related. 

%At the REPP stage, the substructure in the spectrum disappears and the peak height for higher longitudinal momentum value is consistently lower shown in Fig.\ref{FG:2.a}(c). 
%%%%%%it means that less number of particles created with larger momentum which well-known fact.
%%%%%%%%%%%%%%%%%%%%%%%%%%%%%%%%%%%%%%%%%%%%%%%%%%%%%%%%%%%%
%%%%

%%%
%%%%%%%%%%%%%%%%%%%%%%%%%%%%%%%%%%%%%%%%%%%%%%%%%%%%%%%%%%%%%%%%%%%%%%%%%%%%%%%%%%%%%%%%%%%%%%%%%%%%%%%%%%%%%%%%%%%%%%%%%%%%%%%%%%%%%%%%%%%%%%%%%%%%%%%%%%%%%%%
\subsubsection{Approximate expression for  transverse momentum distribution function }
\label{TMSAPPFN}
\begin{figure}[t]
\begin{center}
{
\includegraphics[width =  2.58802in]{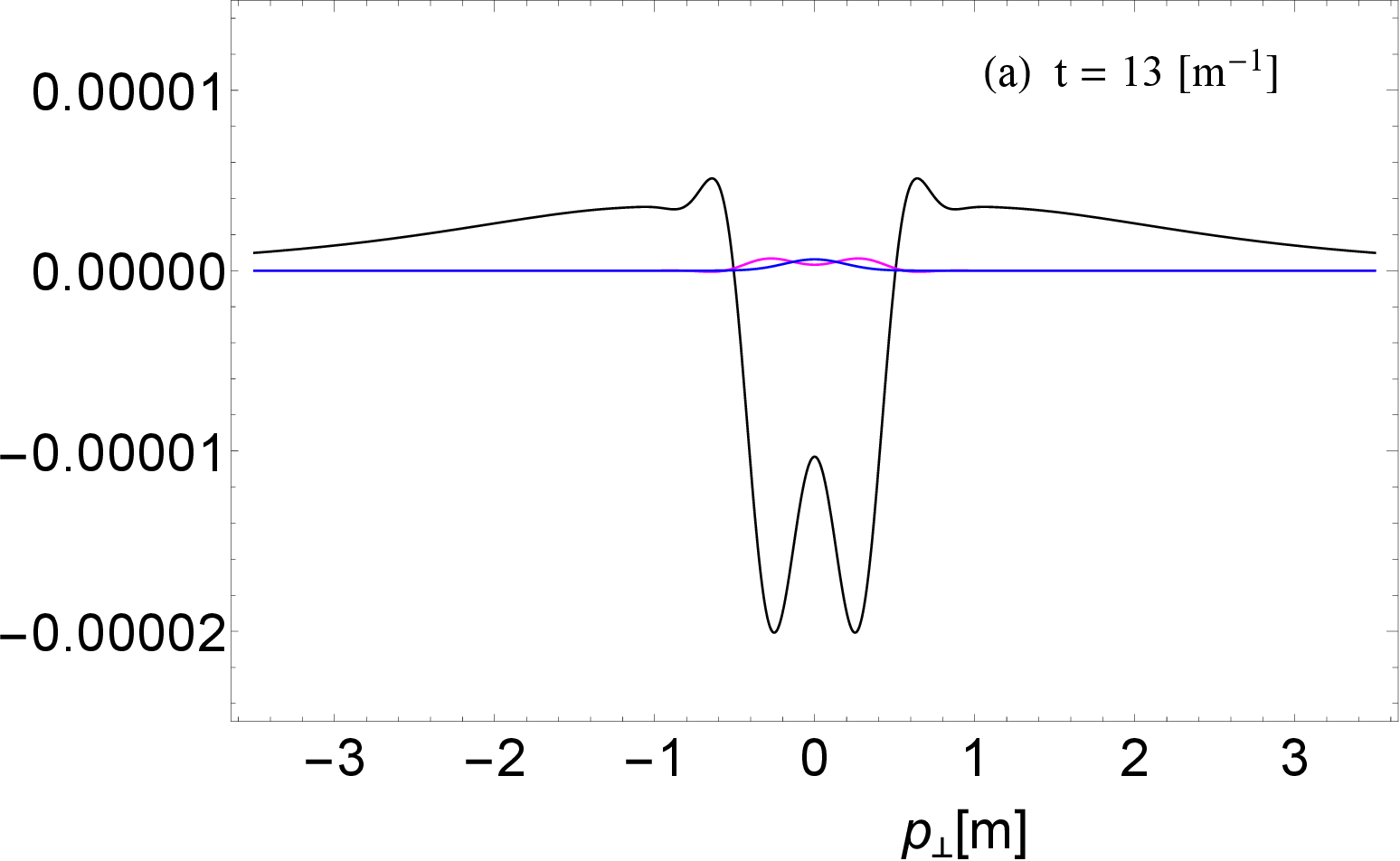}
\includegraphics[width =  2.58802in]{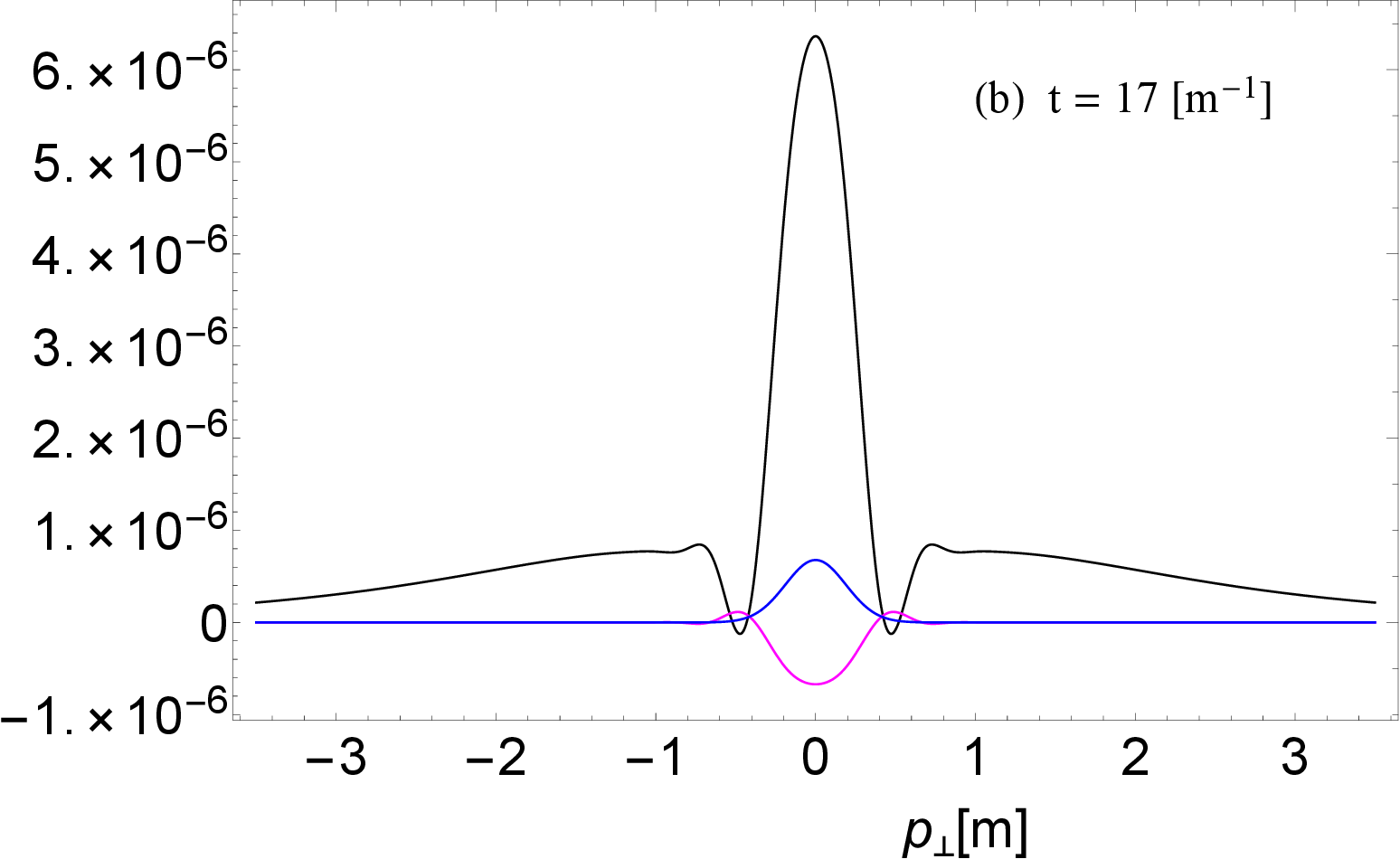}
\includegraphics[width =  2.58802in]{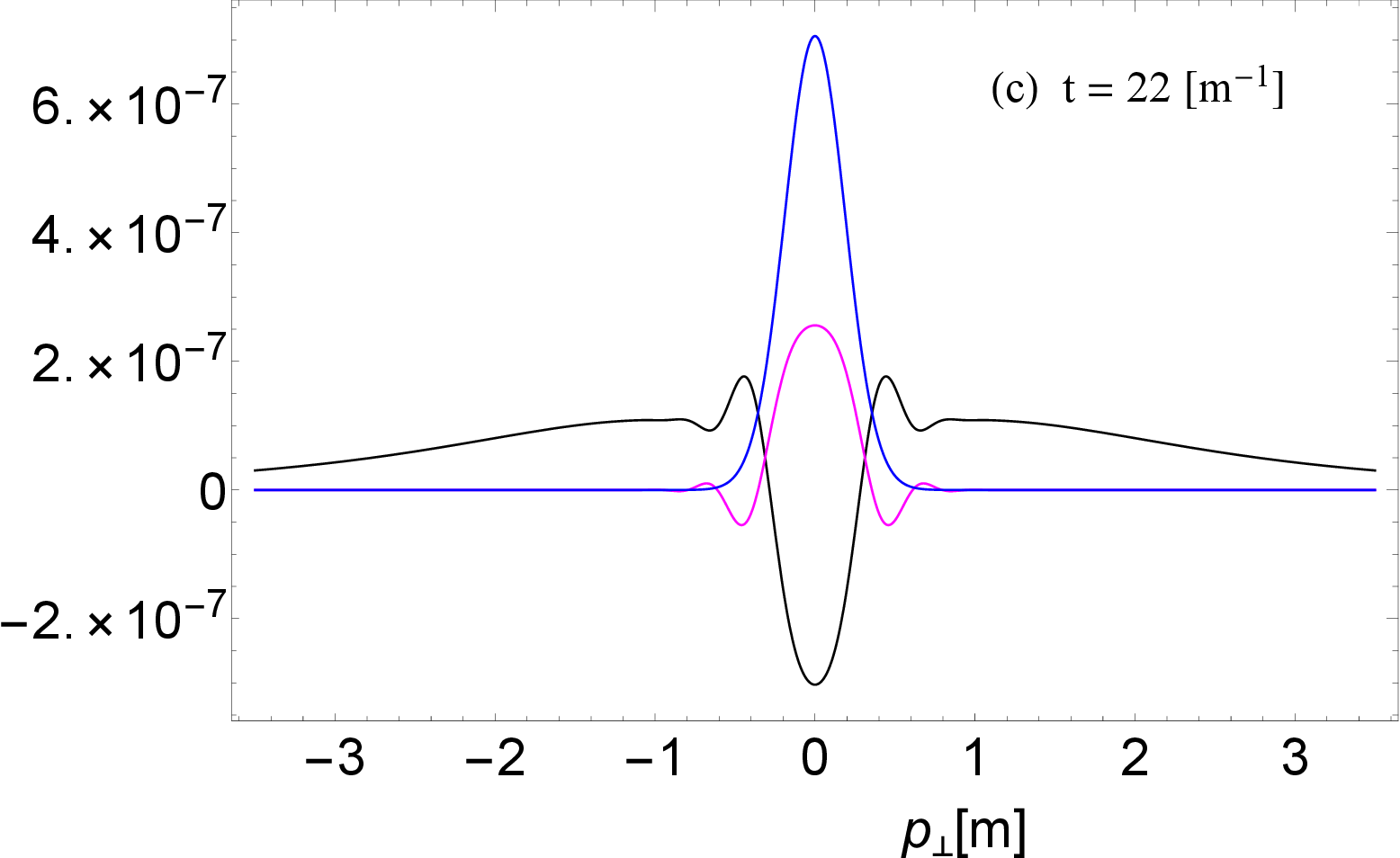}
\includegraphics[width =  2.58802in]{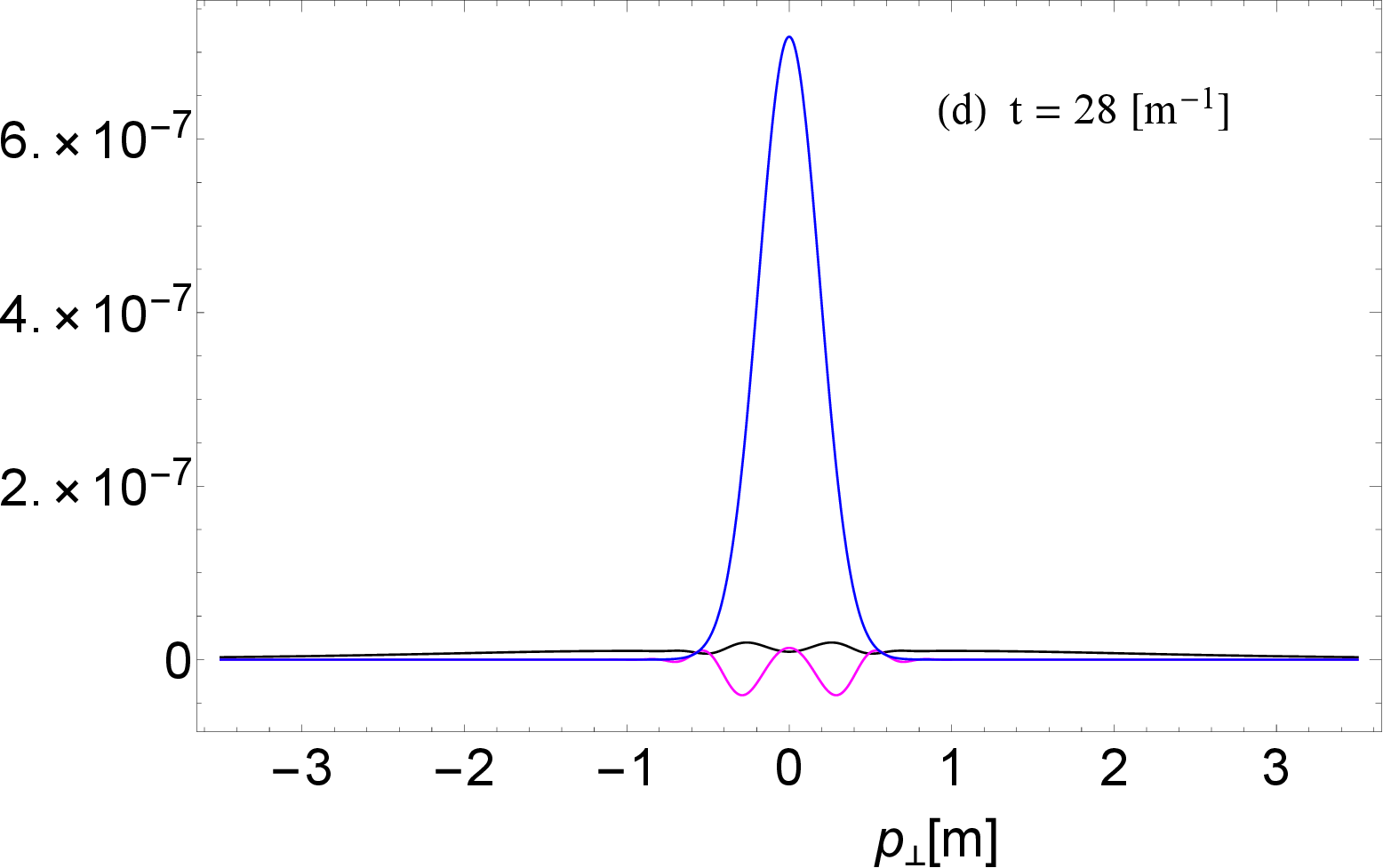}
}
\caption{The zeroth, first, and second-order terms as a function of the transverse momentum for different times. Blue curve: $\mathrm{C}_0(p_\perp, y)$, Magenta curve: $(1-y) \mathrm{C}_1(p_\perp, y)$, and Black curve: $(1-y)^2 \mathrm{C}_2(p_\perp, y)$. The field parameters are $E_0 = 0.2 E_c$ and $\tau = 10 \, [m^{-1}]$.
}
   	\label{allorder_TMS}
\end{center}
\end{figure}

In this subsection, we discuss the mathematical origin of the substructures that appear in the spectrum. Using the approximate analytical expression of the distribution function which is derived in \cref{Approximate analytical}. We find that the approximate expression for the transverse momentum distribution function is given by:
\begin{align}
   f(p_\perp,y) &\approx |N^{(+)}(p_\perp)|^2 \Bigl(\mathrm{C}_0 (p_\perp,y) +  (1-y) \mathrm{C}_1 (p_\perp,y) + (1-y)^2  \mathrm{C}_2 (p_\perp,y)\Bigr),
\label{appTMS}
\end{align}
%%%%
with,
\begin{align}
    |N^{(+)}(p_\perp)|^2 = \frac{1}{2 \omega_0 (\omega_0 - P_0)},
\end{align}
\begin{align}
     \mathrm{C}_0 (p_\perp,y) &= 4 y^2 |\Gamma_2|^2 \Biggl( 1 + p_\perp^2 + E_0 \tau \bigl(E_0  \tau + \sqrt{ 1 + p_\perp^2 + E_0^2 \tau^2} \bigr) \Biggr), \nonumber \\
     \mathrm{C}_1 (p_\perp,y) &= -  \frac{4 y  |\Gamma_1 \Gamma^{*}_2| E_0 \tau  (1 + p_\perp^2)}{ \sqrt{1 + p_\perp^2 + E_0^2 \tau^2}} \cos{(\Upsilon)} ,\nonumber \\ 
         \mathrm{C}_2 (p_\perp,y) &= 4 |\Gamma_1|^2 E_0^2 \tau^2 (1 + p_\perp^2)  \frac{\sqrt{1 + p_\perp^2 + E_0^2 \tau^2} -E_0 \tau }{(1 + p_\perp^2 + E_0^2 \tau^2)^{3/2}} \nonumber \\
     & -  8 |\Gamma_1 \Gamma^{*}_2| E_0 \tau \cos{(\Upsilon)} \Biggl(  \frac{ (1 + p_\perp^2)  ( 1 + p_\perp^2 + E_0 \tau \sqrt{1 + p_\perp^2 + E_0^2 \tau^2})}{ (1 + p_\perp^2 + E_0^2 \tau^2)^{3/2}}
     \nonumber \\ 
     &-  y \Bigl( 5 E_0 \tau - \frac{E_0^2 \tau^2}{\sqrt{1+p_\perp^2 + E_0^2 \tau^2}} 
 + 2 \sqrt{1 + p_\perp^2 + E_0^2 \tau^2} + \frac{E_0^4 \tau^2  [3 + \tau^2 (1  + p_\perp^2 + E_0^2 \tau^2)] }{(1 + p_\perp^2 + E_0^2 \tau^2)^{3/2}}  
 \nonumber \\
 &+  E_0^3 \tau^3 \frac{ (-1 + \tau^2 (1 + p_\perp^2 + E_0^2 \tau^2))}{1 + p_\perp^2 + E_0^2 \tau^2} \Bigr)   -
 \frac{y^2 (1 + E_0^2 \tau^4) }{4 + \tau^2 (1 + p_\perp^2 + E_0^2 \tau^2)} ( 4 + E_0  \tau^3 \sqrt{ 1 + p_\perp^2 + E_0^2 \tau^2})
\nonumber \\
&( E_0 \tau + \sqrt{ 1 + p_\perp^2  +E_0^2 \tau^2})\Biggr).
\label{TMScoef}
\end{align}
%%%%%%%%%%%%%%%%%%%%%%%%at
%%%%%%%%figure%%%\newline%%%%%%%%%%%%%%%%%%%
Fig.~\ref{allorder_TMS} shows the time evolution of individual terms present in Eq.~\eqref{appTMS} as a function of transverse momentum. The profile of $(1-y)^2 \mathrm{C}_2(p_\perp,y)$ exhibits substructure, with its maximum peak located at $p_\perp \approx \pm 0.65 \ [m]$ and a valley near $p_\perp \approx \pm 0.25 \ [m]$, as observed at $t = 13 \ [m^{-1}]$ (see Fig.~\ref{allorder_TMS}(a)). Similarly, the feature shown by $(1-y) \mathrm{C}_1(p_\perp,y)$ has peaks at $p_\perp = \pm 0.25 \ [m]$, but its peak value is much smaller compared to $(1-y)^2 \mathrm{C}_2(p_\perp,y)$.
As time progresses, this substructure can change according to Eq.~\eqref{TMScoef}. Therefore, as seen in Fig.~\ref{allorder_TMS}(b) at $t = 17 \ [m^{-1}]$, the profiles of $(1-y)^2 \mathrm{C}_2(p_\perp,y)$ and $(1-y) \mathrm{C}_1(p_\perp,y)$ change with time, with the peak magnitude of $(1-y)^2 \mathrm{C}_2(p_\perp,y)$ still dominating over the other terms. The change in the shape of $(1-y)^2 \mathrm{C}_2(p_\perp,y)$ and $(1-y) \mathrm{C}_1(p_\perp,y)$ can be understood as $\cos{(\Upsilon)}$ function that presents in the relation Eq.\eqref{TMScoef}.
\par
Furthermore, we can approximate the functions $\mathrm{C}_1(p_\perp,y)$ and $\mathrm{C}_2(p_\perp,y)$ to analyze the qualitative impact on the transverse momentum distribution function. Using the condition  $E_0 \tau > 1 ,$

\begin{align}\mathrm{C}_1 (p_\perp,y)  &\approx \exp\left\{ \frac{p_\perp^2}{2} \tau \left( \frac{2 E_0^2 \tau}{1 + E_0^2 \tau^2} - \frac{E_0 + \pi}{\sqrt{1 + E_0^2 \tau^2}} \right) + \pi \tau \left( E_0 \tau - \sqrt{1 + E_0^2 \tau^2} \right) \right\} \cos(\Upsilon) \nonumber \\ & \times \frac{2 y E_0 \left( 1 + E_0 \tau (E_0 \tau + \sqrt{1 + E_0^2 \tau^2}) \right)}{(1 + E_0^2 \tau^2)^{3/2}}. \label{TMScoef12} \end{align}

\begin{align}
    \mathrm{C}_2 (p_\perp,y) 
	& \approx 4 E_0^2 \tau^2 (1 + p_\perp^2)  \frac{(-E_0 \tau + \sqrt{1 + p_\perp^2 + E_0^2 \tau^2})}{(1 + p_\perp^2 + E_0^2 \tau^2)^{3/2}} |\Gamma_1|^2 
    \nonumber \\&
    + \frac{4 E_0  e^{- \Lambda \frac{p_\perp^2}{2} } \cos{(\Upsilon)}}{\tau(1+ E_0^2 \tau^2)^{7/2}} \Biggl( 3 E_0^4 \tau^5 (3 + 8 E_0^2) + 13 E_0^6 \tau^7 
	\nonumber \\&+ 6 E_0^8 \tau^9 + 12 E_0 \sqrt{1 + E_0^2 \tau^2} + 4 E_0 \tau^2 (1 +7 E_0^2) \sqrt{1 + E_0^2 \tau^2} + 8 E_0^3 \tau^4 (1+ 3 E_0^2) \sqrt{1+ E_0^2 \tau^2}
	\nonumber \\&
	+ 10 E_0^5 \tau^6 \sqrt{1+ E_0^2 \tau^2} + 6 E_0^7 \tau^8 \sqrt{1+ E_0^2 \tau^2}  \Biggr) ,
	\label{TMScoef22}
\end{align}
with,
\begin{align}
	\Lambda &= 
	(\tau(48 E_0^2-1) + E_0^2 \tau^3( 136 E_0^2 -5) + E_0^4 \tau^5(9 + 88 E_0^2) + 29 E_0^6 \tau^7 + 16 E_0^8 \tau^9 + 24 E_0 \sqrt{ 1 + E_0^2 \tau^2}
	\nonumber \\&+ 64 E_0^3 \tau^2 \sqrt{1+ E_0^2 \tau^2} + 8 E_0^3 \tau^4 (11 E_0^2  -1) \sqrt{1+ E_0^2 \tau^2} + 8 E_0^5 \tau^6 \sqrt{1+ E_0^2 \tau^2} 
	\nonumber \\&+ 16 E_0^7 \tau^8 \sqrt{1+ E_0^2 \tau^2}) \Big/ (1 + E_0^2 \tau^2) ( \tau + 16 E_0^2 \tau +  E_0^2 \tau^3 (3 + 40 E_0^2) + 3 E_0^4 \tau^5(3 + 8 E_0^2) + 13 E_0^6 \tau^7
	\nonumber \\&+ 6 E_0^8 \tau^9 + 12 E_0 \sqrt{1+ E_0^2 \tau^2} + 4 E_0 \tau^2 (1+ 7 E_0^2)  \sqrt{1+ E_0^2 \tau^2}  + 8 E_0^3 \tau^4 (1+ 3 E_0^2 )  \sqrt{1+ E_0^2 \tau^2} 
	\nonumber \\&+ 10 E_0^5 \tau^6  \sqrt{1+ E_0^2 \tau^2}  + 6 E_0^7 \tau^8  \sqrt{1+ E_0^2 \tau^2} ). 
\end{align}
%%%%%%%%%%%%%%%%%%%%
From Eq.\eqref{TMScoef12} and \eqref{TMScoef22}, we easily see that $\mathrm{C}_1(p_\perp,y)$ and $\mathrm{C}_2(p_\perp,y)$ shows Gaussian-like profile with the cosine function and its behavior depends on the field parameters. We now analyze the  $\mathrm{C}_2(p_\perp,y)$ function that shows the complicated substructure that changes with time.
From Eq.\eqref{TMScoef22} first term shows a complicated substructure having two peaks at $p_\perp \approx  \pm  1.5 [m] $ and dip at the $p_\perp  \approx 0.$ This structure does not change with time but the second term plays crucial role whose depends the shape of profile on the time since  $\cos{(\Upsilon)}$ factor and its gives multi-peak substructure behavior. 
To further analyze the substructure behavior of the spectra, we reconsider the argument $\Upsilon$ as outlined in \cref{ApprLMSresult}. For this purpose, we approximate $\Upsilon$ by including terms up to $p_\perp^2$ and disregarding higher-order terms.
%%%%%
%We again examine the argument $\Upsilon$, as discussed previously in \cref{ApprLMSresult}, to understand the substructure behavior of the spectra. To do this, we approximate the argument $\Upsilon$. We consider terms up to $p_\perp^2$ and neglect higher-order terms.
%%%%%%%%%%%%%%%%%%%%%%
%%%%%%%%%%%%%%%%%%%%%%%%%%%%%%%%%%%%%%%%%%%%%%%%%%%%%Whereas the $\cos{(\Upsilon)}$argument is approximate as 
\begin{align}
    \Upsilon & \approx   E_0 \tau^2 \ln{(1 + (E_0 \tau + 2 E_0 \tau \sqrt{1 + E_0^2 \tau^2})  )}  + \ln{\Biggl(\frac{(1-y)}{\sqrt{1+ E_0^2 \tau^2}}  \Biggr)} \tau  \sqrt{1+E_0^2 \tau^2}  
     \nonumber \\
     & +  \frac{ p_\perp^2 \tau}{2 \sqrt{1 + E_0^2 \tau^2}} \ln{\Biggl(\frac{(1-y)}{1 + E_0^2 \tau^2} \Biggr)}.
\end{align}
%%%%
It shows that the behavior is symmetric about the origin, which can be understood as the presence of only the $p_\perp^2-$ term and the absence of the linear term in $p_\perp$, in comparison to the longitudinal momentum case as discussed in \cref{ApprLMSresult}.
One more thing we notice is that at $p_\perp = 0$, the behavior of the $\cos{(\Upsilon)}$ function changes drastically with time due to the presence of the time-dependent term $\ln{(1-y)}$. The oscillation frequency also depends on $p_\perp^2$, which means the frequency increases as the transverse momentum value increases. However, the coefficient of $\cos{(\Upsilon)}$ function plays a role here, giving a Gaussian envelope to the cosine function as discussed previously  in ~\cref{ApprLMSresult}. Within the width of the Gaussian profile, the oscillations survive and are seen in the spectrum.
\par
%%%
Fig.\ref{allorder_TMS}(c) shows that the $\mathrm{C}_0(p_\perp,y)$ term dominates over the other terms. We still observe substructure in the terms, $(1-y) \mathrm{C}_1(p_\perp,y)$ and $(1-y)^2 \mathrm{C}_2(p_\perp,y)$. Although their magnitudes decrease drastically, these terms still impact $f(p_\perp, t)$ at this time. The term $\mathrm{C}_0(p_\perp,y)$  shows a smooth Gaussian profile with a peak at $p_\perp = 0$, and its shape does not change with time. At $t = 32 \ [m^{-1}]$, the first and second-order terms are suppressed, and the features of the spectra can be explained by the zeroth-order term alone.
Therefore, by closely analyzing the different terms $\mathrm{C}_0$, $(1-y) \mathrm{C}_1$, and $(1-y)^2 \mathrm{C}_2$ as defined in Eq.~\ref{TMScoef}, we can observe the following: Initially, the TMS behavior is qualitatively explained by the first and second-order terms. Up to $t \approx 2 \tau$, the function $(1-y)^2 \mathrm{C}_2$ continues to qualitatively account for the non-zero peak and substructure near the transient region. This substructure arises from the time-dependent $\cos{(\Upsilon)}$ function. As time progresses, the magnitudes of $(1-y)^2 \mathrm{C}_2$ and $(1-y) \mathrm{C}_1$ decrease and are eventually suppressed as $y \rightarrow 1$. Compared to the other terms in the distribution function, this suppression causes the substructure to nearly disappear (see Fig.~\ref{FG:2}).
%%%
\newline
In the late-time limit, $f(p_\perp, y \rightarrow 1)$ is governed mainly by the zeroth-order term, $\mathrm{C}_0(p_\perp, y)$. As a result, the spectra only show a single peak with a smooth profile. 
To demonstrate this, we can find the asymptotic expression for the transverse momentum distribution function in the limit $y \rightarrow 1$. From Eq.~\eqref{appTMS}, we can write it as:
\begin{align}
        f(p_\perp, y \rightarrow 1) \approx  |N^{(+)}(p_\perp)|^2 \mathrm{C}_0 (p_\perp)
\end{align}
%%%%%%%%%%%%%%%%%%%%%%%%%%%%%%%%%
By lengthy calculation, as we do for the longitudinal momentum case, we found an expression for the transverse momentum distribution function for asymptotic time as follows:
\begin{align} f(p_\perp) &\approx 2 \exp\left\{ 2\pi \tau \left( E_0 \tau - \sqrt{1 + E_0^2 \tau^2} \right) - \pi \tau \frac{p_\perp^2}{\sqrt{1 + E_0^2 \tau^2}} \right\}. \label{asyTMS} \end{align}
The above Eq.~\eqref{asyTMS} clearly shows that in the REPP stage, the transverse distribution function has an invariable Gauss-like distribution whose half-width is defined by the field strength,$E_0$, and pulse duration $\tau.$

%%%%%%%%%%%%%%%%%%%%%%%%
\section{Possible physical interpretation}
\label{sec:physical_interpretation}

%\subsection{Motivation}
%\label{subsec:motivation}

The quasiparticle distribution function \( f(\bm{p},t) \) analyzed in previous sections provides a consistent dynamical description of pair production. However, its physical interpretation at finite times---when the external electric field is still active---requires careful consideration. Because the quasiparticle basis itself is time-dependent, \( f(\bm{p},t) \) is basis-dependent and cannot be directly identified with the number of real, asymptotic particles \cite{Ilderton:2021zej,Dabrowski:2016tsx}. 
%At non-asymptotic times, different adiabatic bases yield different behaviors for the distribution function.

To establish a clear physical interpretation, we adopt a complementary Dirac sea picture (equivalent scattering) approach \cite{Avetissian:2002ucr}. Following Ref.~\cite{Ilderton:2021zej}, we consider an electric field that acts only until a finite time \( T \) and is then switched off abruptly. Within this framework, we solve the Dirac equation as a scattering problem: an initial negative-energy state evolves under the time-dependent field and, after switch-off, becomes a superposition of positive- and negative-energy free states~\cite{Avetissian2016}. The transition amplitude to positive-energy states directly yields the probability \( \mathcal{W}(\bm{p}, T) \) for creating real electron--positron pairs.

Our goal is to prove the key identity:
\begin{equation}
f(\bm{p},T) = \mathcal{W}(\bm{p},T),
\label{eq:goal_identity}
\end{equation}
where \( \mathcal{W}(\bm{p},T) \) is the probability of producing real electron--positron pairs if the field is switched off at time \( T \). This equivalence bridges the formal quasiparticle description with an operationally measurable quantity.
%~\cite{Alvarez-Dominguez:2023zsk}.

\subsection{Setup: Sauter pulse with finite duration}
\label{subsec:setup}

We modify the Sauter-pulse field so that it is nonzero only for \( t < T \):
\begin{equation}
E(t) = \begin{cases}
E_0 \operatorname{sech}^2\left(\frac{t}{\tau}\right), & -\infty < t < T, \\[4pt]
0, & t > T,
\end{cases}
\label{eq:E_finite}
\end{equation}
with the corresponding vector potential
\begin{equation}
A(t) = \begin{cases}
-E_0\tau \tanh\left(\frac{t}{\tau}\right), & t < T, \\[4pt]
-E_0\tau \tanh\left(\frac{T}{\tau}\right), & t > T.
\end{cases}
\label{eq:A_finite}
\end{equation}

The problem naturally divides into three time intervals:
\begin{itemize}
\item \textbf{Region I} (\( t \to -\infty \)): field off, initial negative-energy free state.
\item \textbf{Region II} (\( -\infty < t < T \)): field on, exact Sauter-pulse solution applies.
\item \textbf{Region III} (\( t > T \)): field off, free-particle solutions.
\end{itemize}

\subsubsection{Free Dirac spinors in Regions I and III}
\label{subsubsec:free_spinors}

In the absence of an electric field (\(E(t)=0\)), the Dirac equation admits plane-wave solutions. For a particle with canonical momentum \( \bm{p} = (p_1, p_2, p_\parallel) \) and kinetic momentum \( \Pi_3 = p_\parallel - e A \) (where \( A \) is constant), the energy is \( \mathcal{E} = \sqrt{m^2 + p_\perp^2 + \Pi_3^2} \). The positive- and negative-energy spinors for spin \( r = 1 \) are (see Appendix~\ref{app:spinors} for explicit forms and conventions):
\begin{equation}
\xi_{\bm{p},1}^{(+)}(t) = \frac{1}{\sqrt{2\mathcal{E}(\mathcal{E} - \Pi_3)}}
\begin{pmatrix}
m \\
0 \\
-\mathcal{E} + \Pi_3 \\
p_1 + i p_2
\end{pmatrix} e^{-i\mathcal{E}t},
\qquad
\xi_{\bm{p},1}^{(-)}(t) = \frac{1}{\sqrt{2\mathcal{E}(\mathcal{E} + \Pi_3)}}
\begin{pmatrix}
m \\
0 \\
\mathcal{E} + \Pi_3 \\
p_1 + i p_2
\end{pmatrix} e^{i\mathcal{E}t}.
\label{eq:free_spinors}
\end{equation}
These satisfy the orthonormality condition \( [\xi_{\bm{p},1}^{(\lambda)}(t)]^\dagger \xi_{\bm{p},1}^{(\lambda')}(t) = \delta_{\lambda\lambda'} \) with respect to the Dirac inner product ~\cite{Schweber:1961zz, Avetissian2016}.

In \textbf{Region I} (\( t \to -\infty \)):
\[
\Pi_3 = P_0 \equiv p_\parallel - e A(-\infty) = p_\parallel - E_0\tau,
\quad \mathcal{E} = \omega_0 \equiv \sqrt{m^2 + p_\perp^2 + P_0^2}.
\]

In \textbf{Region III} (\( t > T \)):
\[
\Pi_3 = p_\parallel - e A(T),
\quad \mathcal{E} = \sqrt{m^2 + p_\perp^2 + \Pi_3^2}.
\]

\subsubsection{Exact solution in Region II}
\label{subsubsec:exact_solution}

In Region II, the field is the full Sauter pulse. The mode function satisfies
\begin{equation}
\Big( \partial_t^2 + i E(t) + \omega^2(\bm{p},t) \Big) \psi_{\bm{p}}(t) = 0.
\label{eq:mode_eq_regionII}
\end{equation}
Introducing the variable
\begin{equation}
    y = \frac{1}{2}\left[1 + \tanh\left(\frac{t}{\tau}\right)\right], \quad y \in [0, Y], \quad Y = y(T),
\label{5.1}
\end{equation}

%%\]
this equation transforms into a hypergeometric differential equation. The general solution regular at \( y = 0 \) is
\begin{equation}
    \psi_{\bm{p}}^{(+)}(y) = \mathcal{A} \; y^{-i\tau\omega_0/2} (1-y)^{i\tau\omega_1/2} \, {}_2F_1(a, b; c; y),
\label{5.2}
\end{equation}

where
\[
\begin{aligned}
\omega_0 &= \sqrt{\epsilon_\perp^2(p_\perp) + P_0^2}, \quad \omega_1 = \sqrt{\epsilon_\perp^2(p_\perp) + P_1^2}, \\
%P_0 &= p_\parallel - E_0\tau, \quad P_1 = p_\parallel + E_0\tau, \\
%a &= -iE_0\tau^2 - \frac{i\tau\omega_0}{2} + \frac{i\tau\omega_1}{2}, \\
%b &= 1 + iE_0\tau^2 - \frac{i\tau\omega_0}{2} + \frac{i\tau\omega_1}{2}, \\
%c &= 1 - i\tau\omega_0,
\end{aligned}
\]
and \( \mathcal{A} \) is a normalization constant.

Matching to Region I as \( t \to -\infty \) fixes
\[
\mathcal{A} = \frac{1}{\sqrt{2\omega_0(\omega_0 - P_0)}}.
\]
Thus, the normalized positive-energy mode function in Region II is

\begin{equation}
    \psi_{\bm{p}}^{(+)}(y) = \frac{y^{-i\tau\omega_0/2} (1-y)^{i\tau\omega_1/2}}{\sqrt{2\omega_0(\omega_0 - P_0)}} \; {}_2F_1(a,b;c;y).
\label{5.3}
\end{equation}
The corresponding Dirac spinor for spin \( r = 1 \) in Region II is (see Appendix~\ref{app:spinors} for derivation):
\begin{equation}
\Psi_{\bm{p},1}^{(+)}(t) =
\begin{pmatrix}
m \\
0 \\
i\partial_t + P(p_\parallel,t) \\
p_1 + i p_2
\end{pmatrix}
\psi_{\bm{p}}^{(+)}(t),
\qquad P(p_\parallel,t) = p_\parallel - eA(t).
\label{eq:Psi_regionII}
\end{equation}

%\subsection{Matching at the switching time \( t = T \)}
%\label{subsec:matching}

At \( t = T \), the solution from Region II must match continuously onto a superposition of free states in Region III:
\begin{equation}
\Psi_{\bm{p},1}^{(+)}(T) = \mathcal{R}(\bm{p}) \,\xi_{\bm{p},1}^{(-)}(T) + \mathcal{I}(\bm{p}) \,\xi_{\bm{p},1}^{(+)}(T),
\label{eq:matching_condition}
\end{equation}
where \( \mathcal{R}(\bm{p}) \) is the pair production amplitude (reflection from the Dirac sea) and \( \mathcal{I}(\bm{p}) \) is the transmission amplitude~\cite{Avetissian2016}.

Projecting both sides onto \( \xi_{\bm{p},1}^{(-)}(T) \) using orthonormality yields:
%\[
\begin{equation}
    \mathcal{R}(\bm{p}) = \left[\xi_{\bm{p},1}^{(-)}(T)\right]^\dagger \Psi_{\bm{p},1}^{(+)}(T).
\label{5.4}
\end{equation}

%%\]

Substituting the explicit forms \eqref{eq:free_spinors} and \eqref{eq:Psi_regionII} and simplifying (see Appendix~\ref{app:spinors} for details) gives:
\begin{equation}
\mathcal{R}(\bm{p}) = -i \sqrt{\frac{\mathcal{E} + \Pi_3}{2\mathcal{E}}}
\Big[ \partial_t \psi_{\bm{p}}^{(+)}(T) + i\mathcal{E} \,\psi_{\bm{p}}^{(+)}(T) \Big].
\label{eq:R_explicit}
\end{equation}

The probability to create an electron--positron pair with momentum \( \bm{p} \) and a given spin is \( |\mathcal{R}(\bm{p})|^2 \). Summing over the two independent spin states gives a factor of 2, so the total pair production probability is:
\begin{equation}
\mathcal{W}(\bm{p}, T) = 2 |\mathcal{R}(\bm{p})|^2.
\label{eq:W_def}
\end{equation}

Inserting the expression for \( \psi_{\bm{p}}^{(+)}(t) \) from Eq.~\eqref{5.3} and writing derivatives in terms of the \( y \)-variable leads to:
\begin{equation}
\mathcal{W}(\bm{p}, T) =
\frac{\mathcal{E} + \Pi_3}{2\mathcal{E} \,\omega_0 (\omega_0 - P_0)}
\left|
\frac{2}{\tau} Y(1-Y) \frac{ab}{c} f_1
+ i \big( \mathcal{E} - (1-Y)\omega_0 - Y\omega_1 \big) f_2
\right|^2,
\label{eq:W_final}
\end{equation}
where \( Y = y(T) \), \( f_1 = {}_2F_1(a+1, b+1; c+1; Y) \), and \( f_2 = {}_2F_1(a, b; c; Y) \).

%\subsection{The fundamental equivalence}
%\label{subsec:equivalence}

A direct term-by-term comparison reveals that Eq.~\eqref{eq:W_final} is identical to the quasiparticle distribution function \( f(\bm{p}, y) \) given in Eq.~(72), evaluated at \( y = Y \). Specifically, the prefactor matches \( |N^{(+)}(\bm{p})|^2 \left(1 + \frac{P}{\omega}\right) \) at \( y = Y \), and the expression inside the absolute value is exactly the same as in Eq.~(72).

Therefore,
%we have proven the central result:
\begin{equation}
\boxed{\, f(\bm{p}, T) = \mathcal{W}(\bm{p}, T) \,}
\label{eq:main_identity}
\end{equation}
where \( f(\bm{p}, T) \equiv f(\bm{p}, y(T)) \).

The physical significance of the quasiparticle distribution function within the adiabatic basis---arising from the Bogoliubov transformation---is that it acquires the interpretation of a real particle number if the external electric field is switched off at a finite time \( T \). This occurs because the Bogoliubov transformation captures the essential mixing between negative- and positive-energy states induced by the time-dependent electric field. Once the field is terminated, this mixing freezes, and the quasiparticle excitations evolve into free asymptotic particles.

%Although an instantaneous switch-off is an idealization, it has been shown that a sufficiently rapid decay of the external field leads to the same physical conclusion (see Ref.~\cite{Alvarez-Dominguez:2023zsk}). Therefore, the equality \( f(\bm{p},T) = \mathcal{W}(\bm{p},T) \) provides a robust operational interpretation of the quasiparticle distribution function at finite times.

While an abrupt switch-off represents an idealization, previous studies have shown that the same physical conclusions are obtained when the external field decays rapidly but smoothly (Ref.~\cite{Alvarez-Dominguez:2023zsk}). This ensures that the identification \( f(\bm{p},T) = \mathcal{W}(\bm{p},T) \) provides a physically meaningful and robust interpretation of the finite-time quasiparticle distribution.
%\subsection{Discussion and experimental relevance}
%\label{subsec:discussion}

Equation~\eqref{eq:main_identity} establishes that the quasiparticle distribution function at a finite time \( T \) equals the probability to produce real, observable electron--positron pairs if the electric field is switched off at that instant. This result justifies the detailed analysis of \( f(\bm{p},t) \) presented in Section~\ref{Result}: the oscillatory structures in the momentum spectra and their dependence on longitudinal and transverse momentum reflect the dynamical build-up of real electron--positron pairs, not merely transient quasiparticle features.

Moreover, by linking \( f(\bm{p},t) \) to the probability of observable pair production in a scenario where the electric field is switched off at a finite time, the formalism becomes directly relevant to pulsed-field experiments, such as those anticipated at next-generation high-intensity laser facilities.

%%%%%%%%%%%%%%%%%%%%%%%%%%%%%%%%%%%

%%%%%%%%%%%%%%%%%%%%%%%%%%%%%%%%%%%%%%%%%%%%%%%%%%%%%%%%%%%%%%%%%%%%%%%%%%%%%%%%%%%%%%%%%%%%%%%%%%%%%%%%%%%%%%%%%%%%%%%%%%%%%%%%%%%%%%%%%%%%%%%%%%%%%%%%%%%%%%%%%%%%%%%%%%%%%%%%%%%%%%%%%%%%%%%%%%%%%%%%%%%%%%%%%%%%%%%%%%%%%%%%%%%%%%%%%%%%%%%%%%ConclusionS%%%%%%%%%%%%%%%%%%%%%%%%%%%%%%%%%%%%%%%%%%%%%%%%%%%%%%%%%%%%%%%%%%%%%%%%%%%%%%%%%%%%%%%%%%%%%%%%%%%%%%%%%%%%%%%%%%%%%%%%%%%%%%%%%%%%%%%%%%%%%%%%%%%%%%%%%%%%%%%%%%%%%%%%%%%%%%%%%%%%%%%%%%%%%%%%%%%%%%%%%%%%%%%%%%%%%%%%%%%%%%%%%%%%%%%%%%%%%%%%

%%%%%%%%
\section{Conclusion}
We have conducted a detailed analysis of the electron-positron pair production from the vacuum under the influence of a time-dependent Sauter-pulse electric field. We calculated the particle distribution function using the exact analytic solution for the mode function. We have developed an analytical theory that is valid for finite times $t > \tau.$ In particular, we find an analytical expression for the distribution function in the power series of the small parameter $(1-y).$ The interesting dynamical features of the momentum distribution function at finite times are attributed to the function that appears in this expansion. We have further elaborated on this aspect of the finite-time behavior of the longitudinal and transverse momentum spectra.
\par
%%%%The two-peaked structure in which the central Gaussian-peak structure has onset oscillation, and this quantum interference pattern evolves and fades away. The two-peaked structure in which the central Gaussian-peak structure has onset oscillation, and this quantum interference pattern evolves and fades away. 
We analyzed the temporal evolution of the quasiparticle distribution function. Our investigation revealed that the process of transition from initially virtual particles to real particles occurs in three distinct stages, which crucially depend on the longitudinal and transverse momentum components. Moreover, we have quantified the initiation of the REPP stage, determined by the momentum value, with higher momenta resulting in narrower oscillations in the transient region. We meticulously examined the LMS and TMS to understand the momentum-dependent behavior during these stages. In the LMS, one interesting feature at the beginning of the REPP stage is that the spectrum exhibits an oscillating structure as imprints of a quantum signature at the finite time when the electric field is nearly zero. 
The two-peaked structure, with the central Gaussian-like peak exhibiting onset oscillations, evolves and eventually fades away. Based on this observation, we identified three distinct time scales associated with this behavior. Additionally, we investigated the impact of different electric field strengths ($E_0$) on the duration of the interference pattern's formation and disappearance.
In the multiphoton regime, for $\gamma = 2.5$, we explored the time evolution of  LMS in different stages of pair production, and the spectrum shows the splitting of a smooth uni-modal structure into a multi-modal Gaussian structure near the REPP region, after which it merges into a single peak Gaussian profile at the asymptotic time limit. 
\par
 Utilizing the approximate expressions for the momentum distribution function, we 
 have shown that the LMS structure arises from three distinct functional behaviors. The second-order term dominantly governs the early times, while combining the first and zeroth-order terms leads to the central peak structure with onset oscillation. The first-order term contributes to oscillations, owing to the presence of a Gaussian envelope with oscillatory behavior, but its amplitude diminishes with time, resulting in a smooth spectrum profile at late times. 
%%%
\par
We examined the role of vacuum polarization $u(p_\parallel, t)$, and its counterpart $v(p_\parallel, t)$ by plotting their time evolution. 
In the REPP region, both $u(p_\parallel, t)$ and $v(p_\parallel, t)$ exhibit nearly identical oscillations with the same amplitude, and this amplitude of oscillation decays for higher $p_\parallel$ values.
This observation implies that the qualitative nature of the vacuum polarization function is significantly influenced by the longitudinal momentum. To elucidate this further, we plotted the LMS of $u(p_\parallel, t)$ and $v(p_\parallel, t).$
These plots reveal two distinct patterns emerging at the start of the REPP stage. One pattern exhibits a Gaussian-like structure, while the other displays a deformed oscillatory profile within the Gaussian envelope. These oscillations, exhibiting varying amplitudes, are confined to a small window of longitudinal momentum ($-1 < p_\parallel < 1$).Over time, the regular oscillations of $u(p_\parallel, t)$ and $v(p_\parallel, t)$ became balanced. This balance is explicitly observed in the late-time LMS of the created particles, where oscillations are absent due to the equilibrium between $u(p_\parallel, t)$ and $v(p_\parallel, t).$
As we recognize the pivotal roles of $u(p_\parallel, t)$ and $v(p_\parallel, t)$ functions in accelerating (creation of $e^-$) and decelerating (annihilation of $e^-$) electrons, respectively, the observed oscillations in the $f(p_\parallel, t)$ at finite times can be attributed to the characteristic behavior imprints of these functions, resulting from the acceleration and deceleration of electrons in momentum representation, which manifests as the observed oscillating structure. 
Next, we also discussed the influence of transverse momentum on LMS, which diminishes the value of $f(p_\parallel)$ and smooths out the oscillation for higher $p_\perp$ values. Moreover, we emphasized that as transverse momentum increases, these oscillations in the late REPP stage are smoothed, and the pair formation time is affected.
\par
Finally, we studied the dynamics of TMS, uncovering interesting results that have received limited attention from other authors, particularly in the context of the Sauter pulsed electric field. We observe fluctuating substructures that undergo regular changes in the QEPP and transient region depending on the momentum value. This substructure evolution is further influenced by longitudinal momentum, leading to distinct spectral shapes at given times. This finding indicates that the longitudinal and transverse momentum are related to each other during particle-antiparticle formation quantitatively and qualitatively. At the beginning of the REPP region, this substructure disappears. The shape of the spectrum becomes an invariable Gaussian-like profile and is not affected by the longitudinal momentum, but its peak value still depends on longitudinal momentum during the REPP stage. This contrast highlights the dynamic evolution of substructures in earlier stages, in contrast to the smooth Gaussian-like profile observed in the REPP stage.
%%%
%%%
A similar intriguing behavior in the transverse momentum distribution has been observed in the works by Krajewska et al. \cite{Krajewska:2018lwe}, Bechler et al. \cite{Bechler:2023kjx}, and Majczak et al. \cite{Majczak:2024hmt}, depending on the specifics of the electric field parameters. These studies highlight a fascinating characteristic of the transverse distribution function peak that occurs at non-zero $ p_\perp$ value depending on the frequency. This observation prompts questions about the feasibility of pair formation through tunneling theory, particularly considering the significant distribution function values at non-zero transverse momentum for low frequencies. In our case, the spectrum also shows a peak at non-zero transverse momentum, which signifies pair creation through multiphoton processes at finite times.

%%For our case, the spectrum also shows the occurrence of a peak at non-zero transverse momentum, which signifies the creation of pairs through the multiphoton process at finite times. 
%%%%%%15dec

The finite-time analysis of LMS and TMS reveals not only the critical momentum dependence of the pair creation dynamics but also provides direct insight into observable phenomena. 
Crucially, as established in Sec.\ref{sec:physical_interpretation}, the quasiparticle distribution at time $T$ corresponds to measurable pair production if the field is switched off at that instant. Thus, the oscillatory structures and momentum spectra analyzed here represent genuine physical effects.
%that would imprint on experimentally detectable particles in pulsed-field scenarios.

Moreover, investigating the LMS and TMS of the created quasiparticles at different dynamical stages demonstrates a pronounced dependence on both longitudinal and transverse momenta, while simultaneously uncovering the nontrivial behavior of the vacuum during finite-time pair production processes.
%%%%%%%%%%%%%%%%%%%%%%%%%%%%%%%%%%%%%%%%%%%%%%%%%%%%%%%%%%
%The finite-time analysis of the longitudinal (LMS) and transverse (TMS) momentum spectra reveals not only their critical momentum dependence in the pair-creation dynamics but also provides direct insight into observable phenomena. Crucially, as established in Section V, the quasiparticle distribution at a finite time $T$ corresponds to the measurable probability of real pair production if the external field is switched off at that instant. Consequently, the oscillatory structures and momentum spectra discussed here are genuine physical effects, which would be imprinted on experimentally detectable particles in pulsed-field scenarios.

%Moreover, investigating the LMS and TMS of the created quasiparticles at different dynamical stages demonstrates a pronounced dependence on both longitudinal and transverse momenta, while simultaneously uncovering the nontrivial behavior of the vacuum during finite-time pair production processes.
%%%%%%%%%%%%%%%%%%%%%%%

\section{Acknowledgments}
We are grateful to the anonymous referee for constructive comments that
helped improve the manuscript. 
Deepak gratefully acknowledge the financial support from Homi Bhabha National Institute (HBNI) for carrying out this research work.

\appendix

%\section{Appendixes}

%%%%%%%%
\section{Evaluation approximate analytical expression for the distribution function }
\label{Approximate function appendix}
We have obtained an analytical expression for the distribution function.
\begin{align}
      f(\bm{p},y)  &= |N^{(+)}(\bm{p})|^2 \Biggl( \frac{\omega(\bm{p} ,y) + P(p_\parallel,y )}{\omega(\bm p ,y)}  \Biggr)  \bigg| \frac{2}{\tau} y (1-y)  \frac{a b}{c} f_1 + \ii \biggl(\omega(\bm p ,y)- (1-y) \omega_0 -y \omega_1 \biggr) f_2 \bigg|^2,
\label{A22}
\end{align}
where
\begin{equation}
    f_1 = \Hyper{1+a,1+b,1+c;y},
\end{equation} 
\begin{equation}
    f_2 =\Hyper{a,b,c;y}.
\end{equation}
We aim to derive an approximate expression to study the behavior of the distribution functions in the limit \( t \gg \tau \) (i.e., \( y \rightarrow 1 \)). This is achieved by applying appropriate asymptotic approximations for the Gamma and Gauss-hypergeometric functions, which lead to a simplified analytical form of the particle distribution functions.
\newline
To evaluate the limit \( y \to 1\) of the hypergeometric function \( {}_2F_1(a, b, c; z) \), we apply appropriate linear transformation formulas \cite{abramowitz}
%%%%%%%%%%%%%%%%%%%%%%%%%%%%%%%%%%%%%%%%%%
\begin{multline}
\Hyper{a,b,c;z} = \frac{\Gamma \br{c} \Gamma \br{c-a-b}}{\Gamma \br{c-a} \Gamma \br{c-b}} \Hyper{a,b,a+b-c+1;1-z} \\
  + \br{1-z}^{c-a-b} \frac{\Gamma \br{c} \Gamma \br{a+b-c}}{\Gamma \br{a} \Gamma \br{b}} \Hyper{c-a, c-b, c-a-b+1; 1-z} ,\\
\bet{\textrm{arg}(1- z )} < \pi
\label{1.66appn}
\end{multline}
%%%%%%%%%%%%%%%%%%%%%%%%%%%%
Using the above relation, we approximate the Gauss-hypergeometric functions $f_1$ present in the relation of the distribution function as follows:
 \begin{align}
 f_1  &=  \Hyper{1+a,1+b;1+c;y}  \nonumber \\
  &  =    \frac{ \Gamma (1+ c) \Gamma (c-a-b-1) }{\Gamma (c-a) \Gamma (c-b)} \Hyper{1+a,1+b,2+a+b-c;1-y}
   \nonumber \\
  &  + \frac{ \Gamma (1+ c) \Gamma (1+a+b-c) }{\Gamma (1+a) \Gamma (1+b)} (1-y)^{(c-a-b-1)} \Hyper{c-a,c-b,c-a-b;1-y}   
  \label{hypfun1appna}
\end{align}
%%%
In general Gauss-hypergeometric function,
\begin{align}
    \Hyper{a,b,c;z}  &=  1 +  \frac{ a b }{c } z +  \frac{ a(a+1)  b(b+1) }{c (c+1) } \frac{z^2}{2!} +  \frac{ a(a+1)(a+2)   b(b+1)(b+2) }{c (c+1)(c+2) } \frac{z^3}{3!}+...
    \label{1.69appn}
 \end{align}
 The series continues with additional terms involving higher powers of $z.$ Each term in the series involves the parameters $a, b,$ and $c$ as well as the variable $z$ raised to a specific power.

\begin{align}
 f_1  &=     \Biggl( \frac{ c \Gamma ( c) \Gamma (c-a-b-1) }{\Gamma (c-a) \Gamma (c-b)} \Biggr) \Biggl( 1 +  \frac{ (a+1) (b+1) }{(2+a+b-c) } (1-y) 
  \nonumber \\
 & +  \frac{ (a+1)(a+2)  (b+1)(b+2) }{(2+a+b-c) (2+a+b-c+1) } \frac{(1-y)^2}{2!} + ...  \Biggr)
  \nonumber \\
 &+   (1-y)^{(c-a-b-1)}  
  (a+b-c)   \Biggl(\frac{ c \Gamma ( c) \Gamma (a+b-c) }{ a \Gamma (a)  b \Gamma (b)}  \Biggr) \Biggl( 1 +  \frac{ (c-a) (c-b) }{(c-a-b) } (1-y)  
  \nonumber \\
  &+  \frac{ (c-a)(c-a+1)  (c-b)(c-b+1) }{(c-a-b) (c-a-b+1) } \frac{(1-y)^2}{2!} + ...   \Biggr)
  \label{hypfun1appnb}
\end{align}
Similarly, 
\begin{align}
   f_2  &=  \frac{ \Gamma (c) \Gamma (c-a-b) }{\Gamma (c-a) \Gamma (c-b)} \Hyper{a,b,1-c+a+b;1-y}  
   \nonumber \\ & +  (1-y)^{(c-a-b)}  
  \frac{ \Gamma (c) \Gamma (a+b-c) }{\Gamma (a) \Gamma (b)} \Hyper{c-a,c-b,1+c-a-b;1-y} 
 \nonumber \\
  &= \frac{ \Gamma (c) \Gamma (c-a-b) }{\Gamma (c-a) \Gamma (c-b)} \Biggl( 1 +  \frac{ a b}{1+a+b-c}(1-y) + \frac{(a+1)(b+1))}{(2+a+b-c)} \frac{(1-y)^2}{2 !
  } +...\Biggr)
   \nonumber \\ & +  (1-y)^{(c-a-b)}  
  \frac{ \Gamma (c) \Gamma (a+b-c) }{\Gamma (a) \Gamma (b)} \Biggl(1 + \frac{(c-a)(c-b)}{(1+c-a-b)} (1-y) + ... \Biggr)
  \label{hypfun2appn}
  \end{align}
%%%%%%
Here, $y$ is transformed time variable.
\par
To derive an approximate expression for the distribution function, we truncate the power series of the Gauss-hypergeometric functions \( f_1 \) and \( f_2 \), as presented in Eqs.~\eqref{hypfun1appnb} and \eqref{hypfun2appn}, respectively, up to a specific order. In doing so, we retain only the second-order terms in the expressions for the distribution function, as given by Eq.~\eqref{hypfun1appnb} and \eqref{hypfun2appn}, and neglect higher-order terms of order \( \mathcal{O}((1 - y)^2) \).

%%%%%%%%%%
%To compute an approximate expression for the particle distribution function that depends on finite time, we can truncate the power series of the Gauss-hypergeometric functions $f_1$and $f_2$ given by Eqs.~\eqref{hypfun1appnb} and \eqref{hypfun2appn}  respectively, up to a specific order. 
%The order of truncation will depend on the desired accuracy and the characteristics of the finite-time behavior under consideration.
%\newline
%%%%%%%%%%%%%%%%%%%%%%first term%%%%
%We approximate  the different terms present in expression of the distribution function Eq.~\eqref{hypfun1} and \eqref{hypfun2}, considering only up to the second-order terms and neglecting higher-order terms $\mathcal{O}((1-y)^2)$
%We approximate the various terms in the expressions for the distribution function given by Eq.~\eqref{hypfun1} and \eqref{hypfun2}, considering only up to second-order terms and neglecting higher-order terms of order \( \mathcal{O}((1 - y)^2) \).
%'\newline

%Therefore,
\begin{align}
    \frac{2y}{\tau}(1-y)  \frac{ab}{c} f_1 &\approx \frac{2}{\tau}y(a+b-c)  \Gamma_2 (1-y)^{(c-a-b)} +(1-y) \frac{2 y }{\tau}
\Biggl(a b \Gamma_1 -  (c-a)(c-b)  \Gamma_2 (1-y)^{(c-a-b)}
\Biggr) \nonumber \\
  & + (1-y)^2 \frac{2y}{\tau}\Biggl(  \Gamma_1 \frac{a(1+a) b(1+b)}{(2+a+b-c)}  +  \Gamma_2 (1-y)^{(c-a-b)} \frac{(c-a)(c-b)(c-a+1)(c-b+1)}{(a+b-c-1)}\Biggr)
\end{align}
where we have used 
\begin{align}
     \Gamma_1  &=\frac{ \Gamma ( c) \Gamma (c-a-b-1) }{\Gamma (c-a) \Gamma (c-b)}, \\
   \Gamma_2 &=\frac{ \Gamma ( c) \Gamma (a+b-c) }{\Gamma (a) \Gamma (b)}.
\end{align}
 Similarly,
 \begin{align}
     \Bigl(\omega(\bm{p},y) - (1-y) \omega_0- y \omega_1 \Bigr) f_2 &=   (\omega(\bm{p},y) - (1-y) \omega_0 - y \omega_1) \Biggl(\Gamma_1 (c-a-b-1) 
     \nonumber \\ &
\Bigl(1 + (1-y) \frac{ a b}{(1+a+b-c)} \Bigr) +  (1-y)^{(c-a-b)} \Gamma_2 \Bigl(1 + (1-y) \frac{(c-a)(c-b)}{(1+c-a-b)} \Bigr) \Biggr)
\end{align}
%%%
Also, it is possible to write down the time-dependent quasi-energy $\omega(\bm{p},y)$ as the following series expansion near 
$y \rightarrow 1:$ 
\begin{align}
     \omega(\bm{p},y) &\approx \omega_1 - (1-y) \frac{2 E_0 \tau}{\omega_1} P_1  +  (1-y)^2  \frac{2 E_0^2 \tau^2  }{\omega_1^3}\epsilon^2_\perp(p_\perp)
\end{align}
up to the second order and neglect the other higher order terms.

Therefore,
 \begin{align}
      \Bigl(\omega(\bm{p},y) - (1-y) \omega_0- y \omega_1 \Bigr) f_2 \approx (1-y) \Biggl((\omega_1 - \omega_0) - \frac{2 E_0 \tau }{\omega_1} P_1 \Biggr) \Biggl(\Gamma_1 (c-a-b-1)  + e^{- \ii \tau\omega_1 \ln{(1-y)}} \Gamma_2\Biggr)
      \nonumber \\
      + (1-y)^2 \Biggl( \Gamma_1 (c-a-b-1)  \Biggl(  \frac{ 2 E_0^2 \tau^2  \epsilon^2_\perp }{\omega_1^3}  +     \frac{ a b ( \omega_1(\omega_1 - \omega_0) - 2 P_1 E_0 \tau )}{ \omega_1(1+a+b-c)} \Biggl)
      \nonumber \\
      + \Gamma_2  e^{-\ii \tau \omega_1 \ln{(1-y)}} \Biggl( \frac{ 2 E_0^2 \tau^2 \epsilon^2_\perp }{\omega_1^3} + \Bigl((\omega_1 - \omega_0) - \frac{2 E_0 \tau }{\omega_1} P_1 \Bigr)  \frac{(c-a)(c-b)}{(1+c-a-b)} \Biggr)\Biggr)
 \end{align}
 %%%
 Using the above relation, we get
%%%%%%%Using the above relation we can find out the approximate relation for time-dependent particle distribution function, 
%%%
\begin{align}
   \bigg| \frac{2}{\tau} y (1-y)  \frac{a b}{c} f_1 + \ii (\omega(\bm{p},y) - (1-y) \omega_0 -y \omega_1) f_2 \bigg|^2 
    \nonumber \\
      \approx \Bigg|\frac{2}{\tau} y (a+b-c) \Gamma_2 e^{- \ii \tau \omega_1 \ln{(1-y)}}  + (1-y) \Biggl[\Gamma_1 \Biggl(  \frac{2}{\tau} y a b + \ii (c-a-b-1) ( \omega_1 - \omega_0 - \frac{ 2 E_0 \tau P_1}{\omega_1}) \Biggr)  
  \nonumber \\
  + \Gamma_2  e^{-\ii \tau \omega_1 \ln{(1-y)}}\Biggl(\frac{2}{\tau} y (a+b-c) \frac{(c-a)(c-b)}{(c-a-b)} + \ii ( \omega_1 - \omega_0 - \frac{ 2 E_0 \tau P_1}{\omega_1})   \Biggr) \Biggl] + (1-y)^2 \Biggl[ \Gamma_1 \Biggl(\frac{2}{\tau} y a b 
  \nonumber \\
   \frac{(1+a)(1+b)}{(2+a+b-c)} + \ii  \biggl( \frac{2 E_0^2 \tau^2 \epsilon^2_\perp}{\omega_1^3} + \frac{ a b }{(1+a+b-c)}( \omega_1 - \omega_0  - \frac{ 2 E_0 \tau P_1}{\omega_1})\biggr) (c-a-b-1)  \Biggr)
  \nonumber \\
   + \Gamma_2 e^{-\ii \tau \omega_1 \ln{(1-y)}} \Biggl( \frac{2}{\tau} y (a+b-c) \frac{ (c-a)(c-b)(c-a+1)(c-b+1)}{(c-a-b)(c-a-b+1)}+ \frac{2 E_0^2 \tau^2 \epsilon^2_\perp}{\omega_1^3}  
    \nonumber \\
    +  ( \omega_1 - \omega_0  - \frac{ 2 E_0 \tau P_1}{\omega_1})  \frac{(c-a) (c-b)}{(1+c-a-b)} \Biggr) \Biggr]\Bigg|^2
    \label{absappappn}
 \end{align}

%To calculate the  Eq.\eqref{absapp}
We introduce it here in anticipation of encountering the Gamma function $\Gamma(z)$ in the subsequent content. The Gamma function typically obeys the following relationship:
 \begin{align}
        \Gamma(1+z) = z \Gamma(z), &&
         \Gamma(1-z) \Gamma(z) = \frac{\pi}{\sin(\pi z)},
          \label{GmI1}
 \end{align}
 from which we can derive the  following useful relations,
  \begin{align}
      | \Gamma( \ii z)  |^2= \frac{\pi }{z \sinh{(\pi z)}}, &&
      | \Gamma(1+ \ii z)|^2 = \frac{\pi z}{\sinh{(\pi z)}}, &&
        |\Gamma(\frac{1}{2} + \ii z)|^2 &= \frac{\pi}{\cosh{(\pi z)}}.
        \label{GmI2}
  \end{align}

%%%%%%%%%%%%%%%%%%%%%%%%%%%%%%%%%%%%%%%%%%%%%%%%%%%%%%%%%%%%%%%%%%%%%%%%%%%%%%%%%%%%%%%%%%%%%%%%%%%%%%%%%%%%%%%%%%%%%%%%%%%%%%%%%%%%%%%%%%%%%%%%%%%%%%%%%%%%%%%%%%%%%%%%%%%%%%%%\newline
Using the mathematical identities \eqref{GmI2}, we can compute $|\Gamma_1|^2$ and $|\Gamma_2|^2$ as 
 \begin{align}
       | \Gamma_1|^2 &=  \bet{\frac{ \Gamma ( c) \Gamma (c-a-b-1) }{\Gamma (c-a) \Gamma (c-b)}}^2  \nonumber \\ 
       &= \frac{\omega_0(\omega_1 + \omega_0 + 2 E_0 \tau)}{\omega_1(\omega_1 + \omega_0 - 2 E_0 \tau) (1+ \tau^2 \omega_1^2)}
       \Biggl(  \frac{\sinh{(\frac{\pi \tau}{2} ( \omega_0 +\omega_1 - 2 E_0 \tau))} \sinh{(\frac{\pi \tau}{2} ( \omega_0 +\omega_1 + 2 E_0 \tau))} }{\sinh{(\pi \tau \omega_0)} \sinh{(\pi \tau \omega_1)}}\Biggr),\nonumber \\
         | \Gamma_2|^2 &= \bet{\frac{ \Gamma ( c) \Gamma (a+b-c) }{\Gamma (a) \Gamma (b)}}^2
         \nonumber \\ 
       &= \frac{\omega_0(\omega_0 - \omega_1 + 2 E_0 \tau)}{\omega_1(\omega_1 - \omega_0 + 2 E_0 \tau) (1+ \tau^2 \omega_1^2)}
       \Biggl(  \frac{\sinh{(\frac{\pi \tau}{2} ( \omega_0 -\omega_1 + 2 E_0 \tau))} \sinh{(\frac{\pi \tau}{2} ( -\omega_0 +\omega_1 + 2 E_0 \tau))} }{\sinh{(\pi \tau \omega_0)} \sinh{(\pi \tau \omega_1)}}\Biggr).
       \label{gm1gm2}
 \end{align}

%%%%%
When computing expressions like $\Gamma_1 \Gamma^{*}_2$, approximate methods prove advantageous. A frequently utilized technique entails utilizing Stirling's formula for the Gamma function, offering a simpler yet effective approach to assess the desired expression.
\begin{align}
     \Gamma(z) \approx  z^{z - 1/2} e^{-z}\sqrt{2 \pi }
\end{align}
Then, we derive the set of equations employing Stirling's formula for the Gamma function, which are used to determine the Gamma function in the computation of the particle distribution function.
\begin{align}
     \Gamma(1+ \ii x) &\sim \sqrt{2 \pi}  e^{(\frac{1}{2} ln(x) - \frac{\pi}{2} x )+ \ii ( x (ln(x) -1) + \frac{\pi}{4})}
     \nonumber \\
     \Gamma(-\ii x) &\sim \sqrt{2 \pi}  e^{( \frac{\pi}{2} x -\frac{1}{2} ln(x)  )+ \ii ( x(1- ln(x) ) - \frac{\pi}{4})}
     \nonumber \\
     \Gamma(\ii x) &\sim \sqrt{2 \pi}  e^{( -\frac{\pi}{2} x -\frac{1}{2} ln(x)  )+ \ii ( x(ln(x) -1) - \frac{\pi}{4})}
     \end{align}
 
Subsequently, following certain algebraic manipulations, we obtain :          
 \begin{align}
    \Gamma_1 \Gamma^{*}_2 &\approx |\Gamma_1 \Gamma^{*}_2| e^{\ii \mathrm{\varrho}}
\end{align}

%%%%%%%%%%%%%%%%%%%%%%%%%%%%%%%%%%%%%%%%%%%%%%%%%%%%%%%%%%%%%%%%%%%%%%%%%%%%
     Here,
 \begin{align}
      |\Gamma_1 \Gamma^{*}_2| &=  \frac{\omega_0}{2 \omega_1} \frac{\exp{(\pi \tau/2 (\omega_0 -\omega_1 + 2 E_0 \tau))}}{ \sinh{(\pi \tau \omega_0)}\sqrt{1 + \omega_1^2 \tau^2}}  \sqrt{\frac{((\omega_0 + 2 E_0 \tau)^2 - 
  \omega_1^2)}{(\omega_1^2 - (2 E_0 \tau - \omega_0)^2)}},
    \end{align}
\begin{align}
    \mathrm{\varrho} &=    \frac{\tau}{2} (\omega_0 + \omega_1 -2 E_0 \tau) \ln{(\tau (\omega_0 + \omega_1 -2 E_0 \tau))}  + (-\omega_0 + \omega_1 -2 E_0 \tau) \ln{(\tau (\omega_0 - \omega_1 +2 E_0 \tau))}
    \nonumber\\
& +(-\omega_0 + \omega_1 +2 E_0 \tau) \ln{(\tau (-\omega_0 + \omega_1 +2 E_0 \tau))} +(\omega_0 + \omega_1 +2 E_0 \tau) \ln{(\tau (\omega_0 + \omega_1 +2 E_0 \tau))} 
 \nonumber \\
 &+\pi -  \tan^{-1}(\tau \omega_1 ) - 2 \tau \omega_1  \ln{(2 \omega_1 \tau)}.
\end{align}
%%%%
Now, we use these approximations of the Gamma function to derive an approximate expression for the particle distribution function \eqref{59.1}, which can then be re-expressed using Eq.\eqref{absappappn}  as follows:
  %%%%%%%%%%%%%%%%%%%%%%%%%%% 
  \begin{align}
     f(\bm{p},y)  \approx |N^{(+)}(\bm{p})|^2 \Bigg|\frac{2}{\tau} y (a+b-c) \Gamma_2 e^{- \ii \tau \omega_1 \ln{(1-y)}}  + (1-y) \Biggl[\Gamma_1 \Biggl(  \frac{2}{\tau} y a b + \ii (c-a-b-1) ( \omega_1 - \omega_0 - \frac{ 2 E_0 \tau P_1}{\omega_1}) \Biggr)  
  \nonumber \\
  + \Gamma_2  e^{-\ii \tau \omega_1 \ln{(1-y)}}\Biggl(\frac{2}{\tau} y (a+b-c) \frac{(c-a)(c-b)}{(c-a-b)} + \ii ( \omega_1 - \omega_0 - \frac{ 2 E_0 \tau P_1}{\omega_1})   \Biggr) \Biggl] + (1-y)^2 \Biggl[ \Gamma_1 \Biggl(\frac{2}{\tau} y a b 
  \nonumber \\
   \frac{(1+a)(1+b)}{(2+a+b-c)} + \ii  \biggl( \frac{2 E_0^2 \tau^2 \epsilon^2_\perp}{\omega_1^3} + \frac{ a b }{(1+a+b-c)}( \omega_1 - \omega_0  - \frac{ 2 E_0 \tau P_1}{\omega_1})\biggr) (c-a-b-1)  \Biggr)
  \nonumber \\
   + \Gamma_2 e^{-\ii \tau \omega_1 \ln{(1-y)}} \Biggl( \frac{2}{\tau} y (a+b-c) \frac{ (c-a)(c-b)(c-a+1)(c-b+1)}{(c-a-b)(c-a-b+1)}+ \frac{2 E_0^2 \tau^2 \epsilon^2_\perp}{\omega_1^3}  
    \nonumber \\
    +  ( \omega_1 - \omega_0  - \frac{ 2 E_0 \tau P_1}{\omega_1})  \frac{(c-a) (c-b)}{(1+c-a-b)} \Biggr) \Biggr]\Bigg|^2 \Biggl(1 +  \frac{P (p_\parallel,y)}{\omega(\bm{p},y)}    \Biggr)
    \label{apppdf1appn}
\end{align}
%%%%%%%%%%%%%%%%%%%%%%%%%%%%%%%%%%%%%%%%%%%%%%%%%%%%%%%%%%%%%%%%%%%%
We aim to express $ f(\bm{p},y) $ in a series involving $(1-y)$. We can then consider truncating higher-order terms to simplify the analysis while still capturing essential features.
In this context, we focus exclusively on terms up to the $ (1-y)^2$ order, disregarding higher-order terms in Eq. \eqref{apppdf1appn}. This approach is sufficient to explain the interesting results discussed in the later section \ref{Result}.
%%%
To accomplish this, we again approximate the quantity $\Bigl( 1 + \frac{P(p_\parallel,y)}{\omega(\bm{p},y)} \Bigr)$ using the Taylor series expansion near $y \rightarrow 1$. This yields:
%%%
\begin{align}
   1 + \frac{P(p_\parallel,y)}{\omega(\bm{p},y)}   &\approx \Omega_0 + (1-y) \Omega_1  + (1-y)^2 \Omega_2 ,  
\end{align}
with,
 \begin{align}
     \Omega_0 &= 1 + \frac{P_1}{\omega_1} ,\\
     \Omega_1 &=  \frac{-2 E_0 \tau}{ \omega_1^3} \epsilon^2_\perp (p_\perp),\\
     \Omega_2 &= \frac{-6 E_0^2 \tau^2}{\omega_1^5} P_1 \epsilon^2_\perp(p_\perp).
\end{align}
%%%%%%%%%%%%%%%%%%%%%%%%%%%%%%%%%%%%%%%%%%%%%%%%%%%%%%%%%%%%%%%%%%%%%%%%%%%%%%%%%%%%%%%%%%%%%%%%%%%%%%%%%%%%%%%%%%%%%%%%%%%%%%%%%%%%%%%
%%%%%%%%%%%%%%%%%%%%%%%%%%%%%%%%%%%%%%%%%%%%%%%%%%%%%%%%%%%%%%%
Therefore, we can express the particle distribution function approximately in terms of the small parameter \( (1 - y) \) up to the second order. The expression is given by:

%%%%%%%%%%%%%%%%%%%%%%%%%%%%%%%%%%%%%%%%%%%%%%%%%%%%%%%%%%%%%
\begin{align}
        f(\bm{p},y) &\approx |N^{(+)}(\bm{p})|^2 \Bigl( \mathrm{C}_0(\bm{p},y) +  (1-y) \mathrm{C}_1(\bm{p},y)+ (1-y)^2 \mathrm{C}_2(\bm{p},y) \Bigr),
   \label{appdisfunappn}
\end{align}
%%%
where,
\begin{align}
     \mathrm{C}_0(\bm{p},y) &=  4 y^2 \Omega_0 \omega_1^2 |\Gamma_2|^2,
\label{pdfC0appn}
\end{align}
\begin{align}
    \mathrm{C}_1(\bm{p},y) &=  4 y  \Omega_0  |\Gamma_1 \Gamma^{*}_2| E_0 \tau  (P_1 -\omega_1 )   \cos{(\Upsilon)} +4 y  |\Gamma_2|^2   (2 \Omega_0 ( \omega_1^2 -\omega_0 \omega_1 +(\omega_1 - P_1) E_0 \tau ) +  \Omega_1 \omega_1^2 y) ,
    \label{pdfC1appn}
\end{align}
\begin{align}
    \mathrm{C}_2 (\bm{p},y)   &= \Omega_0 (P_1 - \omega_1)^2 \frac{4 E_0^2 \tau^2}{\omega_1^2} |\Gamma_1|^2 + |\Gamma_2|^2 \Bigl[8 \Omega_1 y (2 E_0 \tau(\omega_1 -P_1) - \omega_0 \omega_1   + \omega_1^2) -8 \Omega_0 y (\omega_1 + P_1 )  
  \nonumber   \\ 
     &(\omega_1-\omega_0 + 2 E_0 \tau) + 4 \Omega_2 y^2 \omega_1^2 + \Omega_0 \biggl( \frac{4}{\omega_1^2} (\omega_1^2 + E_0 \tau (\omega_1 -P_1))^2 + \frac{\tau^2}{4} ((\omega_0 + \omega_1)^2 - 4 E_0^2 \tau^2)^2 y^2 \biggr) 
    \nonumber   \\ 
     &+   \frac{y \Omega_0 }{2 \omega_1^2 (1 + \tau^2 \omega_1^2)} \biggl( 2 \omega_1^2 \tau^2 ((\omega_0 - \omega_1) \omega_1 + 2 E_0 \tau P_1) (\omega_0^2 - (\omega_1 + 2 E_0 \tau)^2)  
     \nonumber   \\ 
     &+ y \omega_1^3 (16 (\omega_0 + \omega_1) + 32 E_0 \tau - 4 (2\omega_0 - 3 \omega_1)(\omega_0 + \omega_1)^2 \tau^2 + 8 E_0 \tau^3 \omega_1 (2 \omega_0 + 3 \omega_1) 
     \nonumber   \\ &
     - \tau^4(\omega_1(\omega_0 + \omega_1)^4 - 16 E_0^2 (2 \omega_0 + \omega_1)) + 32 E_0^3 \tau^5 + 8 E_0^2 \tau^6 \omega_1 (\omega_0+ \omega_1)^2- 16 E_0^4 \tau^8 \omega_1)
     \nonumber   \\ &
     + 16 E_0^2 \tau^2 \epsilon^2_\perp   (1 + \tau^2 \omega_1^2) 
    \biggr)    \Bigr]  +   \frac{|\Gamma_1 \Gamma^{*}_2| \cos{(\Upsilon)}}{2 \omega_1^2 (4 + \omega_1^2 \tau^2)}   \Bigl(-16 E_0^4 \Omega_0 \omega_1^4 \tau^8 y^2 \nonumber   \\ 
     & + 16 E_0^3 \Omega_0 \omega_1^2 \tau^5 y (P_1 (4 + \omega_1^2 \tau^2) - 
     4 \omega_1 y) +  \Omega_0 \omega_1^2 (-\omega_0 + 
     \omega_1) y (2 (-8 (\omega_1 + P_1)  \nonumber   \\ 
     &+ (\omega_0 - 
           \omega_1)^2 \omega_1 \tau^2) (4 + \omega_1^2 \tau^2) +  \omega_1 (32 - 
        4 (\omega_0 - \omega_1) (4 \omega_0 + \omega_1) \tau^2 + (\omega_0 - 
           \omega_1)^3 \omega_1 \tau^4) y) \nonumber   \\ 
     & + 8 E_0^2 \Omega_0 \tau^2 (-2 (\omega_1 - P_1)^2 (4 + \omega_1^2 \tau^2) - (4 + \omega_1^2 \tau^2) (-2 P_1^2 +  \omega_1^2 (2 + \omega_1 (-\omega_0 + \omega_1) \tau^2)) y  \nonumber   \\ 
     & +  \omega_1^3 \tau^2 (-8 \omega_0 + 
        6 \omega_1 + (\omega_0 - \omega_1)^2 \omega_1 \tau^2) y^2) + 
  4 E_0 \omega_1^2 \tau (-4 \Omega_0 (\omega_1 - P_1) (4 + 
        \omega_1^2 \tau^2) 
        \nonumber   \\ 
     &  - (4 + 
        \omega_1^2 \tau^2) (4 \Omega_1 (\omega_1 - P_1) - 
        8 \Omega_0 (\omega_1 + P_1) + 
        \Omega_0 (\omega_0 - \omega_1)^2 P_1 \tau^2) y
    \nonumber   \\ 
     &  - 4 \Omega_0 \omega_1 (4 + (-\omega_0^2 + \omega_1^2) \tau^2) y^2) \Bigr) + 
     \frac{|\Gamma_1 \Gamma^{*}_2| \Omega_0 \tau y}{(\omega_1 (4 + \omega_1^2 \tau^2)))} \Bigl(2 (4 + \omega_1^2 \tau^2) (\omega_1^3 (2 + \omega_1) 
     \nonumber \\
     & - E_0 \omega_1^2 (\omega_1 + 3 P_1) \tau + 
     4 E_0^2 (\omega_1^2 + \omega_1 (-2 + P_1) - P_1^2) \tau^2 + 
     4 E_0^3 (-\omega_1 + P_1) \tau^3)  \nonumber \\
     & + \omega_0^4 \omega_1^2 \tau^2 y - 
  \omega_1^2 (\omega_1 - 2 E_0 \tau) (12 \omega_1 + 8 E_0 \tau + 
     3 \omega_1^3 \tau^2 + 2 E_0 \omega_1^2 \tau^3 - 4 E_0^2 \omega_1 \tau^4 + 
     8 E_0^3 \tau^5) y  \nonumber \\
     &+ 4 \omega_0 \omega_1^2 (4 + \omega_1^2 \tau^2) (-2 + \omega_1 y) + 
  2 \omega_0^2 ((\omega_1 (2 + \omega_1) + E_0 (\omega_1 - P_1) \tau) (4 + 
        \omega_1^2 \tau^2) \nonumber \\
        & -y \omega_1^2 (10 + 3 \omega_1^2 \tau^2 - 2 E_0 \omega_1 \tau^3 + 
        4 E_0^2 \tau^4) ) \Bigr) \sin{(\Upsilon)}.
         \label{pdfC2appn1}
\end{align}
Let us denote
\begin{align}
 \Upsilon &= \mathrm{\varrho} + \tau \omega_1 \ln{(1-y)}, \\
\sigma_0 &=\bigg(16 (\omega_1^2 + E_0 (\omega_1 - P_1) \tau)^2 (1 + \omega_1^2 \tau^2) -   \nonumber\\
&4 \Big(-8 (\omega_0 - \omega_1) \omega_1^2 (\omega_1 + P_1) + 
16 E_0 \omega_1^2 (\omega_1 + P_1) \tau +   \nonumber\\
&\big(8 E_0^2 (-\omega_1^2 + P_1^2) - (\omega_0 - \omega_1) \omega_1^3 (\omega_0^2 + 
\omega_1 (7 \omega_1 + 8 P_1))\big) \tau^2 +   \nonumber\\
&2 E_0 \omega_1^2 (\omega_0 + 3 \omega_1) (2 \omega_1^2 - \omega_0 P_1 + 3 \omega_1 P_1) \tau^3 +   \nonumber\\
&4 E_0^2 \omega_1^2 \big((\omega_0 - 3 \omega_1) \omega_1 + 2 \omega_1 P_1 + 2 P_1^2\big) \tau^4 + 
8 E_0^3 \omega_1^2 P_1 \tau^5\Big) y +   \nonumber\\
&\omega_1^2 \Big(32 \omega_1 (\omega_0 + \omega_1) + 
64 E_0 \omega_1 \tau + (\omega_0 + \omega_1)^2 (\omega_0^2 - 
14 \omega_0 \omega_1 + 25 \omega_1^2) \tau^2 +   \nonumber\\
&16 E_0 \omega_1^2 (2 \omega_0 + 3 \omega_1) \tau^3 - 
\big(\omega_1^2 (\omega_0 + \omega_1)^4 + 
8 E_0^2 (\omega_0^2 - 6 \omega_0 \omega_1 - 3 \omega_1^2)\big) \tau^4 +   \nonumber\\
&64 E_0^3 \omega_1 \tau^5 + 
8 E_0^2 \big(2 E_0^2 + \omega_1^2 (\omega_0 + \omega_1)^2\big) \tau^6 - 
16 E_0^4 \omega_1^2 \tau^8\Big) y^2\bigg),
\end{align}
\begin{align}
    \sigma_1 &= 16 E_0^3 \Omega_0 \omega_1^2 \tau^5 y (P_1 (4 + \omega_1^2 \tau^2) - 
     4 \omega_1 y)  -16 E_0^4 \Omega_0 \omega_1^4 \tau^8 y^2 +  \Omega_0 \omega_1^2 (-\omega_0 + 
     \omega_1) y (2 (-8 (\omega_1 + P_1)  \nonumber   \\ 
     &+ (\omega_0 - 
           \omega_1)^2 \omega_1 \tau^2) (4 + \omega_1^2 \tau^2) +  \omega_1 (32 - 
        4 (\omega_0 - \omega_1) (4 \omega_0 + \omega_1) \tau^2 + (\omega_0 - 
           \omega_1)^3 \omega_1 \tau^4) y) \nonumber   \\ 
     & + 8 E_0^2 \Omega_0 \tau^2 (-2 (\omega_1 - P_1)^2 (4 + \omega_1^2 \tau^2) - (4 + \omega_1^2 \tau^2) (-2 P_1^2 +  \omega_1^2 (2 + \omega_1 (-\omega_0 + \omega_1) \tau^2)) y  \nonumber   \\ 
     & +  \omega_1^3 \tau^2 (-8 \omega_0 + 
        6 \omega_1 + (\omega_0 - \omega_1)^2 \omega_1 \tau^2) y^2) + 
  4 E_0 \omega_1^2 \tau (-4 \Omega_0 (\omega_1 - P_1) (4 + 
        \omega_1^2 \tau^2) 
        \nonumber   \\ 
     &  - (4 + 
        \omega_1^2 \tau^2) (4 \Omega_1 (\omega_1 - P_1) - 
        8 \Omega_0 (\omega_1 + P_1) + 
        \Omega_0 (\omega_0 - \omega_1)^2 P_1 \tau^2) y
    \nonumber   \\ 
     &  - 4 \Omega_0 \omega_1 (4 + (-\omega_0^2 + \omega_1^2) \tau^2) y^2),   \\ 
    \sigma_2 &= 2 (4 + \omega_1^2 \tau^2) (\omega_1^3 (2 + \omega_1) - E_0 \omega_1^2 (\omega_1 + 3 P_1) \tau 
     \nonumber \\
     & + 4 E_0^2 (\omega_1^2 + \omega_1 (-2 + P_1) - P_1^2) \tau^2 + 
     4 E_0^3 (-\omega_1 + P_1) \tau^3)  \nonumber \\
     & + \omega_0^4 \omega_1^2 \tau^2 y - 
  \omega_1^2 (\omega_1 - 2 E_0 \tau) (12 \omega_1 + 8 E_0 \tau + 
     3 \omega_1^3 \tau^2 + 2 E_0 \omega_1^2 \tau^3 - 4 E_0^2 \omega_1 \tau^4 + 
     8 E_0^3 \tau^5) y  \nonumber \\
     &+ 4 \omega_0 \omega_1^2 (4 + \omega_1^2 \tau^2) (-2 + \omega_1 y) + 
  2 \omega_0^2 ((\omega_1 (2 + \omega_1) + E_0 (\omega_1 - P_1) \tau) (4 + 
        \omega_1^2 \tau^2) \nonumber \\
        & -y \omega_1^2 (10 + 3 \omega_1^2 \tau^2 - 2 E_0 \omega_1 \tau^3 + 
        4 E_0^2 \tau^4) ) .
        \label{sigm2ap}
\end{align}
Therefore,
\begin{align}
    \mathrm{C}_2 (\bm{p},y)   &= \Omega_0 (P_1 - \omega_1)^2 \frac{4 E_0^2 \tau^2}{\omega_1^2} |\Gamma_1|^2 + |\Gamma_2|^2 \Bigl[8 \Omega_1 y (2 E_0 \tau(\omega_1 -P_1) - \omega_0 \omega_1   + \omega_1^2) 
  \nonumber   \\ 
     &+ 4 \Omega_2 y^2 \omega_1^2  + \frac{\Omega_0 }{4 \omega_1^2 ( 1 + \tau^2 \omega_1^2)}\sigma_0 \Bigr]  + \frac{|\Gamma_1 \Gamma^{*}_2| }{2 \omega_1^2 (4 + \omega_1^2 \tau^2)} ( \cos{(\Upsilon)} \sigma_1 + \sin{(\Upsilon)} \sigma_2  ).
     \label{pdfC2appn}
\end{align}

%This characteristic makes this case especially convenient for physical interpretations.
%%%%%%%%%%%%%%%%%%%%%%%%%%%%%%%%%%%%%%%%%%%%%%%%%%%%%%%%%%%%%%%%%%%%%%%%%%%%%%%%%%%%%%%%%%%
%%%%%%%%%%%%%%%%%%%%%%%%%%%%%%%%%%%%%%%%%%%%%%%%%%%%%%%%%%%%%%%%%%%%%%%%%%%%%%%%%%%%%%%%%%%%%%%%%%%%%%%%%%%%%%%%%%
%%%%%%%%%%%%%%%%%%%%%%%%%%%%%%%%%%%%%%%%%%%%%%%%%%%%%%%%%%%%%%%%%%%%%%%%%%%%%%%%%%%%%%%%%%%%%%%%%%%%%%%%%%%%%%%%%%%%%%%%%%%%%%%%%%%%%%%%%%%%%%%%%%%%%%%%%%%%%%%%%%%%%%%%%%%%%%%%%%%%%%%%%%%%%%%%%%%%%%%%%%%%%%%%%%%%%%%%%%%%%

%%%%

%%%
\begin{figure}
    \centering
    \includegraphics[width=1.0\linewidth]{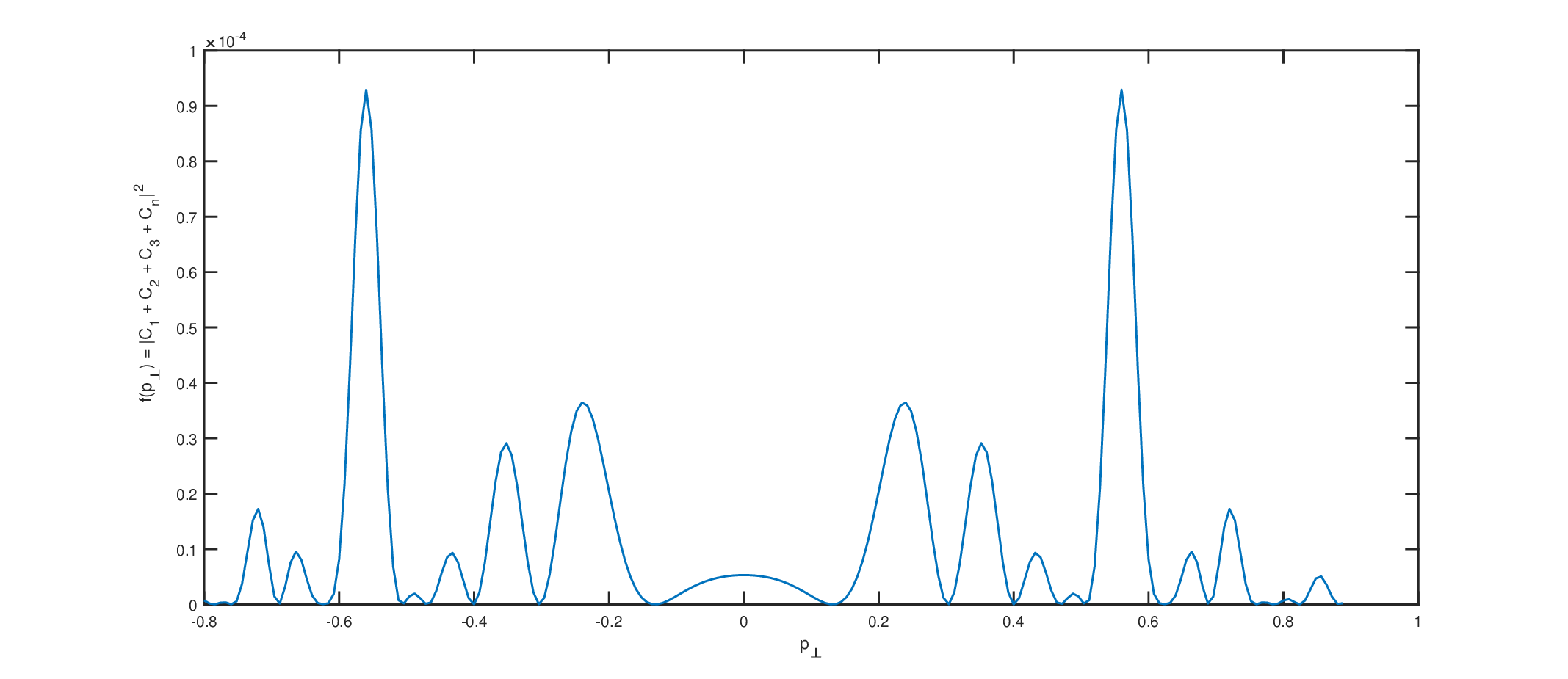}
    \caption{Total Probability of pair production at time $t =15 [m^{-1}],$ where $-3 < \nu < 3, E_0 =0.2 E_c,   $ and $\tau =10 [m^{-1}].$}
    \label{fig:fourier}
\end{figure}

\section{Fourier analysis }
\label{fourier}
 We decompose the time-dependent electric field $E(t)$ into cosine and sine components to analyze the TMS at finite time using the Fourier analysis. 
\begin{align}
     E(t) &=  \frac{1}{(2 \pi)} \int \Tilde{E}(\nu) e^{i \nu t}  d\nu
\end{align}
where,
\begin{align}
    \Tilde{E}(\nu) &=\int_{t_{in}}^{t_{out}}    E(t)  e^{-i \nu t}  dt
    \label{finitefreuqn}
\end{align}
We replace the continuous integration with a discrete one to analyze the behavior of the electric field in a two-level system of pair production \cite{Krajewska:2018lwe}.
\begin{align}
     E(t) &= \sum_{-n}^{n}  \Tilde{E}(\nu_n) e^{i \nu_n t}  
\end{align}

We calculate the frequency \( \nu\) using the Eq.~\eqref{finitefreuqn} at time $t$, and the total probability can be compute by absolute sum of different amplitude of pair production due to different $\nu_n$ at the finite time \( t = t_{\text{out}} \) and plot the total probability as the function of $p_\perp$ as shown in Fig.\ref{fig:fourier}.

%%%%%%%%%%%%%%%%%

%%%%%%%%%%%%%%%%%%%%%%%%%%%%%%%%%%%%%%%%%%%%%%%
% APPENDIX A.3
%%%%%%%%%%%%%%%%%%%%%%%%%%%%%%%%%%%%%%%%%%%%%%%

\section{Explicit Spinor solutions for free and Sauter-pulse fields}
\label{app:spinors}

%This appendix provides the explicit forms of the Dirac spinors used in ~\textbf{Section~\ref{sec:physical_interpretation}},  the three time regions and to derive the measurable pair production probability \( \mathcal{W}(\bm{p},T) \). All expressions are given in the Weyl basis introduced in Eq.~(2), with natural units \( \hbar = c = m = 1 \).

This appendix provides the explicit forms of the Dirac spinor solutions used in Section ~\ref{sec:physical_interpretation} for the free-field regions (I and III) and the Sauter-pulse region (II), as well as the details of the matching procedure at the switch-off time $t=T.$

\subsubsection{Free-field spinors (Regions I and III)}
\label{app:free_spinors}

In regions where the electric field vanishes, the Dirac equation admits plane-wave solutions of the form \cite{Avetissian2016}
\[
\Psi_{\bm{p},r}^{(\lambda)}(\bm{x},t) = \xi_{\bm{p},r}^{(\lambda)}(t) e^{i\bm{p}\cdot\bm{x}}, \quad \lambda = \pm,
\]
where \( \lambda = + \) corresponds to positive-energy (electron) states and \( \lambda = - \) to negative-energy (positron) states. The canonical momentum \( \bm{p} = (p_1, p_2, p_\parallel) \) is conserved, and the kinetic momentum along the field direction is \( \Pi_3 = p_\parallel - eA \) with \( A \) constant. The corresponding energy is
\[
\mathcal{E} = \sqrt{m^2 + p_\perp^2 + \Pi_3^2}, \qquad p_\perp^2 = p_1^2 + p_2^2.
\]

For spin projection \( r = 1 \), the normalized free spinors are
\begin{align}
\xi_{\bm{p},1}^{(+)}(t) &= \frac{1}{\sqrt{2\mathcal{E}(\mathcal{E} - \Pi_3)}}
\begin{pmatrix}
m \\[2pt]
0 \\[2pt]
-\mathcal{E} + \Pi_3 \\[2pt]
p_1 + i p_2
\end{pmatrix} e^{-i\mathcal{E}t}, \\
\xi_{\bm{p},1}^{(-)}(t) &= \frac{1}{\sqrt{2\mathcal{E}(\mathcal{E} + \Pi_3)}}
\begin{pmatrix}
m \\[2pt]
0 \\[2pt]
\mathcal{E} + \Pi_3 \\[2pt]
p_1 + i p_2
\end{pmatrix} e^{i\mathcal{E}t}.
\end{align}
For \( r = 2 \), the corresponding spinors are
\begin{align}
\xi_{\bm{p},2}^{(+)}(t) &= \frac{1}{\sqrt{2\mathcal{E}(\mathcal{E} + \Pi_3)}}
\begin{pmatrix}
p_1 - i p_2 \\[2pt]
\mathcal{E} + \Pi_3 \\[2pt]
0 \\[2pt]
m
\end{pmatrix} e^{-i\mathcal{E}t}, \\
\xi_{\bm{p},2}^{(-)}(t) &= \frac{1}{\sqrt{2\mathcal{E}(\mathcal{E} - \Pi_3)}}
\begin{pmatrix}
-p_1 + i p_2 \\[2pt]
\mathcal{E} - \Pi_3 \\[2pt]
0 \\[2pt]
m
\end{pmatrix} e^{i\mathcal{E}t}.
\end{align}
%These satisfy the orthonormality condition
%\[
%\big[ \xi_{\bm{p},r}^{(\lambda)}(t) \big]^\dagger \xi_{\bm{p}',r'}^{(\lambda')}(t) = \delta_{\lambda\lambda'}\delta_{rr'}(2\pi)^3\delta(\bm{p} - \bm{p}')\]with respect to the Dirac inner product.

\subsubsection{Sauter-pulse spinors (Region II)}
\label{app:sauter_spinors}

In Region II, the electric field is given by the Sauter pulse \( E(t) = E_0 \operatorname{sech}^2(t/\tau) \). The Dirac spinor is constructed from the scalar mode function \( \psi_{\bm{p}}^{(\pm)}(t) \) obtained in Section~\ref{Result}. Using the constant spinor basis vectors \( R_r \) defined in Eq.~(12), the positive-energy solution is
\begin{equation}
\Psi_{\bm{p},r}^{(+)}(t) =
\big[ i\gamma^0\partial_t + \bm{\gamma}\cdot\bm{p} - e\gamma^3 A(t) + m \big]
\psi_{\bm{p}}^{(+)}(t) R_r \,.
\label{eq:Psi_plus_general}
\end{equation}
For \( r = 1 \), this takes the explicit form
\begin{equation}
\Psi_{\bm{p},1}^{(+)}(t) =
\begin{pmatrix}
m \\
0 \\
i\partial_t + P(p_\parallel,t) \\
p_1 + i p_2
\end{pmatrix}
\psi_{\bm{p}}^{(+)}(t), \qquad P(p_\parallel,t) = p_\parallel - eA(t),
\label{eq:Psi_plus_explicit}
\end{equation}
with the mode function \( \psi_{\bm{p}}^{(+)}(t) \) given by Gauss-hypergeometric expression. The corresponding negative-energy spinor \( \Psi_{\bm{p},1}^{(-)}(t) \) is obtained by replacing \( \psi_{\bm{p}}^{(+)}(t) \) with \( \psi_{\bm{p}}^{(-)}(t) \) and changing the sign of the energy component.

\subsubsection{Matching at \( t = T \) and the reflection amplitude}
\label{app:matching}

In Section~\ref{sec:physical_interpretation}, the field is switched off abruptly at \( t = T \). The solution in Region II must match continuously onto a superposition of free states in Region III:
\begin{equation}
\Psi_{\bm{p},1}^{(+)}(T) = \mathcal{R}(\bm{p}) \, \xi_{\bm{p},1}^{(-)}(T) + \mathcal{I}(\bm{p}) \, \xi_{\bm{p},1}^{(+)}(T),
\label{eq:matching_app}
\end{equation}
where \( \mathcal{R}(\bm{p}) \) is the pair-production amplitude (reflection from the Dirac sea) and \( \mathcal{I}(\bm{p}) \) is the transmission amplitude. Projecting both sides onto \( \xi_{\bm{p},1}^{(-)}(T) \) using orthonormality yields
\[
\mathcal{R}(\bm{p}) = \big[ \xi_{\bm{p},1}^{(-)}(T) \big]^\dagger \Psi_{\bm{p},1}^{(+)}(T).
\]

Substituting the explicit forms of \( \xi_{\bm{p},1}^{(-)} \) and \( \Psi_{\bm{p},1}^{(+)} \) leads to Eq.~\eqref{eq:R_explicit} in the main text, which after simplification gives the probability \( \mathcal{W}(\bm{p},T) = 2|\mathcal{R}(\bm{p})|^2 \).

The identity \( f(\bm{p},T) = \mathcal{W}(\bm{p},T) \) established in Section~\ref{sec:physical_interpretation} follows from comparing this expression with the quasiparticle distribution function \( f(\bm{p},t) \) derived in Section~\ref{PDF}. 

%The explicit spinor forms provided here ensure the matching is performed consistently and the result is independent of the chosen spin basis.

%\newpage
%\bibliographystyle{unsrt}
\bibliography{main}

\end{document}